\documentclass[aps,prd,twocolumn,showpacs,tightenlines,floatfix,nofootinbib,superscriptaddress]{revtex4-1}
\usepackage{natbib}
\usepackage{graphicx,subfigure}
\usepackage{caption}
\usepackage{tabularx,dcolumn,booktabs}
\usepackage{multirow}
\usepackage{float}
\usepackage{mathtools}
\usepackage{multirow}
\usepackage{bm}
\usepackage{hyperref}
\usepackage{setspace}
\usepackage{caption}
\newcommand{\Chi}{{\large{\chi}}}
\begin{document}

\title{Open-source QCD analysis of nuclear parton distribution functions at NLO and NNLO}

\author{Marina Walt}
 \email{marina.walt@uni-tuebingen.de}
\affiliation{Institute for Theoretical Physics, University of T\"ubingen, Auf der Morgenstelle 14, 72076 T\"ubingen, Germany}

\author{Ilkka Helenius}
\email{ilkka.m.helenius@jyu.fi}
\affiliation{University of Jyvaskyla, Department of Physics, P.O. Box 35, FI-40014 University of Jyvaskyla, Finland}%
\affiliation{Helsinki Institute of Physics, P.O. Box 64, FI-00014 University of Helsinki, Finland}%

\author{Werner Vogelsang}
\email{werner.vogelsang@uni-tuebingen.de}
\affiliation{Institute for Theoretical Physics, University of T\"ubingen, Auf der Morgenstelle 14, 72076 T\"ubingen, Germany}%

\date{\today}

\begin{abstract}
We present new sets of nuclear parton distribution functions (nPDFs) at next-to-leading order and next-to-next-to-leading order. Our analyses are based on deeply inelastic scattering data with charged-lepton and neutrino beams on nuclear targets. In addition, a set of proton baseline PDFs is fitted within the same framework with the same theoretical assumptions. The results of this global QCD analysis are compared to existing nPDF sets and to the fitted cross sections. Also, the uncertainties resulting from the limited constraining power of the included experimental data are presented. The published work is based on an open-source tool, \textsc{xFitter}, which has been modified to be applicable also for a nuclear PDF analysis. The required extensions of the code are discussed as well.
\end{abstract}

\maketitle
{
\tableofcontents
}
\section{Introduction}

Phenomenology based on collinear factorization \cite{Collins:1989gx} has proven extremely successful in the era of the Large Hadron Collider (LHC). In this approach the parton distribution functions (PDFs) \cite{Kovarik:2019xvh}, describing the number distributions of quarks and gluons in the colliding hadrons, are factorized from the hard partonic scattering. The latter can be calculated within perturbative quantum chromodynamics (QCD) \cite{Dokshitzer:1991wu, Brock:1993sz} but the PDFs have to be determined in a global analysis using experimental data and Dokshitzer-Gribov-Lipatov-Altarelli-Parisi (DGLAP) equations that provide the scale evolution of the PDFs \cite{Lipatov:1974qm, Gribov:1972ri, Altarelli:1977zs, Dokshitzer:1977sg}. The most precise constraints for PDFs come from high-energy deeply inelastic scattering (DIS) experiments where the hadron structure is probed with a highly-virtual photon or a massive electroweak boson. The data available from the HERA collider \cite{Abramowicz:2015mha}, combined with older fixed-target measurements, provide plenty of data points with a broad kinematic reach that can be used to constrain the proton PDFs in the kinematic region relevant at the LHC \cite{Gao:2017yyd}. However, the data suitable for nuclear PDF (nPDF) analyses are far more sparse \cite{Kovarik:2019xvh}. In these analyses the nuclear modifications, first observed in DIS experiments with nuclear targets, are assumed to be nonperturbative and absorbed into the PDFs obeying the same scale evolution equations as free protons \cite{Armesto:2006ph}.

The global nPDF analyses rely heavily on nuclear DIS data. Compared to the HERA data available for proton PDF fits, the kinematic reach of the fixed-target nuclear data is quite limited. Such data provide direct constraints for quarks, but the gluon distributions are probed only at higher orders in perturbative QCD (pQCD) and via scale-evolution effects. There is potential for improvement by including neutrino-nucleus DIS data that have additional sensitivity to the flavour decomposition of the PDFs due to the different coupling to up- and down-type quarks. Most of the recent next-to-leading order (NLO) nPDF analyses, e.g. DSSZ \cite{deFlorian:2011fp}, include data for Drell-Yan (DY) dilepton production which provide additional constraints for antiquarks. Some further constraints for gluons have been obtained from pion-production data in d+Au collisions at the Relativistic Heavy-Ion Collider (RHIC) \cite{Adler:2006wg} which were first used in the EPS08 analysis \cite{Eskola:2008ca}, and later also in EPS09 \cite{Eskola:2009uj} and nCTEQ15 \cite{Kovarik:2015cma}. The EPPS16 \cite{Eskola:2016oht} was the first analysis to use measurements from the LHC by incorporating data for $Z$ and $W^{\pm}$ bosons \cite{Khachatryan:2015hha, Khachatryan:2015pzs, Aad:2015gta} and dijet production \cite{Chatrchyan:2014hqa} in p+Pb collisions. These provide further constraints for the flavour decomposition and gluon nuclear modifications, but the statistics of the Run I data for these observables was still quite limited. The more precise data from Run II from the LHC will bring more constraints especially for the gluon nPDFs \cite{Eskola:2019dui}. Furthermore, the existing heavy-meson data from LHCb demonstrate promising potential to directly measure the gluon shadowing down to $x \sim 10^{-5}$ \cite{Zenaiev:2015rfa, Cacciari:2015fta, Gauld:2016kpd, deOliveira:2017lop, Eskola:2019bgf}. A complementary channel to study small-$x$ gluons would be direct photon production at forward rapidities \cite{Helenius:2014qla}. Such a measurement with the required forward instrumentation could be performed with the proposed FoCal upgrade in ALICE \cite{Peitzmann:2018kie}.

Another way to improve the PDFs is to increase the perturbative precision of the analysis. For proton PDFs the current standard is next-to-next-to-leading order (NNLO) \cite{Harland-Lang:2014zoa, Dulat:2015mca, Abramowicz:2015mha, Ball:2017nwa, Alekhin:2017kpj} in pQCD. The splitting functions required for the DGLAP evolution have been available at the required NNLO accuracy for a while already \cite{Vogt:2004mw, Moch:2004pa}, but despite the recent progress \cite{Boughezal:2016wmq, Currie:2016bfm, Campbell:2016lzl, Gaunt:2015pea, Czakon:2016ckf, Grazzini:2017mhc, Chen:2019zmr, Caola:2019nzf}, the number of processes for which a full NNLO calculation is publicly available is still limited. For nPDFs first works in this direction have been performed recently. The first one, KA15 \cite{Khanpour:2016pph}, includes data from different fixed-target DIS experiments with a lepton beam. The fit is performed in the zero-mass variable-flavour-number-scheme (ZM-VFNS) where all quarks are assumed massless for energies above their mass threshold. The more recent analysis nNNPDF1.0 \cite{AbdulKhalek:2019mzd}, applies the NNPDF methodology \cite{Forte:2002fg, Ball:2008by} in which the resulting nPDFs are determined by a neural network. Also there, the applied data were restricted to neutral-current DIS, but a more realistic general-mass variable-flavour-number-scheme (GM-VFNS) was applied.

In this work we introduce two new nPDF analyses, performed at NLO and NNLO in pQCD, respectively, which we refer to as TUJU19. The presented work is based on the open-source \textsc{xFitter} package \cite{Zenaiev:2016jnq,Bertone:2017tig} (formerly known as \textsc{HeraFitter} \cite{Alekhin:2014irh}) that has been modified in order to also accommodate data from nuclear collisions and suitable PDF parameterizations. In addition to neutral-current DIS with a lepton beam, we also include charged-current neutrino DIS ($\nu$DIS) data with nuclear targets that are sensitive to the flavour decomposition of nonisoscalar nuclei. For a free proton baseline we fit new PDF sets mainly based on the combined HERA I and II data, providing baseline fits consistent with our assumptions and kinematical cuts made for the nPDF analyses. Furthermore, the required extensions of the code will be published, providing a first open-source tool to analyze nuclear PDFs. In this paper we describe the theoretical framework in section \ref{sec:theory}, then discuss the analysis procedure in section \ref{sec:analysis} and the selection of experimental data in section \ref{sec:expdata}. The results of the analysis are presented in section \ref{sec:results} and the work is then summarized in section \ref{sec:summary}, where an outlook towards future developments is also presented.

\section{Theoretical framework}
\label{sec:theory}

\subsection{Deeply inelastic scattering}
\label{sec-crosssections}

In this analysis, neutral-current (NC) and charged-current (CC) DIS processes are considered: NC in case of electron(positron)-nucleus ($\mathrm{e} A$) and CC for (anti)neutrino-nucleus ($\nu A$) scattering. For these processes the differential cross section is given by 
\begin{equation}
\frac{\mathrm{d}^2 \sigma}{\mathrm{d}x\,\mathrm{d}y} = N^{\, l}\,\left[ y^2 x F^{\,l}_1+(1-y)\,F^{\,l}_2 \mp\,\left( y-\frac{y^2}{2} \right)\,xF^{\,l}_3  \right]\,,
\label{neutrinoDIS}
\end{equation}
where $x$ and $y$ are the standard kinematic variables in DIS. The coupling factor $N^{\, l}$ depends on the scattering type and $F^{\, l}_{1,2,3}$ denote the structure functions \cite{Roberts:1990ww,sterman_handbook_2001,Patrignani:2016xqp} for the scattering of lepton $l$. In equation (\ref{neutrinoDIS}), the index $l$ covers different beams including $l=\nu,\,\bar{\nu},\,e^+,\,e^-,\,\mu^+,\,\mu^-$. For the nuclear data used in this work two normalization factors are relevant: one for NC DIS in the case of unpolarized leptons and one for the CC DIS of incoming (anti)neutrinos. In the case of unpolarized leptons the normalization factor $N^{\, l,\,NC}$ for NC DIS is \cite{Patrignani:2016xqp}
\begin{equation}
N^{\,l,\,\mathrm{NC}} = \frac{4 \pi \alpha^2}{x y Q^2} \,,
\end{equation}
\noindent and in the case of incoming (anti)neutrinos the normalization factor $N^{\, \nu,\,CC}$ for CC DIS is 
\begin{equation}
N^{\, \nu, \,\mathrm{CC}} = \frac{G_{\mathrm{F}}^2\,M_{W}^4\,Q^2}{4\pi\,x\,y\,\left( Q^2 + M_W^2 \right)^2}\,,
\end{equation}
where $Q^2$ is the virtuality of the intermediate boson that provides the scale at which the nucleons are probed. For the CC processes $G_F$ is the Fermi coupling constant and $M_W$ is the mass of the $W^{\pm}$ boson. When combining the structure functions into the differential cross sections in alignment with~(\ref{neutrinoDIS}), the sign before $F_{3}$ is positive for $\nu$ and $e^+$, and negative for $\bar{\nu}$ and $e^-$. 

In QCD the structure functions $F_i$, as introduced in eq. (\ref{neutrinoDIS}), are related to the scale-dependent parton distribution functions $f_j(x,\,Q^2)$, with $j=g$ or $j=q\,,\bar{q}$, via
\begin{equation}
F_i\left(x,\,Q^2\right) = \sum_j C_i^j\left(x,\,\alpha_s(\mu^2),\,\mu^2/Q^2\right)\otimes f_j\left(x,\,\mu^2\right)\,,
\label{eq-convolution}
\end{equation}
\noindent where one typically chooses $\mu=Q$. The symbol $\otimes$ in equation (\ref{eq-convolution}) denotes a convolution between the parton distribution functions and the Wilson coefficients $C_i^j$ (see e.g. Refs.~\cite{vanNeerven:1999ca,vanNeerven:2000uj,Vermaseren:2005qc} for $C_2^j$ at NLO and NNLO). For example, the structure functions for neutrinos and antineutrinos are given at leading order (LO) by \cite{Tanabashi:2018oca}
\begin{align}
\label{eq:sF-neutrino}
F_1^{\,\nu} &= d + s + b + \bar{u} + \bar{c} + \bar{t} \,, \notag \\
F_2^{\,\nu} &= 2x\,(d+s+b+\bar{u} +\bar{c} +\bar{t}) \,, \\
F_3^{\,\nu} &= 2\,(d+s+b-\bar{u}-\bar{c}-\bar{t}) \,, \notag
\end{align}
and
\begin{align}
\label{eq:sF-antin}
F_1^{\,\bar{\nu}} &= u + c + t + \bar{d} + \bar{s} + \bar{b} \,, \notag \\
F_2^{\,\bar{\nu}} &= 2x\,(u+c+t+\bar{d} +\bar{s} +\bar{b}) \,, \\
F_3^{\,\bar{\nu}} &= 2\,(u+c+t-\bar{d}-\bar{s}-\bar{b}) \,. \notag
\end{align}
\noindent As can be seen from relations (\ref{eq:sF-neutrino}) and (\ref{eq:sF-antin}), there is an added value to the analysis by including the neutrino-nucleus DIS data that have additional sensitivity to flavour decomposition of the PDFs due to the different coupling to up- and down-type quarks.

The factorization of the partonic scattering process and the nonperturbative PDFs \cite{Collins:1989gx}, as well as the perturbative treatment are valid at sufficiently high energy scales $(Q^{\,2}\gtrsim 1~\text{GeV}^2)$. In this work we have selected kinematic cuts $Q^{\,2} > 3.5\,\mathrm{GeV^2}$, the Bjorken variable $x<0.7$, and the invariant mass of the hadronic final state $W^2>12\,\mathrm{GeV^2}$. The latter can be expressed in terms of the other invariant variables as
\begin{align}
W^2 \approx Q^2\left(\frac{1}{x}-1\right)\,.
\label{eq-W2}
\end{align}
Some of the experimental data sets do not specify the invariant $y$, but when the collision energy is known, it can be derived from the relation
\begin{equation}
Q^2 \approx y \, x \, s.
\end{equation}

\subsection{PDF parameterization}

A global DGLAP-based analysis requires a nonperturbative input for the PDFs at the initial scale of the fit.
In this analysis parton distributions of a free proton and of a nucleon bound in a nucleus are parameterized as
\begin{equation}
xf^{p/A}_i\left(x,Q_0^2 \right) = c_0\,x^{c_1} (1-x)^{c_2} \left(1+c_3\,x + c_4\,x^2 \right)
\label{pdf-parameterization}
\end{equation}
for parton flavour $i=g,\,d_{\mathrm{v}},\,u_{\mathrm{v}},\,\bar{u},\,\bar{d},\,\bar{s}$, at the initial scale $Q_0^2=1.69\,\mathrm{GeV^2}$. This form of PDF parameterization is similar to the functional form used in the HERAPDF2.0 analysis \cite{Abramowicz:2015mha} and is motivated by the fact that the main constraints for the free proton PDF baseline come from the same DIS data. To keep the framework consistent we use the same form for the nuclear PDFs. 

The main focus of this work is on the nuclear PDFs for which the fit parameters $c_k$ in equation (\ref{pdf-parameterization}) are re-parameterized to be dependent on the nuclear mass number $A$ as
\begin{equation}
c_k\,\rightarrow c_k(A) = c_{k,0}+c_{k,1}\left( 1 - A^{-c_{k,2}} \right)
\label{coeff-A}
\end{equation}
where $k={0,\dots,4}$. A similar form was also successfully used in the nCTEQ15 analysis \cite{Kovarik:2015cma}. At the same time, if $A=1$ the $A$-dependent right-hand part of equation (\ref{coeff-A}) becomes zero and the free proton PDFs are recovered by default. The explicit $A$ dependence of the nuclear PDFs also allows us to make predictions for nuclei which were not part of the actual analysis, but are possibly interesting for future experiments.

As discussed, $xf^{p/A}_i$ given in equation (\ref{pdf-parameterization}) defines the parton distribution in a proton bound to a nucleus $A$. In addition there are also neutrons in a nucleus which we denote by $f_i^{\,n/A}$. The full PDF for a nucleon inside a nucleus can be obtained by averaging over the number of protons and neutrons in nuclei:
\begin{equation}
f_i^{\,N/A} \left( x,Q^{\,2} \right) = \frac{Z\cdot f_i^{\,p/A}+ (A-Z)\cdot f_i^{\,n/A}}{A}\,.
\label{nucleon}
\end{equation} 
The PDFs of neutrons are not separately fitted, but are determined from the proton PDFs based on isospin symmetry. In addition to this symmetry, we have assumed $s=\bar{s}$ and $s=\bar{s}=\bar{u}=\bar{d}$ as the included DIS data are not sensitive enough to constrain the strange-quark content, nor the sea quark flavour decomposition. In particular, even though the neutrino DIS data are sensitive to the separation of up- and down-type quarks, the kinematic region covered by the incorporated data ($x\gtrsim 0.01$)is where the valence quarks dominate the cross section. 
 
The heavy quarks are treated within the general-mass variable-flavour-number-scheme (GM-VFNS), see Ref.~\cite{Accardi:2016ndt} for a recent overview. There are several options for GM-VFNS implemented in {\textsc{xFitter}}\footnote{Also fixed-flavour (FF) mass schemes, like e.g. the ABM scheme \cite{ABM-scheme,Alekhin:2010sv}, are available in \textsc{xFITTER}.}, including (S)ACOT schemes \cite{Aivazis:1993pi,Kramer:2000hn,Tung:2001mv,Kretzer:2003it}, RT and RT optimal schemes \cite{Thorne:1997ga, Thorne:2006qt, Thorne:2012az}, as well as a FONLL scheme \cite{Cacciari:1998it, Forte:2010ta} for GM-VFNS. In this work we apply the FONLL-A scheme for the NLO analysis and the FONLL-C at NNLO, implemented in the \textsc{APFEL} package \cite{Bertone:2013vaa}. The heavy-quark masses are fixed to $m_{\mathrm{charm}}=1.43~\mathrm{GeV}$ and $m_{\mathrm{bottom}}=4.50~\mathrm{GeV}$. The strong coupling constant $\alpha_S$ is set to $\alpha_{\mathrm{S}}(M_{\mathrm{Z}})=0.118$ for both the NLO and the NNLO fits.


For the parton distribution functions $xf^{p/A}_i$ as defined in equation (\ref{pdf-parameterization}), we assume the baryon number sum rules and the momentum sum rule satisfied by every nucleon in the nucleus,
\begin{equation}
\int_0^1 \mathrm{d}x f_{u_{\mathrm{v}}}^{p/A}\left(x,\,Q_0^2 \right) = 2\,,	
\label{eqsumruleuv}
\end{equation}
\begin{equation}
\int_0^1 \mathrm{d}x f_{d_{\mathrm{v}}}^{p/A}\left(x,\,Q_0^2 \right) = 1\,,
\label{eqsumruledv}
\end{equation}
\begin{equation}
\int_0^1 \mathrm{d}x \sum_i xf^{p/A}_i\left(x,\,Q_0^2 \right) = 1\,.
\label{eqmomsumrule}
\end{equation}
Strictly speaking, for nuclear parton distribution functions the sum rules are approximations that might not hold for individual nucleons in a nucleus in general, but which are reasonable at the available level of precision in regard to the experimental uncertainties. In this work, equation (\ref{eqsumruleuv}) is used to fix the normalization of $u_{\mathrm{v}}$ quarks in a proton and equation (\ref{eqsumruledv}) defines the normalization of $d_{\mathrm{v}}$ quarks in a proton, nucleus per nucleus. The momentum sum rule (\ref{eqmomsumrule}) is used to constrain the normalization of the sea quarks. The remaining unconstrained normalization coefficient $c^g_0$ in the gluon PDF is treated as a regular free parameter during the fitting procedure. Alternatively, the gluon normalization could have been fixed by the momentum sum rule as done in many earlier analyses, e.g. Ref. \cite{Eskola:2016oht}.


As described above, we parameterize and fit the PDFs of a proton in a nucleus, and the neutron PDFs are determined based on SU(2) symmetry. In particular, the distributions of $u$ and $d$ quarks are exchanged: $u^p \rightarrow d^n$ and $d^p \rightarrow u^n$, which is valid for valence and sea quarks. For completeness we mention that this interchange requires the validity of charge symmetry, and in Ref.~\cite{Boros:1998es} it has been suggested that some charge symmetry violation (CSV) could take place in the small-$x$ region. However, in the $x$ region covered by the nuclear DIS data we use, such effects should be negligible.
Besides the DIS experiments, CSV effects can be studied in experiments measuring asymmetries in $W$ boson production. Further experiments and tests of CSV in parton distributions were suggested in Refs.~\cite{Saito:2000fx,Londergan:2005ht}. Similarly, possible isospin symmetry violations have been studied in Refs.~\cite{Martin:2003sk,Wang:2015msk}. In this work, however, we assume that the charge and isospin symmetries hold.

\section{Analysis procedure}
\label{sec:analysis}

\subsection{Fitting procedure}

The optimal values for the parameters are obtained by minimizing $\Chi^2$ defined as

\begin{equation}
\Chi^2 = \sum_i \frac{\left( \mu_i - \hat{m}_i \right)^2}{\Delta^2_i} + \sum_{\alpha} b^2_{\alpha}
\label{eq-chi2}
\end{equation}
\noindent with
\begin{equation}
\hat{m}_i = m_i + \sum_{\alpha} \Gamma_{i\alpha} b_{\alpha}.
\label{eq-theory}
\end{equation}
Here, $\mu_i$ is the value of the measured data point for a given observable, $\Delta_i$ is the uncorrelated experimental error, whereas the sum over correlated systematic errors is given by the term $\sum_{\alpha} b^2_{\alpha}$ in equation (\ref{eq-chi2}). The theoretical predictions for each data point $i$ are represented by $\hat{m}_i$, defined in equation (\ref{eq-theory}). There, $m_i$ is the actual theoretical value calculated using DGLAP-evolved PDFs with given parameters $\{c_k\}$, $\Gamma_{i\alpha}$ are the correlated errors and $b_{\alpha}$ are the so-called nuisance parameters. A nuisance parameter quantifies the strength of the correlated error source $\alpha$, whereas $\Gamma_{i\alpha}$ quantifies the sensitivity of the $i^{\mathrm{th}}$ measurement to the correlated systematic error source $\alpha$. The quality of the fit can be estimated from the resulting $\Chi^2/N_{\mathrm{dp}}$ ratio, where $N_{\mathrm{dp}}$ is the number of data points. A value $\Chi^2/N_{\mathrm{dp}} \approx 1$ indicates that the agreement between the theoretical prediction and the measured observable is on average at the level of the experimental uncertainties.

There are several ways to take into account the correlated and uncorrelated uncertainties and to combine the statistical and uncorrelated systematic uncertainties in the $\Chi^2$ definition in \textsc{xFitter} used in this analysis. Here we use the following definition
\begin{align}
\Chi^2\left(\textbf{m},\,\textbf{b}\right) =& \sum_i \, \frac{\left[ m_i - \Sigma_{\alpha}\,\gamma^i_{\alpha}\,\mu_i\,b_{\alpha} - \mu_i \right]^2}{\left( \delta_{i,\mathrm{stat}} \sqrt{\mu_i m_i} \right)^2 + \left( \delta_{i,\mathrm{uncorr}} m_i \right)^2 } \notag \\
+& \sum_{\alpha} b^2_{\alpha}.
\label{eq-chi2-scaled}
\end{align}
The variables have been introduced in (\ref{eq-chi2}) with $\Gamma^i_{\alpha}= \gamma^i_{\alpha}\,\mu_i$, and $\delta_{i,\mathrm{stat}}$ and $\delta_{i,\mathrm{uncorr}}$ are the relative statistical and uncorrelated systematic uncertainties, respectively. The above form in (\ref{eq-chi2-scaled}) corresponds to a Poisson-like scaling for the statistical experimental uncertainties, whereas the systematic uncorrelated and correlated uncertainties are scaled linearly. This choice for $\Chi^2$ is similar to the one used in the HERAPDF2.0 analysis incorporating the combined H1 and ZEUS DIS data \cite{Abramowicz:2015mha}, on which our proton PDF baseline is mainly based. For consistency the same form has also been used for the nuclear PDF analysis.

\subsection{Uncertainty analysis}
\label{sec-uncertainties}

The minimization of $\Chi^2$ provides a central set of PDFs with the parameter values providing the best description of the used data. However, the experimental data always contain several uncertainties of different types, such as statistical, systematic or correlated, like e.g. normalization errors. To study how well the experimental data actually constrain the fitted distributions, a separate error analysis needs to be performed. Such an analysis quantifies how much (according to the given criteria) room there is for the parameters to vary so that the resulting cross sections are still in agreement within the experimental uncertainties. The distribution functions resulting from the uncertainty analysis are typically provided as part of the PDF sets.

There are two established methods which can be used for the error analysis, the Hessian method \cite{Pumplin:2001ct,Pumplin:2000vx} or the Monte Carlo (MC) method \cite{Giele:1998gw,Giele:2001mr,Hartland:2019bjb}. The former relies on quadratic approximation of $\Chi^2$ with respect to the parameters $\{c_k\}$ near the minimum. In the latter method, the data is varied within the given uncertainties, and for each variation a PDF replica set is fitted. Therefore this method is less sensitive to the form of the parameterization but numerically more demanding. Also the Lagrange multiplier method \cite{Pumplin:2000vx, Stump:2001gu} has been used to study the uncertainties (see e.g. Ref. \cite{Martin:2002aw}), but there the error propagation to an observable becomes more involved. In this QCD analysis the Hessian method is used for the analysis of the uncertainties. 

The Hessian error analysis is performed assuming a quadratic expansion of the function $\Chi^2 = \Chi_0^2 + \Delta \Chi^2$ around its global minimum. Here, $\Chi_0^2$ is the value of the function at the global minimum (with the best-fit parameters $\{k_0\}$) and $\Delta \Chi^2$ is the displacement from the minimum \cite{Pumplin:2001ct,Pumplin:2000vx}. The Hessian matrix $H$ is constructed from the second derivatives of $\Chi^2$ at the minimum. The matrix elements $H_{ij}$ are defined as
\begin{equation}
H_{ij} = \frac{1}{2}\left( \frac{\delta^2\Chi^2}{\delta y_i\,\delta y_j} \right)\,,
\label{eq-hesse-matrix}
\end{equation}
with $y_{i}$ being the displacement of the parameter $a_i$ from its value $a_0$ at the minimum. For the analysed function $\Chi^2$ one writes
\begin{equation}
\Chi^2 = \Chi_0^2 + \sum_{i,j} H_{ij}\,y_i\,y_j\,.
\end{equation}
The Hessian matrix is symmetric and thus has a complete set of orthonormal eigenvectors $v_{ij}$. The eigenvectors and the eigenvalues $\epsilon_{j}$ of the Hessian matrix are used to transform the displacements ${y_i}$ into a new set of parameters ${z_i}$
\begin{equation}
y_i=\sum_j v_{ij} \sqrt{\frac{1}{\epsilon_j}} z_j\,,
\end{equation}
leading to a simplified relation
\begin{equation}
\Delta \Chi^2 = \Chi^2 - \Chi_0^2 = \sum_i z_i^2\,.
\end{equation}
This representation has the advantage that the surfaces of constant $\Chi^2$ are spheres in ${z_i}$ space with $\Delta \Chi^2$ being the squared distance from the minimum. The varied parameters ${a_i}$ from which the resulting error sets are defined can then be written as
\begin{equation}
a_i = a_0 \pm \Delta a_i = a_0 \pm \Delta \Chi^2 \sum_j \frac{v_{ij}^2}{\epsilon_j}\,,
\label{eq-ai}
\end{equation}
where $\Delta \Chi^2$ defines the tolerance criterion determining the allowed growth of $\Chi^2$. The relation in equation (\ref{eq-ai}) shows that the parameters which correspond to the eigenvectors of the Hessian matrix with large eigenvalues are well determined since their $\Delta a_i$ is small, whereas the weakly determined parameters correspond to small eigenvalues. The uncertainties for a given observable $X$ can be calculated via
\begin{align}
\label{eq-obsX}
(\Delta X^{\pm})^2 = \sum_{i=1}^{n_{\mathrm{param}}} \bigg\{ {\substack{\text{max}\\ \text{min}}} \bigg[ &X(S_i^+)-X(S_0), \notag \\
&X(S_i^-)-X(S_0),\, 0 \bigg] \bigg\}^2\,,
\end{align}
where $X(S_0)$ is the observable calculated with the central parameter set and the $S_i^{\pm}$ correspond to the error sets in the positive and negative direction determined from the diagonalized parameter $z_i$.

In an ideal case one would choose the tolerance criterion so that $\Delta \Chi^2 = 1$. However, since we consider several different data sets which are not necessarily in a mutual agreement with one another, such a choice would underestimate the underlying uncertainty. In this work the tolerance for $\Delta \Chi^2$ is based on the statistically motivated method as discussed e.g. in Refs.~\cite{Hirai:2007sx, Eskola:2016oht}.  For the proton baseline with $13$ free fit parameters it becomes $\Delta \Chi^2 = 20$ at the 90\% confidence level. This choice has also been validated by comparing to the error bands generated with the MC method, though the $\Delta \Chi^2$ value preferred by the MC method is quite flavour- and kinematics-dependent. 
Previous nPDF studies have shown that such a statistically motivated method would not fully cover the experimental uncertainties in the nuclear data \cite{Kovarik:2015cma, Eskola:2016oht}. Thus for the nuclear PDF error analysis we increase the tolerance from the statistically motivated value and choose $\Delta \Chi^2 = 50$ for our 16 free parameters.

\subsection{The fitting framework}
\begin{center}
\begin{figure*}[tb!]
\includegraphics[width=0.8\textwidth]{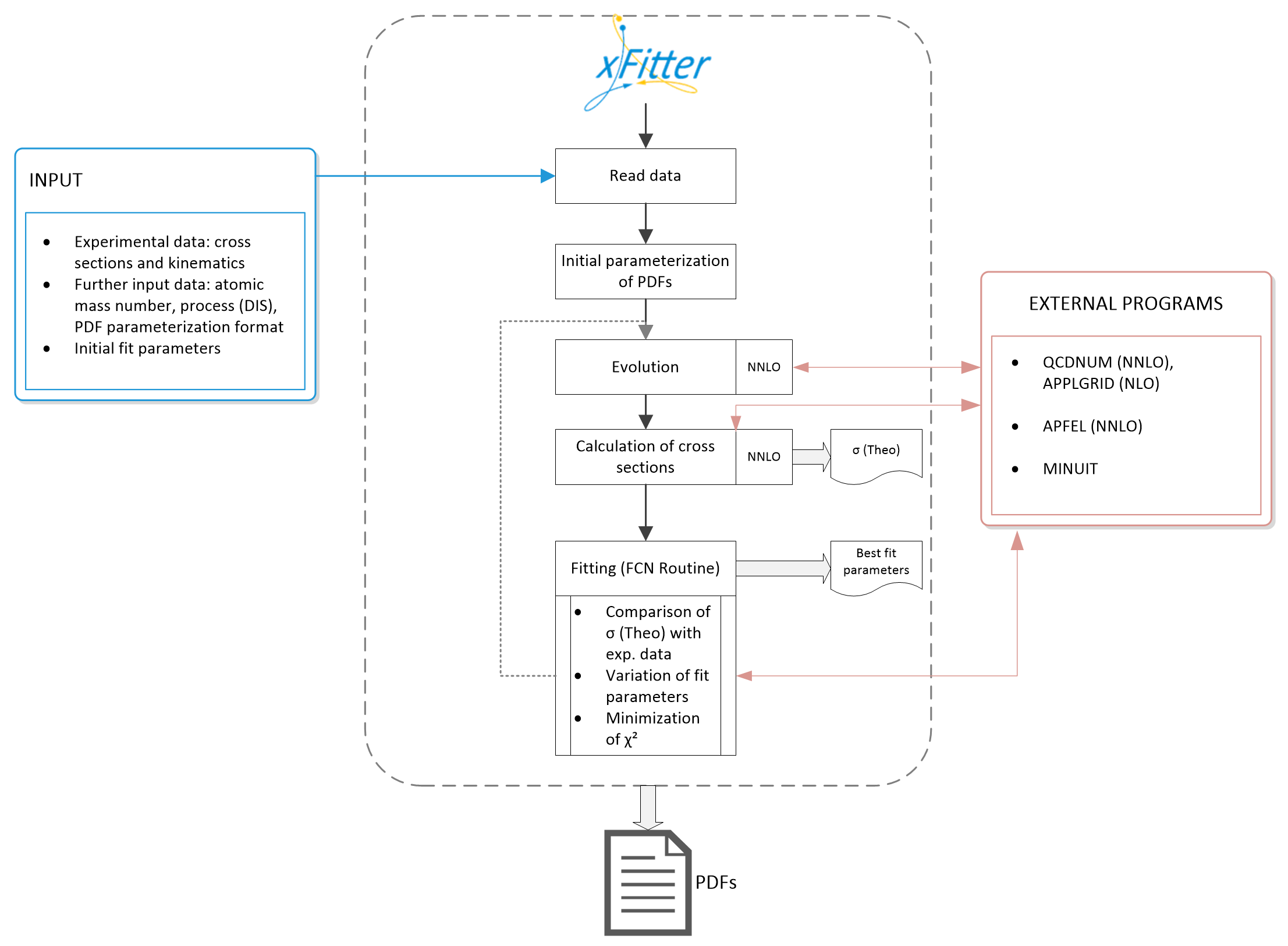}
\caption{Schematic view of the high-level \textsc{xFitter} functionalities. The \textsc{xFitter} logo is taken from Ref. \cite{xfitter_link}.}
\label{fig-xfitter}
\end{figure*}
\end{center}
The global analyses of the baseline proton and nuclear PDFs were performed with the \textsc{xFitter} \cite{Zenaiev:2016jnq,Bertone:2017tig} tool. The main goal of the \textsc{xFitter} project is to provide an open-source tool to fit proton PDFs with different theoretical assumptions. A schematic view of the fitting procedure and relations to different external programs are shown in figure \ref{fig-xfitter}. Being an open-source tool it is available to everyone and makes the research process fully transparent, which is important in order to establish a common knowledge base and a deep understanding of the opportunities and limitations. The released version covers various options like different PDF parameterization forms, mass schemes, etc. Furthermore, \textsc{xFitter} provides interfaces programmed in Fortran or C++ to the commonly used tools like \textsc{MINUIT} \cite{James:1975dr,Lazzaro:2010zza}, \textsc{QCDNUM} \cite{Botje:2010ay}, \textsc{APPLGrid} \cite{Carli:2010rw} or \textsc{APFEL} \cite{Bertone:2013vaa}, etc. The DGLAP evolution routine and the calculation of DIS cross sections are implemented at NNLO. Further functionalities in regards to the future potential and alternative fitting approaches, including dipole models \cite{Luszczak:2013rxa, Luszczak:2016bxd} and small-$x$ resummation \cite{Ball:2017otu, Abdolmaleki:2018jln}, are available in the released version.

In order to perform a nuclear PDF analysis several modifications of the code were required. First, the PDF parameterization had to be adapted for the purpose of nuclear PDFs. Thus, new parameters $c_k(A)$ dependent on the nuclear mass number $A$ as per equation (\ref{coeff-A}) were introduced. In order to reflect the new nPDF parameterization, the form of the steering file, as well as the file containing the initial parameters for \textsc{MINUIT}, and the according interpretation routine were adapted. As the next step, the mass number $A$ and proton number $Z$ of a given nucleus for the nucleon decomposition (cf. eq.~(\ref{nucleon})) of the up and down quarks were included. The possible combinations of data sets for different mass numbers $A$ and proton numbers $Z$ were kept flexible in order to deal with data for ratios between different nuclei. The information on $A$ and $Z$ depends on the data set and thus needs to be provided inside the data files. Therefore, the form and the routine to read the experimental data files were extended accordingly. Additionally, the overall minimization routine \textsc{FCN} was modified so that the DGLAP evolution can evolve nuclear PDFs covering different combinations of $A$ and $Z$ individually. Next, the calculation of sum rules (equations \ref{eqsumruleuv}, \ref{eqsumruledv}, \ref{eqmomsumrule}) had to be adapted in order to reflect the flexibility of an $A$-dependent normalization. Additionally, to keep the form of the PDF parameterization flexible, a new numerical integration routine for the calculation of sum rules was implemented. 

Besides that, the cross section calculation routine was enhanced for the treatment of various isoscalar modifications, as described in subsection \ref{isoscalar-corr}. Three flags identifying the 'NMC', 'EMC' or 'SLAC' forms of the corrections were implemented. Furthermore, experimental nuclear data is often provided in terms of ratios $\sigma(A_1)/\sigma(A_2)$ or $F_2(A_1)/F_2(A_2)$ for two different nuclei $A_1$ and $A_2$. Thus, we had to extend the \textsc{xFitter} mechanisms for the consideration of these ratios by the implementation of a two-step loop. The underlying PDF flavour decomposition for a proton was modified so that in the case of a nucleus the PDF decomposition is applied for a nucleon of the form (\ref{nucleon}). 

Moreover, charged-current (CC) processes for the neutrino DIS data\footnote{The implemented calculation of the reduced CC DIS cross section by using the FONLL scheme in \textsc{xFITTER} was customized to the HERA framework. Specifically, a factor $1/2$, e.g., in equation (8) of Ref. \cite{Adloff:2003uh}, is included in the reduced cross section to account for the factor 2 in the normalization of the structure functions used there. For the neutrino-nucleus DIS data, we use the nonreduced differential cross section with the prefactor from \cite{Tanabashi:2018oca}, and especially with the structure functions defined in equations (\ref{eq:sF-neutrino}) and (\ref{eq:sF-antin}). Thus, for the calculation of the CC DIS cross sections, a division by a factor of $2$ was removed from \textsc{xFitter} in the case of neutrino beams.}  were incorporated in \textsc{xFitter} according to the differential cross section described in subsection \ref{sec-crosssections}. 
Finally, the uncertainty analysis routine{\footnote{In this work the option 'DoBands' has been used to generate asymmetric error bands, which is based on the 'iterate' method by John Pumplin \cite{Pumplin:2000vx}. Its advantage is that, if necessary, the iteration routine will add a positive value $X$ to all eigenvalues to force the matrix to be positive definite, which is as close as possible to the actual $\hat{H}$. The positive definiteness of the Hessian matrix relies on the second derivatives, which is a difficult computation and is numerically often approximate. One reason is that the minimized function $\Chi^2$ is not exact, but given by a second-order polynomial in the space of the fit parameters. Thus, if some fit parameters are not well constrained by the data, higher-order polynomial terms of $\Chi^2$ might become relevant \cite{Eskola:2019dui}. Another point is that the function $\Chi^2$ might not be as smooth as necessary due to the limited numerical precision at which the DGLAP equations are solved and due to the finite accuracy of the integrals.}} \cite{Pumplin:2001ct,Pumplin:2000vx} was modified so that scaling of the error bands (cf. eq. \ref{eq-ai}) could also be performed for $\Delta \Chi^2 > 1.0$. The modifications described in this section will be published as a part of the package later on. 

\section{Experimental data}
\label{sec:expdata}

\subsection{Charged-lepton DIS data}

\begin{figure}[tb!]
     \begin{center}
      \subfigure{        
              \includegraphics[width=0.48\textwidth]{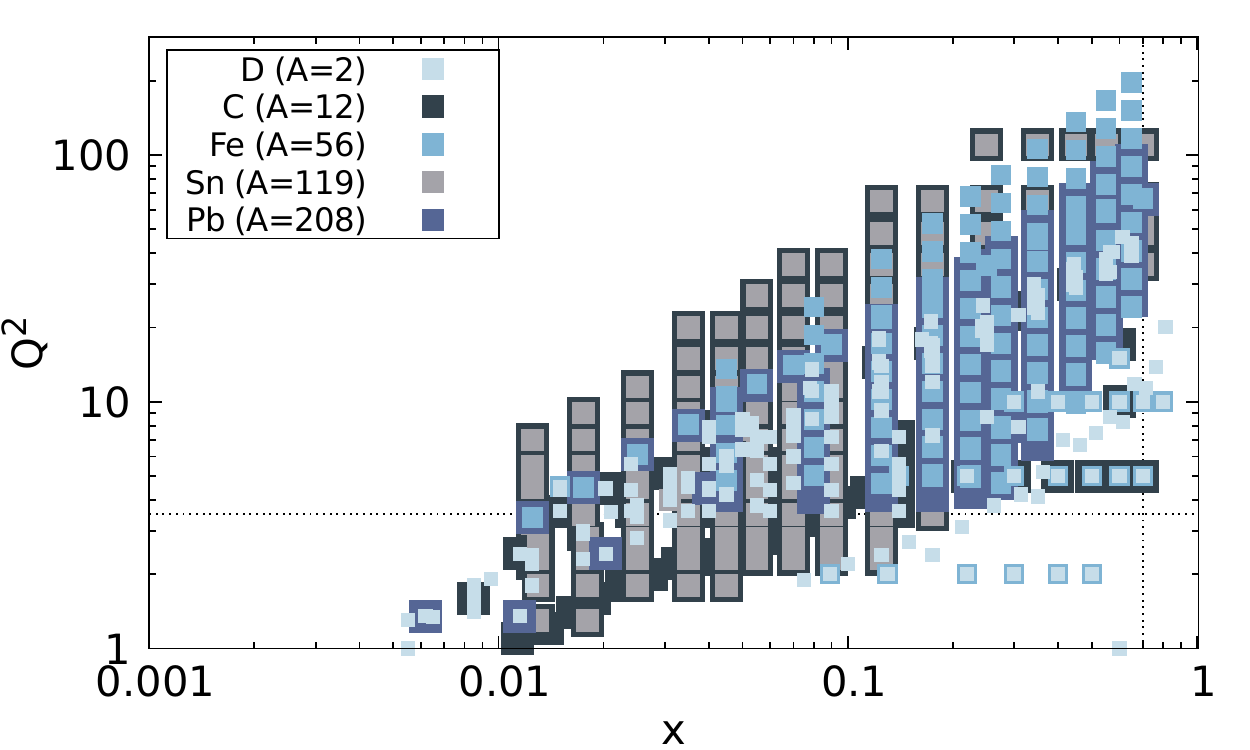}} 
          \subfigure{        
              \includegraphics[width=0.48\textwidth]{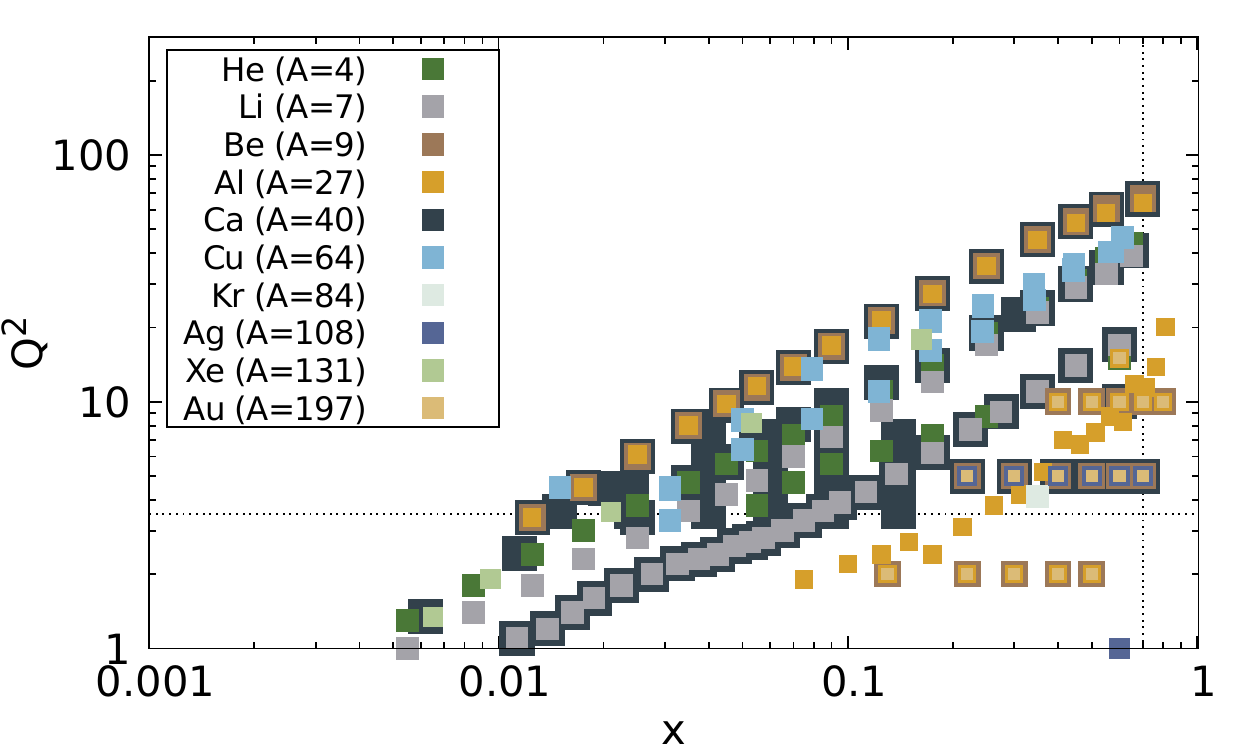}} 

          \end{center} 
\caption{Kinematic reach of experimental DIS data in the $(x,\,Q^2)$ plane used to constrain the nuclear PDFs.}
\label{fig-xQ2A-plane}    
    \end{figure} 
The QCD analyses presented here have been performed by including the experimental data from DIS measurements. The free proton baseline was fitted using data from the HERA \cite{Abramowicz:2015mha}, BCDMS \cite{Benvenuti:1989rh} and NMC \cite{Arneodo:1996qe} experiments, as listed in table \ref{tab-expdata-proton}. The fixed-target DIS data with lepton and neutrino beams used to determine the nuclear parton distribution functions are summarized in table \ref{tab-expdata}. The kinematic reach of the included experimental DIS data in the $(x,Q^2)$ plane is shown in figure \ref{fig-xQ2A-plane} for the different nuclear targets. The applied kinematic cuts $x \leq 0.7$ and $Q^2 \geq 3.5\,\mathrm{GeV^2}$ are illustrated by the dotted lines in the plots. The number of available data points varies for different nuclei. A large number of data points are available for deuteron (D) and the heavier nuclei of carbon (C), iron (Fe), tin (Sn) and lead (Pb), as shown in the upper panel of figure \ref{fig-xQ2A-plane}. These data points are provided either in the form of absolute cross sections, or as ratios where D is usually used as the reference (denominator). Also calcium (Ca) has been intensively used in the relevant experiments. For the other nuclei (lower panel in figure \ref{fig-xQ2A-plane}), only a few data points are available. Therefore the nuclei predominantly present in the included data are expected to be better constrained than the nuclei with fewer data points. 
\begin{center}
\renewcommand{\arraystretch}{1.25}
\begin{table}[tb]
\caption{{ Summary of experimental DIS data used to determine proton PDFs. In the last two columns the resulting $\Chi^2$ values at NLO and NNLO obtained in our analysis are provided.}}
\label{tab-expdata-proton}
\scriptsize
\begin{tabular}{l|l|c|r|c|r|r}
\toprule
  Exp. & Data set & Year & Ref. & $N_{\mathrm{dp}}$  & $\Chi^2$ NLO  & $\Chi^2$ NNLO  \\ \hline \hline
  BCDMS & F2p 100GeV & 1996 & \cite{Benvenuti:1989rh} & 83 & 88.88 & 90.98\\
        & F2p 120GeV & & &						        90   &  69.97 & 67.75 \\  
        & F2p 200GeV & & &						        79   &  89.46 &  85.91 \\  
        & F2p 280GeV  & & &						        75   &  66.97 & 68.73 \\ \hline
  HERA 1+2 & NCep 920 & 2015 & \cite{Abramowicz:2015mha} & 377 & 455.15 & 475.14 \\ 
           & NCep 820 & & &						        70   &  72.47 & 73.84 \\
           & NCep 575 & & &						        254   &  225.24 & 228.97 \\
           & NCep 460 & & &						        204   &  223.23 & 223.95 \\ 
           & NCem & & &						        159   &  233.55 & 229.42 \\
           & CCep & & &						        39   &   42.19 & 44.41 \\
           & CCem & & &						        42   &  65.94 & 68.99 \\ \hline                                                              
  NMC-97 & NCep & 1997 & \cite{Arneodo:1996qe}  & 100& 124.56 & 111.64 \\ \hline \hline            
   \multicolumn{4}{r|}{\textbf{In total:}} & 1559 & 1845.99 &  1909.08  \\ 
\toprule  
\end{tabular} 
\end{table}
\end{center}
\begin{center}
\renewcommand{\arraystretch}{1.25}
\begin{table}[tb!]
\caption{{Summary of experimental DIS data used to determine the nuclear PDFs. In the last two columns the resulting $\Chi^2$ values at NLO and NNLO obtained in our analysis are provided.}}
\label{tab-expdata}
\scriptsize
\begin{tabular}{l|l|c|r|c|r|r}
\toprule
  Nucleus & Exp. & Year & Ref. & $N_{\mathrm{dp}}$  & $\Chi^2$ NLO  & $\Chi^2$ NNLO  \\   
\hline
\hline
  D & NMC 97 & 1996  & \cite{Arneodo:1996qe} &  120 & 124.85  & 118.66  \\ 
       & EMC 90 & 1989  & \cite{Arneodo:1989sy} & 21 & 29.23  & 31.73  \\ \hline
  He/D  & HERMES  & 2002    & \cite{Airapetian:2002fx} & 7 &  54.64 & 37.99   \\ 
      & NMC 95, re.  & 1995   & \cite{Amaudruz:1995tq} &  13 & 12.44  & 12.98  \\   
      & SLAC E139 & 1994  & \cite{Gomez:1993ri} & 11 & 7.21  & 4.68 \\ \hline
   Li/D  & NMC 95 & 1995  & \cite{Arneodo:1995cs} &  12 & 7.06  & 5.93  \\ \hline
   Be/D  &  SLAC E139 & 1994  & \cite{Gomez:1993ri} & 10 & 7.84  & 7.83  \\ \hline    
   Be/C   & NMC 96 & 1996  & \cite{Arneodo:1996rv} & 14 & 14.80  & 16.19  \\ \hline
   C    & EMC 90 & 1989  & \cite{Arneodo:1989sy} & 17 &  11.01 & 10.05  \\
   C/D  & FNAL E665 & 1995   & \cite{Adams:1995is} & 3  & 5.12  & 5.91 \\ 
        & SLAC E139 & 1994   & \cite{Gomez:1993ri} & 6 &  15.12 & 17.16  \\
        & EMC 88 & 1988  & \cite{Ashman:1988bf} & 9 & 4.49  & 3.50  \\  
        & NMC 95, re.  & 1995    & \cite{Amaudruz:1995tq} & 13 &  38.08 & 36.52  \\ \hline
   C/Li  & NMC 95, re. & 1995    & \cite{Amaudruz:1995tq} & 10 & 17.27  & 13.90 \\ \hline
   N/D  & HERMES &  2002    & \cite{Airapetian:2002fx} & 1 &  2.20 & 0.97  \\ \hline
   Al/D     & SLAC E139 & 1994  & \cite{Gomez:1993ri} &  10 & 11.20  &  14.22 \\\hline    
   Al/C    & NMC 96 &  1996   & \cite{Arneodo:1996rv} & 14   & 6.51  & 6.55   \\\hline 
   Ca &    EMC 90  & 1989  & \cite{Arneodo:1989sy} & 19 & 13.17  & 12.56  \\ 
   Ca/D  & NMC 95, re. & 1995    & \cite{Amaudruz:1995tq} & 12 & 29.61  & 31.12 \\ 
     &    FNAL E665 &  1995  & \cite{Adams:1995is} & 3  &  4431 & 6.01  \\
     &    SLAC E139  &  1994  & \cite{Gomez:1993ri} & 6  & 8.44  & 9.34 \\                   
   Ca/Li  & NMC 95, re. & 1995  & \cite{Amaudruz:1995tq} &  10 & 7.36  &  5.16  \\ \hline
   Ca/C  & NMC 95, re. & 1995  & \cite{Amaudruz:1995tq}  & 10  & 6.47  & 6.70  \\ 
     &  NMC 96 &  1996  & \cite{Arneodo:1996rv} & 14  & 7.14  &  6.99  \\ \hline 
   Fe  &    SLAC E140  & 1993 & \cite{Dasu:1993vk} & 2  & 0.05  & 0.05 \\
   Fe/D  &    SLAC E139  & 1994  & \cite{Gomez:1993ri} & 14 & 34.08  & 34.18  \\
   Fe/C  & NMC 96 & 1996  & \cite{Arneodo:1996rv} & 14  &  9.82 & 9.96 \\ \hline     
   $\nu$ Fe    & CDHSW  & 1991 & \cite{Berge:1989hr}  &  464  & 347.74  &  365.14  \\ 
   $\bar{\nu}$ Fe    & CDHSW  & 1991  & \cite{Berge:1989hr} & 462  & 423.06  &  398.25  \\ \hline
   Cu/D  & EMC 93 & 1993   & \cite{Ashman:1992kv} & 19  & 18.12  &  17.45 \\ 
        & EMC 88  &  1988 & \cite{Ashman:1988bf}  & 9  & 5.59  & 7.22  \\ \hline   
   Kr/D  & HERMES  & 2002   & \cite{Airapetian:2002fx}  & 1   & 2.02  &  2.02 \\ \hline   
   Ag/D  & SLAC E139  & 1994  & \cite{Gomez:1993ri} & 6  & 16.24  & 18.81  \\ \hline   
   Sn/D  & EMC 88 &  1988 & \cite{Ashman:1988bf}  & 8 & 14.56 & 9.24 \\ \hline   
   Sn/C  & NMC 96 & 1996 & \cite{Arneodo:1996rv}  & 14 & 12.90  & 7.61 \\
        & NMC 96, $Q^2$ dep. & 1996  & \cite{Arneodo:1996ru} & 134 & 94.7  & 79.85  \\ \hline     
   Xe/D   & FNAL E665 & 1992  & \cite{Adams:1992nf} &  3 & 2.13  & 2.53 \\ \hline
   Au/D  & SLAC E139  & 1994  & \cite{Gomez:1993ri} &  11 & 16.64   &  19.80  \\ \hline 
   Pb/D  & FNAL E665  & 1995  & \cite{Adams:1995is} & 2  &  12.24 & 13.32   \\ \hline 
   Pb/C   & NMC 96 &  1996  & \cite{Arneodo:1996rv} & 14   & 9.94  &  6.77  \\ \hline 
   $\nu$ Pb   & CHORUS  & 2005 & \cite{Onengut:2005kv} &  405   & 229.11  &  243.85 \\ 
   $\bar{\nu}$ Pb   & CHORUS  & 2005  & \cite{Onengut:2005kv} & 405  & 361.35  & 328.28  \\ \hline \hline 
   \multicolumn{4}{r|}{\textbf{In total:}} & 2336 & 2072.29 & 2014.02 \\ 
\toprule
\end{tabular}
\end{table}
\end{center}
\subsection{Neutrino DIS data}
Neutrino data were included in the analysis by using the measured cross sections for neutrino and antineutrino beams. The advantage compared to the isospin-averaged structure functions $F_2$ and $F_3$ utilized in DSSZ \cite{deFlorian:2011fp} is that the sensitivity to the flavour decomposition is retained in the cross sections. Another approach was used in EPPS16 \cite{Eskola:2016oht} where normalized (anti)neutrino cross sections were considered. This increases the sensitivity to the shape of the nuclear modifications. In order to extract complete, i.e. without isospin averaging, information from the incorporated neutrino data sets, the absolute cross sections are exploited here. 

The data from the CDHSW $\nu \mathrm{Fe}$ experiment \cite{Berge:1989hr} and the CHORUS $\nu \mathrm{Pb}$ experiment \cite{Onengut:2005kv} have been included in this analysis. In addition there is more neutrino scattering data available, e.g. measured cross sections with an $\mathrm{Fe}$ target by the the NuTeV collaboration \cite{Tzanov:2005kr}, and also data from the CCFRR collaboration \cite{MacFarlane:1983ax}. The data from the CCFRR experiment were excluded from our analysis for two reasons. First, the quantities $x$ and $Q^2$, required for the analysis procedure, were not publicly available for the cross sections. Second, only the averaged structure functions $F_2$ and $F_3$ for $\nu \mathrm{Fe}$ and $\bar{\nu} \mathrm{Fe}$ were available, which lose the sensitivity to flavour decomposition. In regard to NuTeV data, an early study in Ref.~\cite{Owens:2007kp} found that these data could be accommodated together with the CHORUS neutrino DIS data when constraining the $d/u$ ratio but with the applied nuclear corrections some tension with other DIS and DY data were observed. Later on, the analyses in Refs.~\cite{Schienbein:2007fs, Schienbein:2009kk} found some unresolved tension between the NuTeV neutrino DIS data and lepton-nucleus data. In a following work, a similar tension was also found when taking into account neutrino DIS data from the CHORUS and CCFRR experiments in Ref.~\cite{Kovarik:2010uv}. Simultaneously, a study presented in Ref.~\cite{Paukkunen:2010hb} concluded that the tension with other data was specifically due to the NuTeV data at certain energies, whereas CDHSW and CHORUS data were well compatible with the existing lepton-nucleus DIS data. This tension was further studied in Ref.~\cite{Paukkunen:2013grz} where again the NuTeV data were found to be incompatible with the other considered data. Only by normalizing the differential data with the integrated cross section at each energy bin was an acceptable agreement achieved. Due to the demonstrated tension, we have not included the NuTeV neutrino DIS data in this analysis.

\subsection{Isoscalar modifications and experimental uncertainties}
\label{isoscalar-corr}
%
%
Some experimental analyses of charged-lepton DIS have modified the measured structure functions to achieve isospin symmetry for nonsymmetric nuclei such as iron or lead. According to the relations summarized in Ref.~\cite{Eskola:2016oht}, an isoscalar structure function of a nucleus with the mass number $A$ is defined as
\begin{equation}
\hat{F}^A_2 \equiv \frac{1}{2}\,F_2^{p,A}+\frac{1}{2}\,F_2^{n,A}
\label{F2isoscalar}
\end{equation}
with $F_2^{p,A}$ and $F_2^{n,A}$ representing the structure functions of the bound protons and neutrons. By definition, the isoscalar structure function contains an equal number of protons and neutrons, which holds only for specific nuclei. A general structure function for a nucleus with $Z$ protons and $N=A-Z$ neutrons can be written as
\begin{equation}
F^A_2=\frac{Z}{A}\,F_2^{p,A}+\frac{N}{A}\,F_2^{n,A}\,,
\label{F2general}
\end{equation}
\noindent which is \textit{not} isoscalar if $Z \neq N \neq A/2$. The relation between the isoscalar structure function (\ref{F2isoscalar}) and the general structure function (\ref{F2general}) is given by
\begin{equation}
\hat{F}^A_2 = \beta\,F^A_2\,,
\label{F2relation}
\end{equation}
where
\begin{equation}
\beta = \frac{A}{2}\,\left( 1 + \frac{F_2^{n,A}}{F_2^{p,A}} \right) / \left( Z + N \frac{F_2^{n,A}}{F_2^{p,A}} \right)\,.
\label{F2beta}
\end{equation}
Furthermore, it is assumed that the ratio $ F_2^{n,A}/F_2^{p,A}$ for any nucleus is unchanged compared to that for unbound nucleons, so that the relation
\begin{equation}
\frac{F_2^{n,A}}{F_2^{p,A}} = \frac{F_2^{n}}{F_2^{p}}
\end{equation}
can be used in (\ref{F2beta}) to describe the isoscalar modifications. The ratio $F_2^{n}/F_2^{p}$ for the isoscalar ``correction'' is parameterized in a different way by each experiment,
\begin{itemize}
\item EMC \cite{Ashman:1992kv}:
\begin{equation}
\frac{F_2^{n}}{F_2^{p}}=0.92 - 0.86 x
\end{equation}
\item SLAC \cite{Gomez:1993ri}:
\begin{equation}
\frac{F_2^{n}}{F_2^{p}}=1 - 0.8x
\end{equation}
\item NMC \cite{Amaudruz:1991nw}:
\begin{equation}
\frac{F_2^{n}}{F_2^{p}}= A(x)\,\left( Q^2/20\,\mathrm{GeV}^2 \right)^{B(x)}\,\left( 1+x^2\,\mathrm{GeV^2}/Q^2 \right)
\end{equation}
\begin{align}
\mathrm{with}\,\,\,A(x)= &0.979 - 1.692x + 2.979 x^2-4.313x^3 \notag \\
&+3.075x^4 \notag \\
\mathrm{and}\,\,\,B(x)=&-0.171x+0.244x^2\,. \notag
\end{align}
\end{itemize}

In this work the general form of the structure function (\ref{F2general}) is used to calculate the theoretical predictions. In case isoscalar modifications were applied to the measured quantities, for consistency the same modifications are applied to the obtained theoretical results by using relation (\ref{F2relation}).

%
Some of the experiments provide normalization uncertainties on top of the systematic and statistical errors. In this work normalization uncertainties have been treated as \textit{correlated} errors as discussed in Refs. \cite{Stump:2001gu,Pumplin:2001ct,Pumplin:2000vx}. 
The correlated uncertainties are treated as implemented in \textsc{xFitter}, described further in Refs.~\cite{Stump:2001gu,Botje:2001fx,Aaron:2009bp}. The same procedure applies if any \textit{overall} uncertainties are provided in addition to the point-to-point uncertainties, e.g. for the SLAC data \cite{Gomez:1993ri,Bodek:1983ec,Dasu:1993vk}. 

\section{Results}
\label{sec-results}
\label{sec:results}

\subsection{Proton baseline}
\label{sec-protonpdfs}

Analyses of nuclear PDFs have often been performed by using an existing proton PDF set as a baseline for the nuclear modifications. In this work, however, we have fitted the proton PDFs using the same setup as for the nuclear PDFs. This ensures that all assumptions like sum rules, parton flavour decomposition, etc., as well as all parameters like coupling constants and quark masses, and also further settings like e.g. the heavy flavour mass scheme, are applied in a consistent way. Furthermore, this paves the way for a future combined proton and nuclear PDF analysis.
\begin{figure*}[tb!]
     \begin{center}
          \subfigure{
              \includegraphics[width=0.237\linewidth]{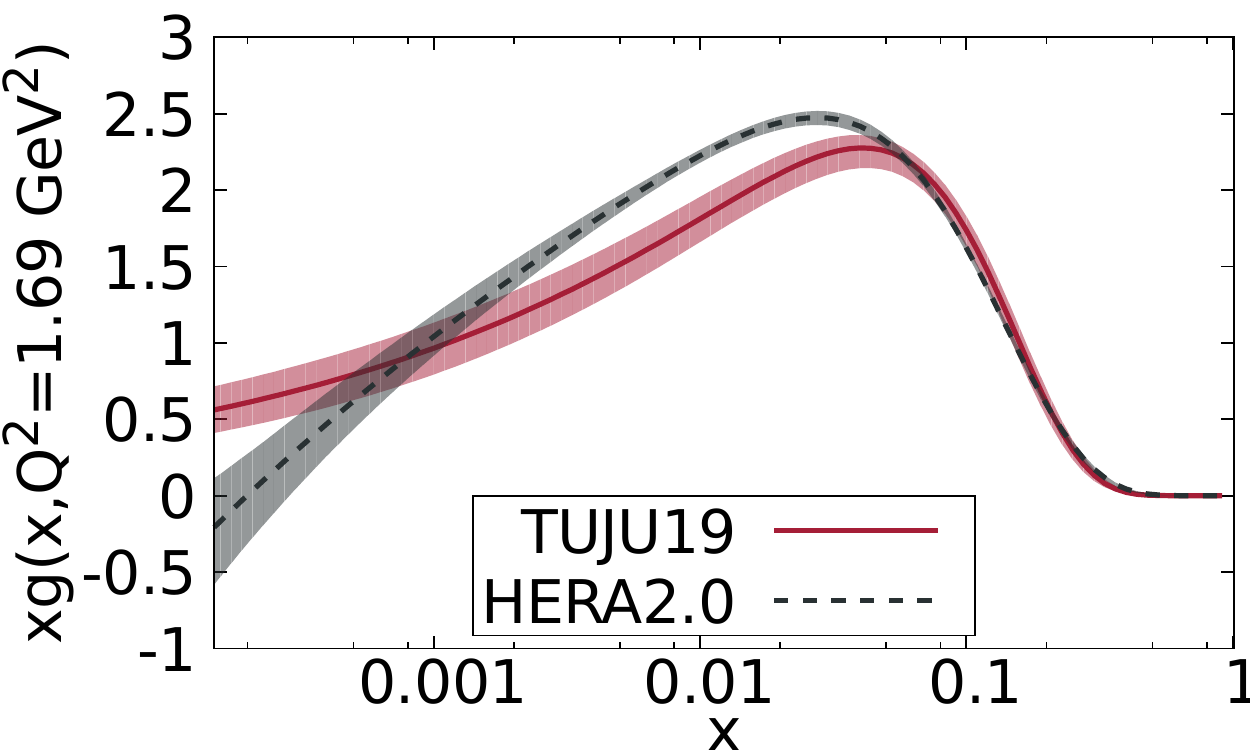}} 
          \subfigure{    
              \includegraphics[width=0.237\linewidth]{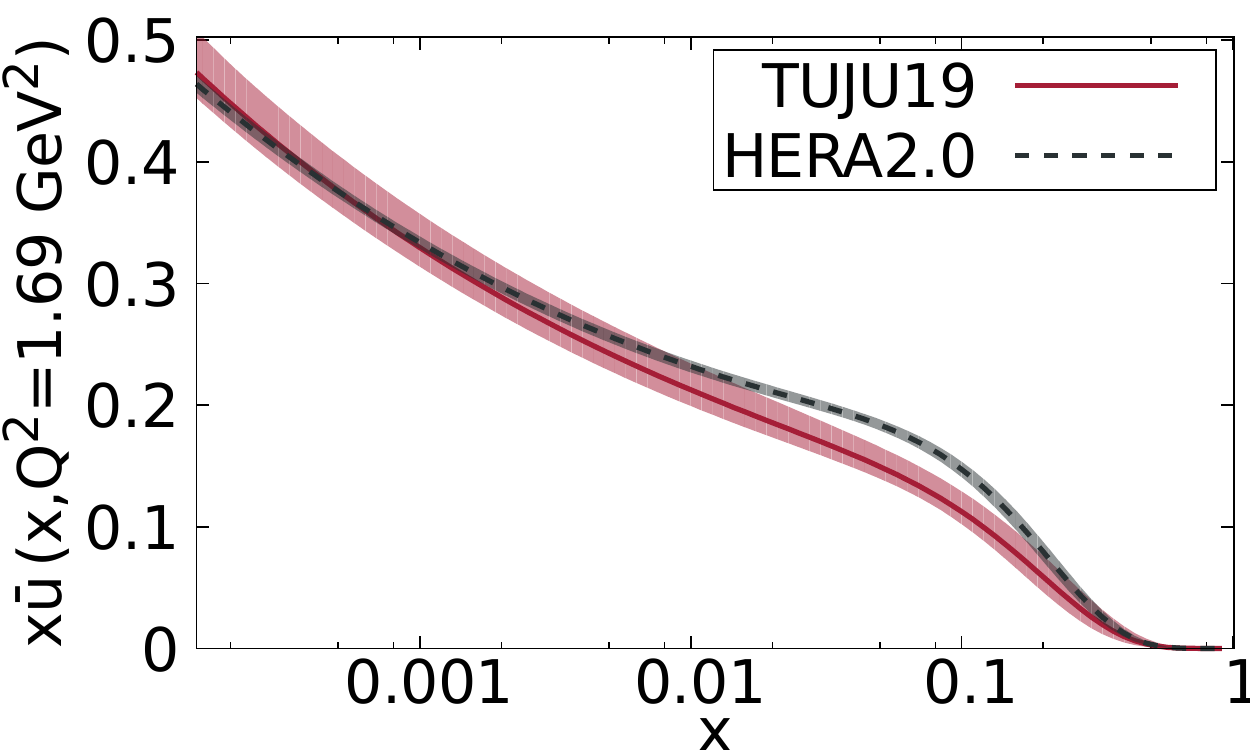}} 
          \subfigure{                           
              \includegraphics[width=0.237\linewidth]{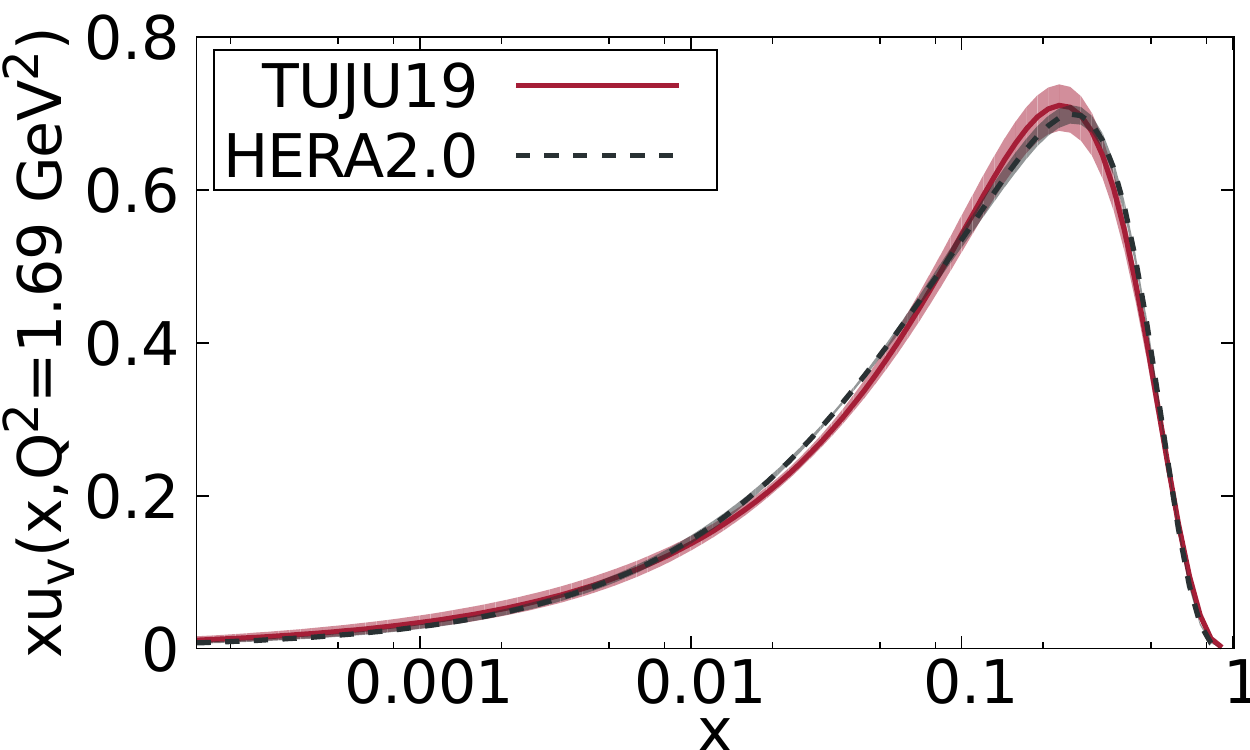}} 
          \subfigure{                                
              \includegraphics[width=0.237\linewidth]{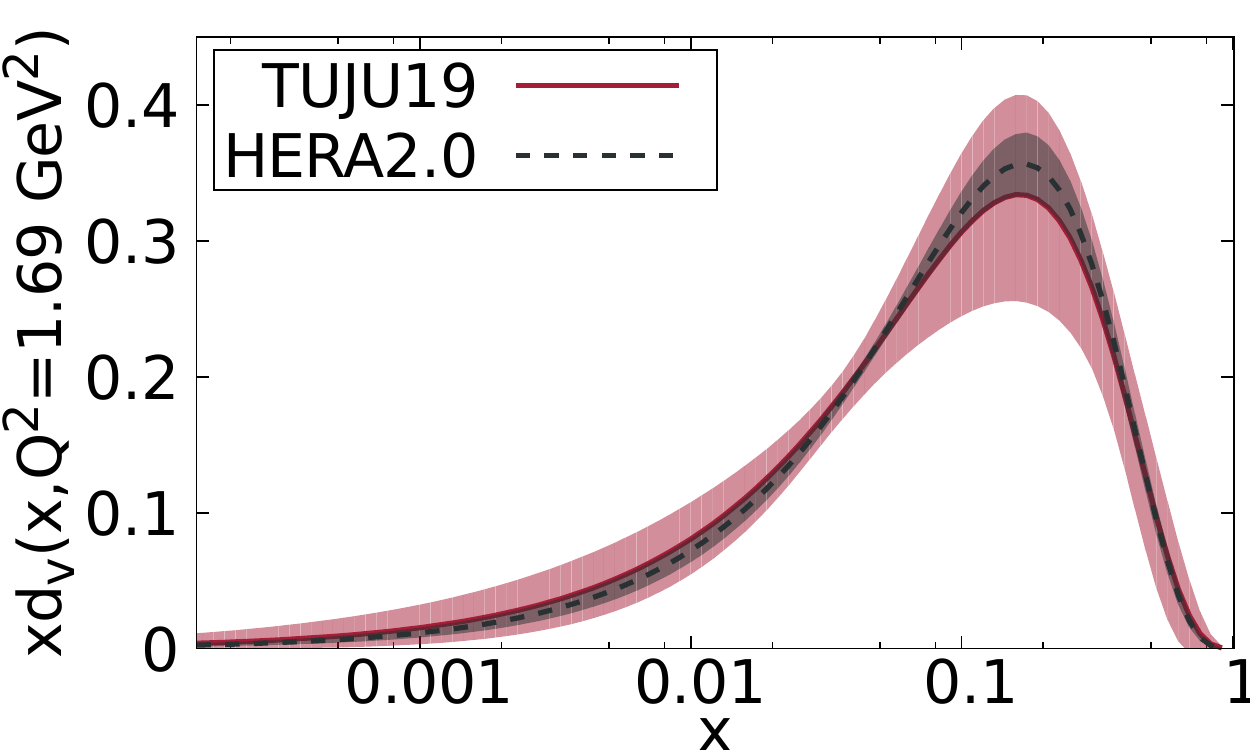}} 
          \subfigure{              
              \includegraphics[width=0.237\textwidth]{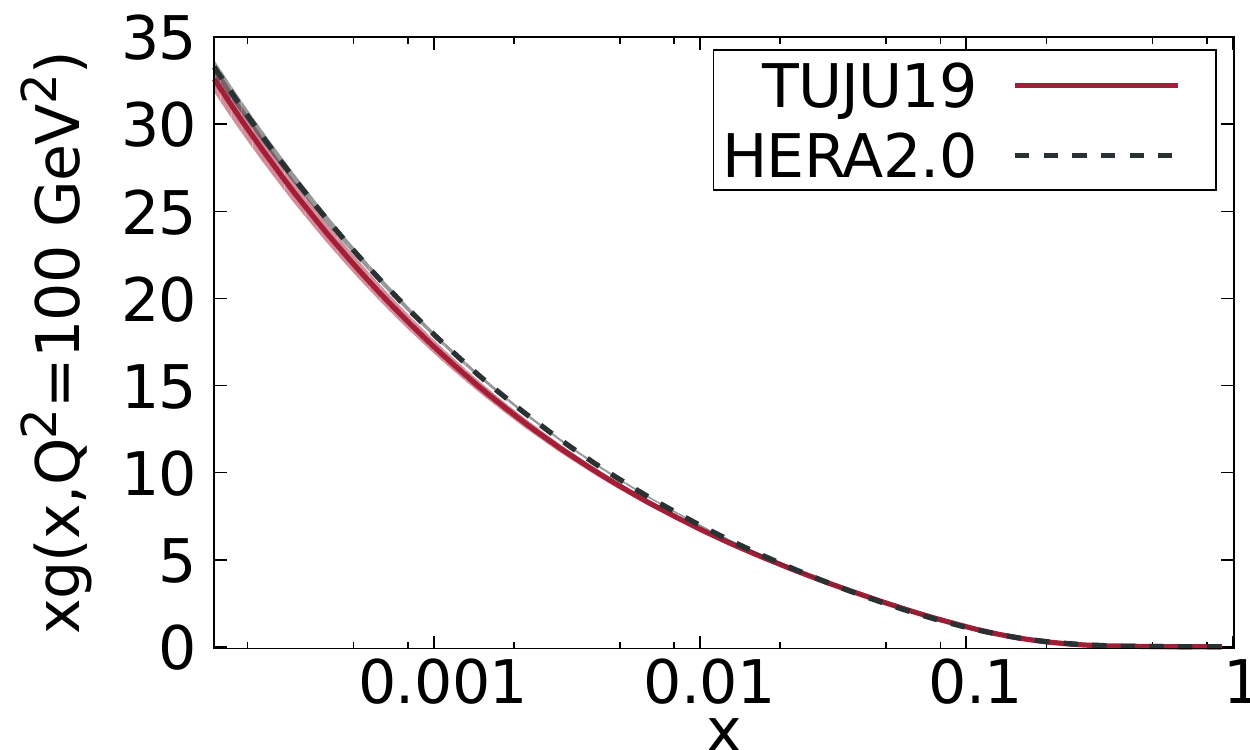}} 
          \subfigure{              
              \includegraphics[width=0.237\textwidth]{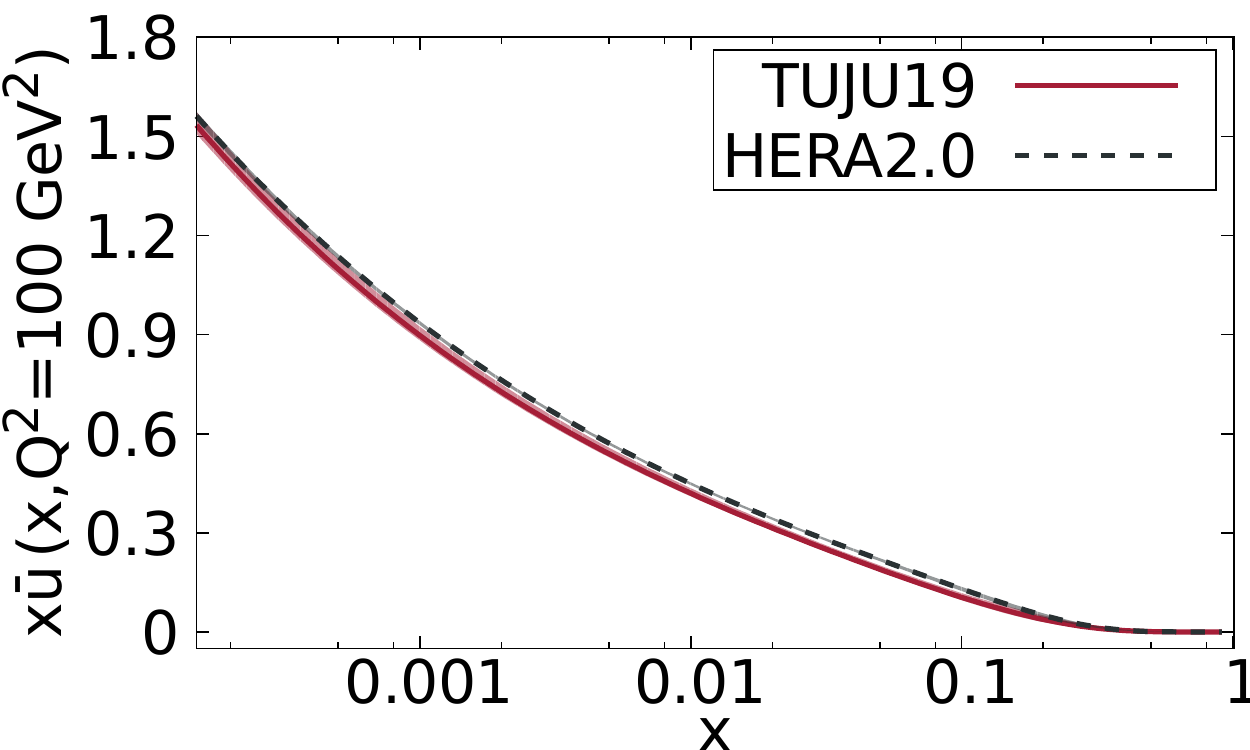}} 
          \subfigure{                        
              \includegraphics[width=0.237\textwidth]{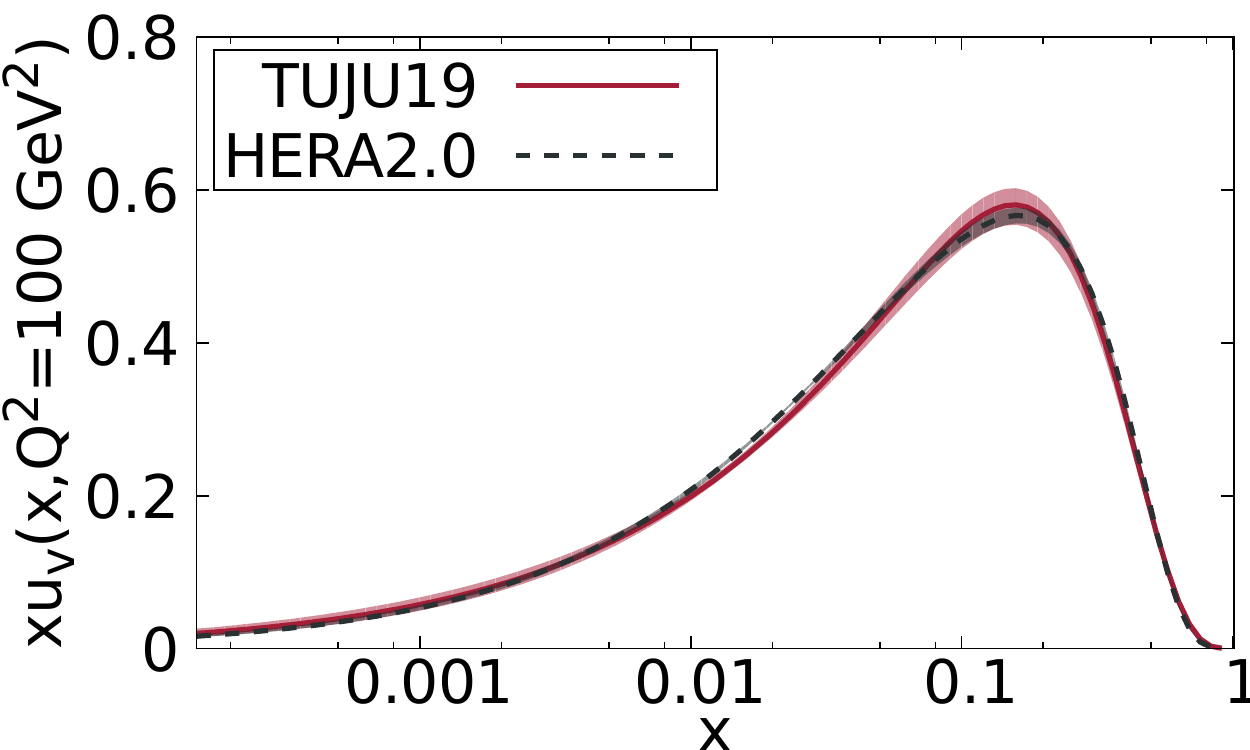}} 
          \subfigure{                                   
              \includegraphics[width=0.237\textwidth]{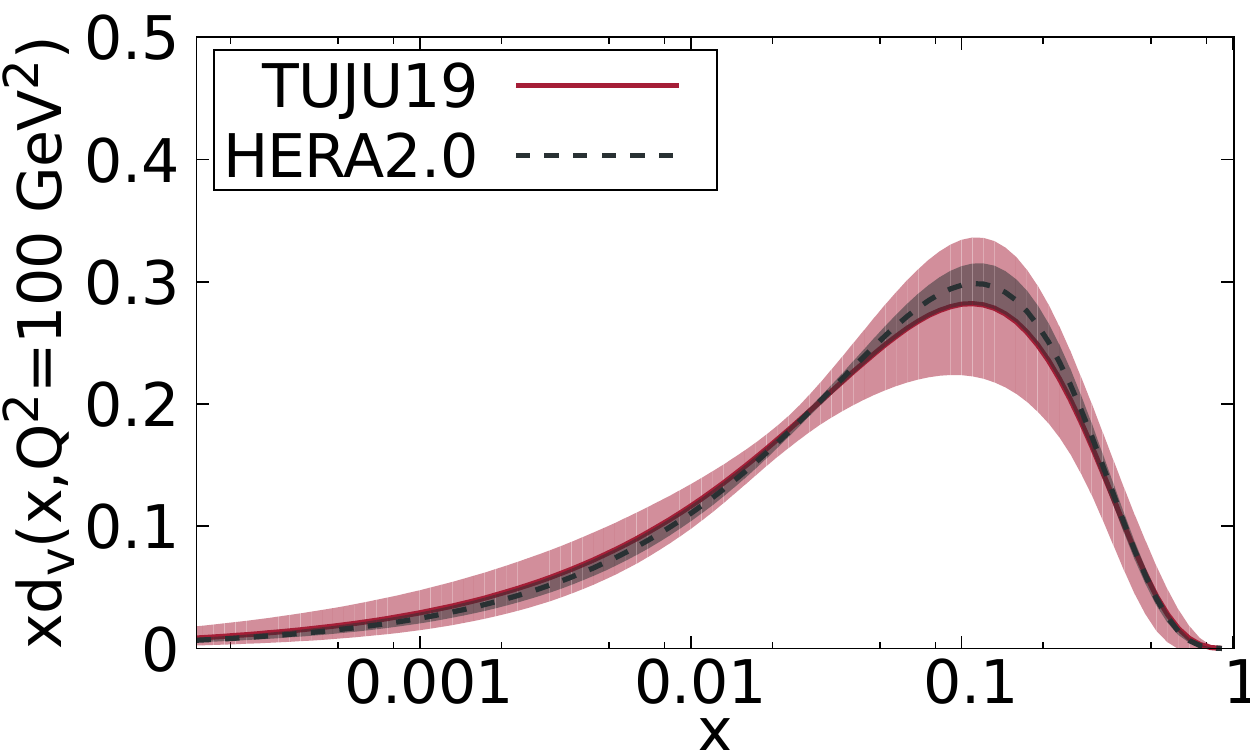}} 
          \end{center} 
    \caption{Proton baseline PDFs TUJU19 at NLO compared to the HERA2.0 results, shown at the initial scale $Q_0^2=1.69\,\mathrm{GeV}^2$ and at $Q^2=100\,\mathrm{GeV}^2$ after DGLAP evolution.}
\label{figprotonNLO}    
    \end{figure*}

\begin{figure*}[tb!]
     \begin{center}
          \subfigure{
              \includegraphics[width=0.237\textwidth]{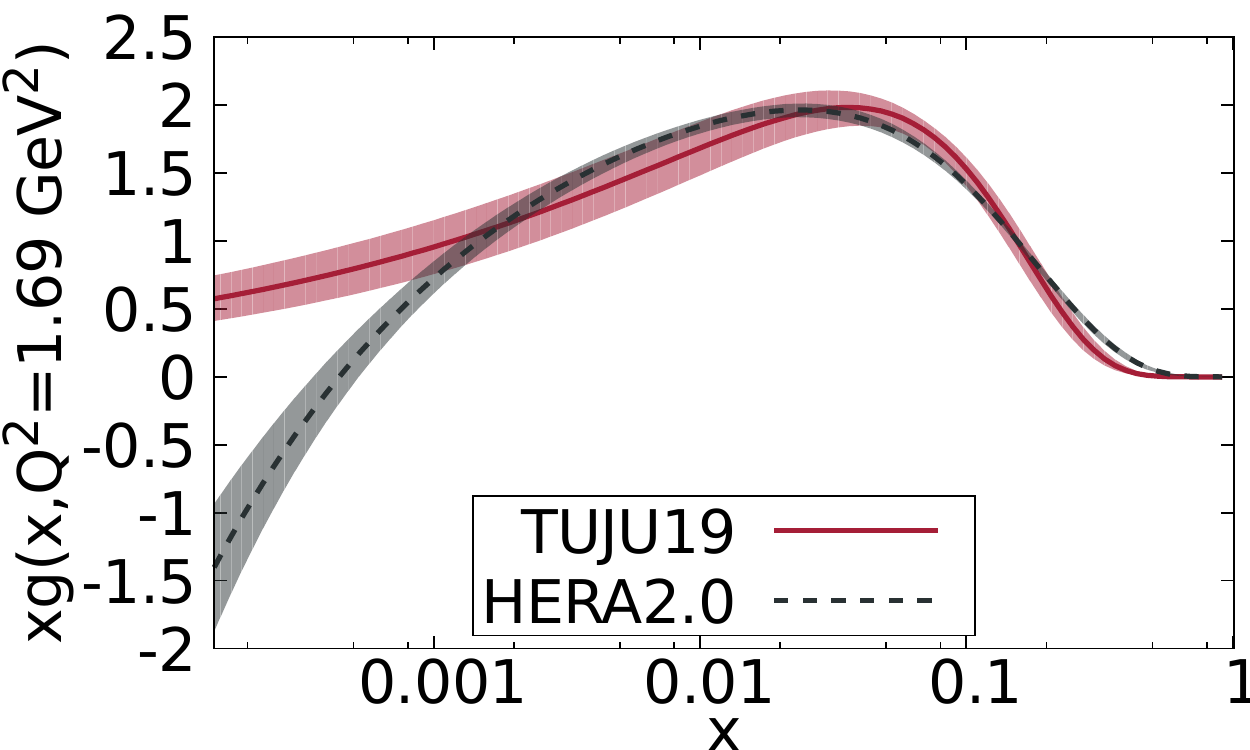}} 
          \subfigure{    
              \includegraphics[width=0.237\textwidth]{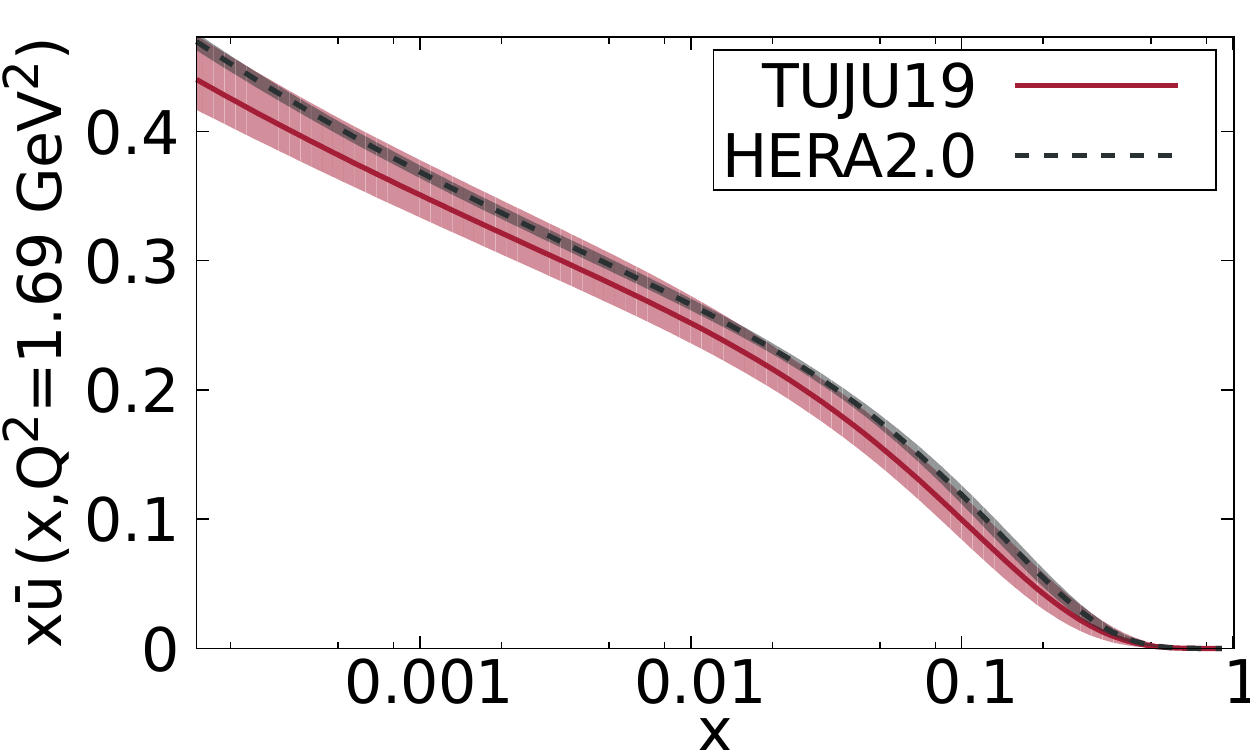}} 
          \subfigure{                           
              \includegraphics[width=0.237\textwidth]{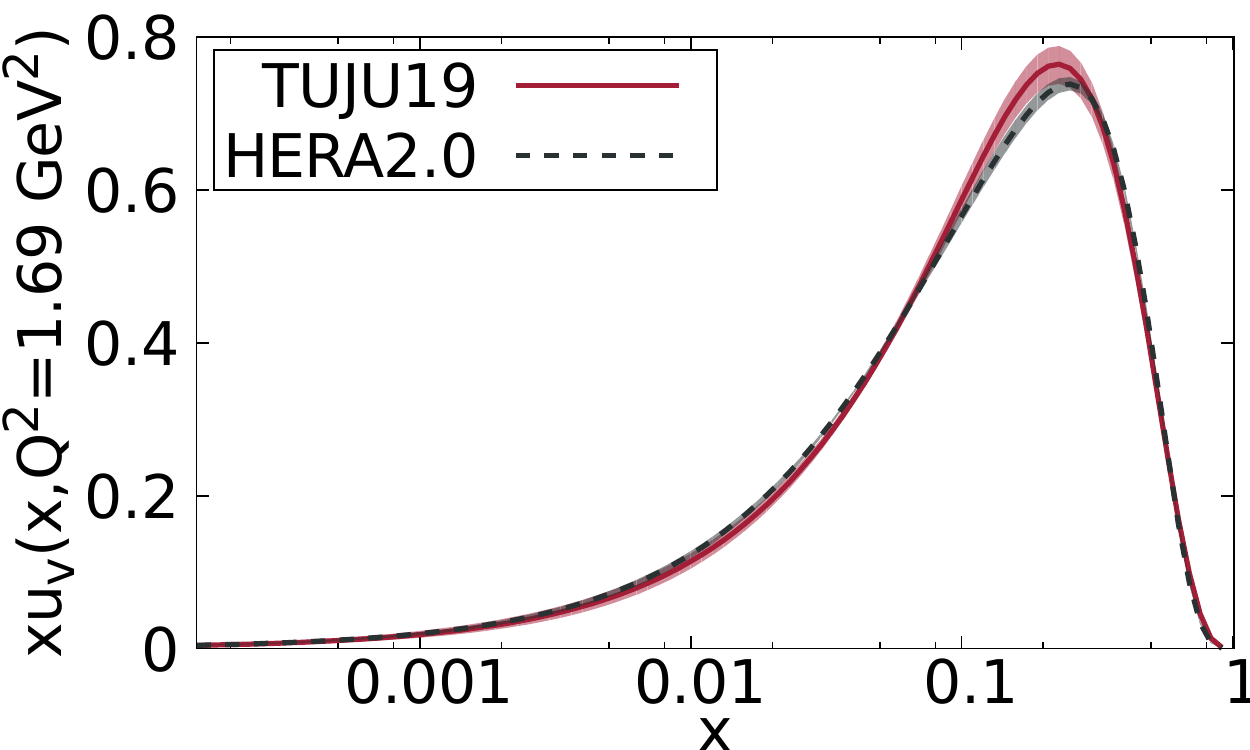}} 
          \subfigure{                                
              \includegraphics[width=0.237\textwidth]{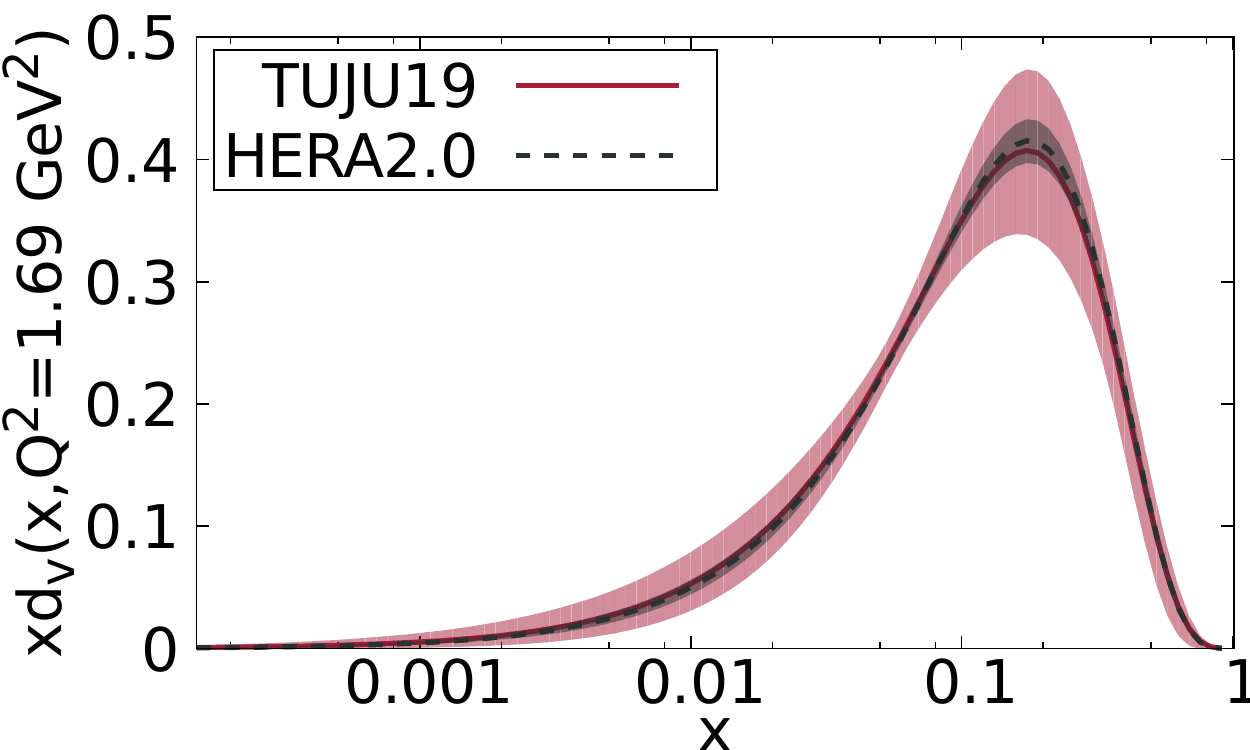}} 
          \subfigure{              
              \includegraphics[width=0.237\textwidth]{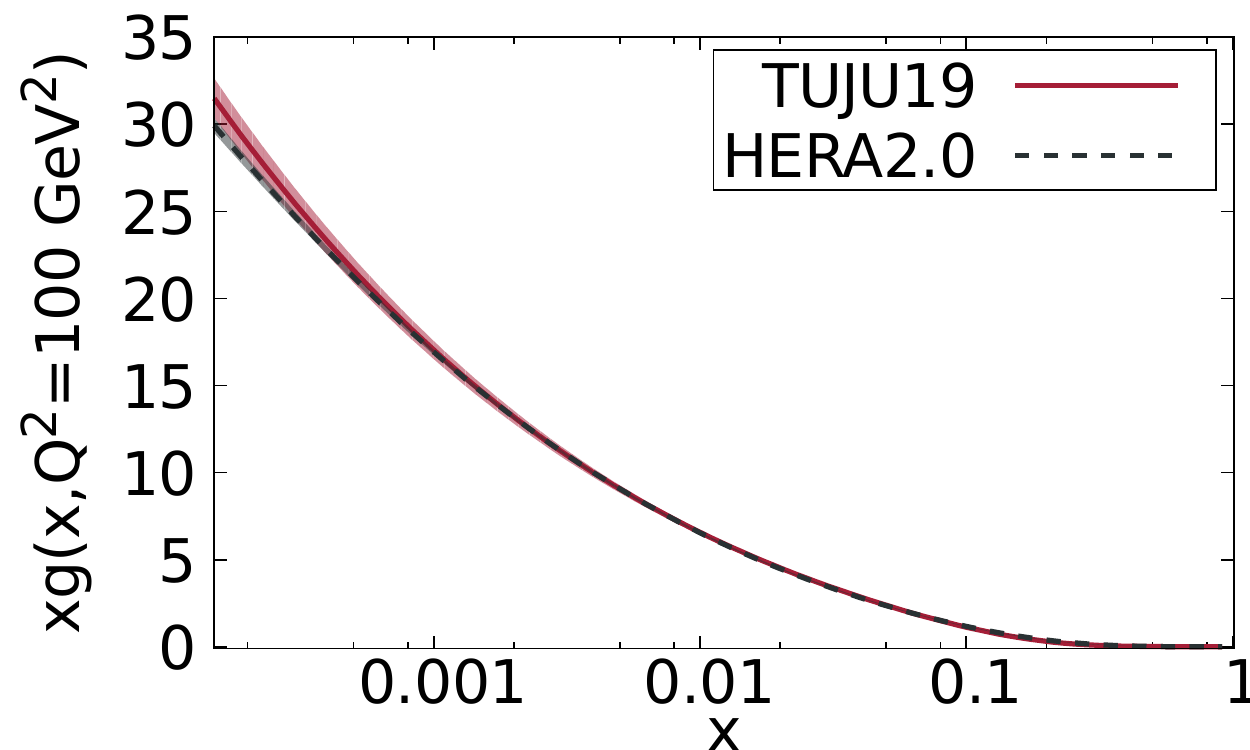}} 
          \subfigure{              
              \includegraphics[width=0.237\textwidth]{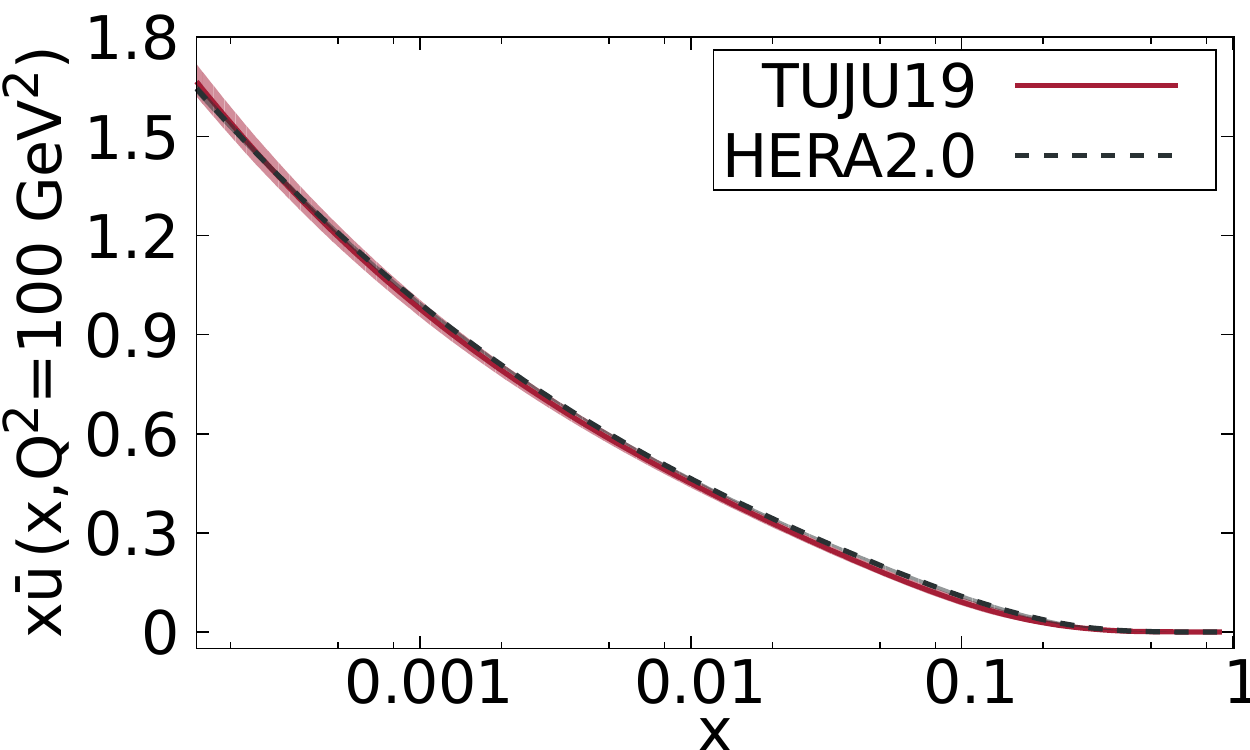}} 
          \subfigure{                        
              \includegraphics[width=0.237\textwidth]{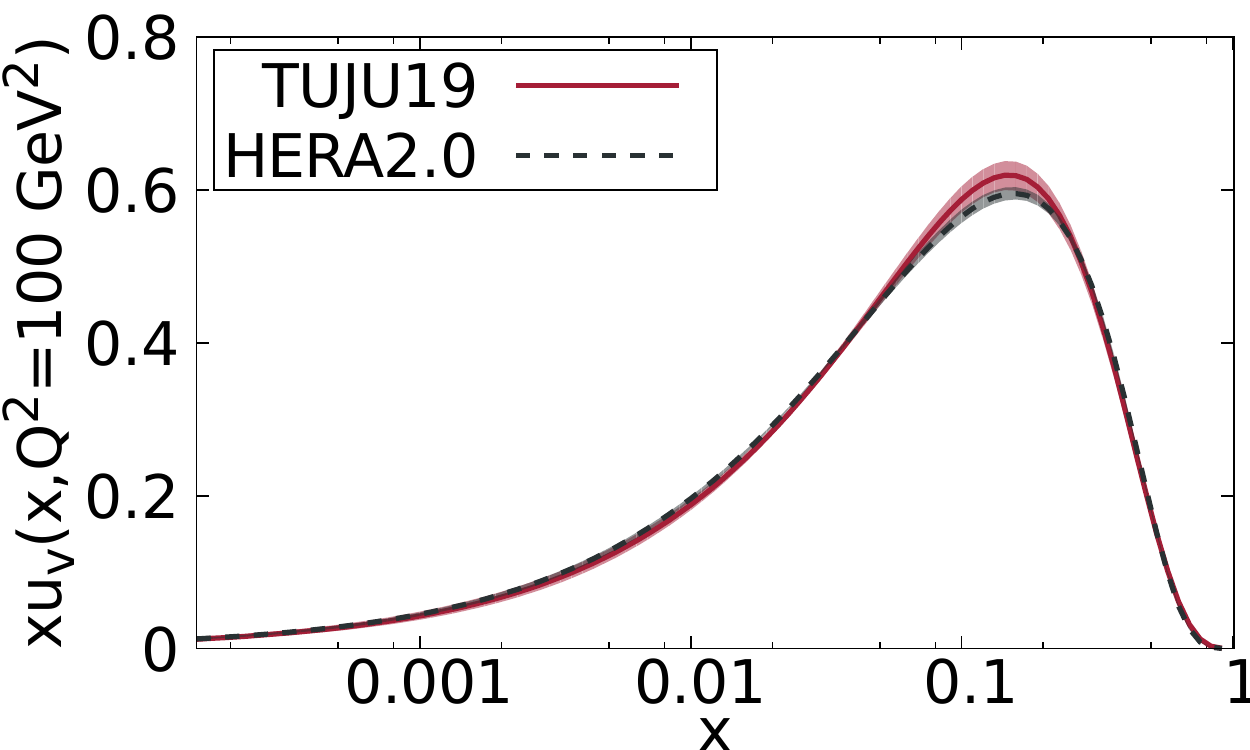}} 
          \subfigure{                                   
              \includegraphics[width=0.237\textwidth]{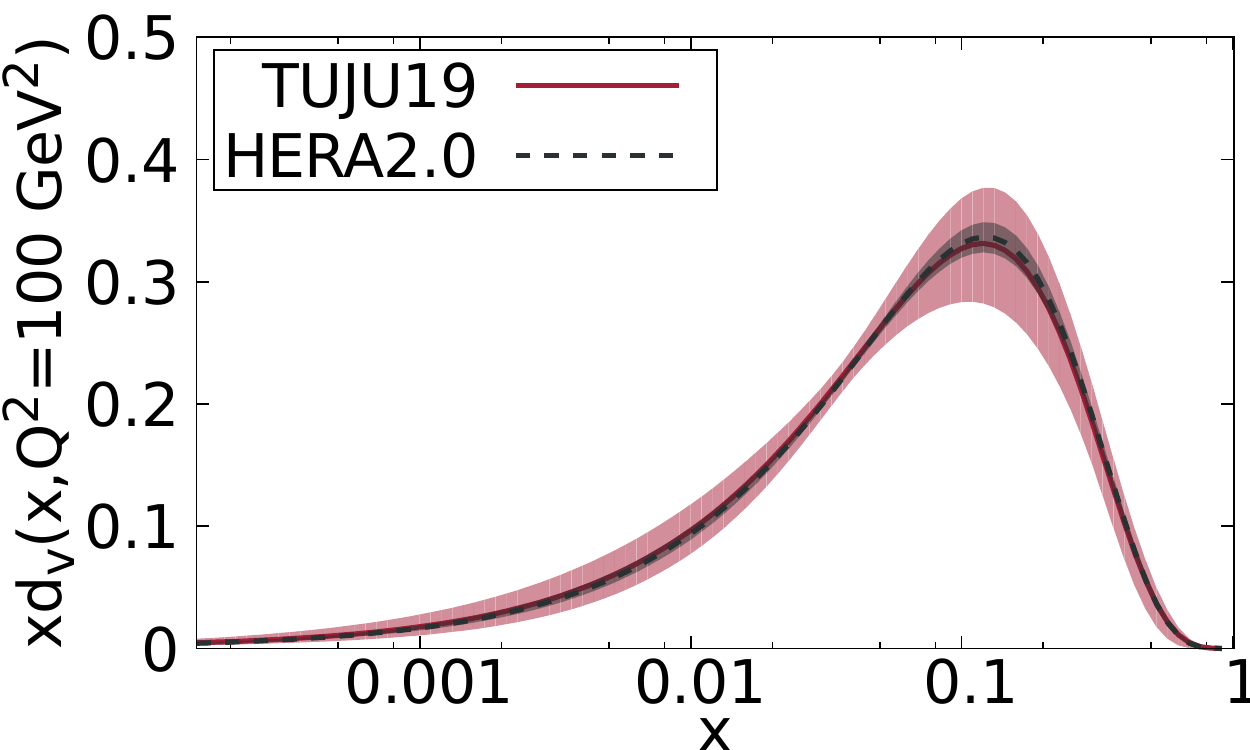}} 
          \end{center} 
    \caption{Same as for figure \ref{figprotonNLO}, but at NNLO.}
\label{figprotonNNLO}    
    \end{figure*}

As we use \textsc{xFitter} as our analysis framework, the baseline proton PDFs are derived with a very similar setup as for the HERA2.0 PDFs \cite{Abramowicz:2015mha}. However, in addition to the combined HERA DIS data we also include data from other experiments (cf. table \ref{tab-expdata-proton}). Another difference is that we use the parameterization in eq. (\ref{pdf-parameterization}) whereas the HERA2.0 analysis includes additional terms for the gluon at the initial scale of the analysis. The obtained parton distribution functions are compared to the HERA2.0 PDFs \cite{Abramowicz:2015mha} in figure \ref{figprotonNLO} at NLO and in figure \ref{figprotonNNLO} at NNLO. As shown in Ref.~\cite{Abramowicz:2015mha}, the HERA2.0 PDFs are well compatible with other state-of-the-art proton PDFs in the kinematic region considered in this work, and since the main focus of this work is on nuclear PDFs we do not present further comparisons to other proton PDF sets. As expected, because of our use of the same fitting framework with similar data and similar definition of $\Chi^2$, the agreement with the HERA2.0 PDFs is very good both at NLO and NNLO. The observed difference of the gluon PDFs at small $x$ can be traced back to the different parameterization applied. Since we also include data from experiments other than HERA, we have used a larger $\Delta \Chi^2$ value, namely $\Delta \Chi^2=20$ (see section \ref{sec-uncertainties}), which results in larger uncertainties than those quoted for the HERA2.0 PDFs.

\subsection{Nuclear PDFs}

\begin{figure*}[b!]
     \begin{center}
          \subfigure{
              \includegraphics[width=0.237\textwidth]{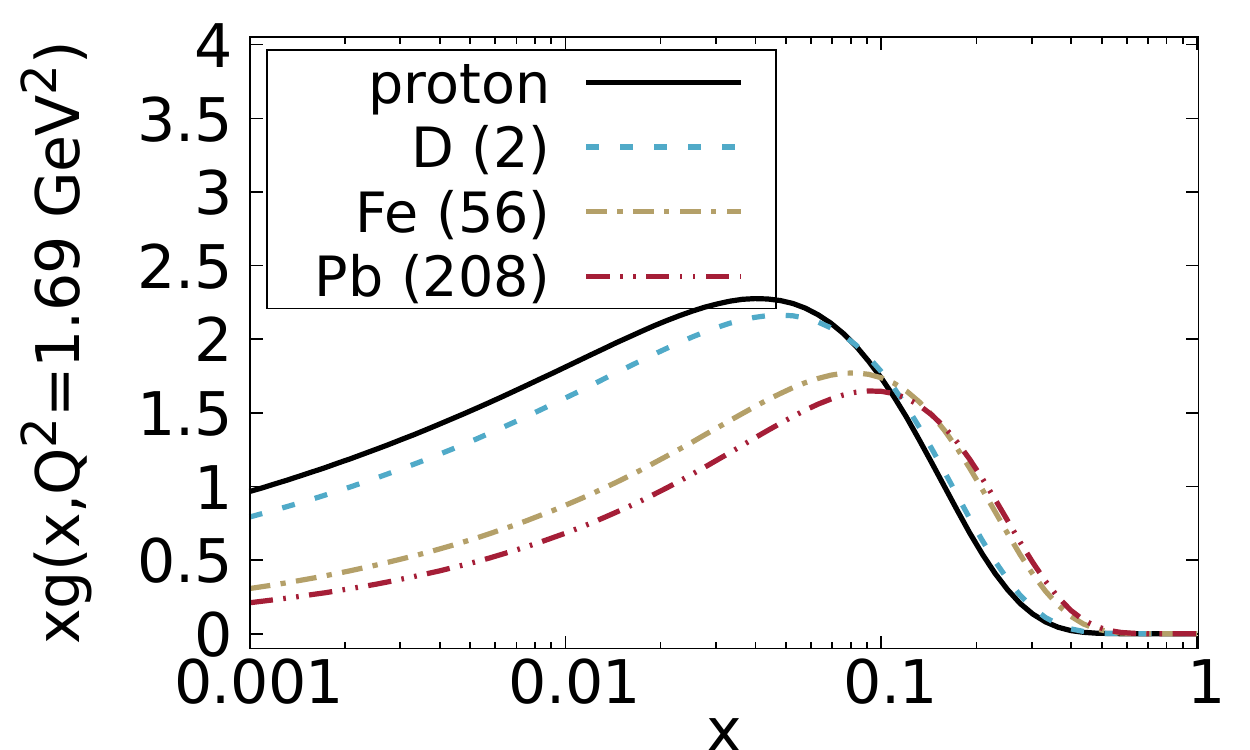}} 
          \subfigure{    
              \includegraphics[width=0.237\textwidth]{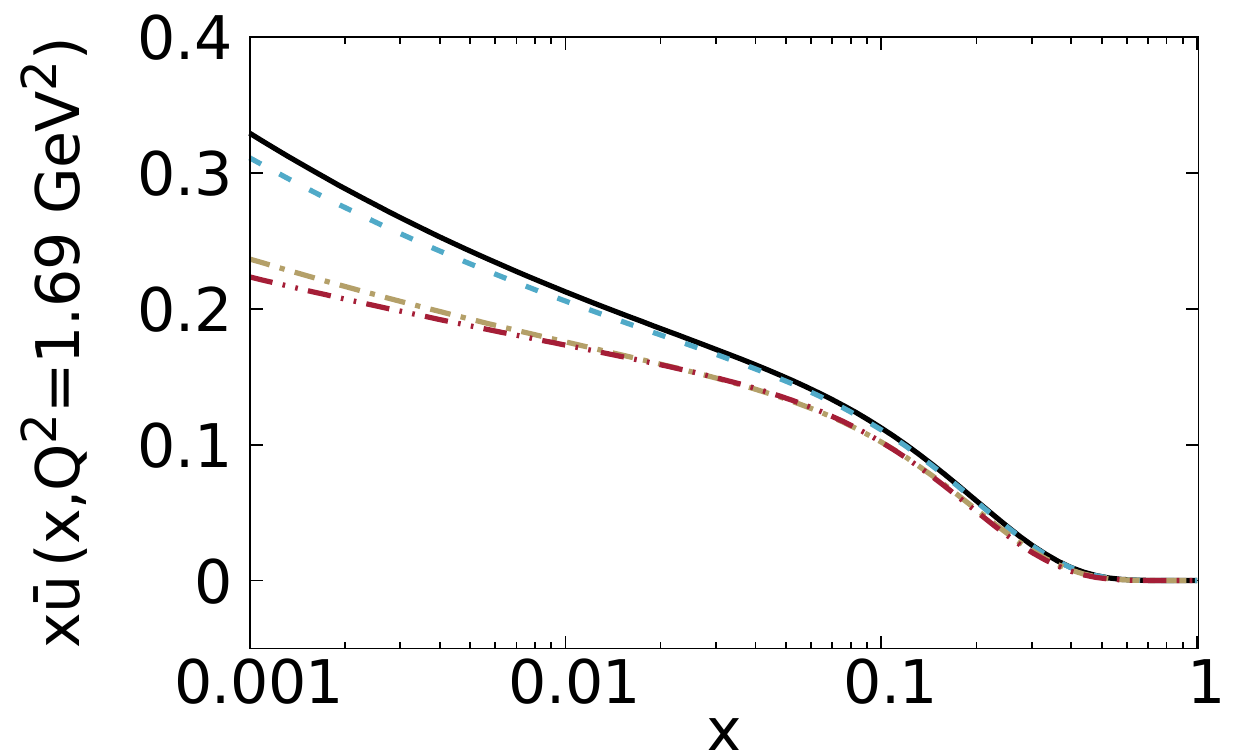}} 
          \subfigure{                           
              \includegraphics[width=0.237\textwidth]{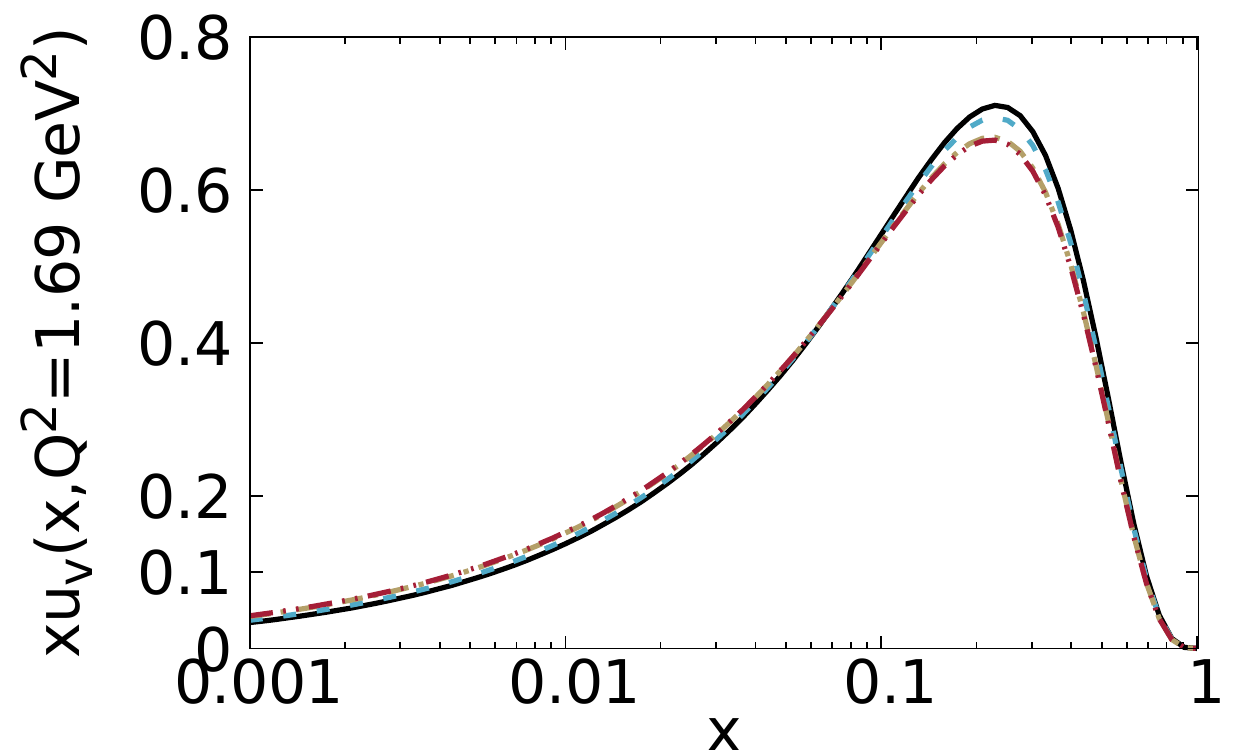}} 
          \subfigure{                                
              \includegraphics[width=0.237\textwidth]{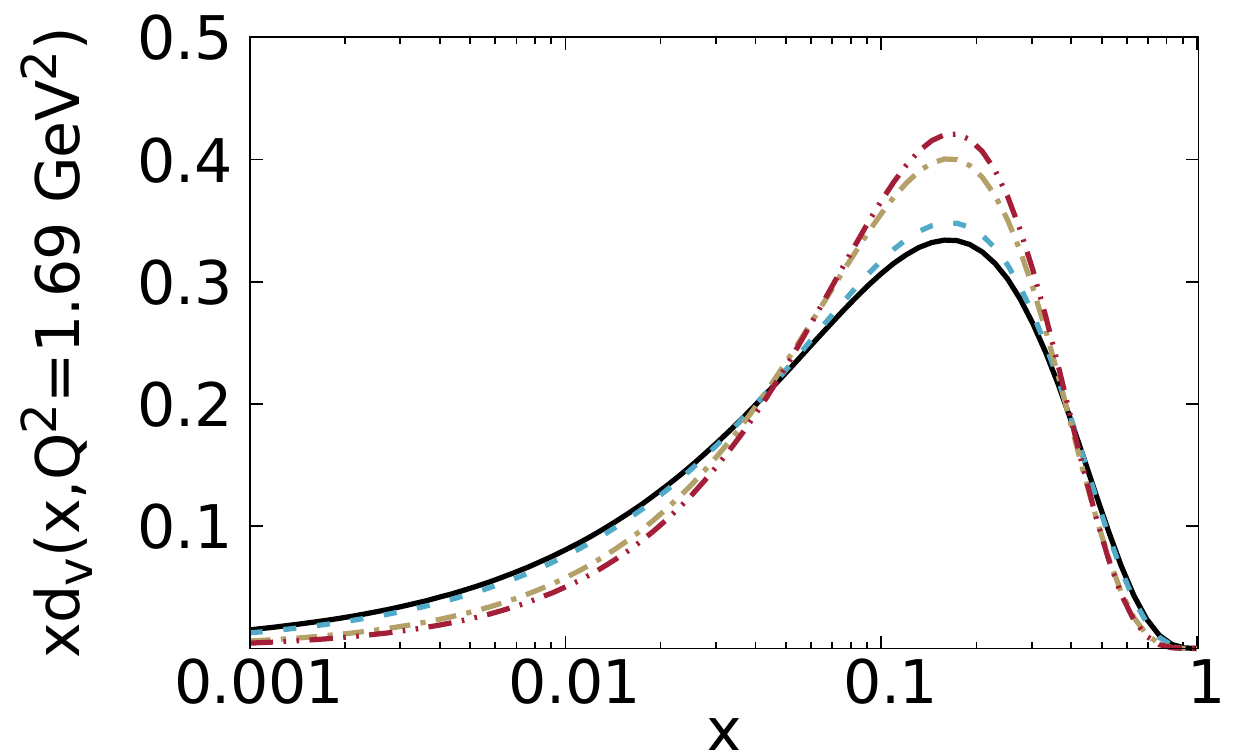}} 
          \subfigure{              
              \includegraphics[width=0.237\textwidth]{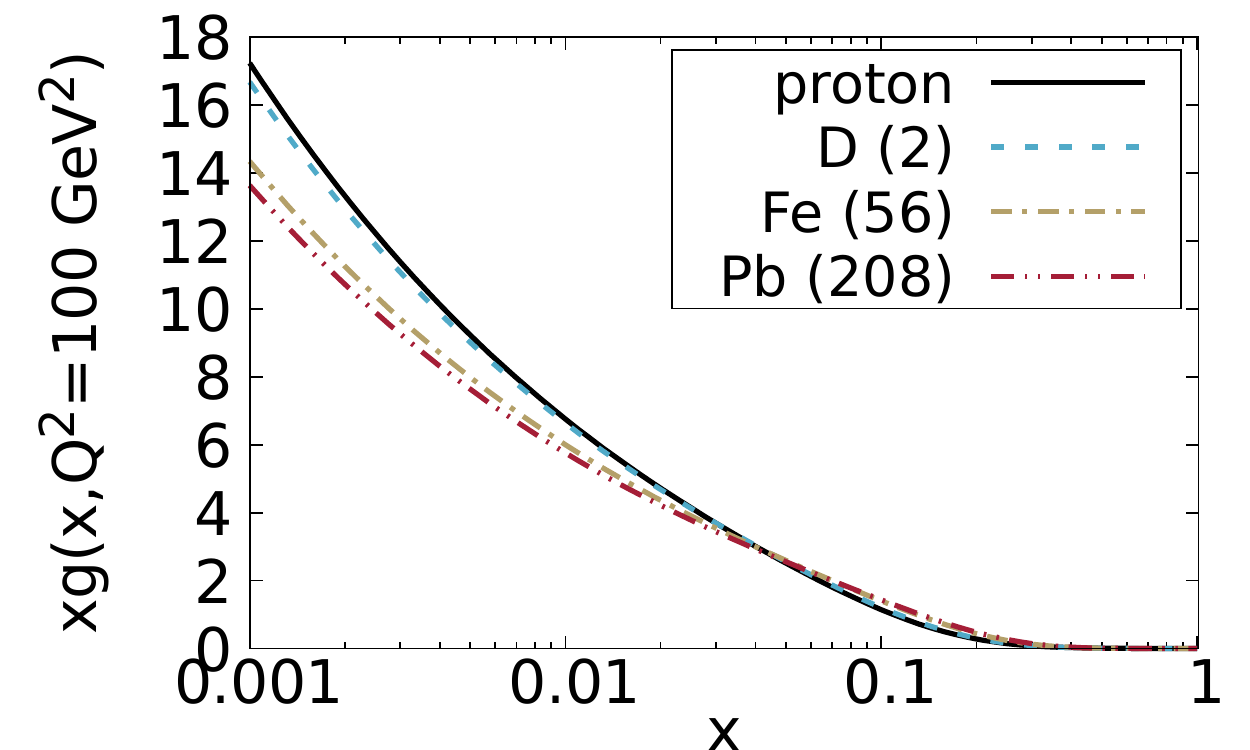}} 
          \subfigure{              
              \includegraphics[width=0.237\textwidth]{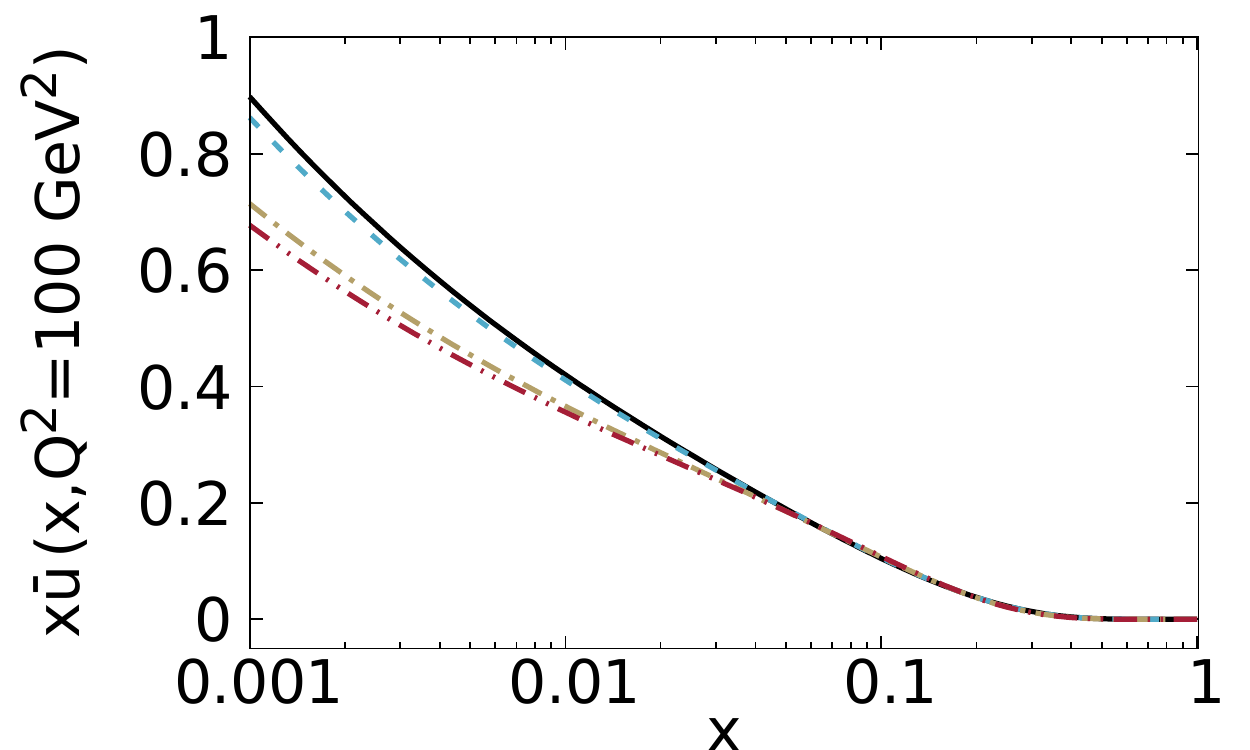}} 
          \subfigure{                        
              \includegraphics[width=0.237\textwidth]{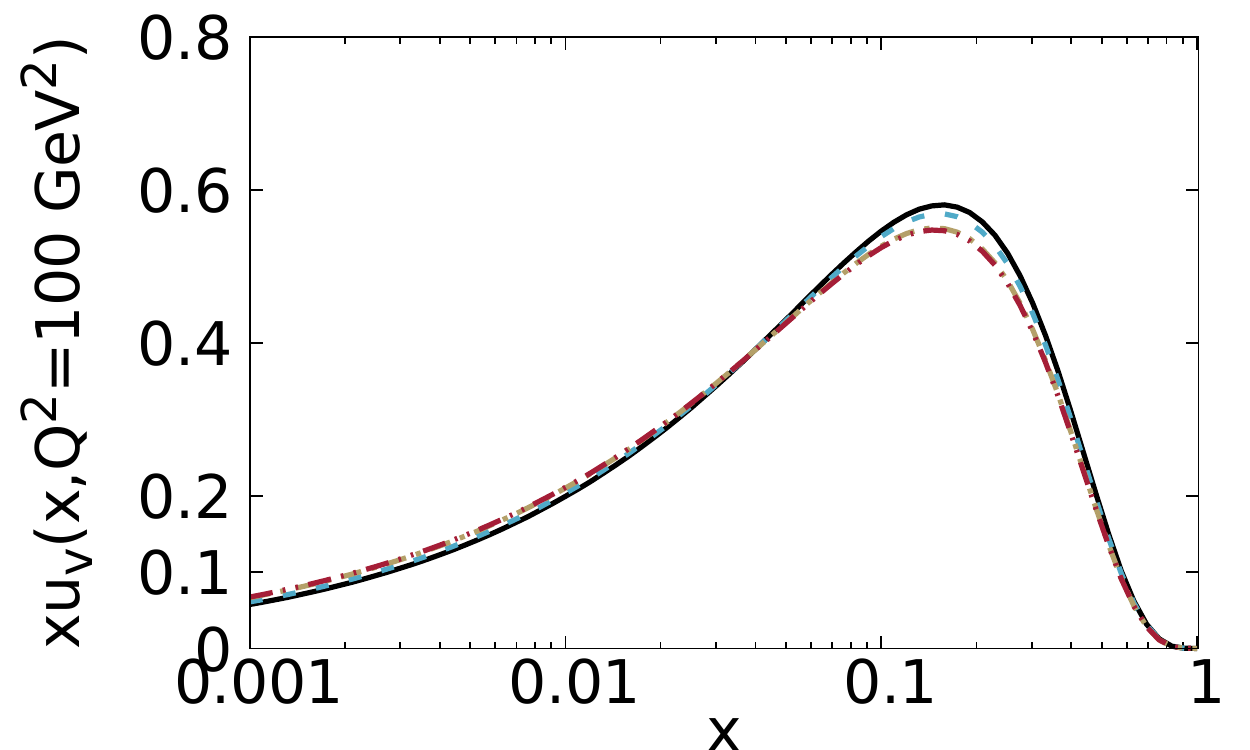}} 
          \subfigure{                                   
              \includegraphics[width=0.237\textwidth]{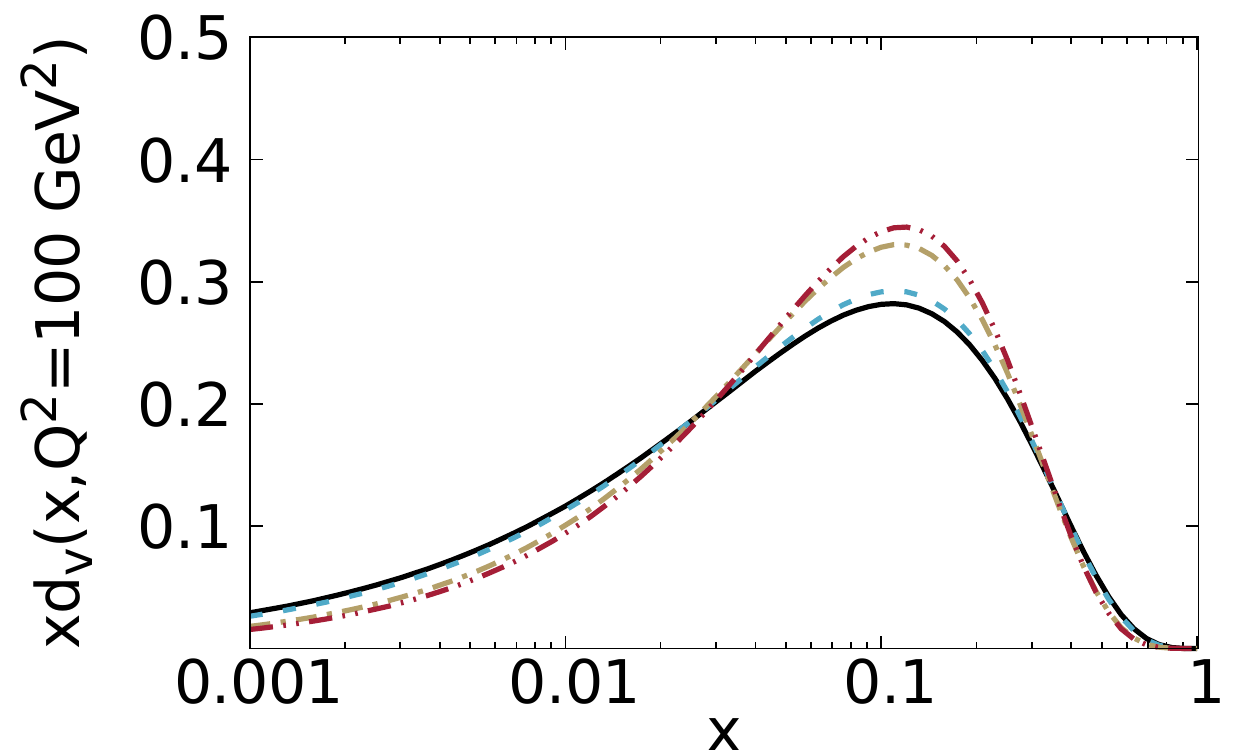}} 
          \end{center} 
    \caption{Nuclear parton distributions functions TUJU19 in different nuclei with the mass number $A$ at NLO, shown at the initial scale $Q_0^2=1.69\,\mathrm{GeV}^2$ and at $Q^2=100\,\mathrm{GeV}^2$ after DGLAP evolution.}
\label{fig-nPDF-A-NLO}    
    \end{figure*}

\begin{figure*}[tb!]
     \begin{center}
          \subfigure{
              \includegraphics[width=0.237\textwidth]{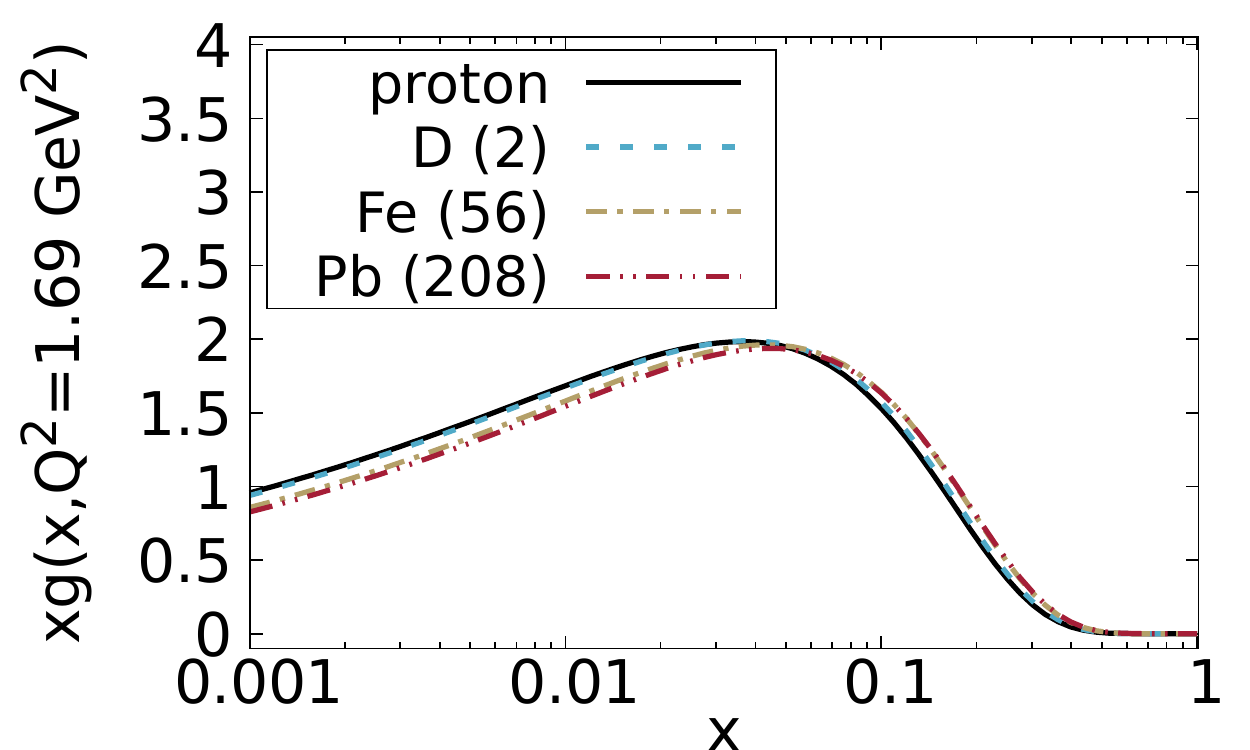}} 
          \subfigure{    
              \includegraphics[width=0.237\textwidth]{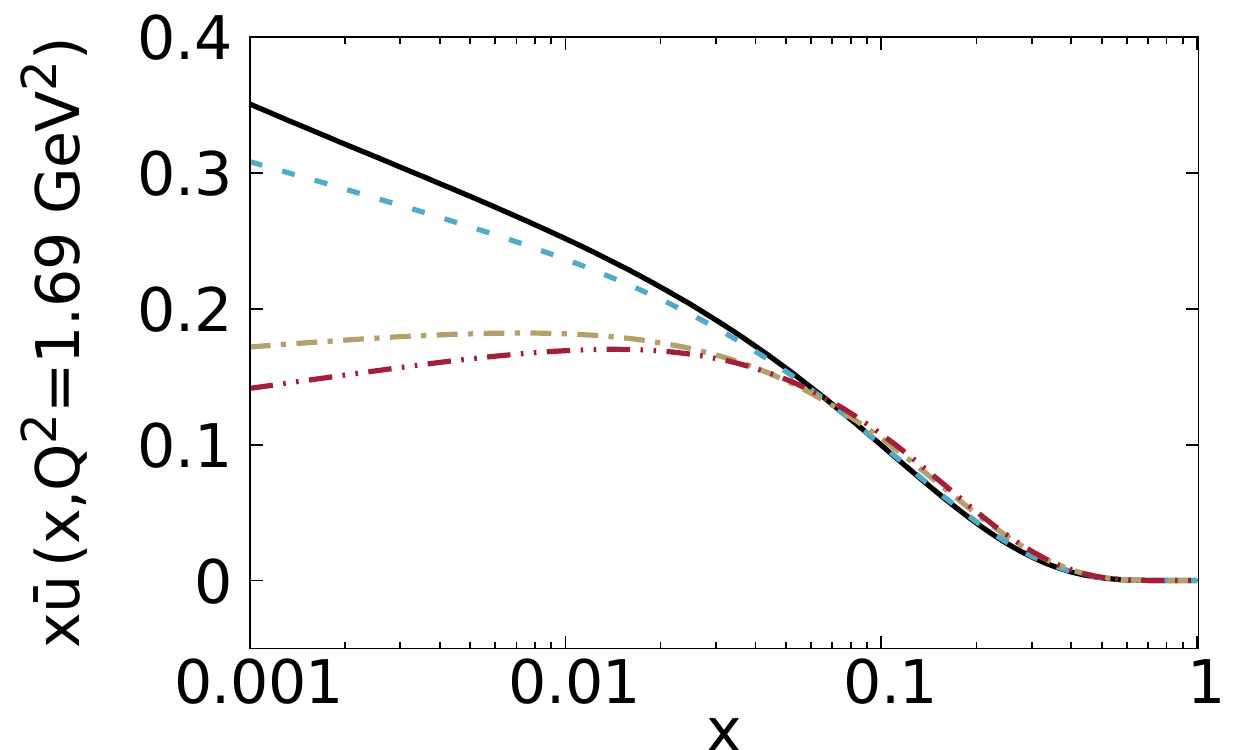}} 
          \subfigure{                           
              \includegraphics[width=0.237\textwidth]{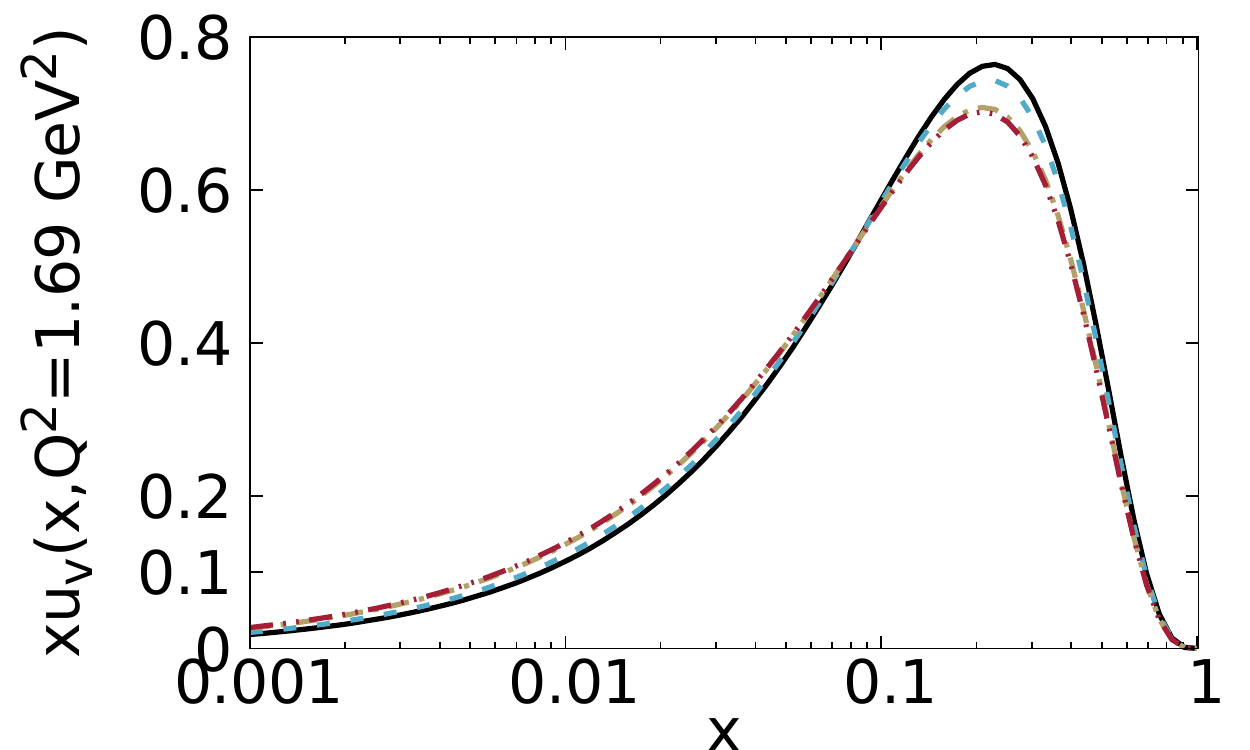}} 
          \subfigure{                                
              \includegraphics[width=0.237\textwidth]{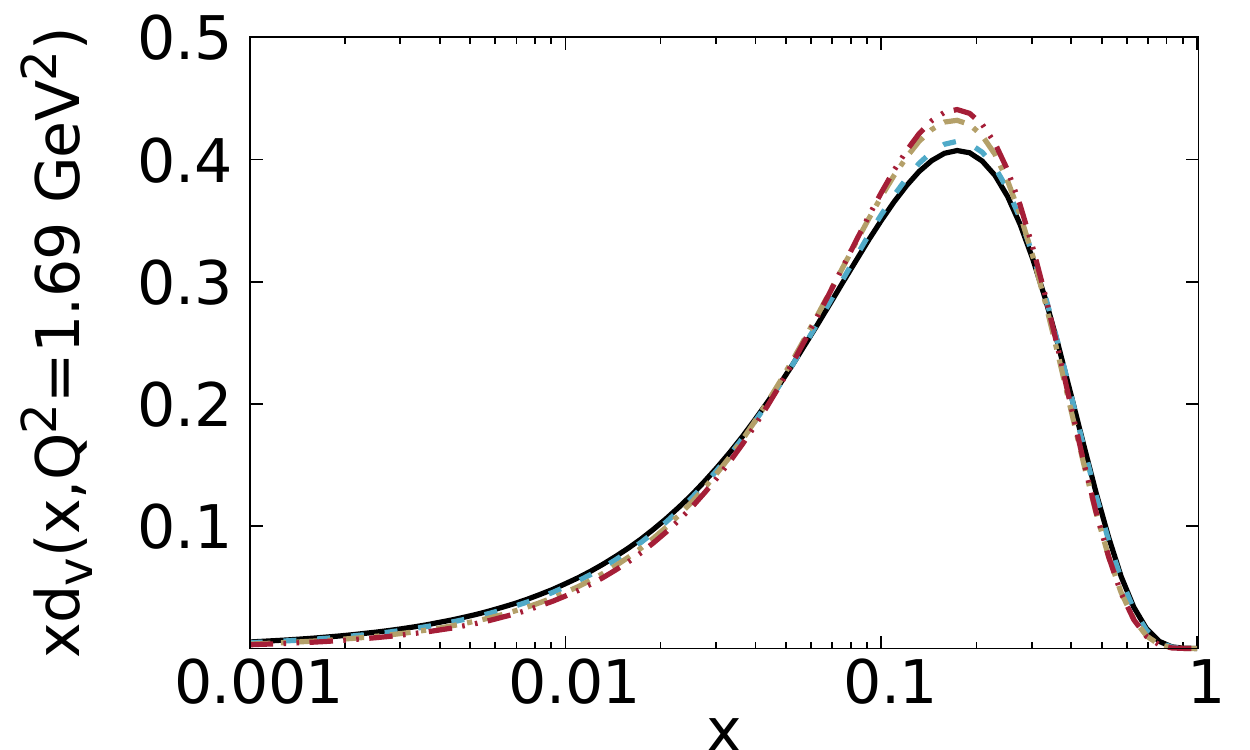}} 
          \subfigure{              
              \includegraphics[width=0.237\textwidth]{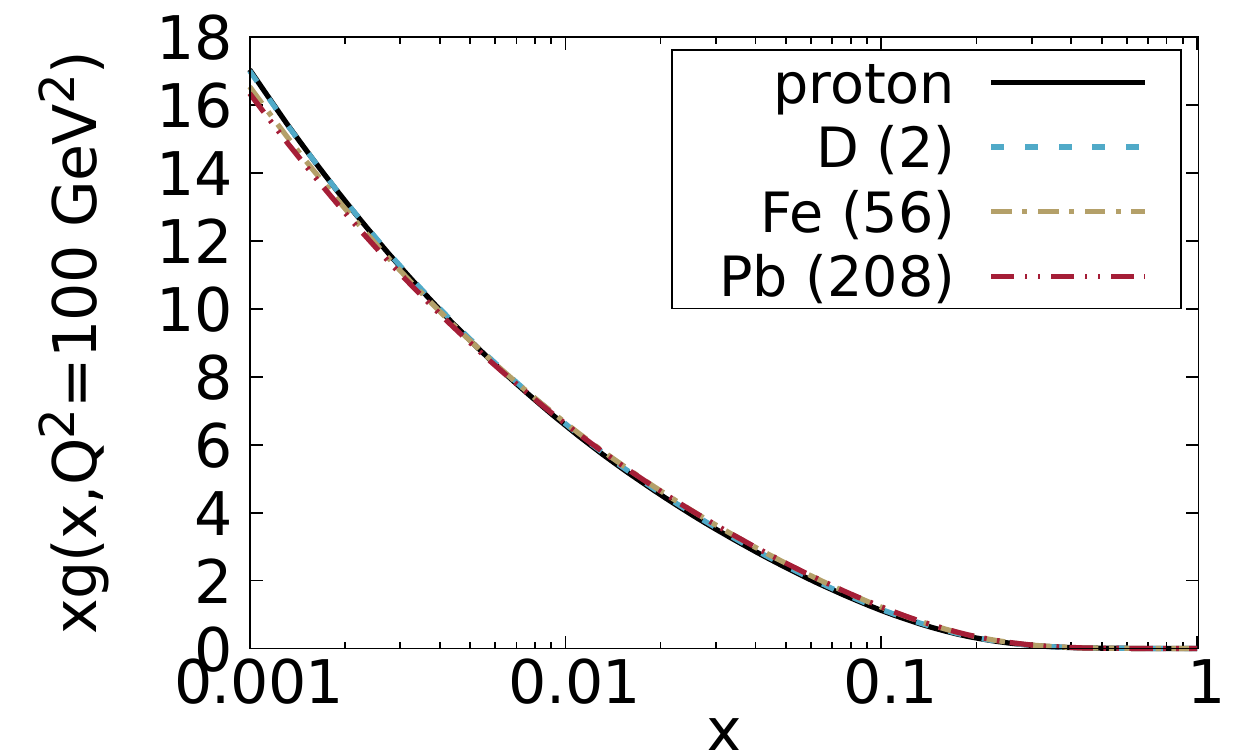}} 
          \subfigure{              
              \includegraphics[width=0.237\textwidth]{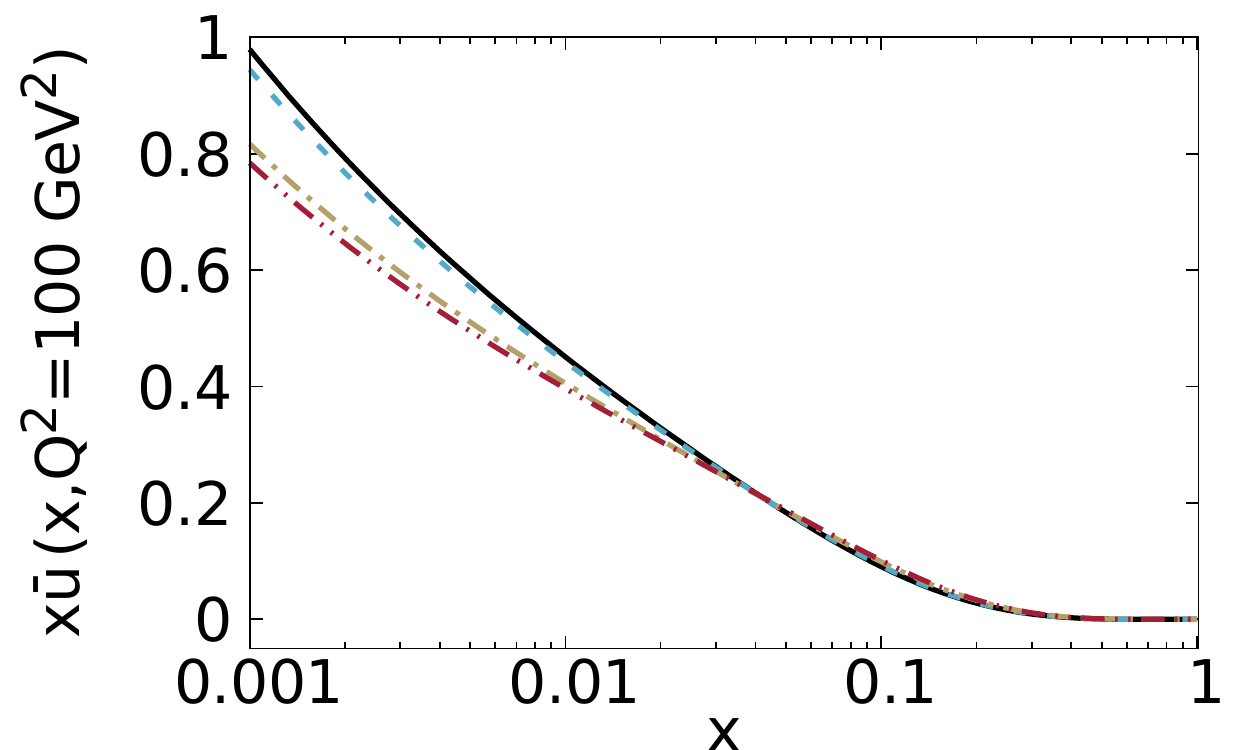}} 
          \subfigure{                        
              \includegraphics[width=0.237\textwidth]{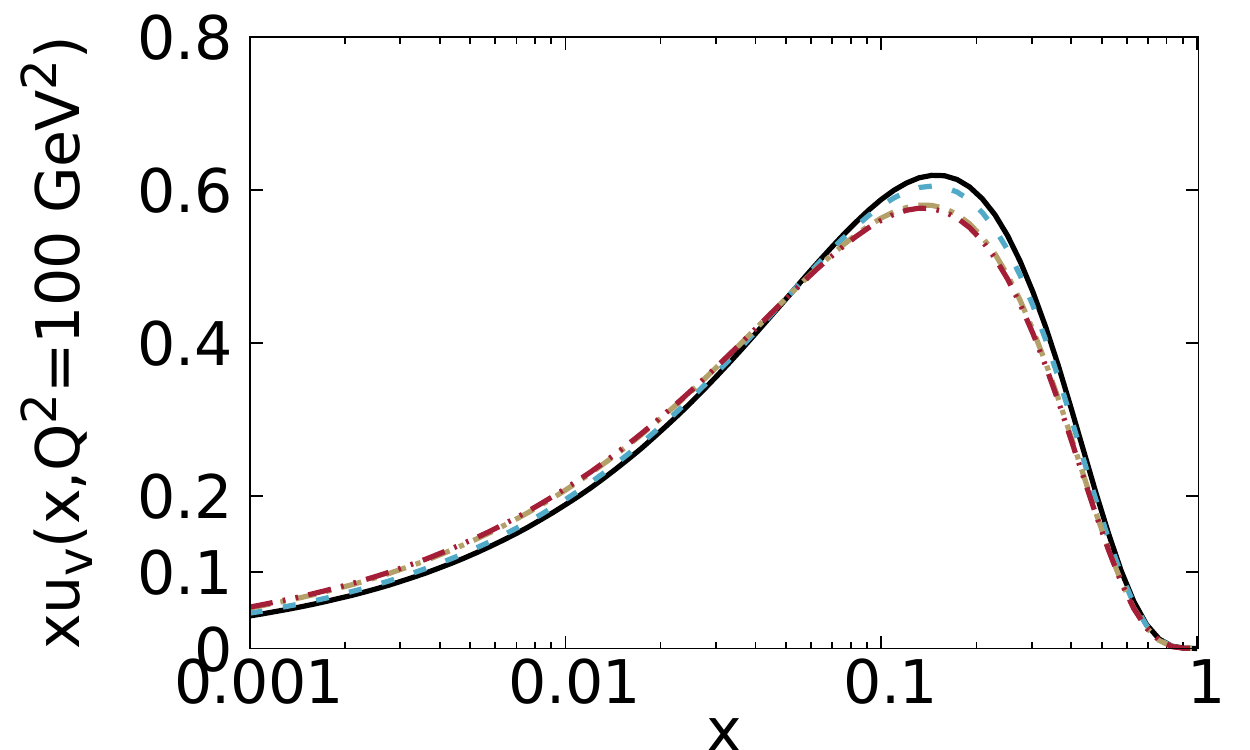}} 
          \subfigure{                                   
              \includegraphics[width=0.237\textwidth]{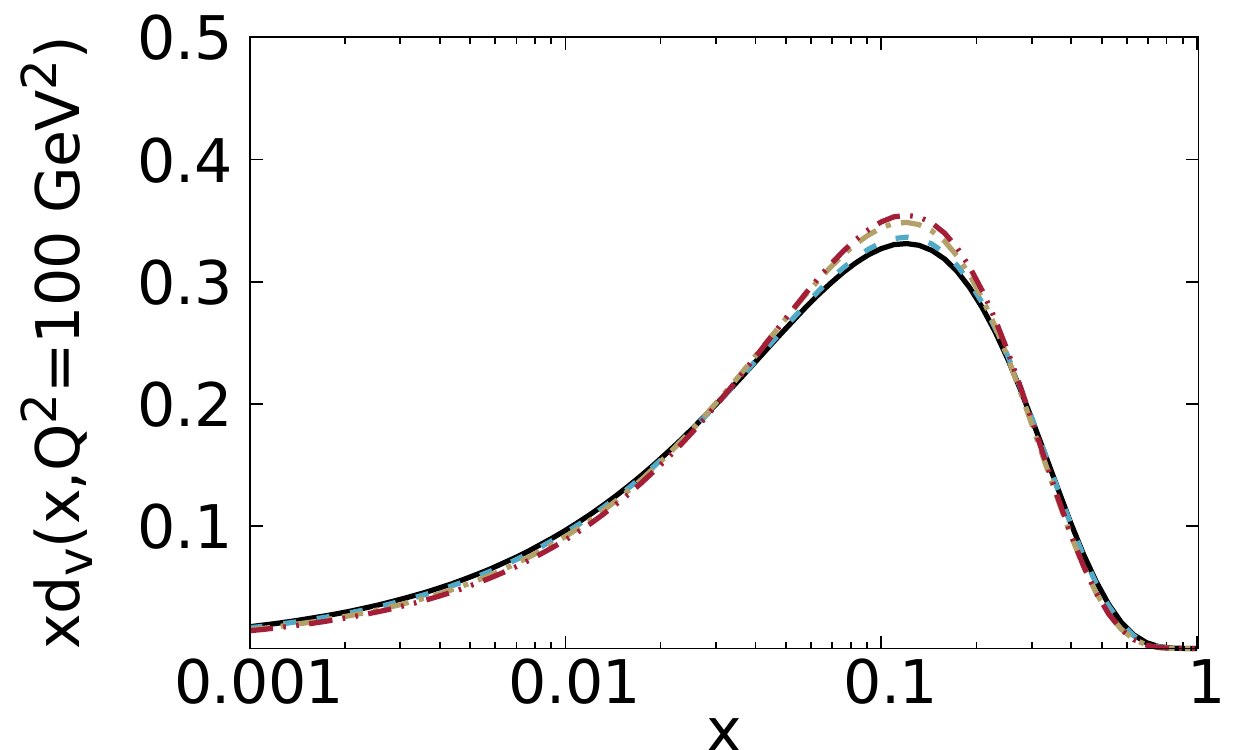}} 
          \end{center} 
    \caption{Same as for figure \ref{fig-nPDF-A-NLO}, but at NNLO.}
\label{fig-nPDF-A-NNLO}    
    \end{figure*}

The resulting nuclear PDFs, referred to as TUJU19, are presented in figure \ref{fig-nPDF-A-NLO} at NLO and in figure \ref{fig-nPDF-A-NNLO} at NNLO for a few different nuclei, together with the fitted proton baseline PDFs. As the sea-quark nPDFs have been assumed to be flavour independent, i.e. $s=\bar{s}=\bar{u}=\bar{d}$, the $x\bar{u}\left(x,\,Q^2\right)$ distribution represents all sea quarks in the plots. Many earlier analyses have assumed that the nuclear modifications for the deuteron are negligible and constructed its PDFs from the free proton ones using isospin symmetry. In this work we, instead, treat the deuteron as a nucleus in the fitting procedure. Small deviations from the proton PDFs are found for the proton in a deuteron, as shown in figures \ref{fig-nPDF-A-NLO} and \ref{fig-nPDF-A-NNLO}. The deviation from the proton PDFs becomes larger with increasing $A$, and significant effects are found in heavy nuclei such as iron and lead. The optimal parameters according to the chosen parameterizations in eqs.~(\ref{pdf-parameterization}) and (\ref{coeff-A}) are listed in appendix \ref{app-pdf-params}.

\begin{figure*}[bt!]
     \begin{center}
     \subfigure{\includegraphics[width=0.237\textwidth]{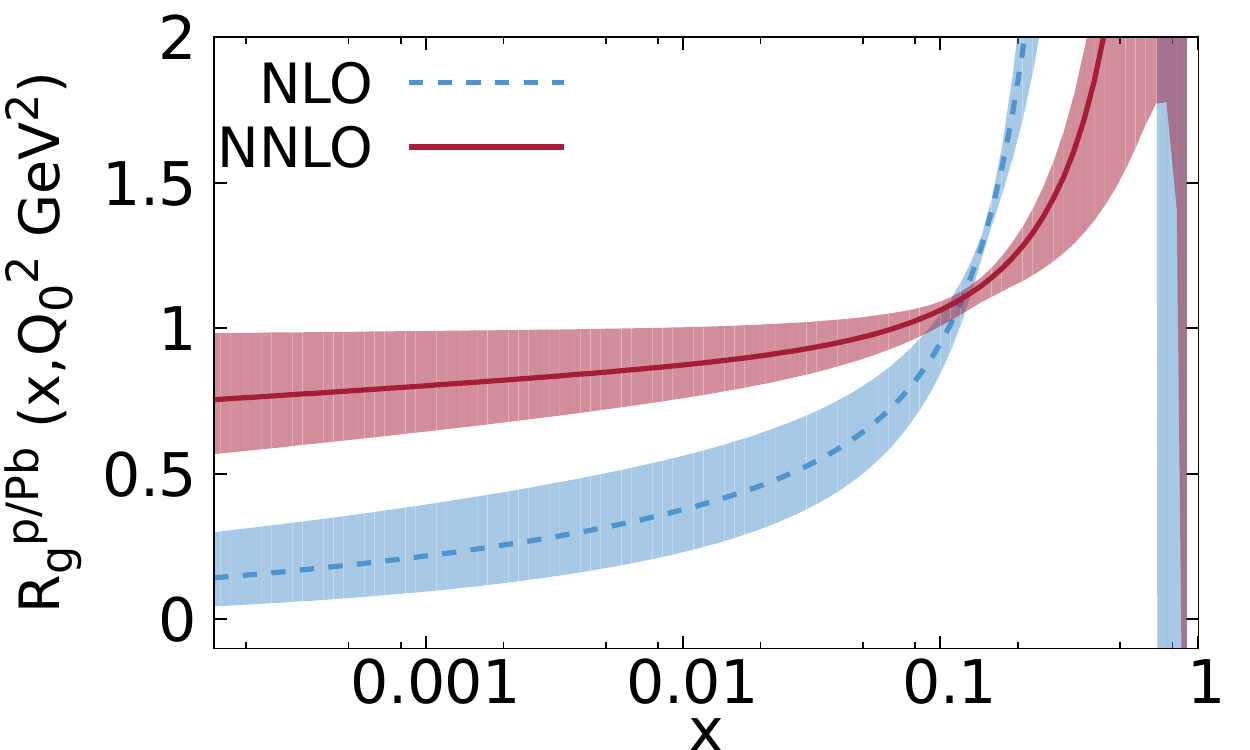}} 
     \subfigure{\includegraphics[width=0.237\textwidth]{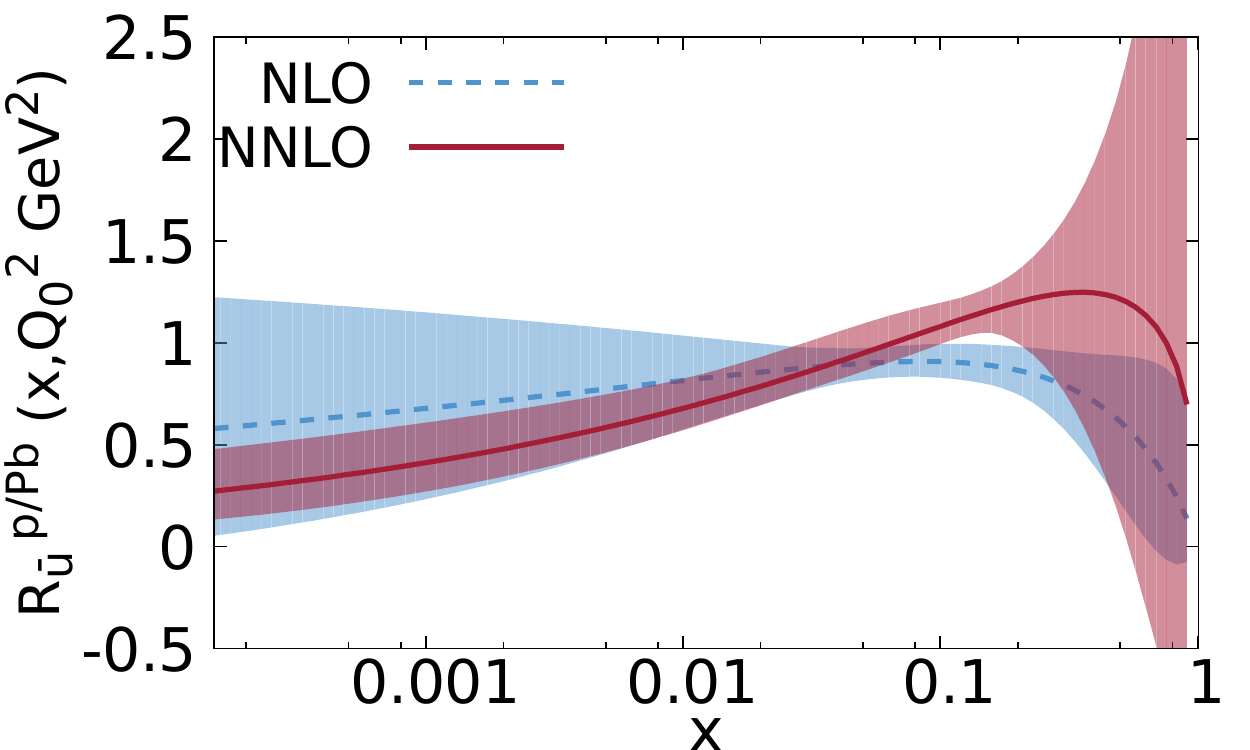}} 
     \subfigure{\includegraphics[width=0.237\textwidth]{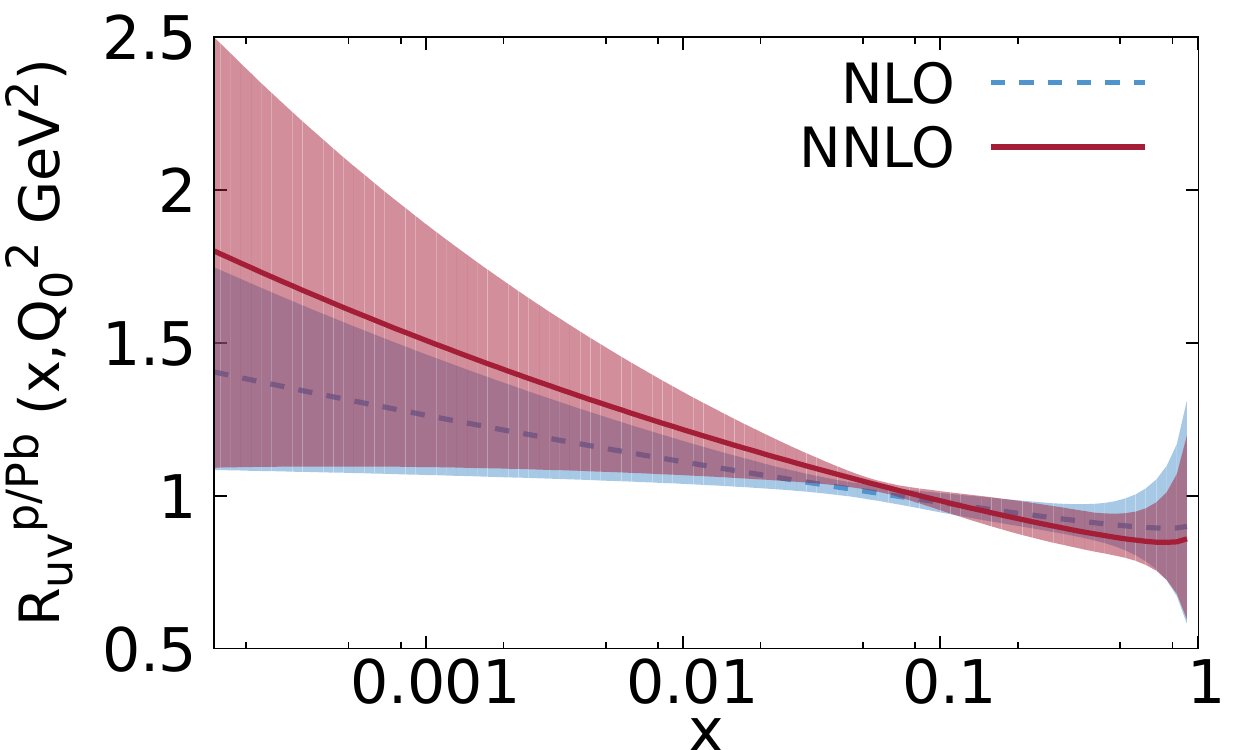}} 
     \subfigure{\includegraphics[width=0.237\textwidth]{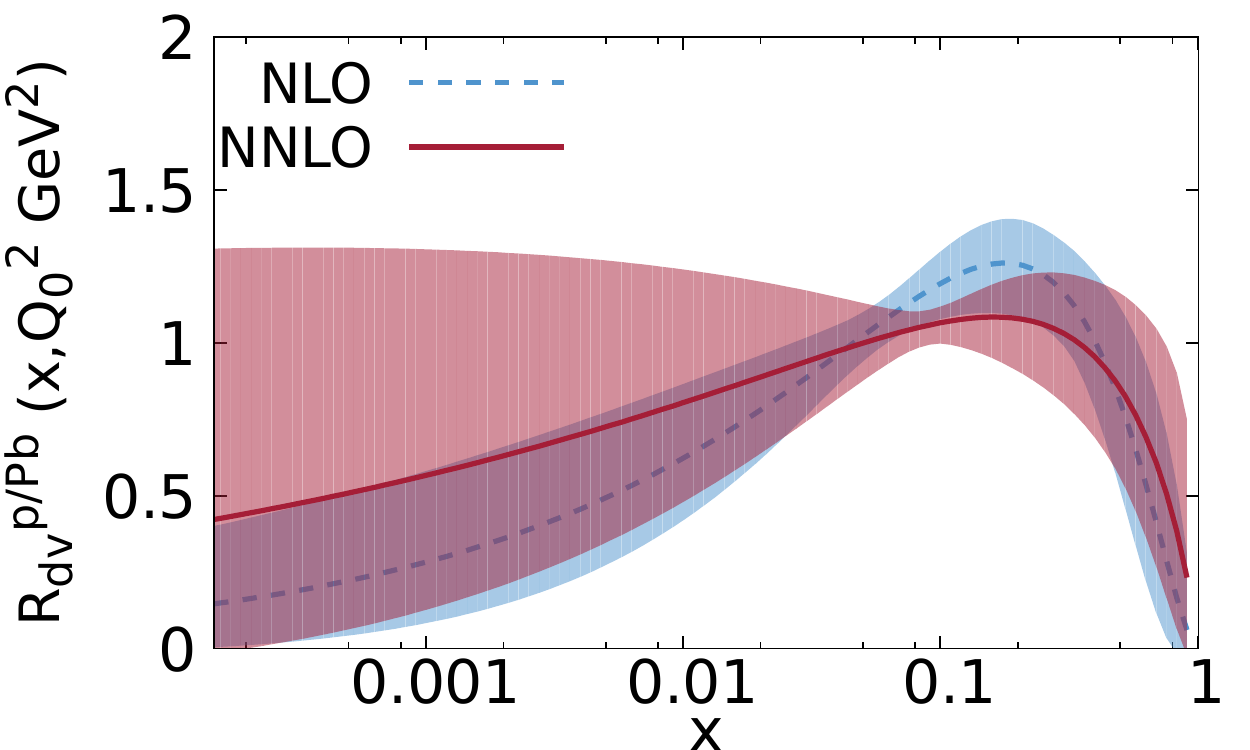}} 
     \subfigure{\includegraphics[width=0.237\textwidth]{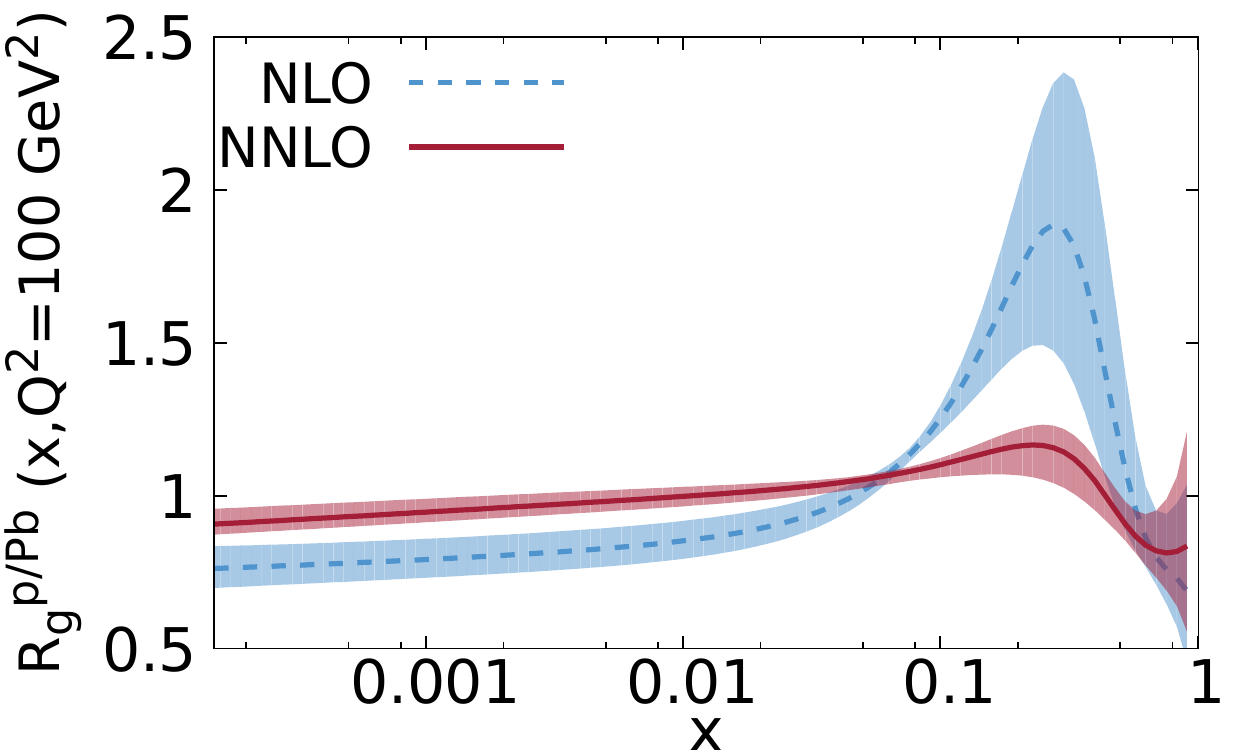}} 
     \subfigure{\includegraphics[width=0.237\textwidth]{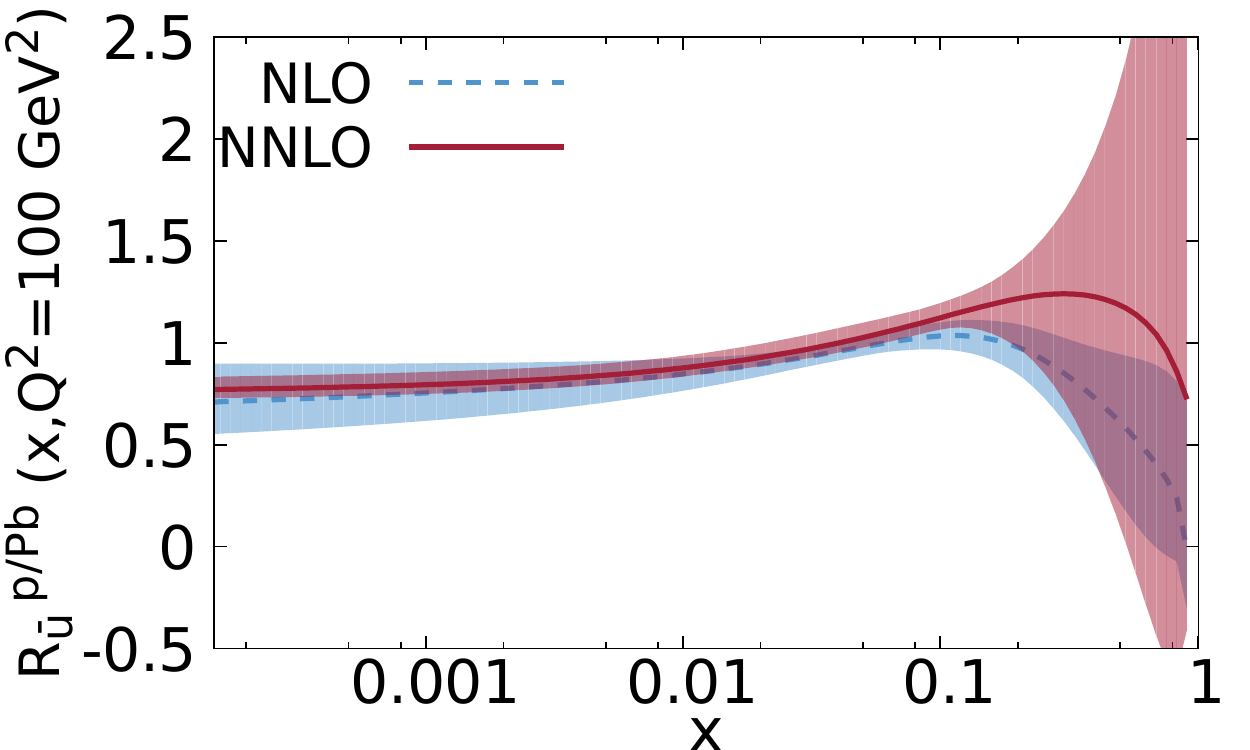}} 
     \subfigure{\includegraphics[width=0.237\textwidth]{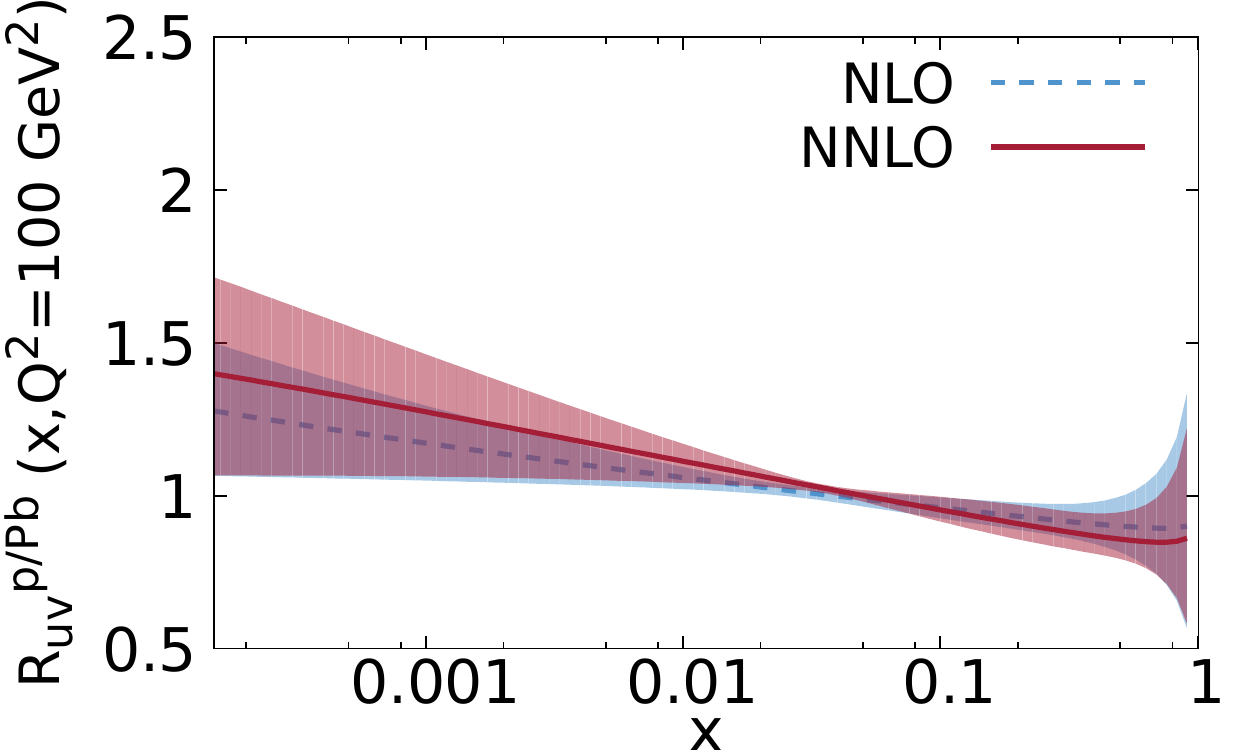}} 
     \subfigure{\includegraphics[width=0.237\textwidth]{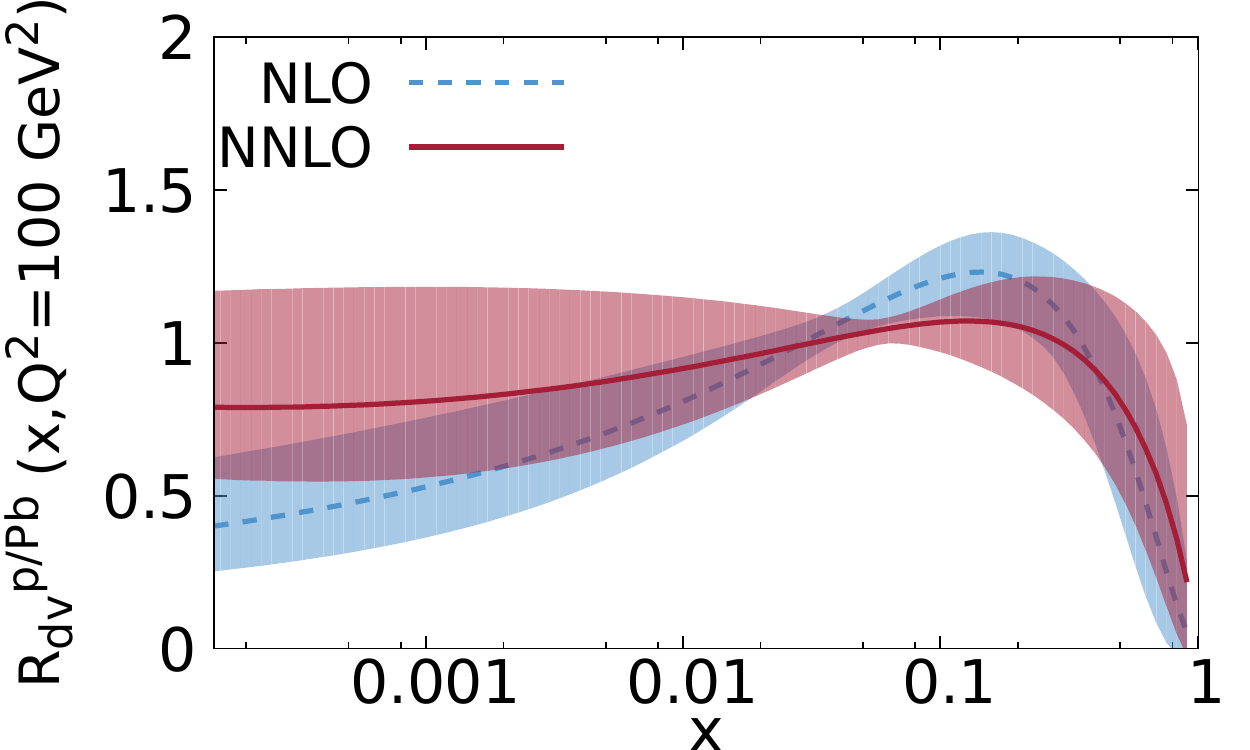}} 
          \end{center} 
    \caption{Ratios $R_i^{p/Pb}$ of parton distribution functions per parton flavour $i$ in a proton bound in lead compared to a free proton $p$. The obtained ratios are shown at NLO and NNLO, both at the initial scale $Q_0^2=1.69\,\mathrm{GeV}^2$ and at a higher scale $Q^2=100\,\mathrm{GeV}^2$.}
\label{fig-nPDF-R-NLO-vs-NNLO}    
    \end{figure*}    
   
The nuclear modifications of the PDFs for the lead nucleus, defined as 
\begin{equation}
R_i^{\mathrm{p/Pb}}= \frac{f_i^{\mathrm{p/Pb}}\left(x,\,Q^2\right)}{f_i^{\mathrm{p}}\left(x,\,Q^2\right)}\,,
\label{eq:Ri}
\end{equation}
where $f_i^{\mathrm{p/Pb}}(x,\,Q^2)$ and $f_i^{\mathrm{p}}(x,\,Q^2)$ are the PDFs for the bound and the free proton, respectively, are shown in figure~\ref{fig-nPDF-R-NLO-vs-NNLO}. The NLO and NNLO modifications are compared at the initial scale of the analysis ($Q^2=1.69~\text{GeV}^2$) and at a higher scale ($Q^2=100~\text{GeV}^2$) after DGLAP evolution. In both cases, the ratio of gluon PDFs shows some low-$x$ shadowing and a rapid rise with increasing $x$. This behaviour is similar to what was observed in the HKN07 analysis \cite{Hirai:2007sx}, but in our case the enhancements are moderated at higher scales and a recognizable antishadowing peak develops around $x\sim 0.3$. For the sea quarks the typical nuclear modifications -- shadowing, antishadowing and EMC suppression -- are visible already at the initial scale. However, especially the magnitude of the small-$x$ shadowing differs at different orders, NNLO favoring a stronger effect. Since for gluons the behaviour is opposite we conclude that these differences arise due to the fact that the sea-quark and gluon evolution are coupled and the applied DIS data is not sensitive enough to fully separate the contributions. At higher scales the sea quark modifications are in better agreement though some difference still persists at large $x$. The valence-quark parameters were allowed to be flavour-dependent also for nPDFs and the resulting nuclear effects indeed become rather different for bound $u_v$ and $d_v$ distributions. For $d_v$ again the typical features including shadowing and antishadowing are well visible but for $u_v$ we find that some amount of low-$x$ enhancement is preferred. We note that a similar behaviour was observed in the nCTEQ15 analysis \cite{Kovarik:2015cma}, although no neutrino DIS data were included there that would provide additional flavour sensitivity especially for the valence sector. One should keep in mind that the full nPDF for an average nucleon will be the sum of those for protons and neutrons, so the opposite behaviour will cancel out to a certain extent.

The uncertainty bands for the nPDFs provided in this work have been generated with $\Delta \Chi^2=50$ as described in subsection \ref{sec-uncertainties}. The resulting uncertainty bands do, however, depend also on the flexibility of the applied parameterization. Due to the limited sensitivity of the applied data to the gluon and sea quark nPDFs, we had to limit the number of $A$-dependent parameters in order to achieve numerical convergence of the fits. Therefore the provided uncertainty bands for the gluon distribution likely underestimate the true uncertainty to some extent, which should be kept in mind when comparing to previous works. In the future, by adding more data providing further constraints one could consider admitting more parameters and therefore allowing for a larger flexibility of the parameterization.

\subsection{Comparison to data}

\begin{figure}[htb!]
     \begin{center}
          \subfigure{        
              \includegraphics[width=0.23\textwidth]{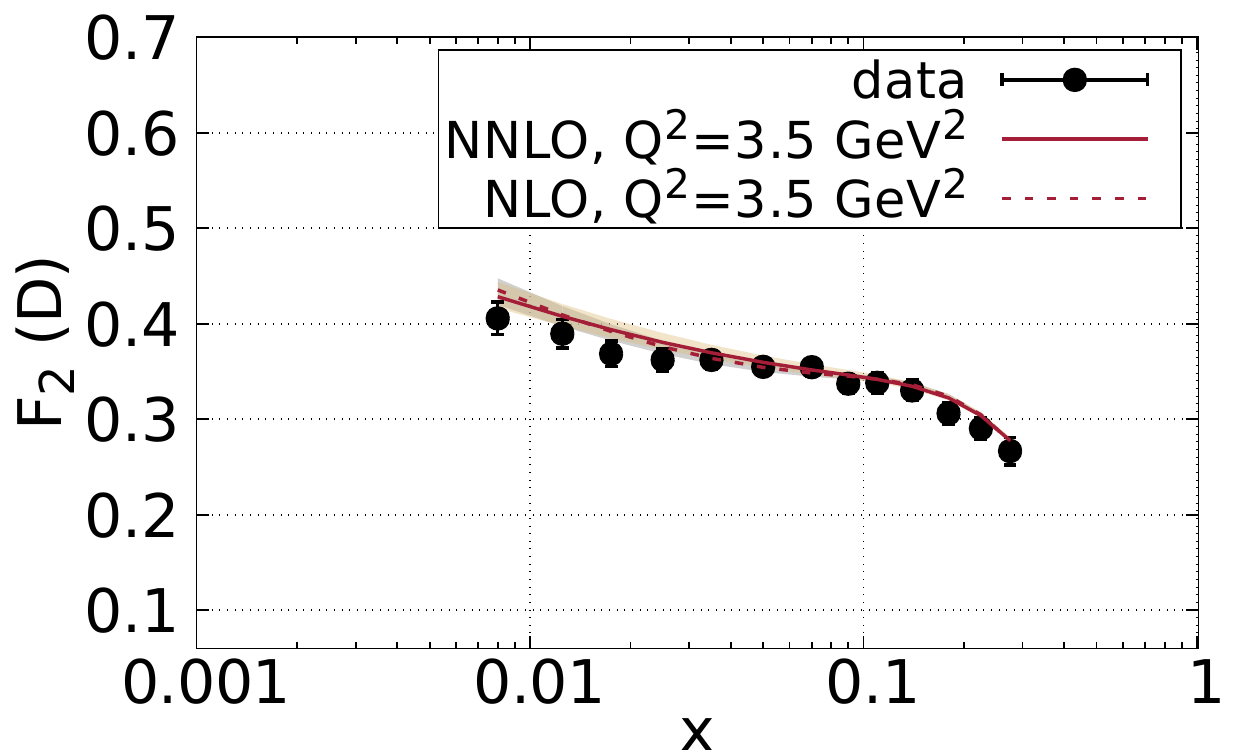}} 
         \subfigure{        
              \includegraphics[width=0.23\textwidth]{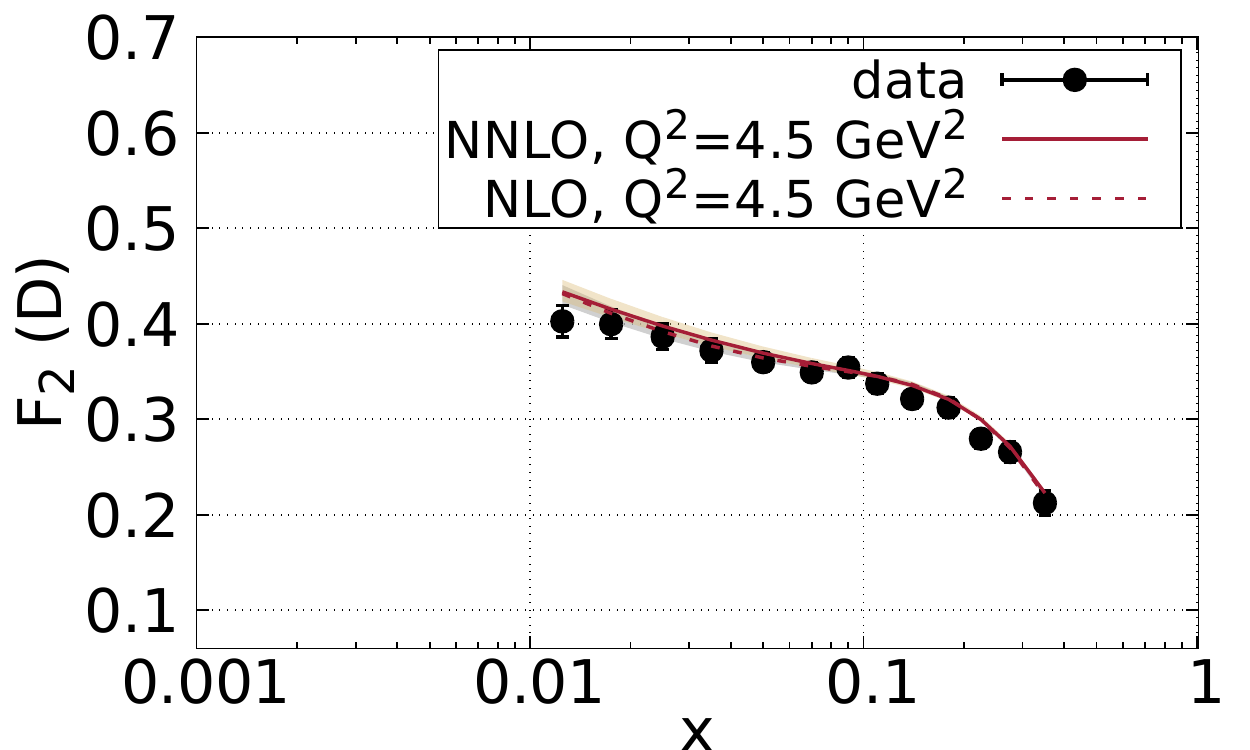}} 
         \subfigure{        
              \includegraphics[width=0.23\textwidth]{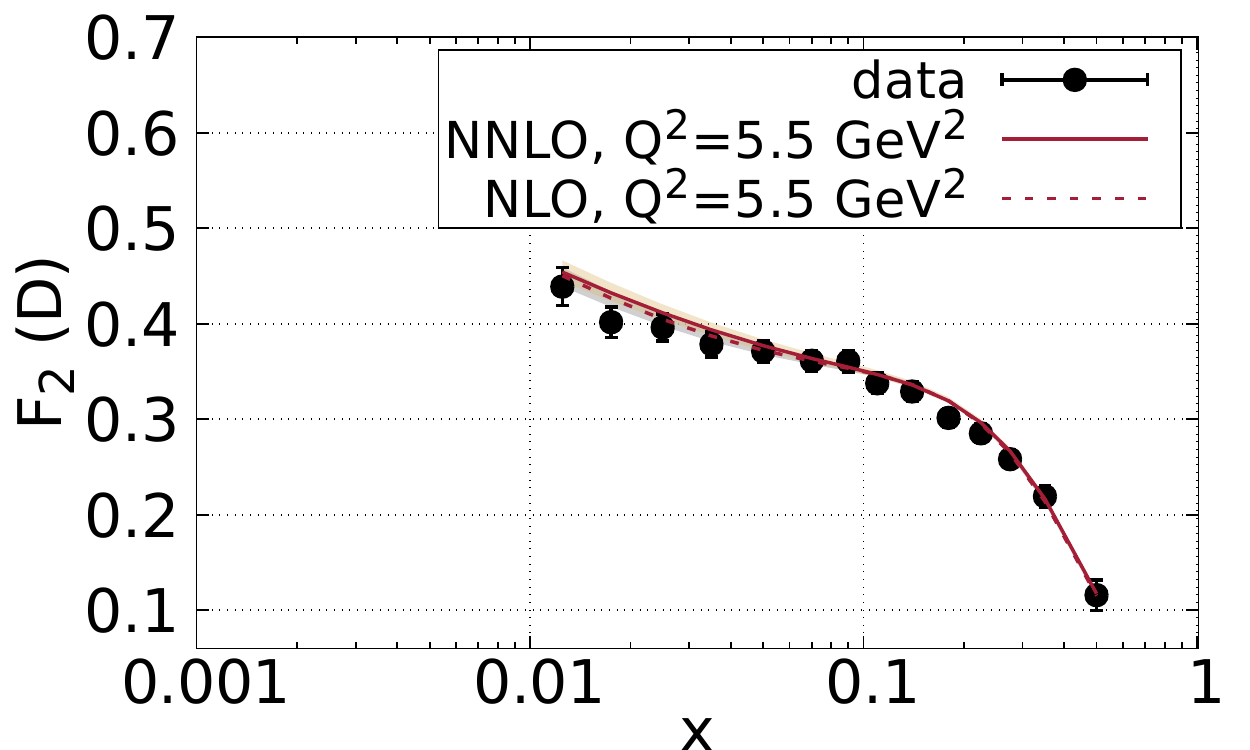}} 
         \subfigure{        
              \includegraphics[width=0.23\textwidth]{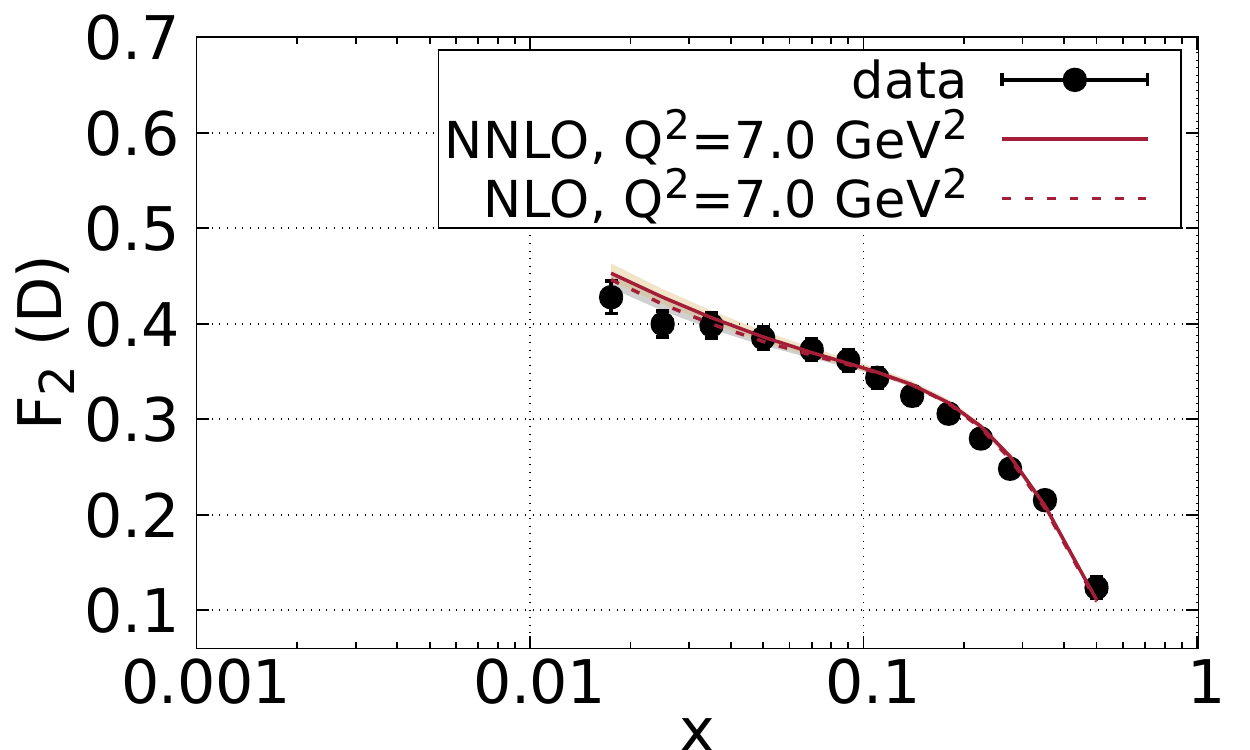}} 
         \subfigure{        
              \includegraphics[width=0.23\textwidth]{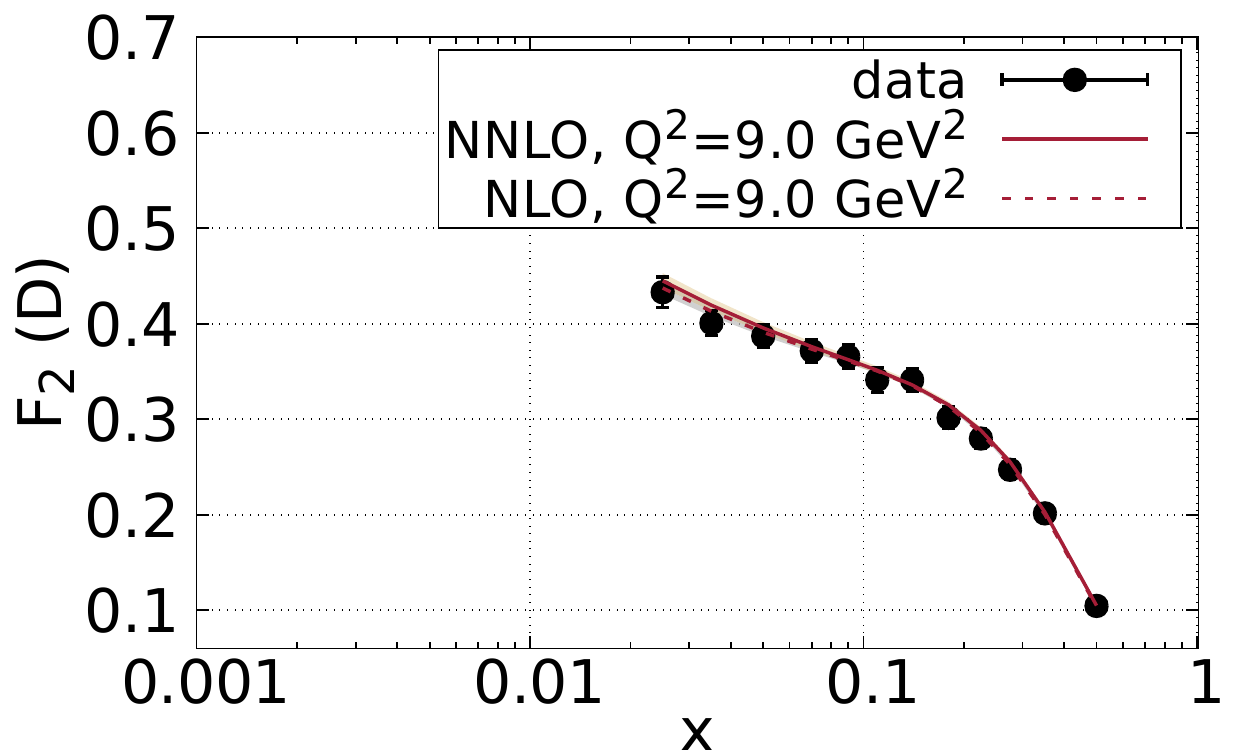}} 
         \subfigure{        
              \includegraphics[width=0.23\textwidth]{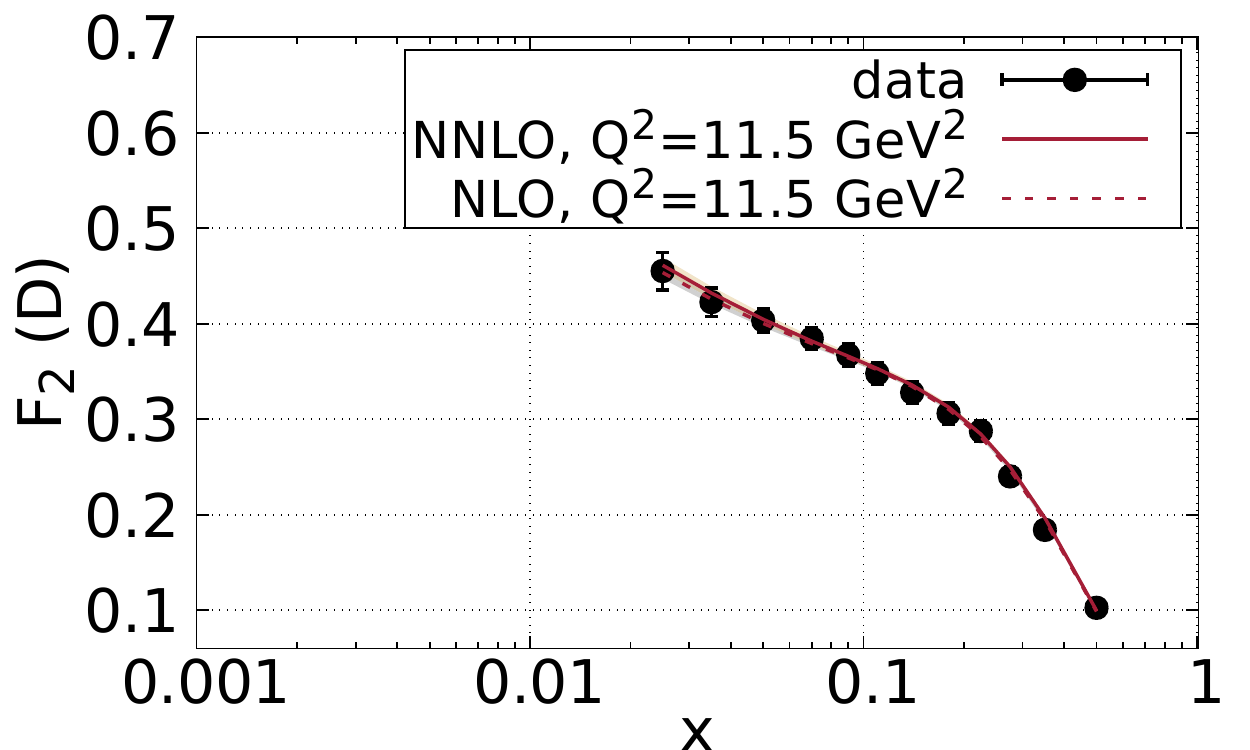}}                             
         \subfigure{        
              \includegraphics[width=0.23\textwidth]{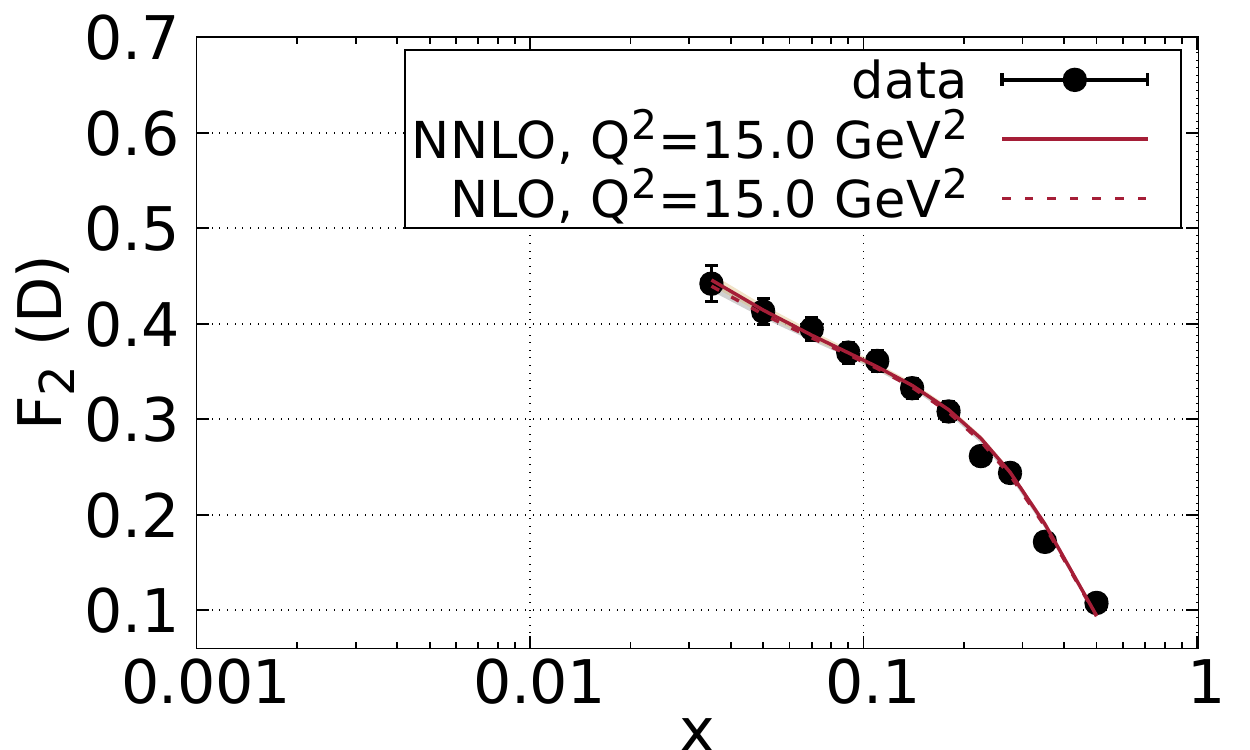}}                             
         \subfigure{        
              \includegraphics[width=0.23\textwidth]{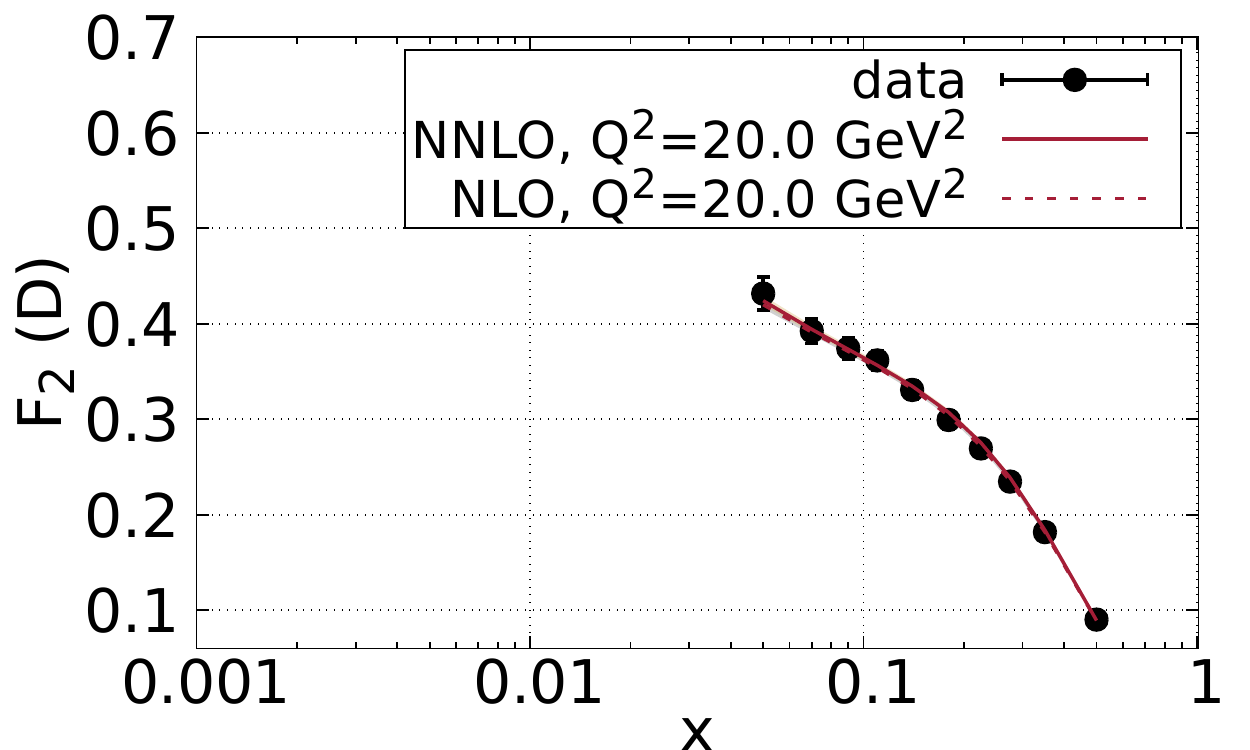}}                             
         \subfigure{        
              \includegraphics[width=0.23\textwidth]{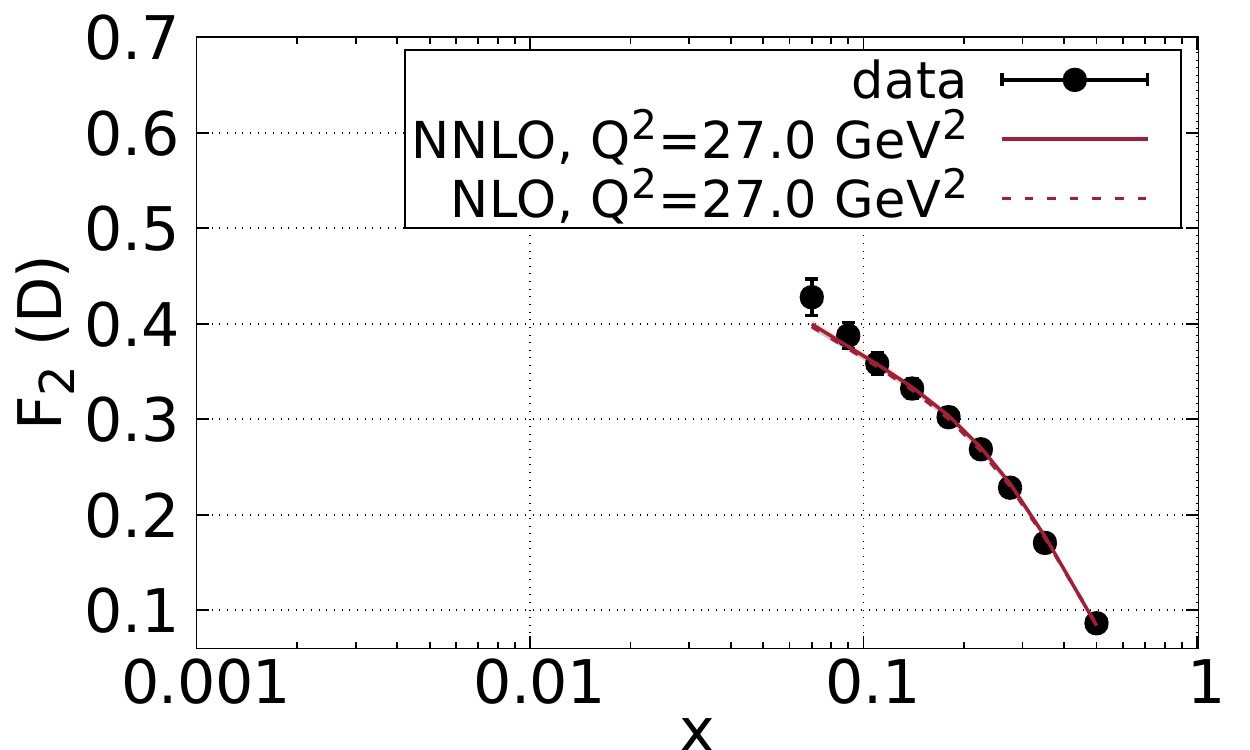}}                             
         \subfigure{        
              \includegraphics[width=0.23\textwidth]{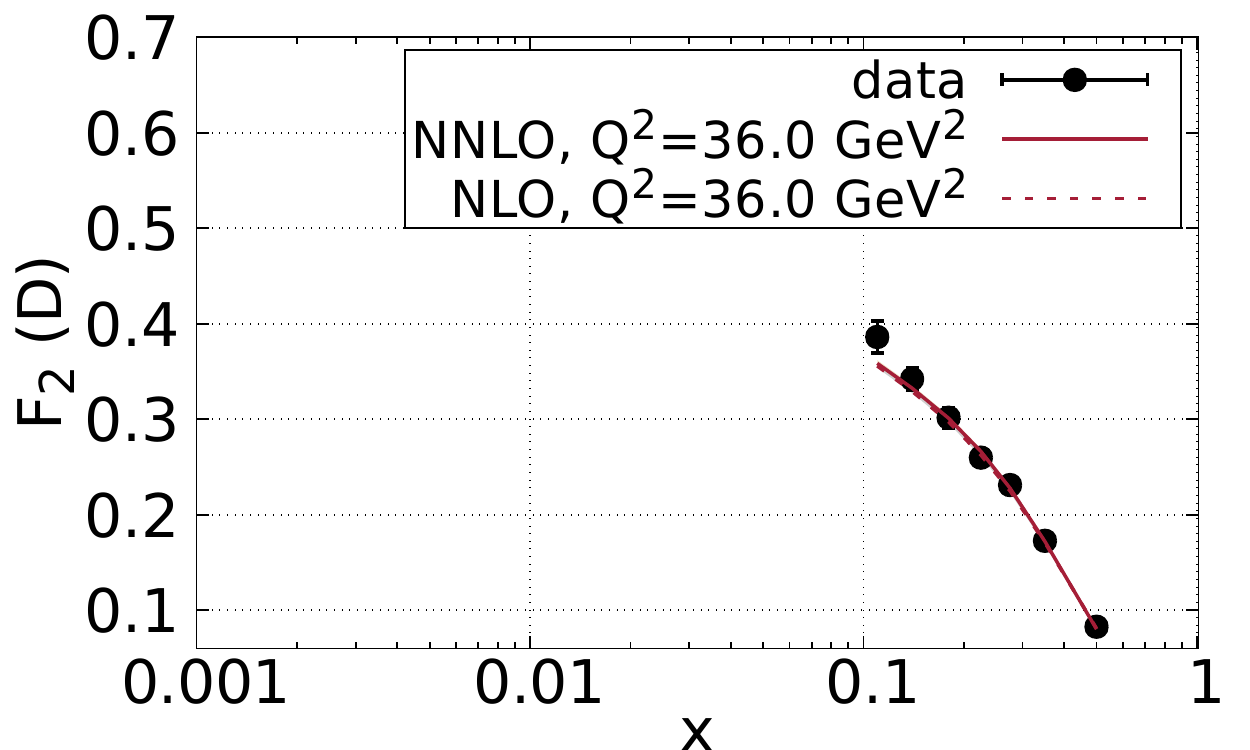}}                             
         \subfigure{        
              \includegraphics[width=0.23\textwidth]{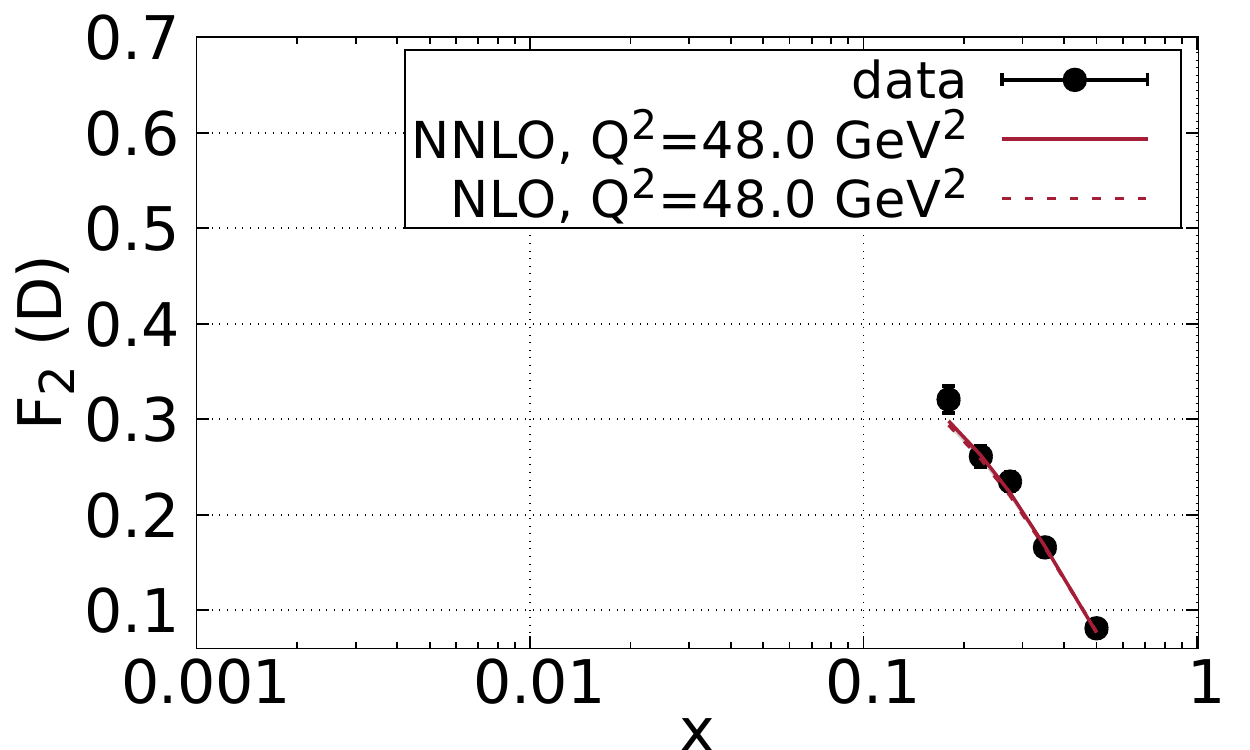}}                             
          \end{center} 
    \caption{Comparison to NMC $F_2(\mathrm{D})$ data at different values of $Q^2$ at NLO (dashed line, grey error bands) and NNLO (solid line, golden-coloured error bands).}
\label{figNMCd}    
    \end{figure}

\begin{figure*}[hbt]
\begin{center}
\includegraphics[width=\textwidth]{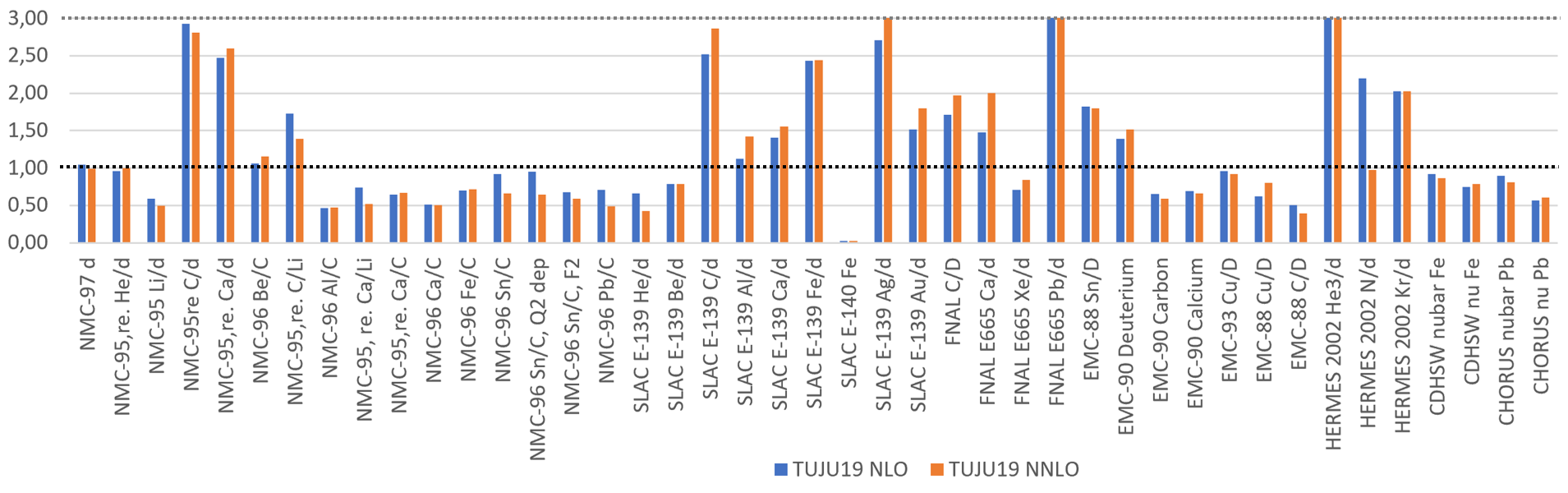}
\caption{Comparison of $\Chi^2$ values divided by the individual number of data points per data set $N_{\mathrm{dp}}$ at NLO and NNLO. The ``ideal'' value $\Chi^2/N_{\mathrm{dp}}\,=\,1.0$ is marked by the horizontal black dotted line. The bars in the diagram corresponding to $\Chi^2/N_{\mathrm{dp}}\,> \,3.0$ have been truncated for the purpose of a clearer representation, which is symbolised by the dashed light-grey line. }
\label{fig-Chi2}
\end{center}
\end{figure*}

\begin{figure*}[htb!]
     \begin{center}
          \subfigure{        
              \includegraphics[width=0.237\textwidth]{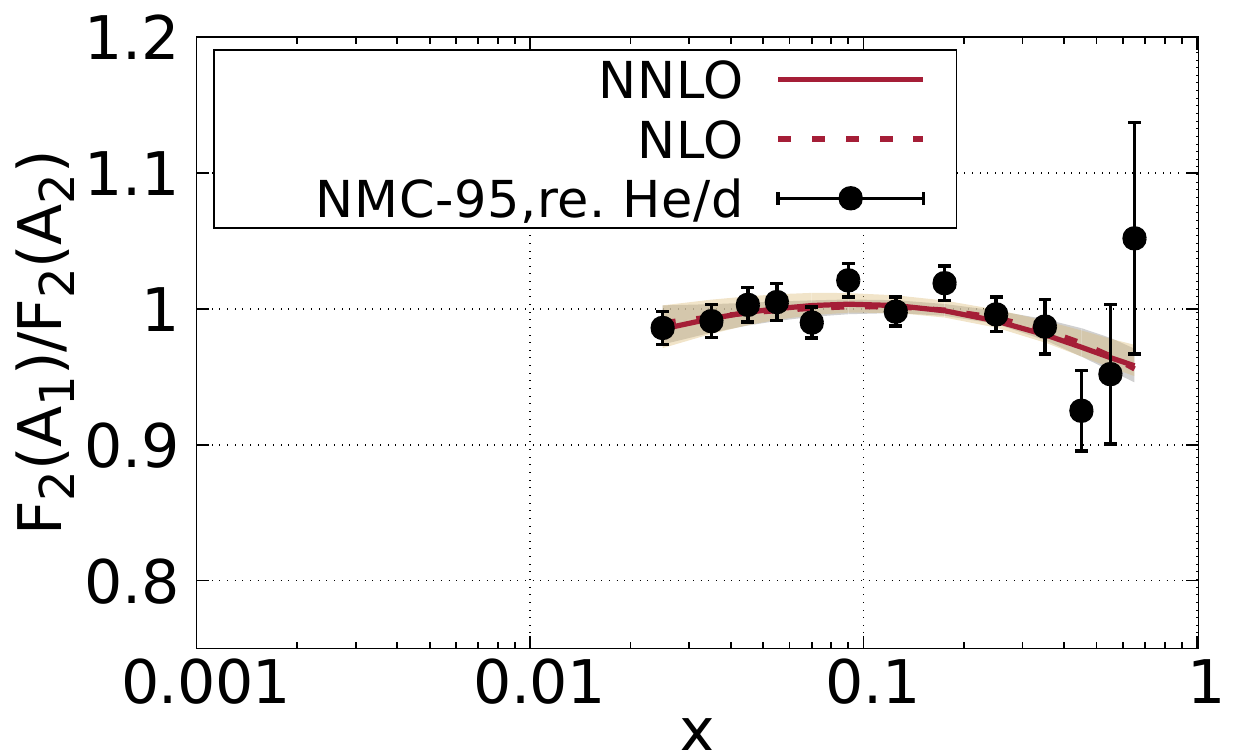}} 
         \subfigure{        
              \includegraphics[width=0.237\textwidth]{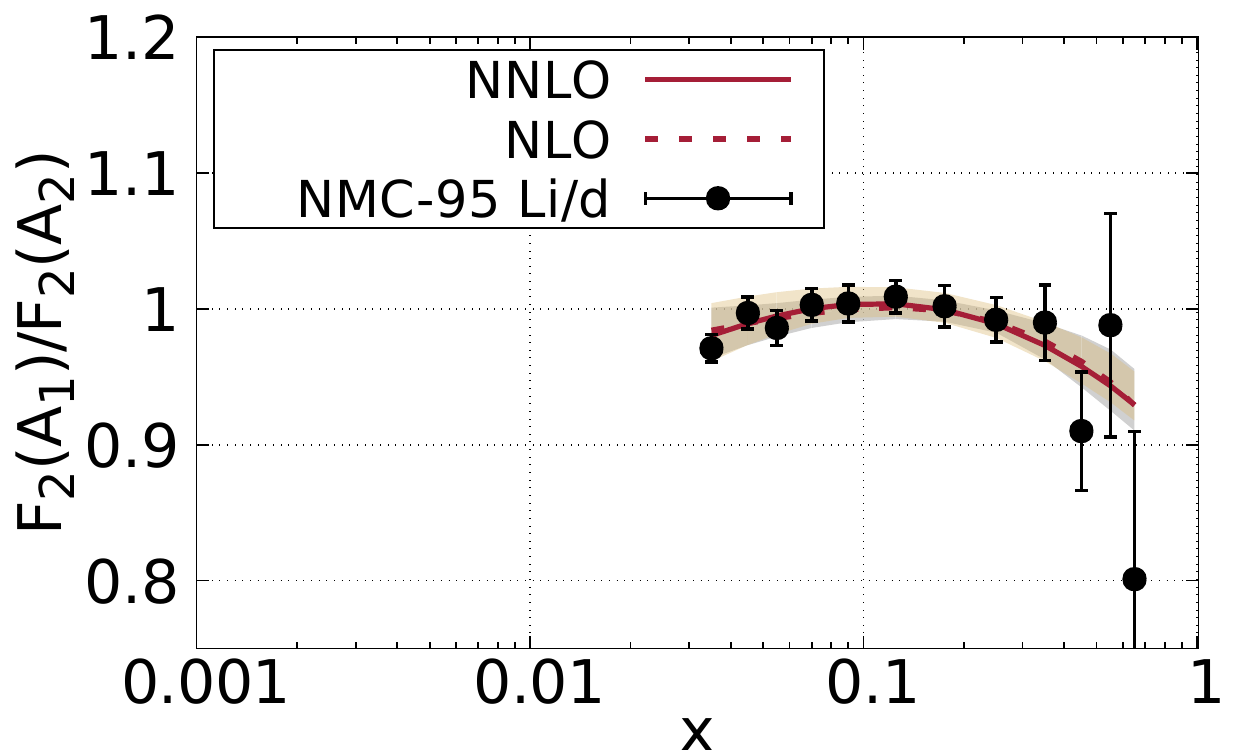}} 
         \subfigure{        
              \includegraphics[width=0.237\textwidth]{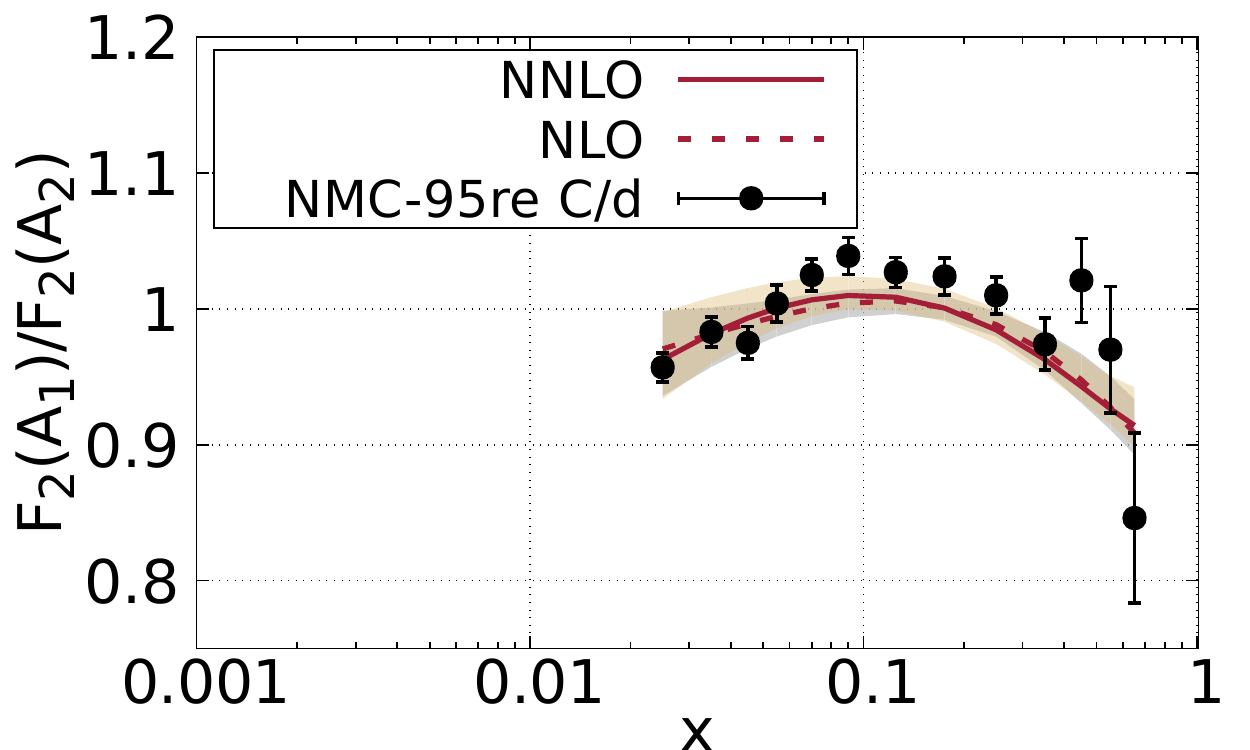}} 
         \subfigure{        
              \includegraphics[width=0.237\textwidth]{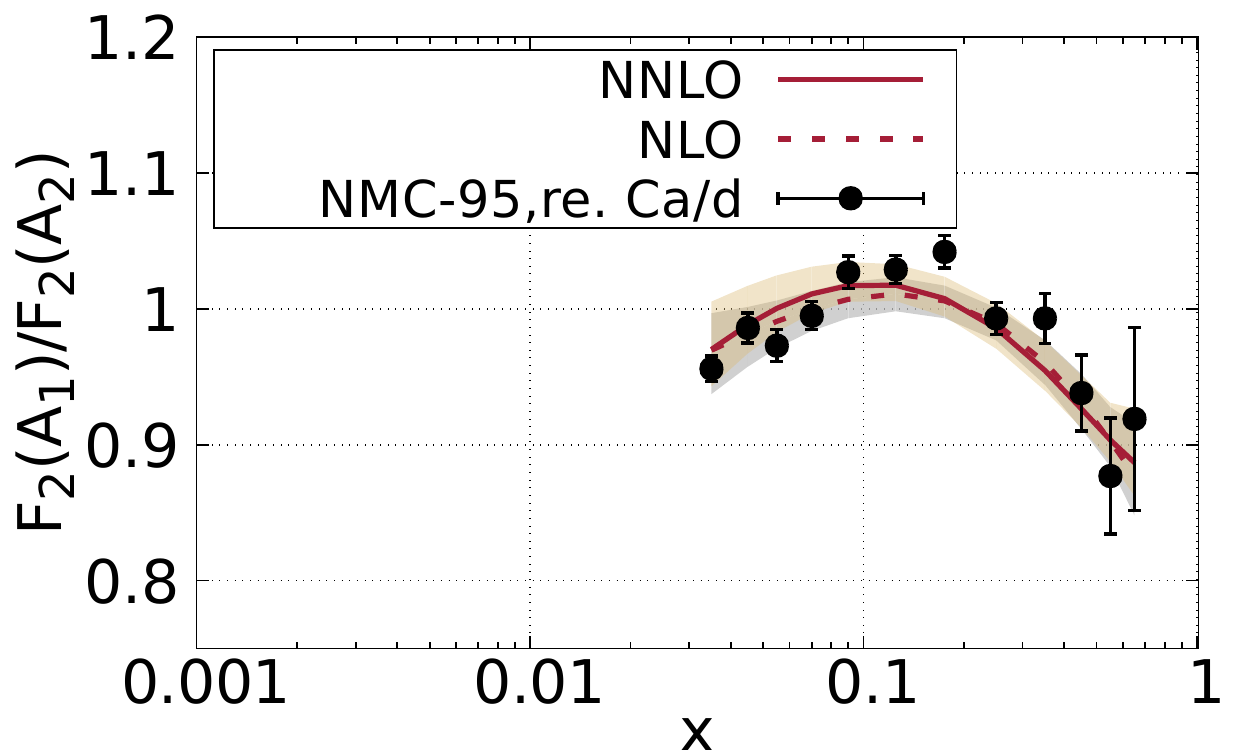}} 
         \subfigure{        
              \includegraphics[width=0.237\textwidth]{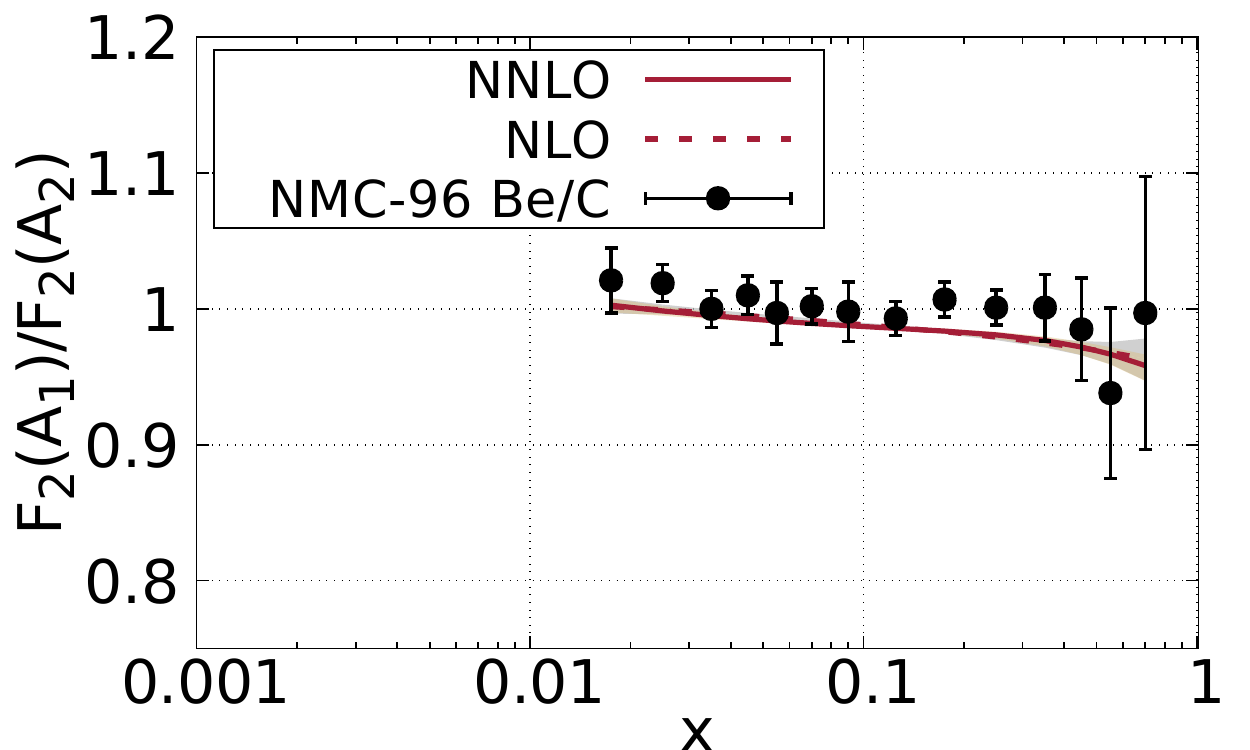}} 
         \subfigure{        
              \includegraphics[width=0.237\textwidth]{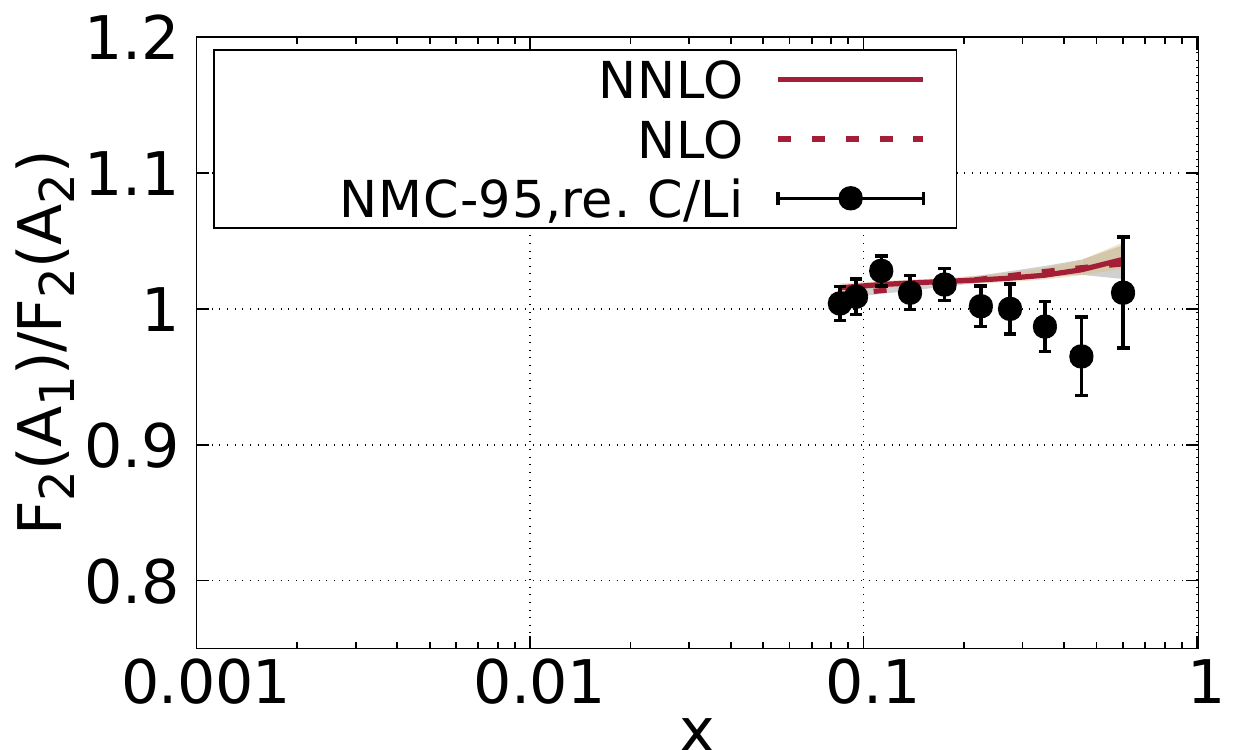}}                             
         \subfigure{        
              \includegraphics[width=0.237\textwidth]{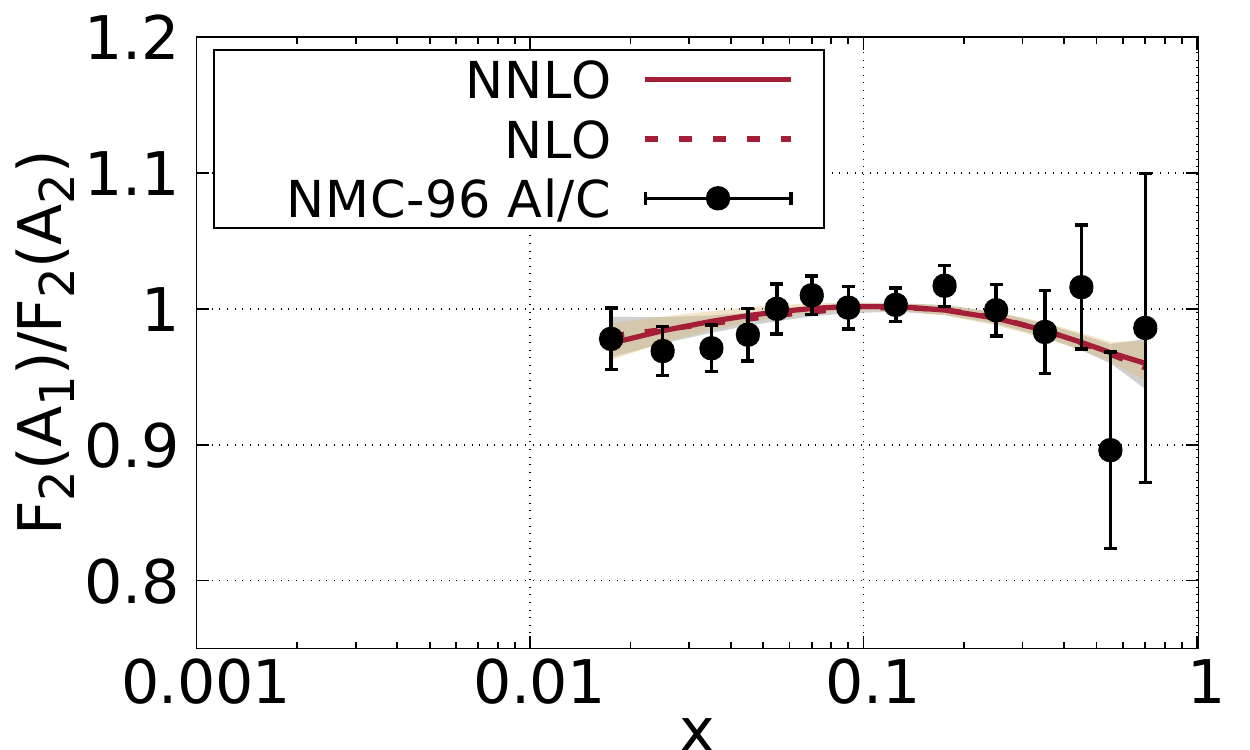}}                             
         \subfigure{        
              \includegraphics[width=0.237\textwidth]{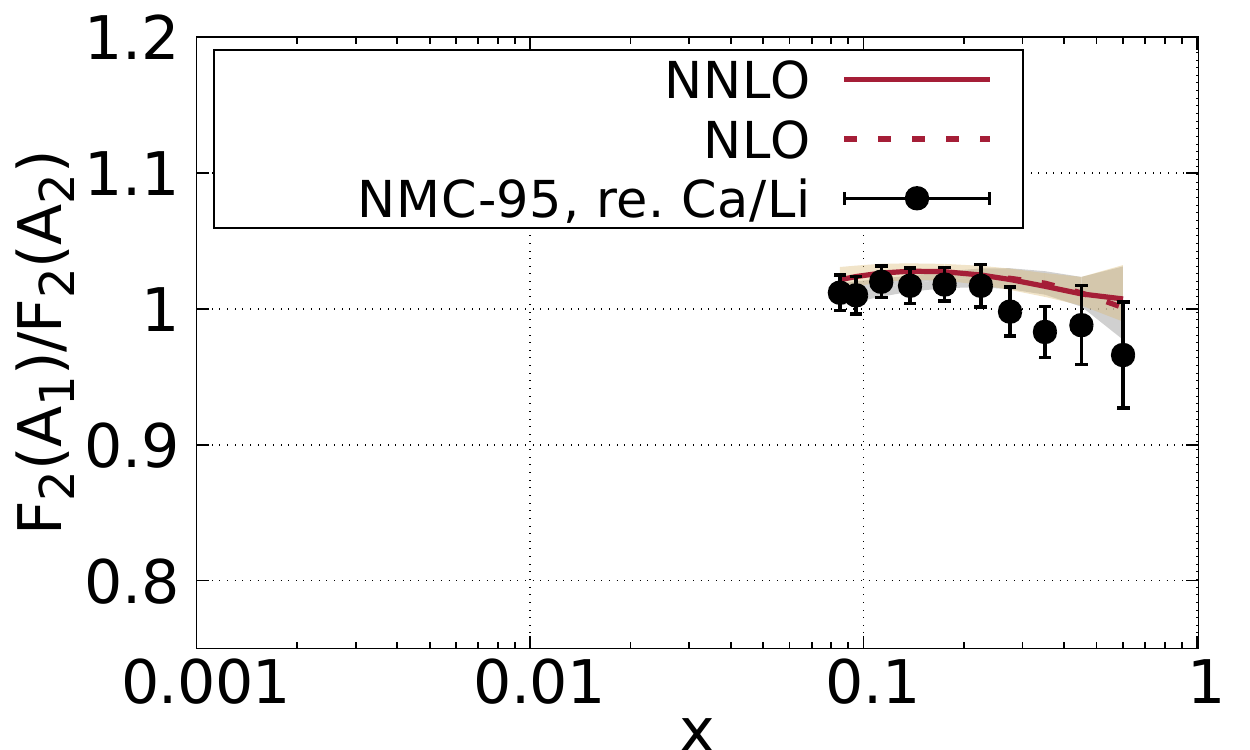}}                             
         \subfigure{        
              \includegraphics[width=0.237\textwidth]{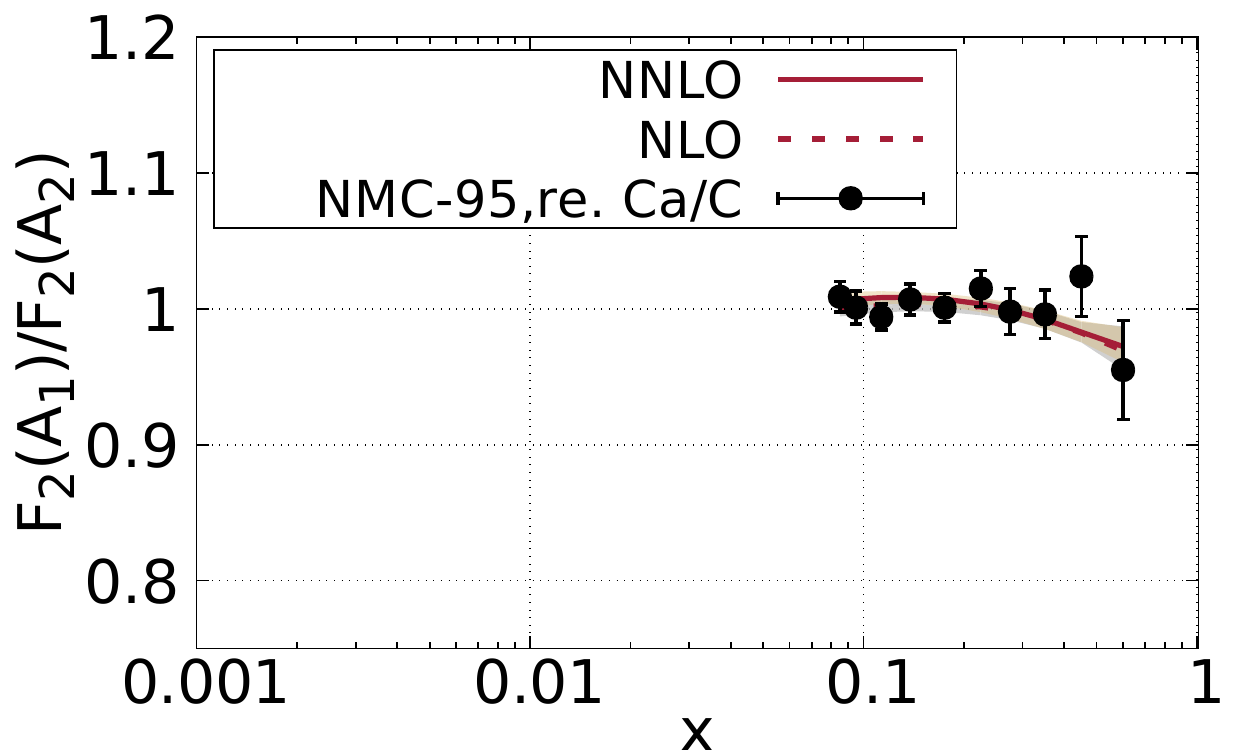}}                             
         \subfigure{        
              \includegraphics[width=0.237\textwidth]{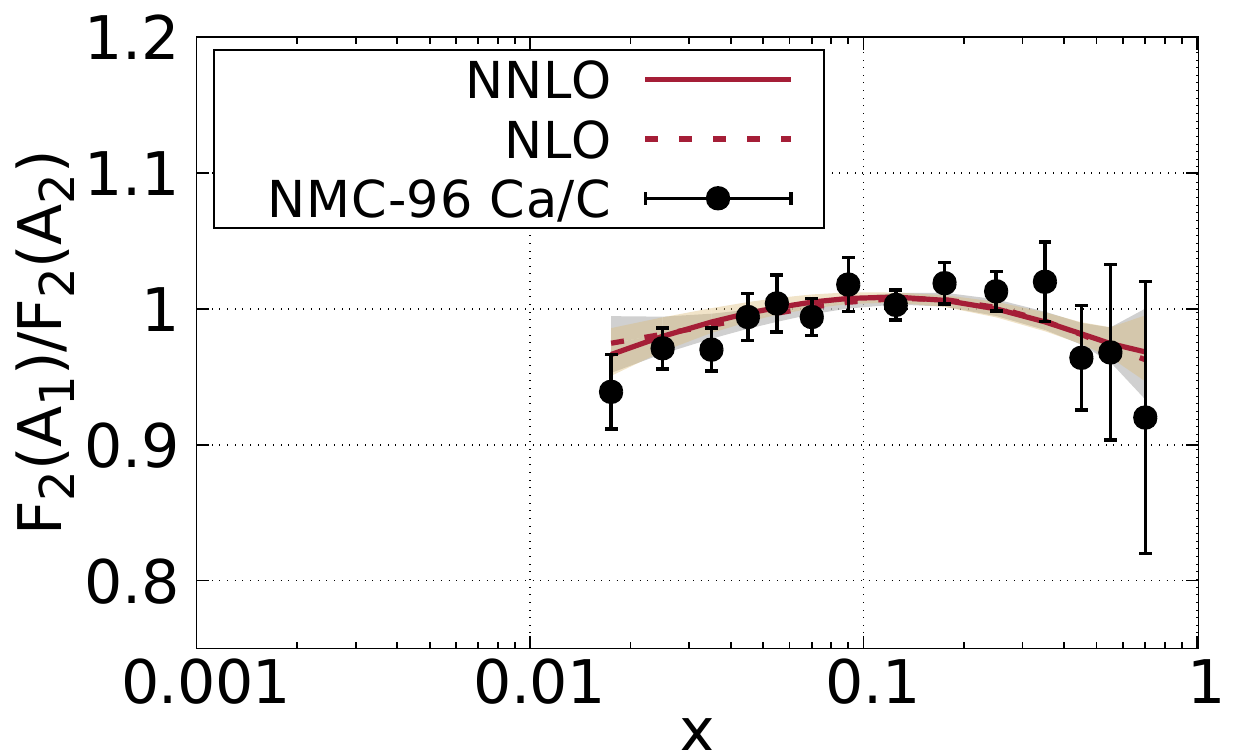}}                             
         \subfigure{        
              \includegraphics[width=0.237\textwidth]{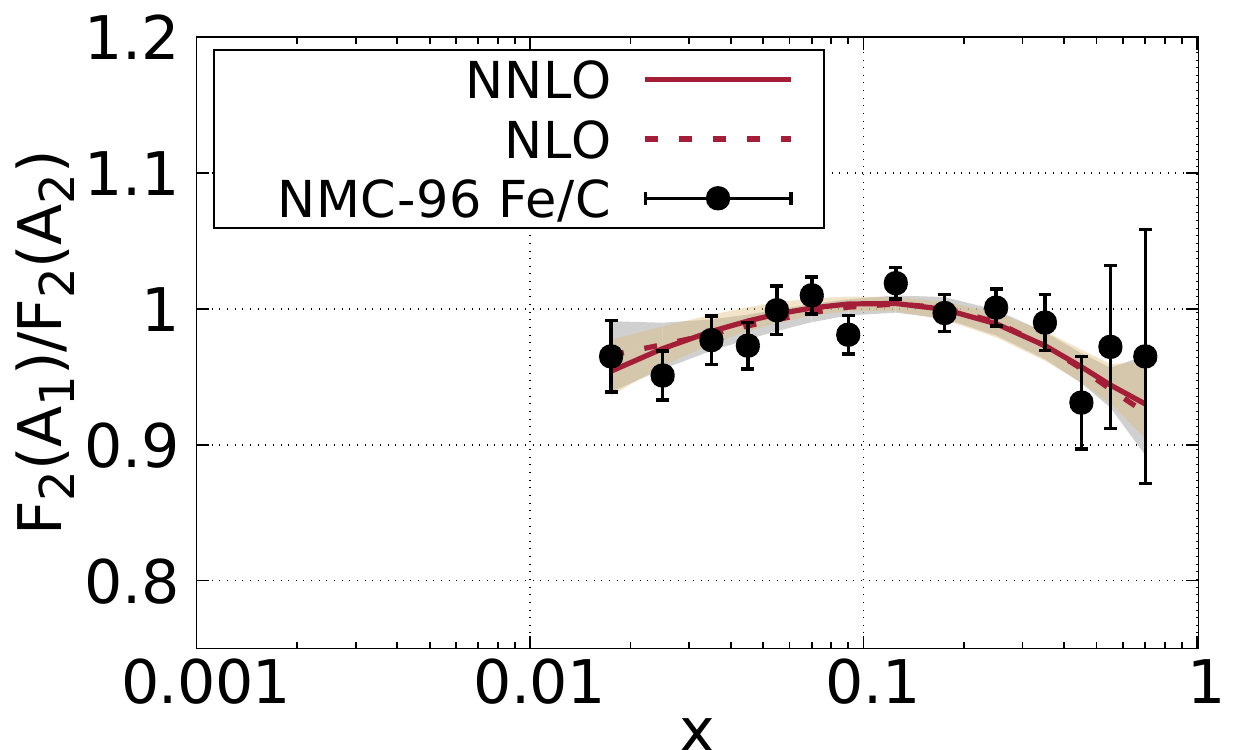}}                             
         \subfigure{        
              \includegraphics[width=0.237\textwidth]{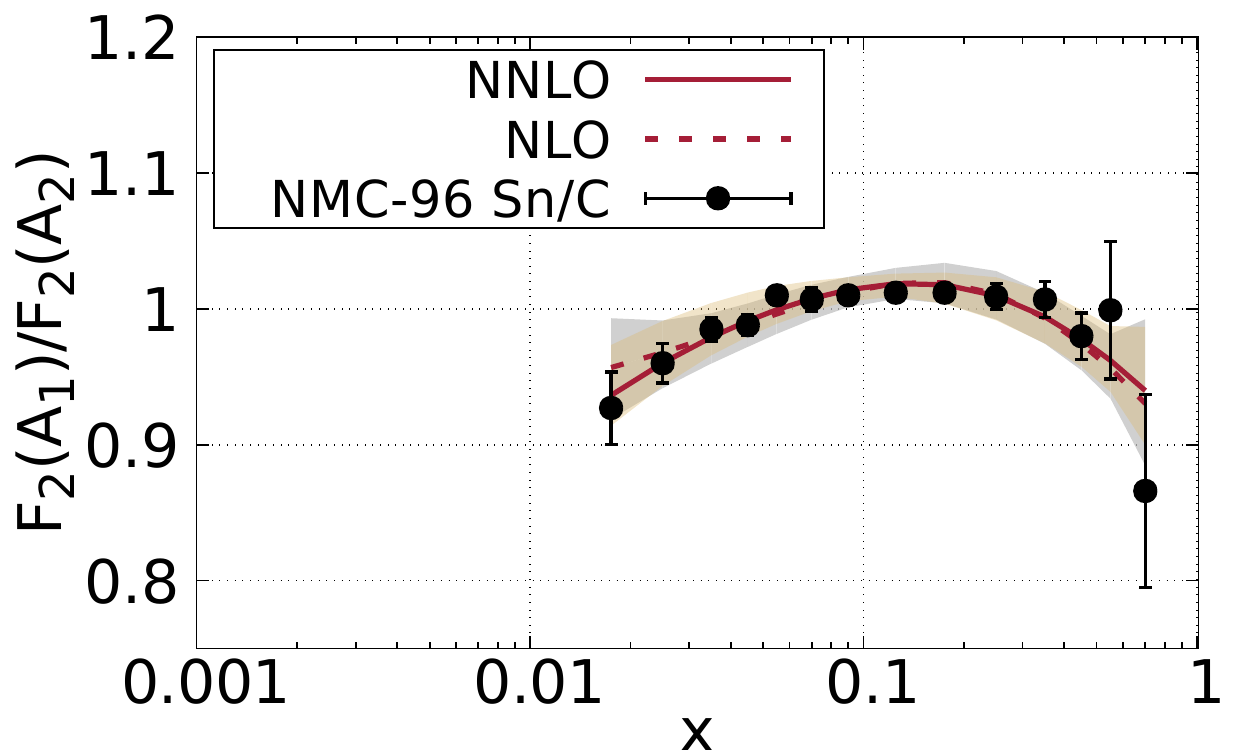}}                             
         \subfigure{        
              \includegraphics[width=0.237\textwidth]{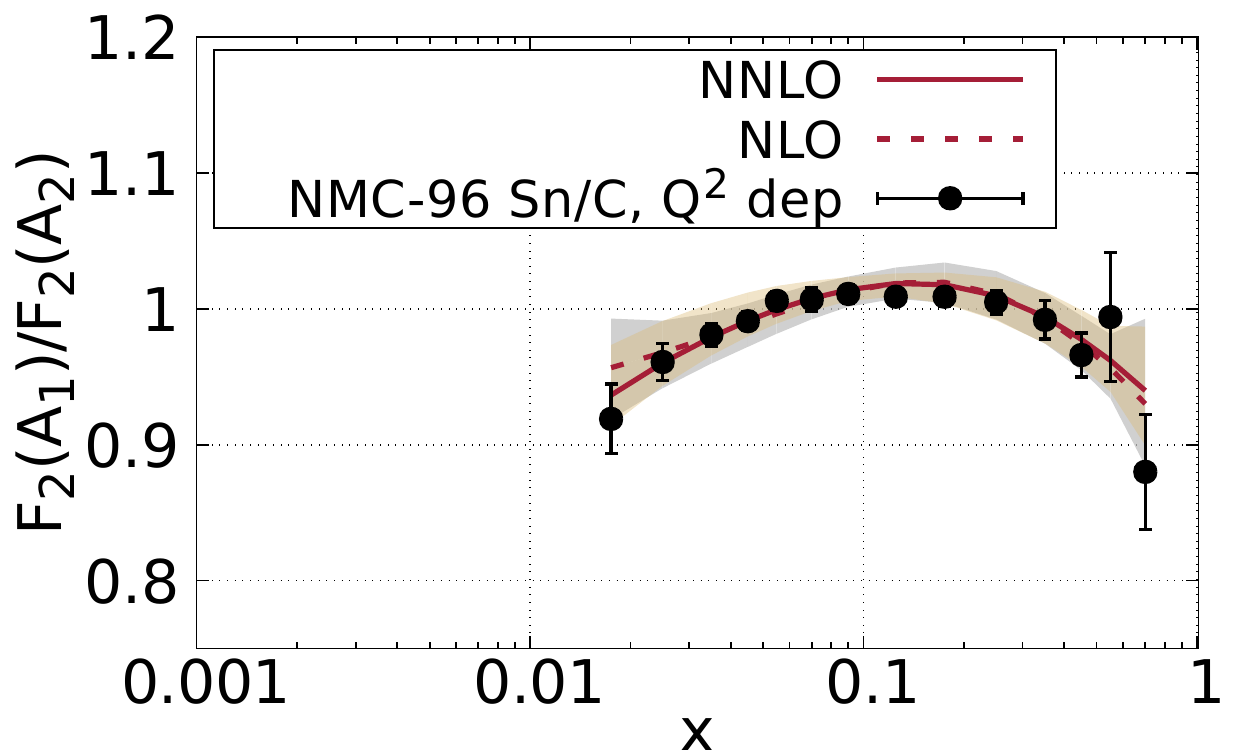}}                             
         \subfigure{        
              \includegraphics[width=0.237\textwidth]{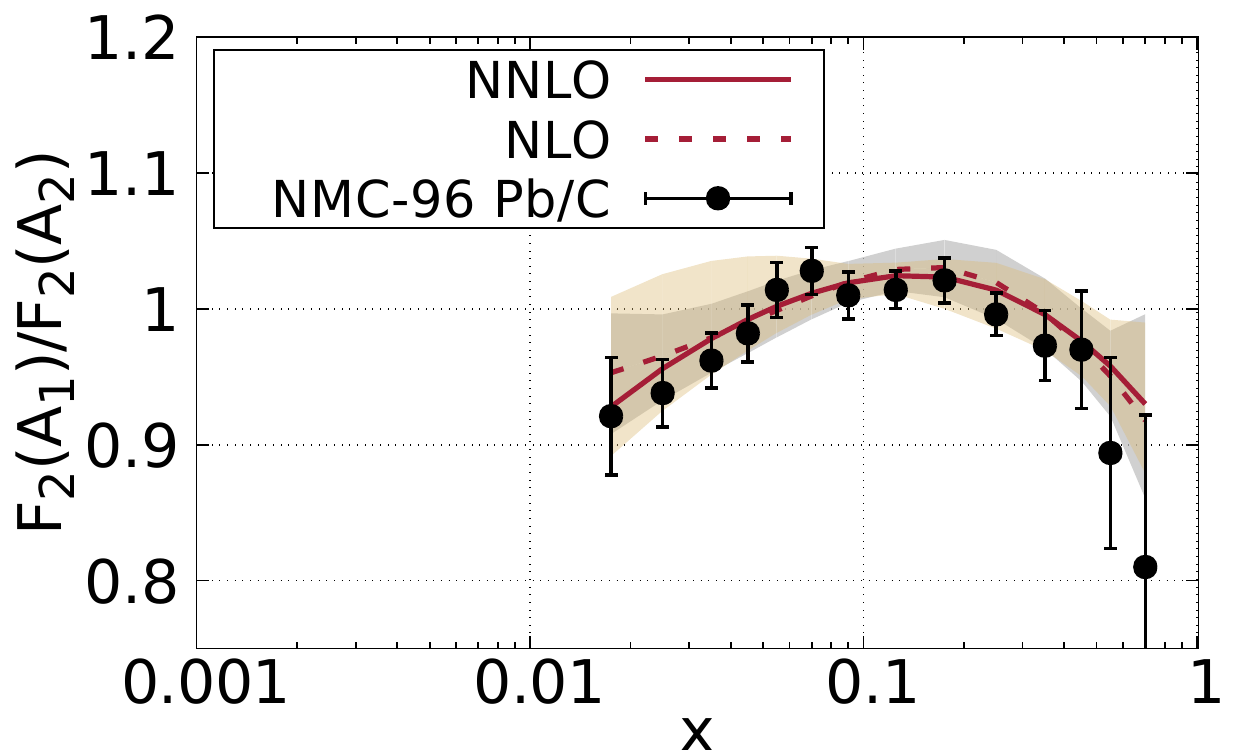}}                                                                                                                                             
          \end{center} 
    \caption{Comparison to NMC $F_2(A_1)/F_2(A_2)$ data measured for different combinations of nuclei with mass numbers $A_1$ and $A_2$, at NLO (dashed line, grey error bands) and NNLO (solid line, golden-coloured error bands).}
\label{figNMC}    
    \end{figure*}

\begin{figure*}[hbt!]
     \begin{center}
          \subfigure{        
              \includegraphics[width=0.237\textwidth]{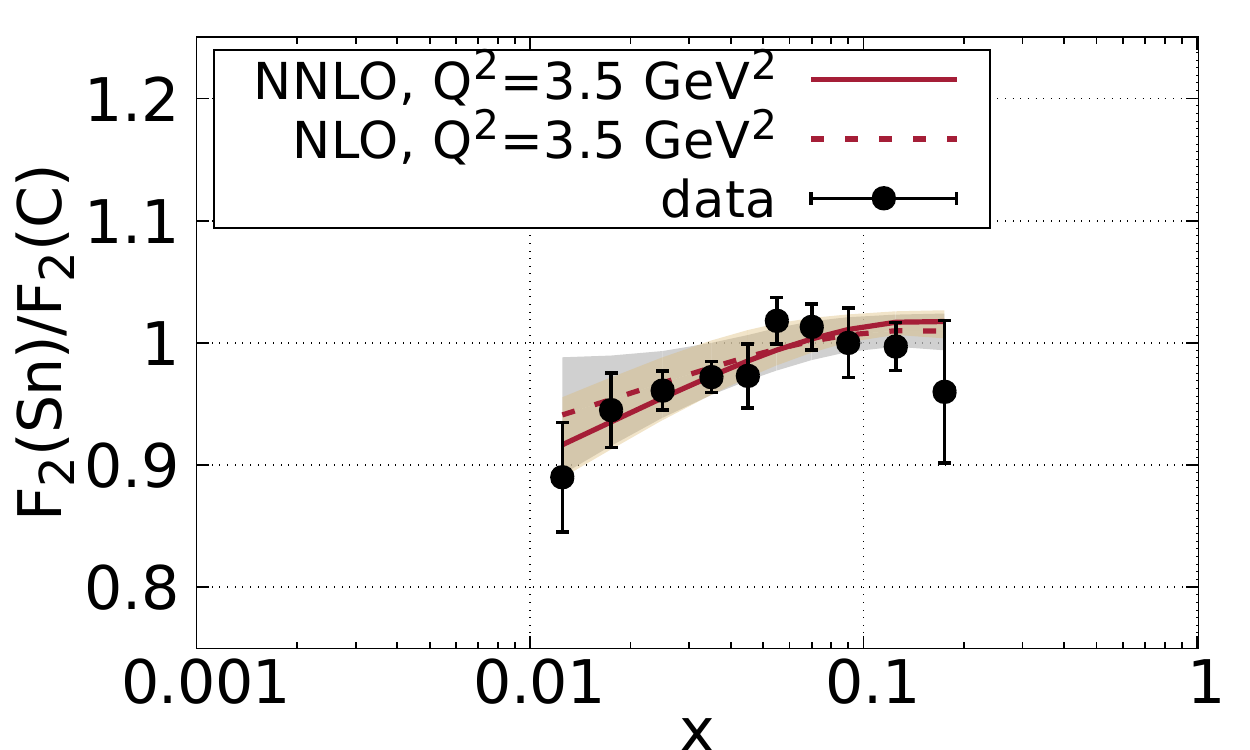}} 
         \subfigure{        
              \includegraphics[width=0.237\textwidth]{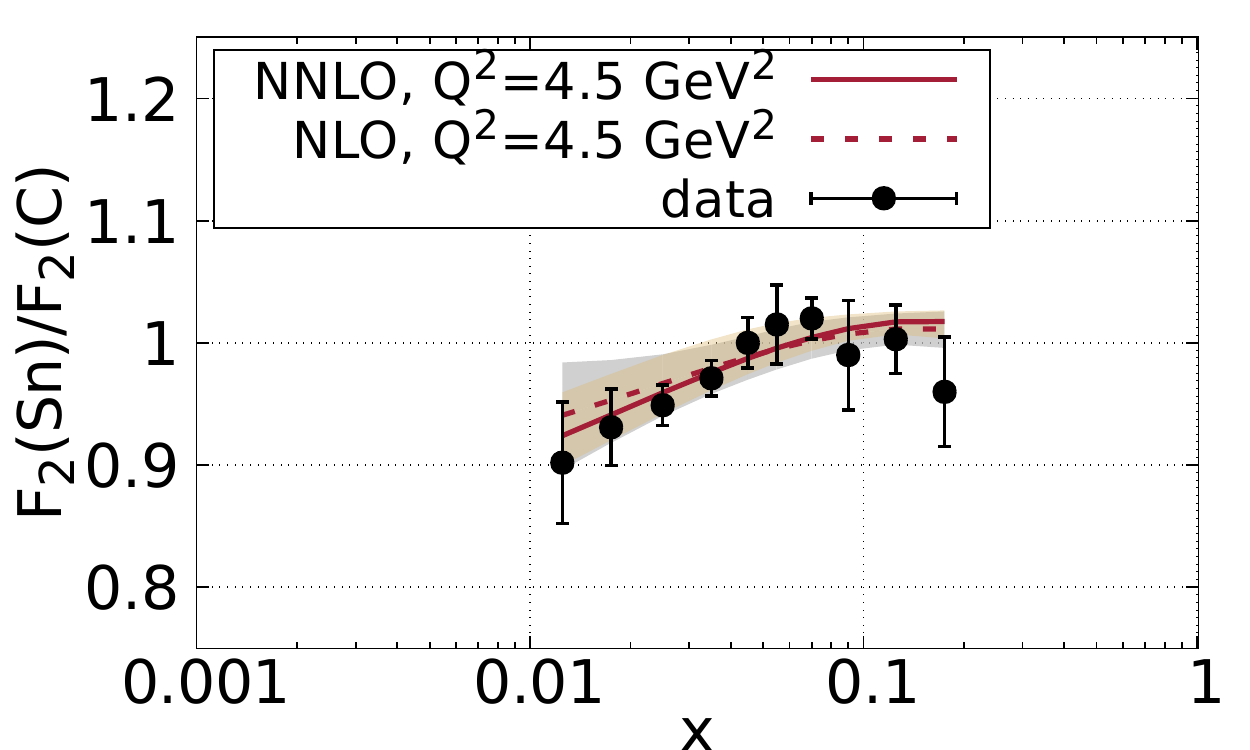}} 
         \subfigure{        
              \includegraphics[width=0.237\textwidth]{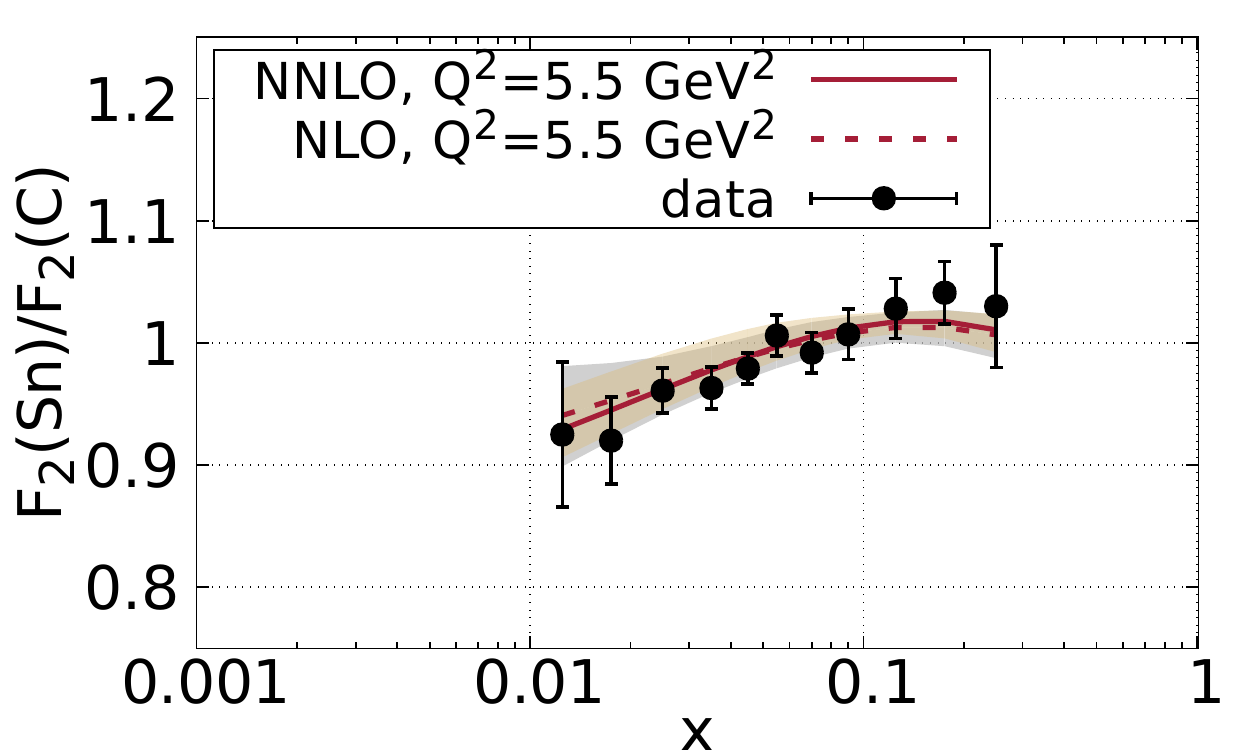}} 
         \subfigure{        
              \includegraphics[width=0.237\textwidth]{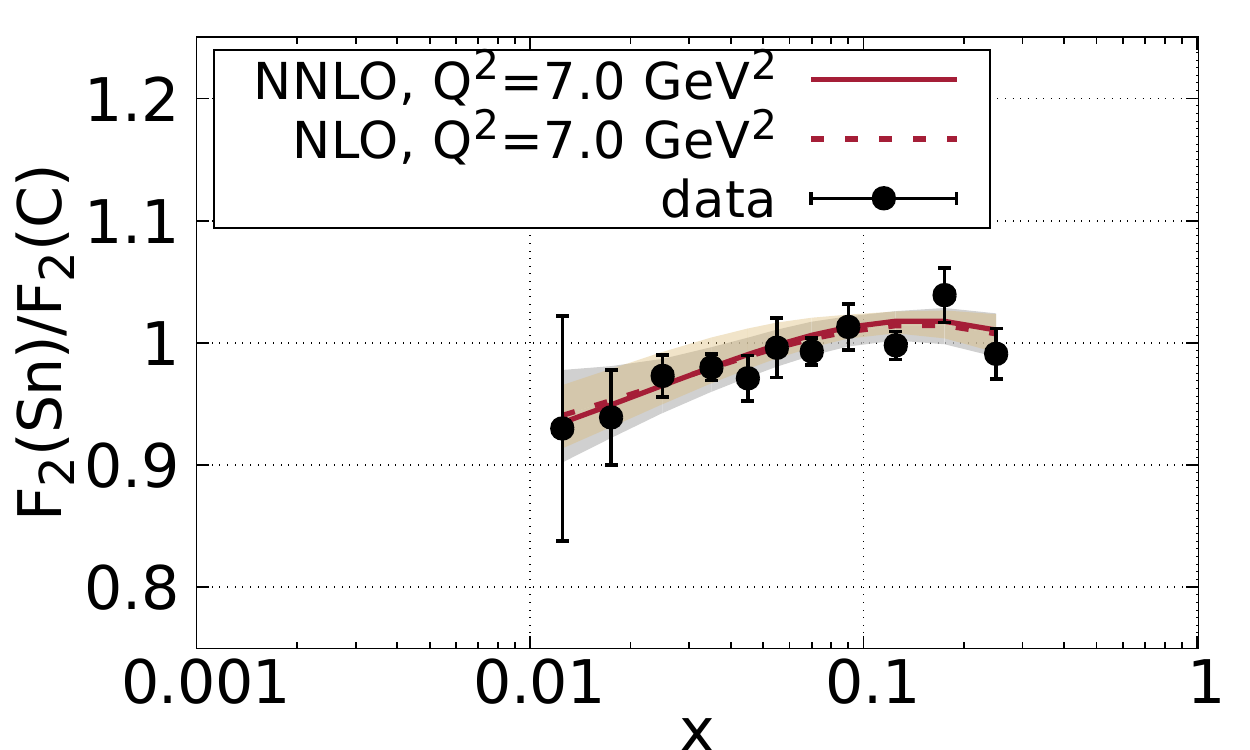}} 
         \subfigure{        
              \includegraphics[width=0.237\textwidth]{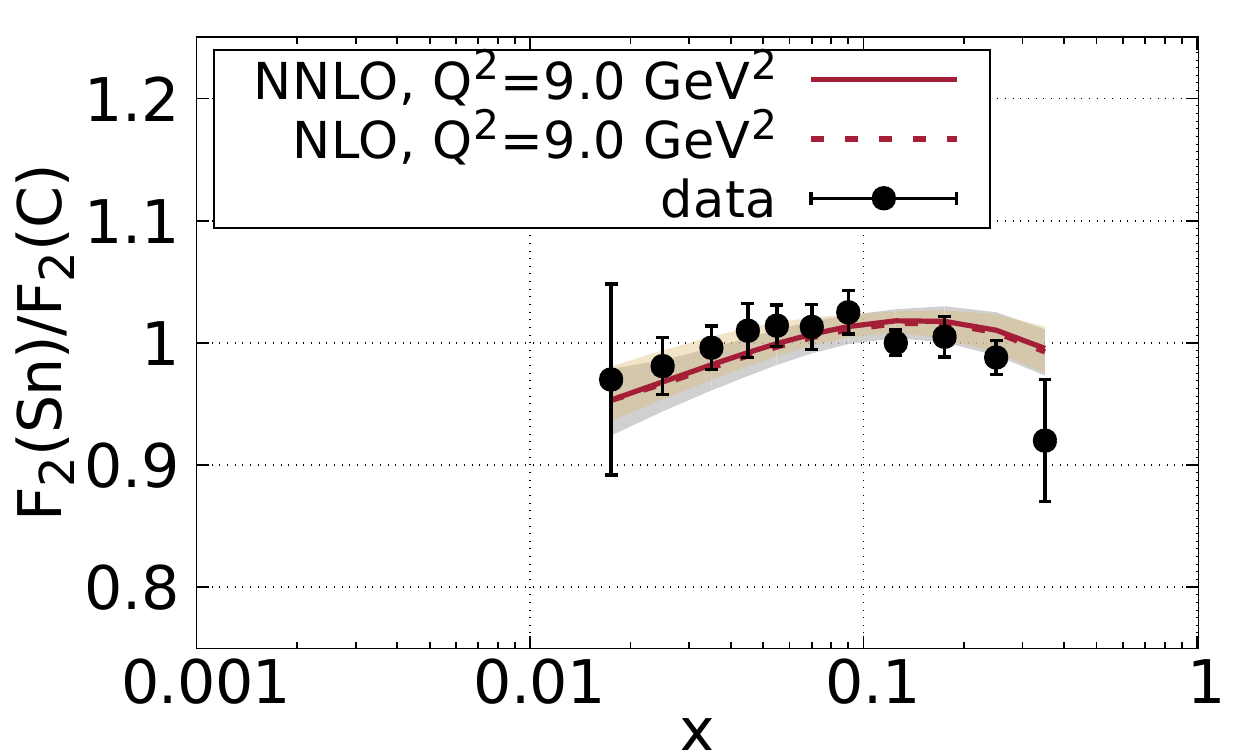}} 
         \subfigure{        
              \includegraphics[width=0.237\textwidth]{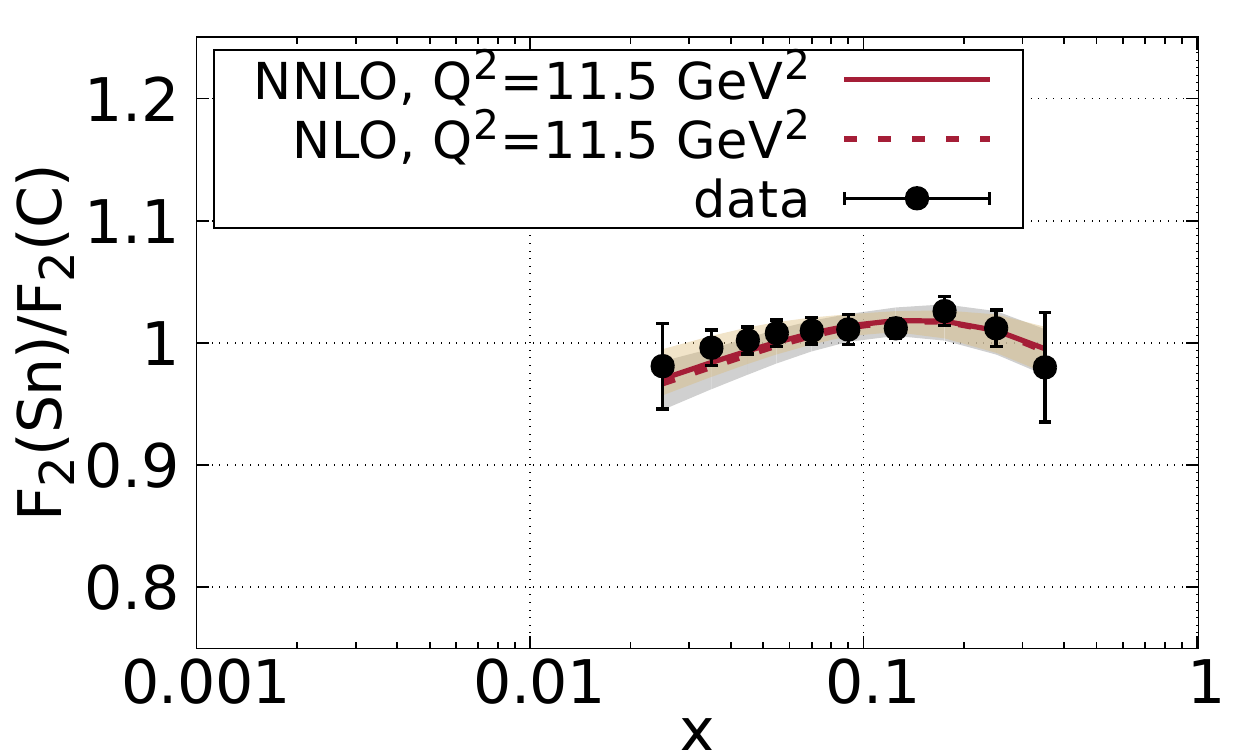}}                             
         \subfigure{        
              \includegraphics[width=0.237\textwidth]{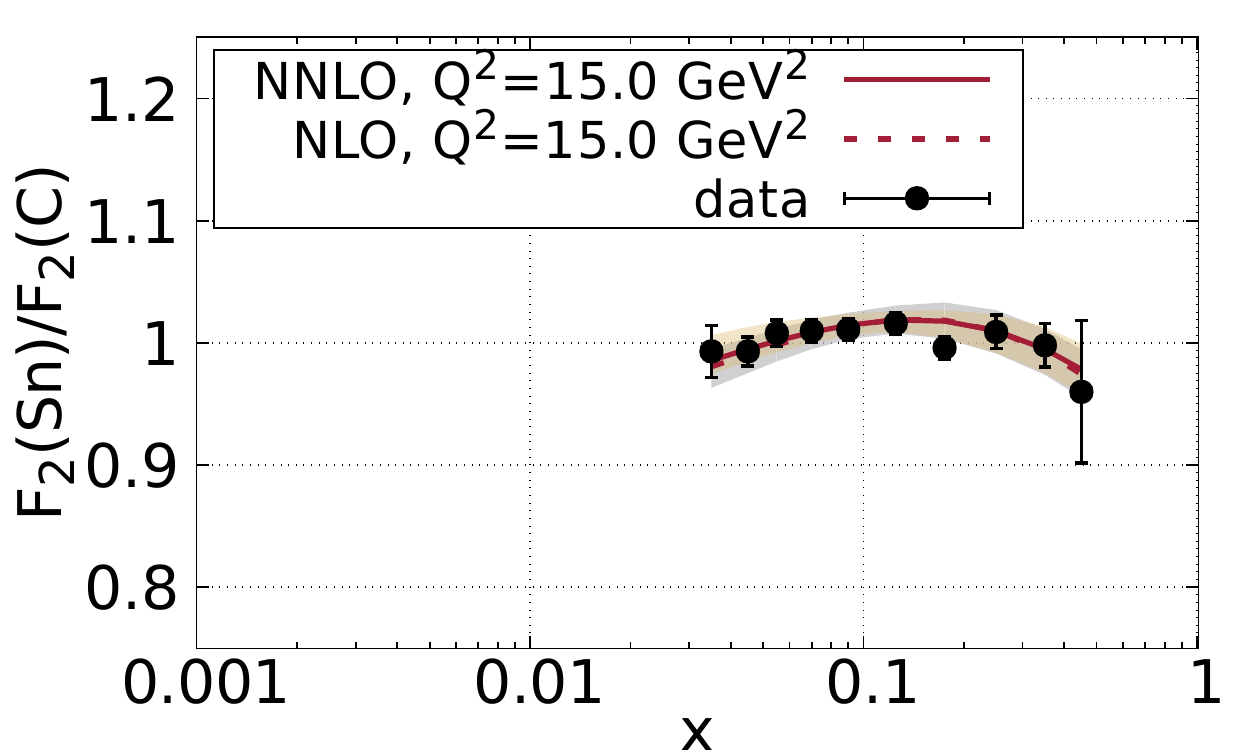}}                             
         \subfigure{        
              \includegraphics[width=0.237\textwidth]{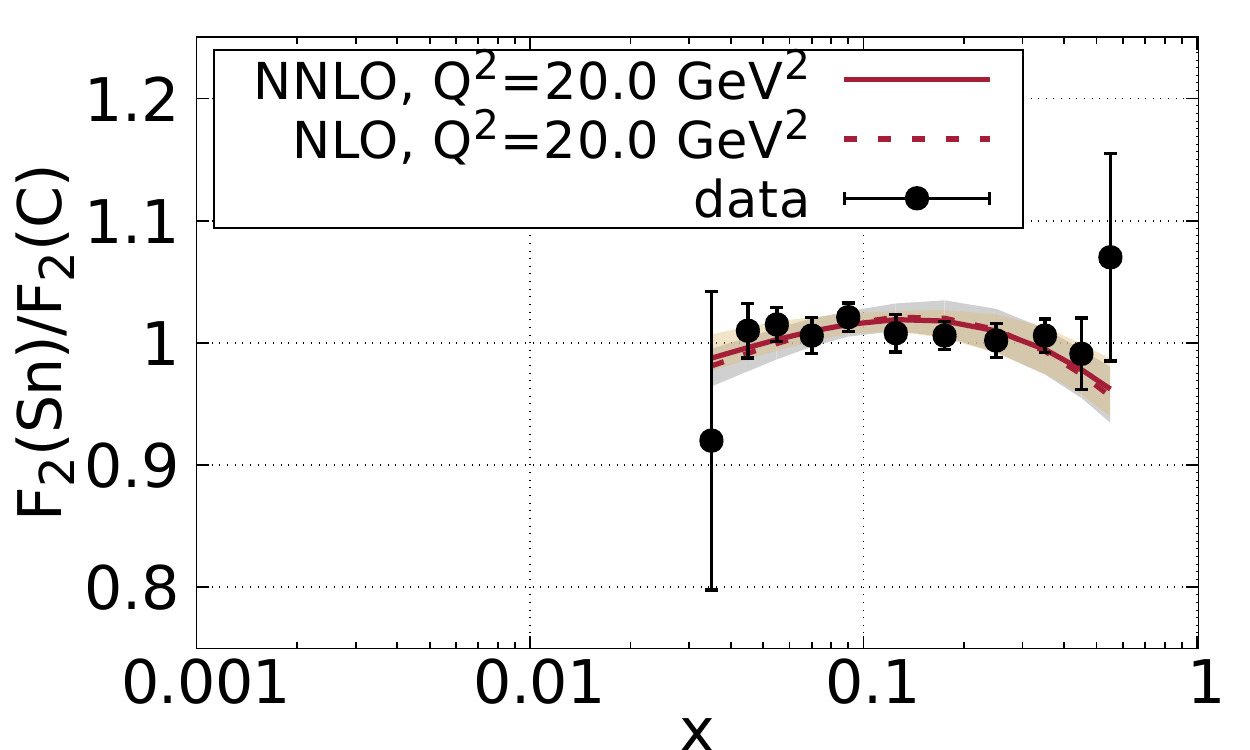}}                             
         \subfigure{        
              \includegraphics[width=0.237\textwidth]{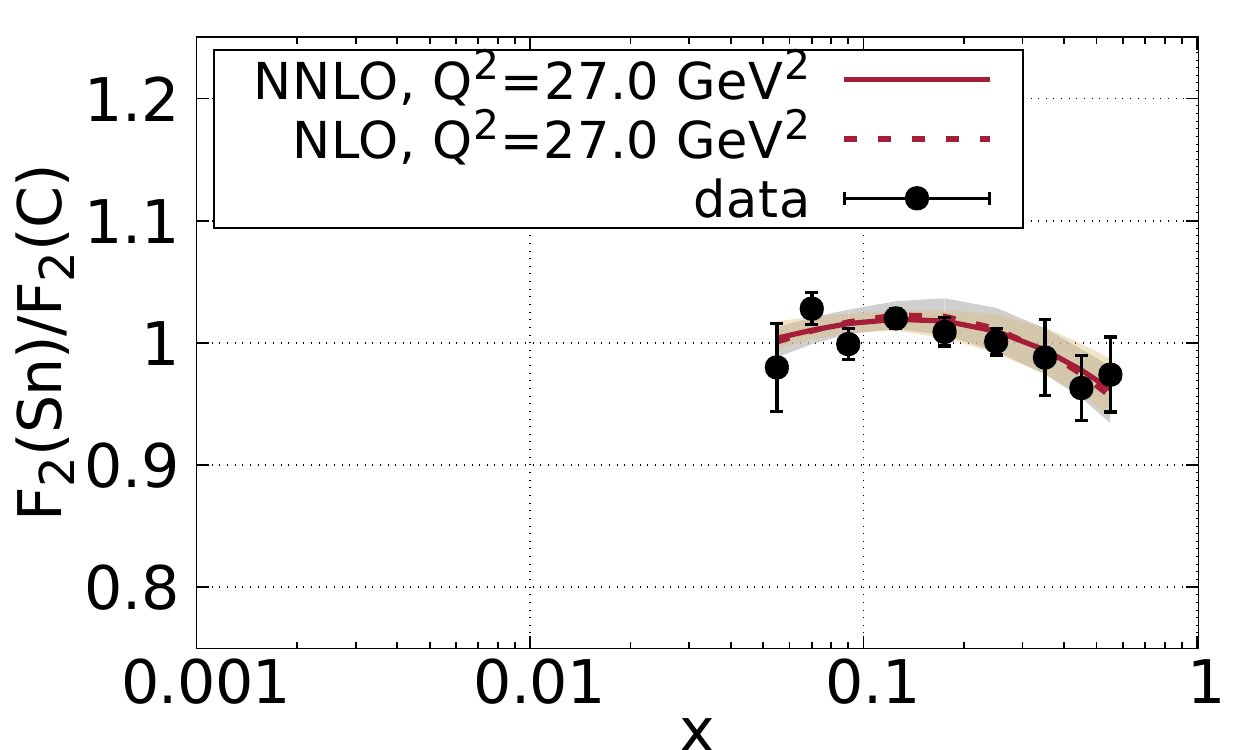}}                             
         \subfigure{        
              \includegraphics[width=0.237\textwidth]{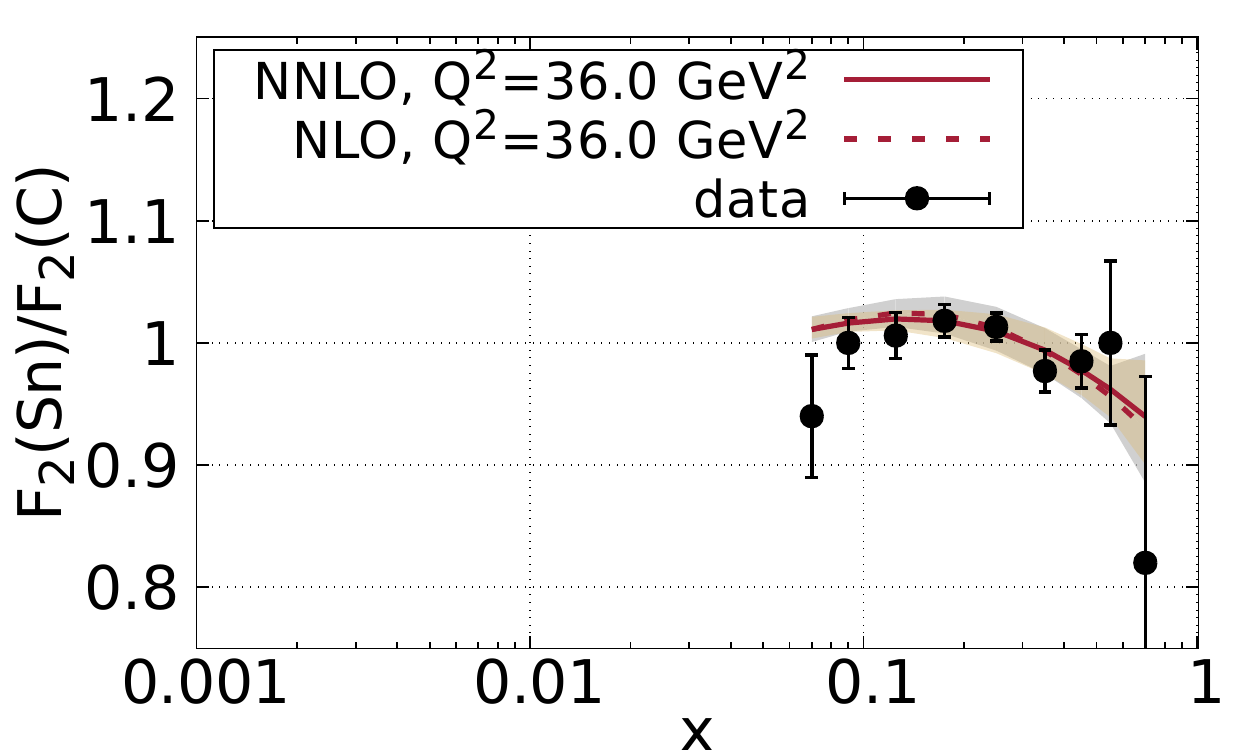}}                             
         \subfigure{        
              \includegraphics[width=0.237\textwidth]{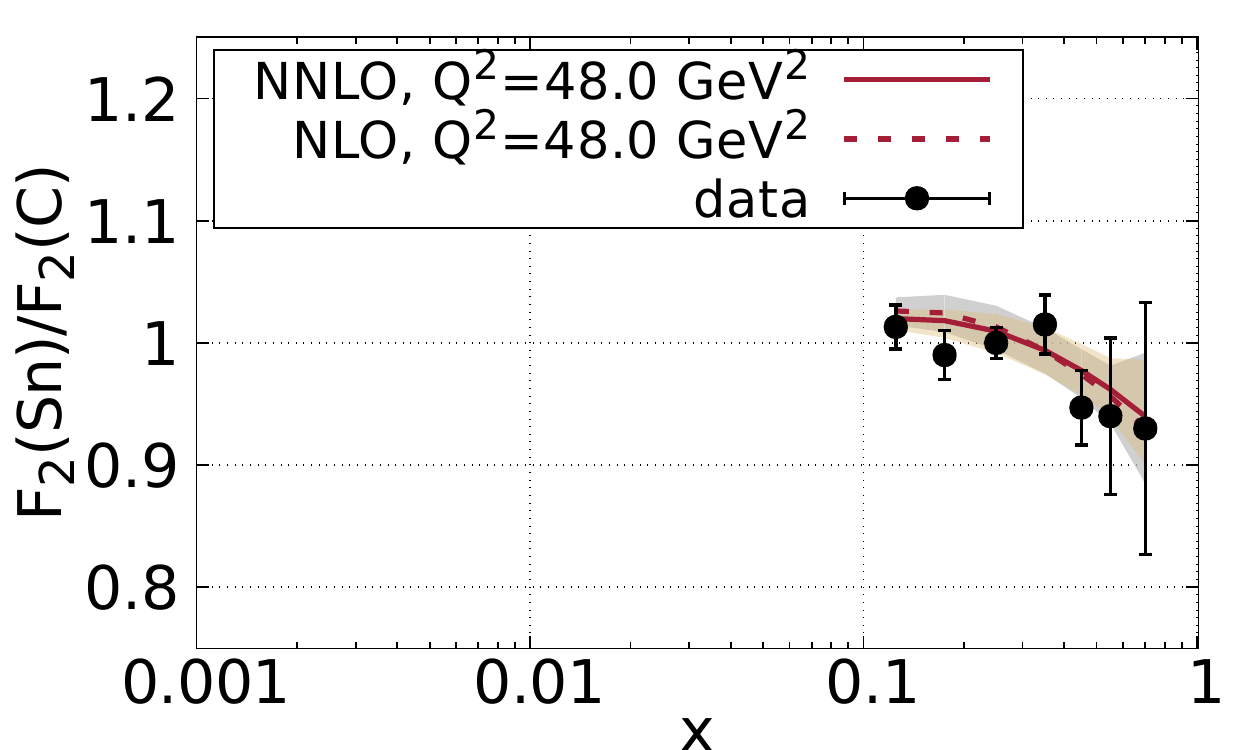}}                             
         \subfigure{        
              \includegraphics[width=0.237\textwidth]{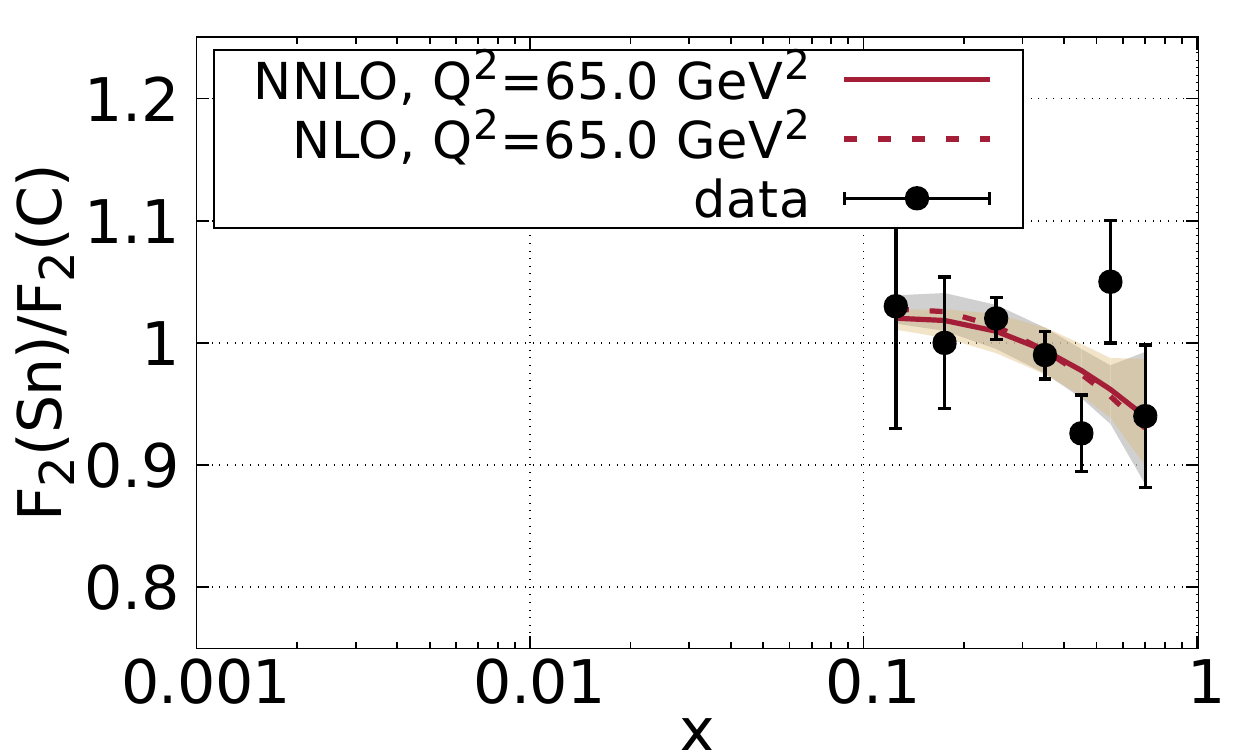}}                             
         \subfigure{        
              \includegraphics[width=0.237\textwidth]{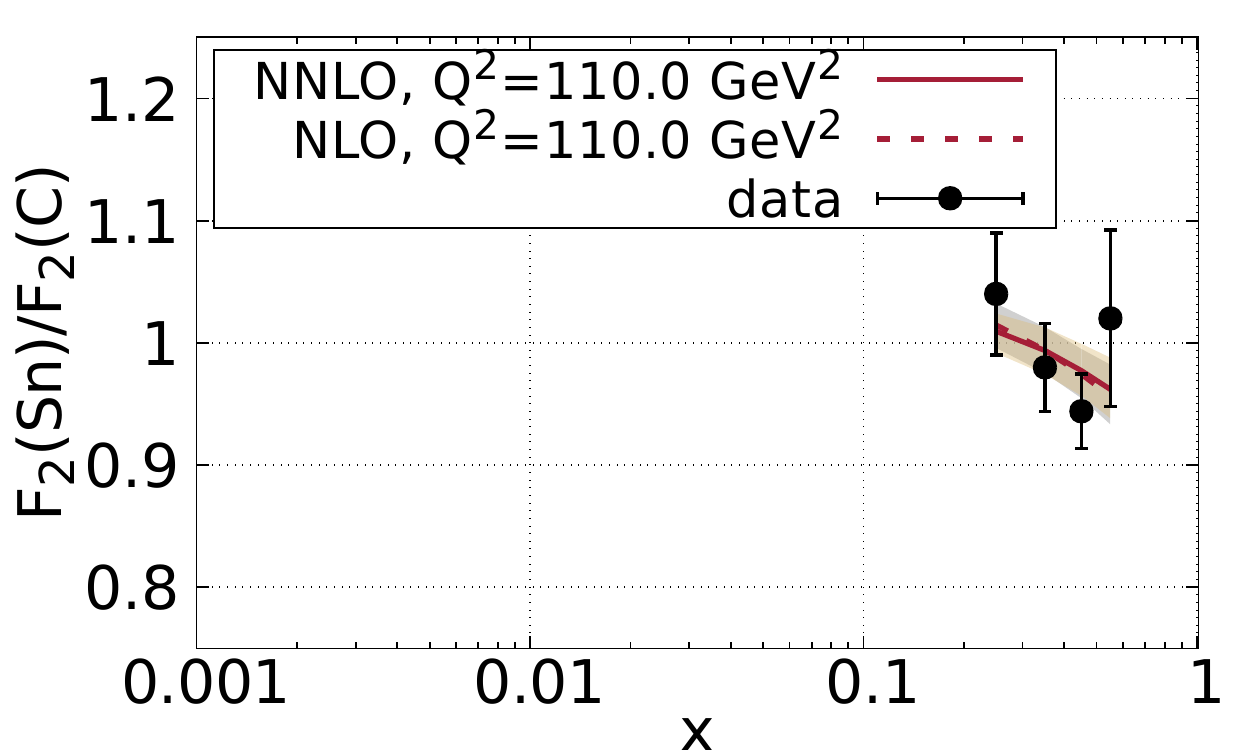}}                             
          \end{center} 
    \caption{Comparison to NMC $F_2(\mathrm{Sn})/F_2(\mathrm{C})$ data at different values of $Q^2$ at NLO (dashed line, grey error bands) and NNLO (solid line, golden-coloured error bands).}
\label{figNMCSnC}    
    \end{figure*}

\begin{figure*}[htb!]
     \begin{center}
          \subfigure{        
              \includegraphics[width=0.237\textwidth]{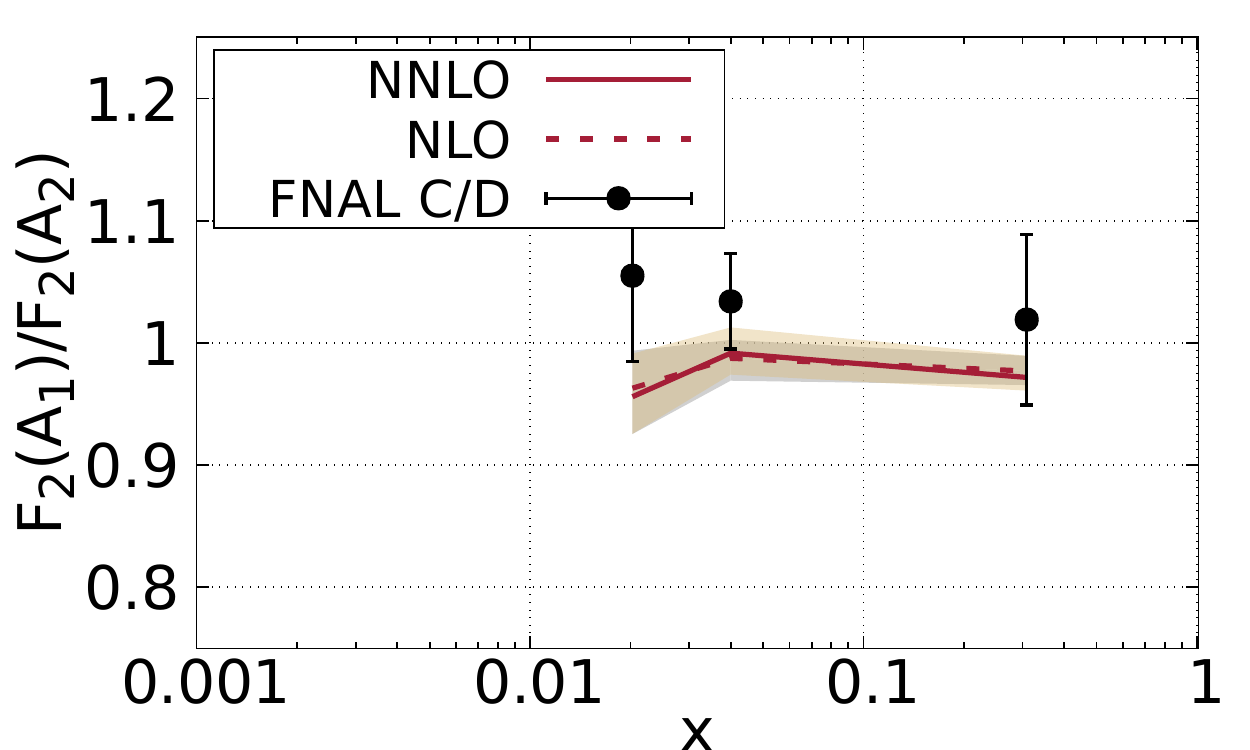}} 
         \subfigure{        
              \includegraphics[width=0.237\textwidth]{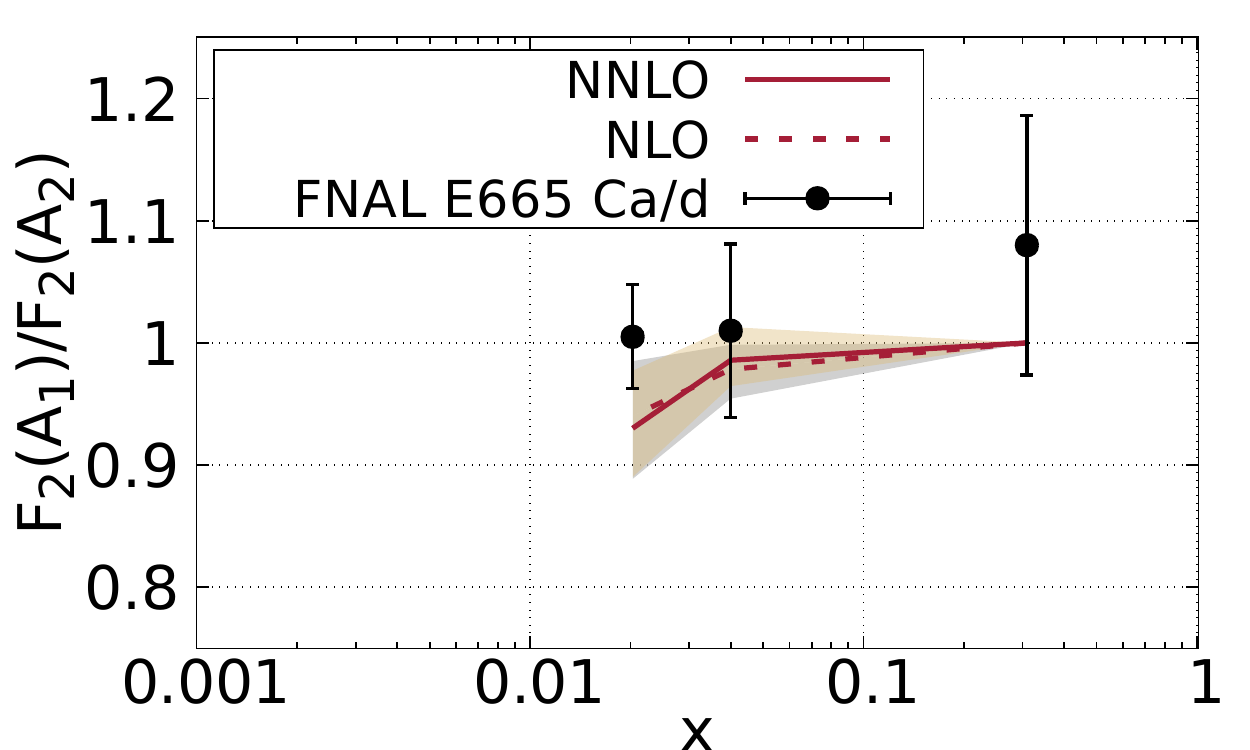}} 
         \subfigure{        
              \includegraphics[width=0.237\textwidth]{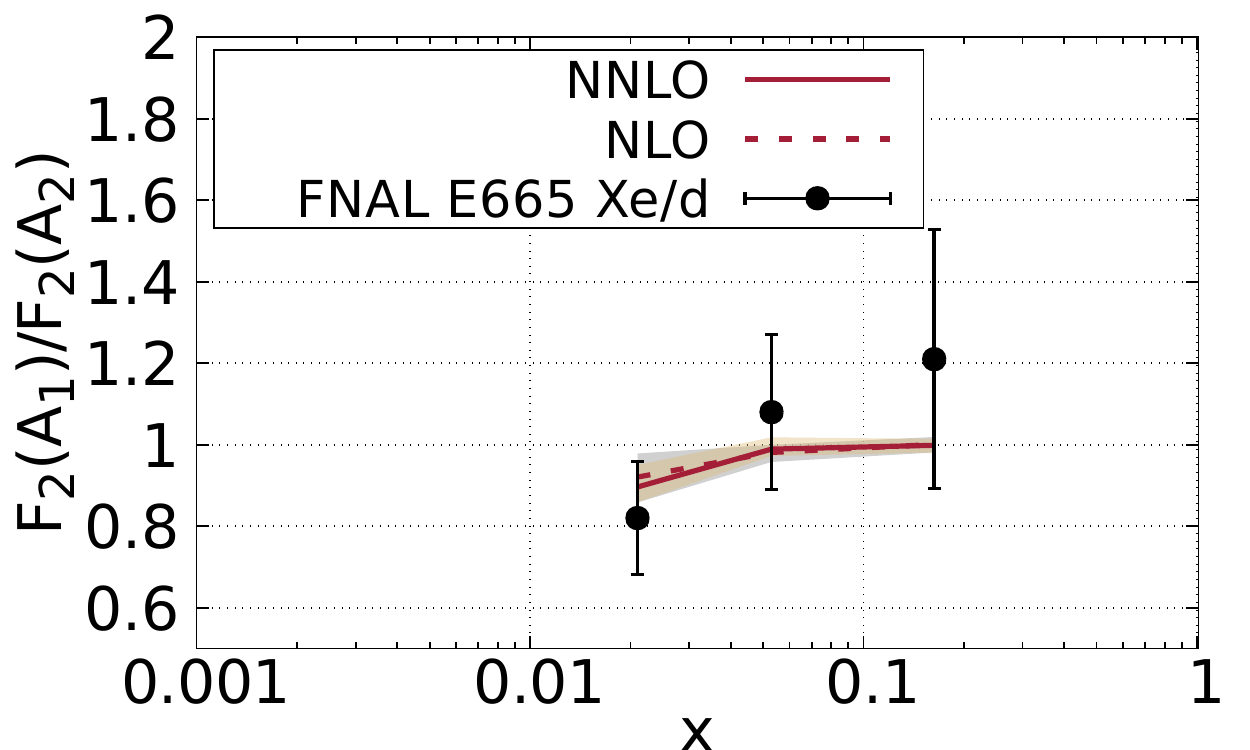}} 
         \subfigure{        
              \includegraphics[width=0.237\textwidth]{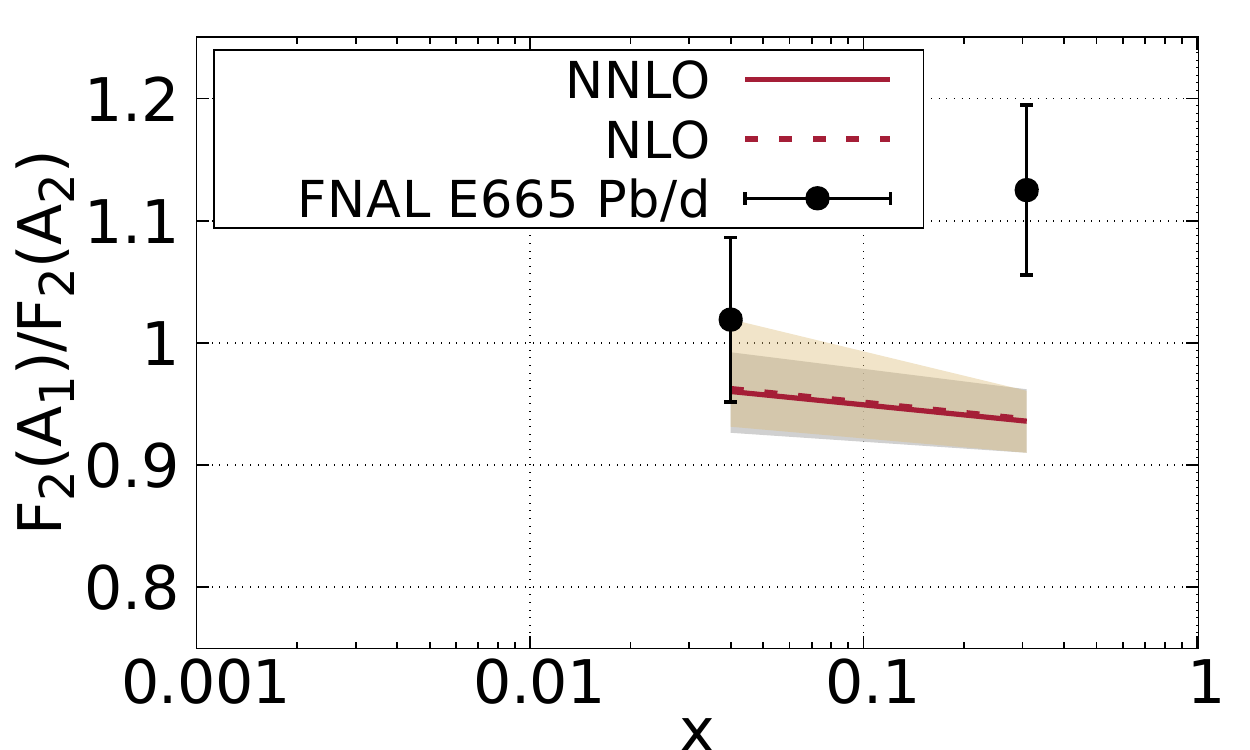}} 
          \end{center} 
    \caption{Comparison to FNAL E665 data for different ratios $F_2(A_1)/F_2(A_2)$ for nuclei with mass numbers $A_1$ and $A_2$, at NLO (dashed line, grey error bands) and NNLO (solid line, golden-coloured error bands).}
\label{figFNAL}    
    \end{figure*}

\begin{figure*}[htb!]
     \begin{center}
          \subfigure{        
              \includegraphics[width=0.237\textwidth]{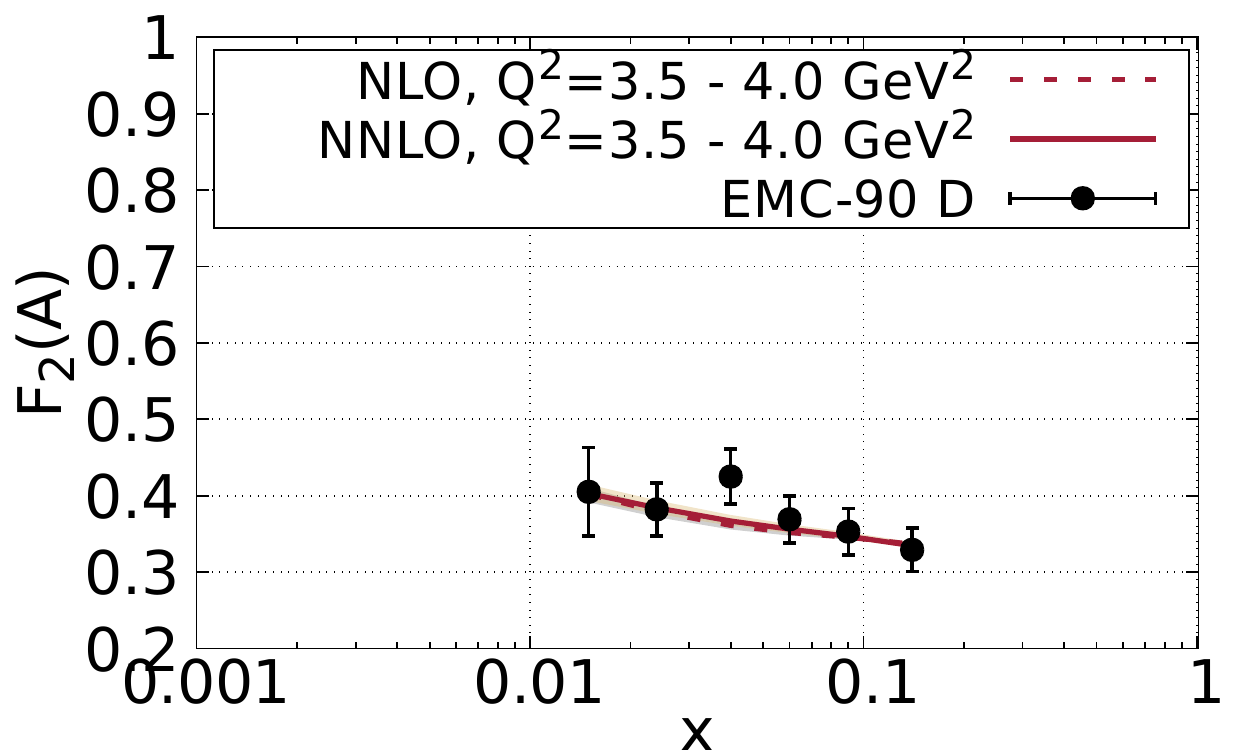}} 
         \subfigure{        
              \includegraphics[width=0.237\textwidth]{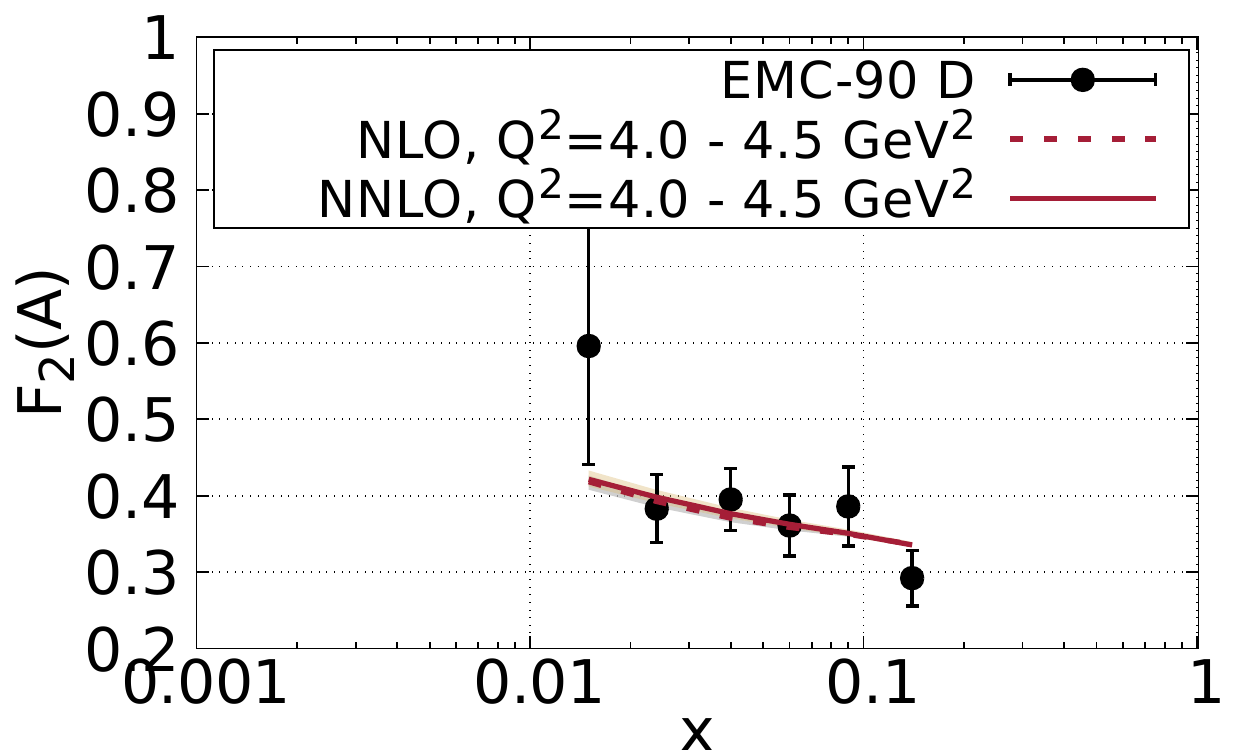}} 
         \subfigure{        
              \includegraphics[width=0.237\textwidth]{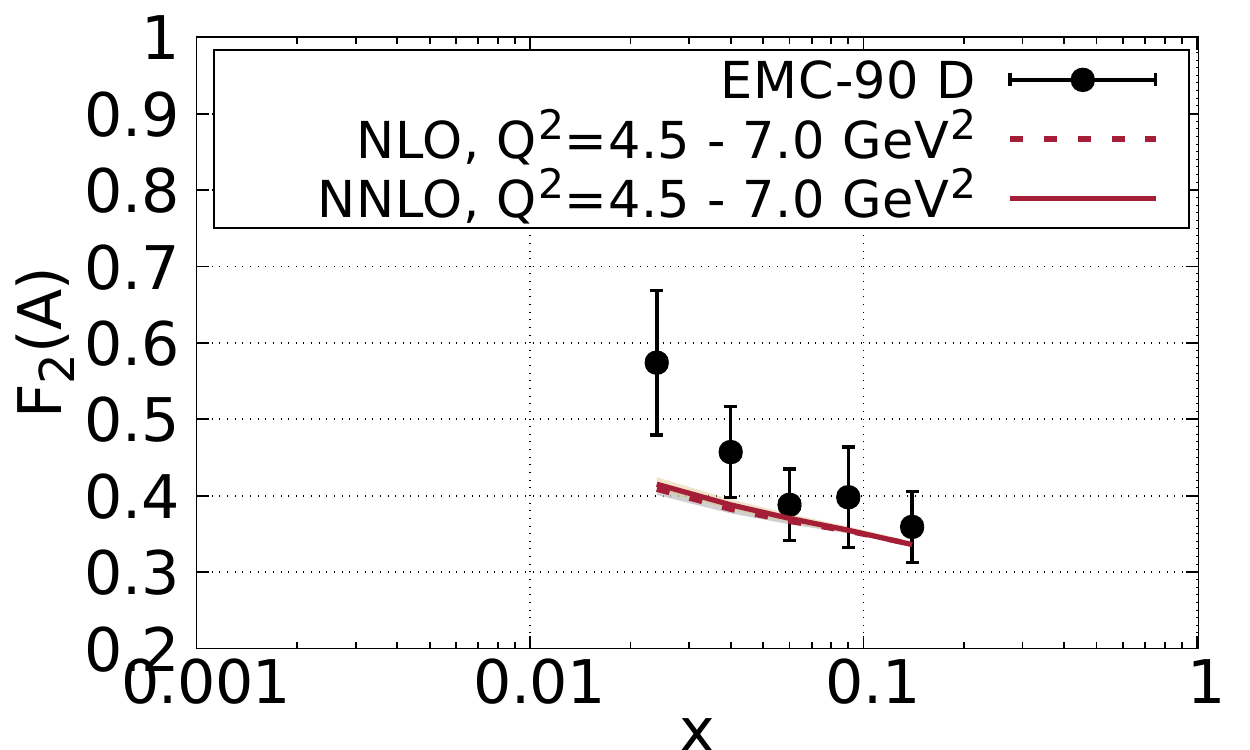}} 
         \subfigure{        
              \includegraphics[width=0.237\textwidth]{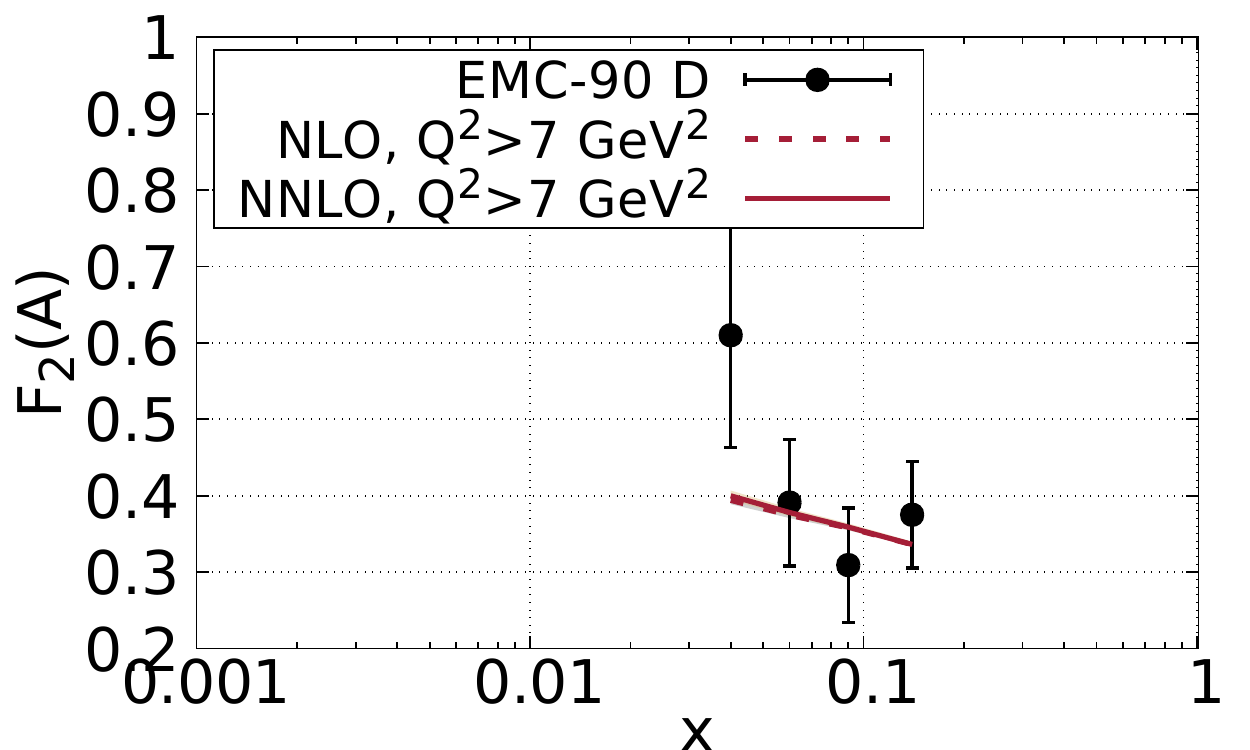}} 
          \subfigure{        
              \includegraphics[width=0.237\textwidth]{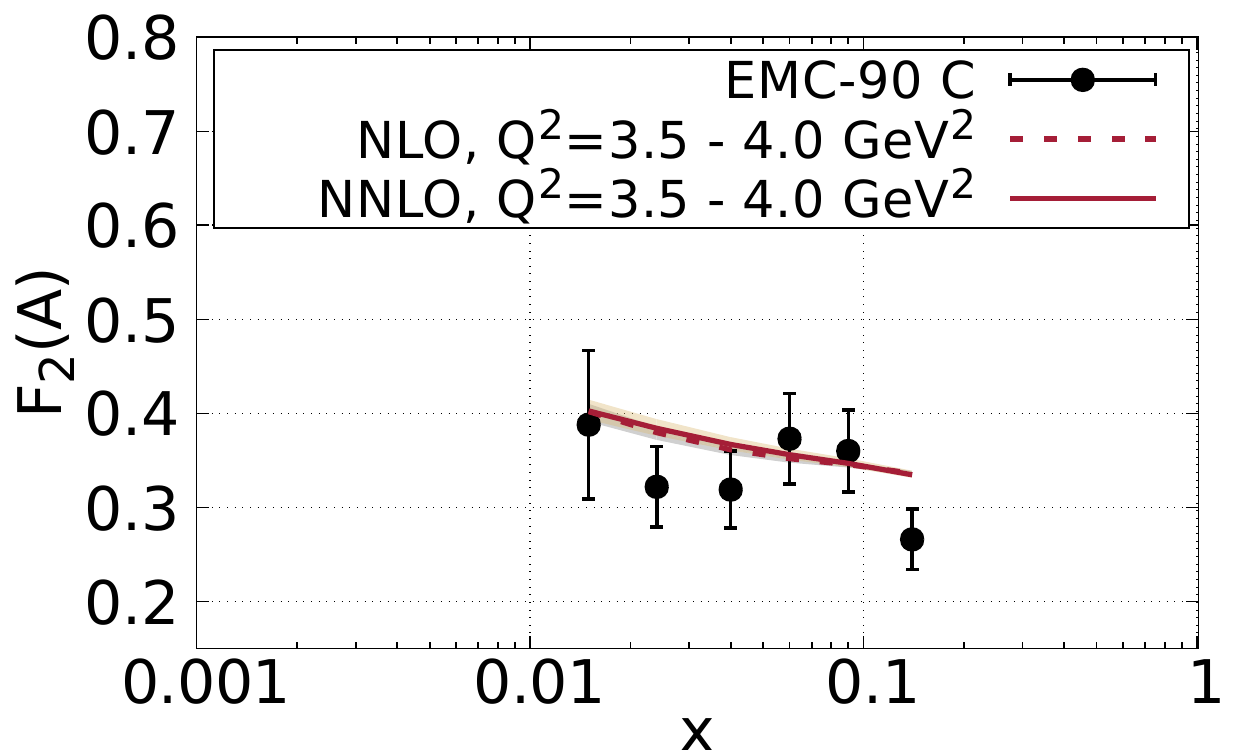}} 
         \subfigure{        
              \includegraphics[width=0.237\textwidth]{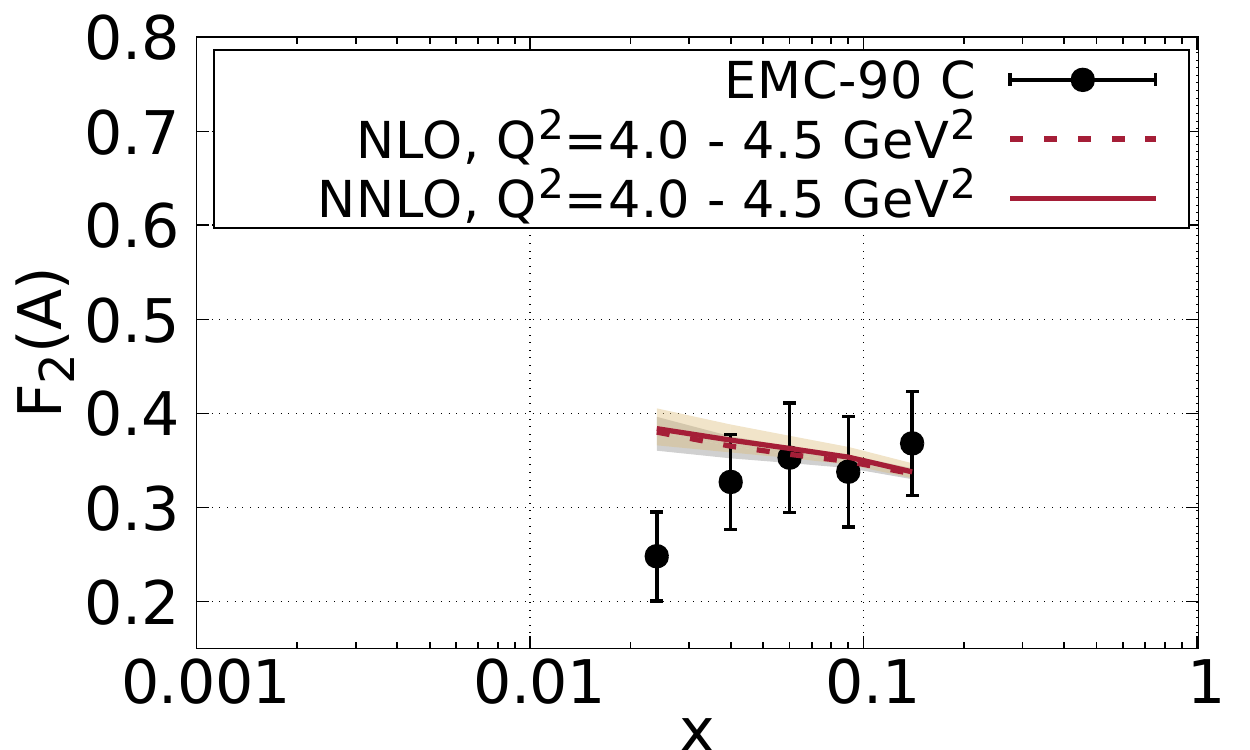}} 
         \subfigure{        
              \includegraphics[width=0.237\textwidth]{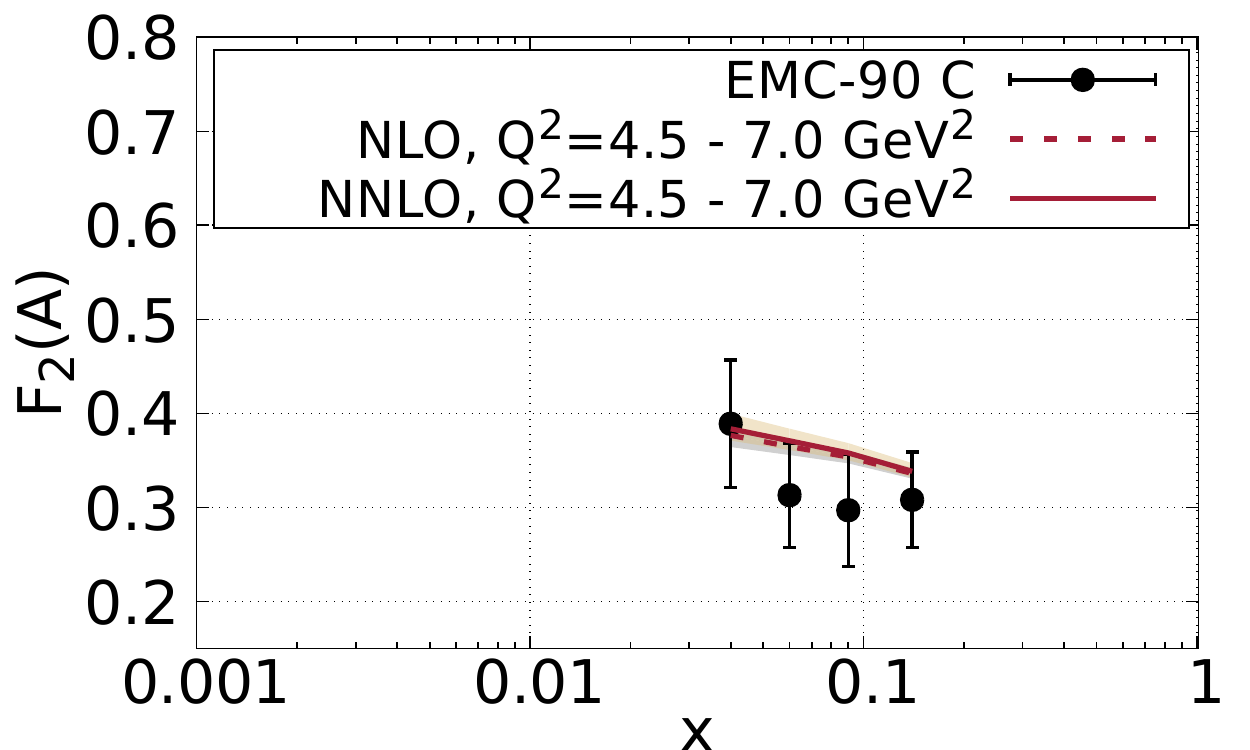}} 
         \subfigure{        
              \includegraphics[width=0.237\textwidth]{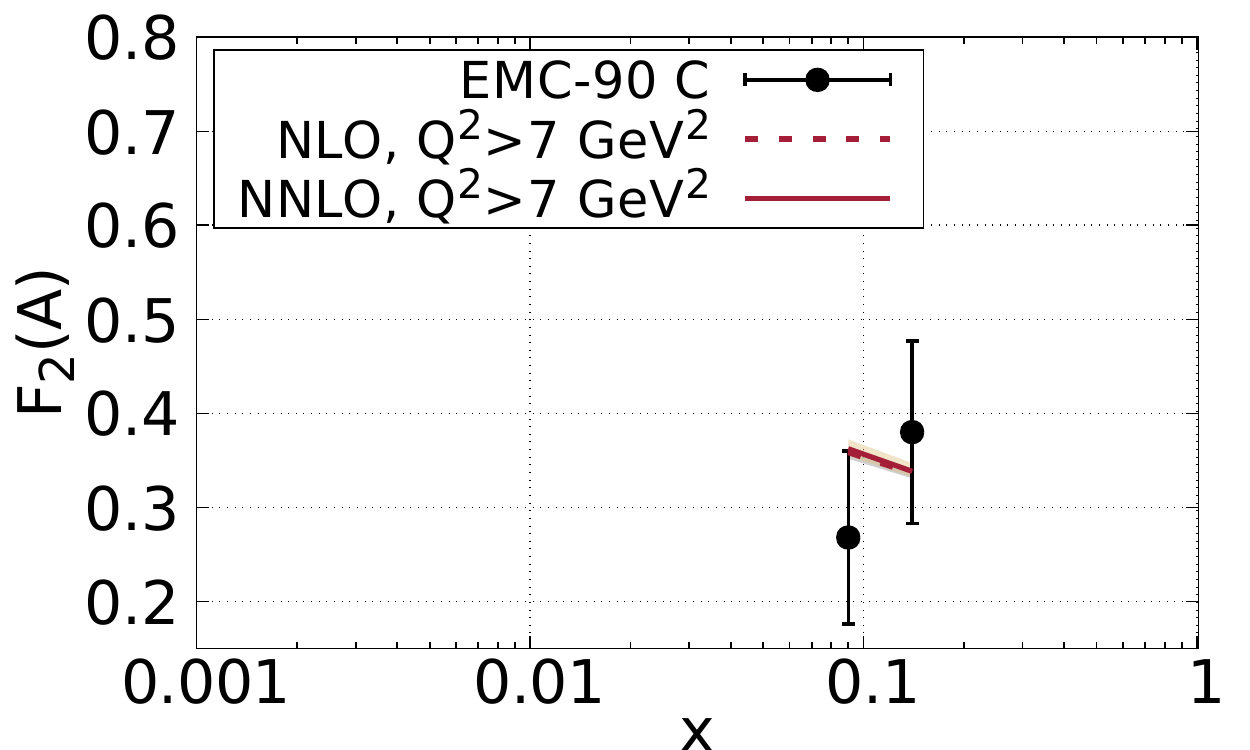}} 
          \subfigure{        
              \includegraphics[width=0.237\textwidth]{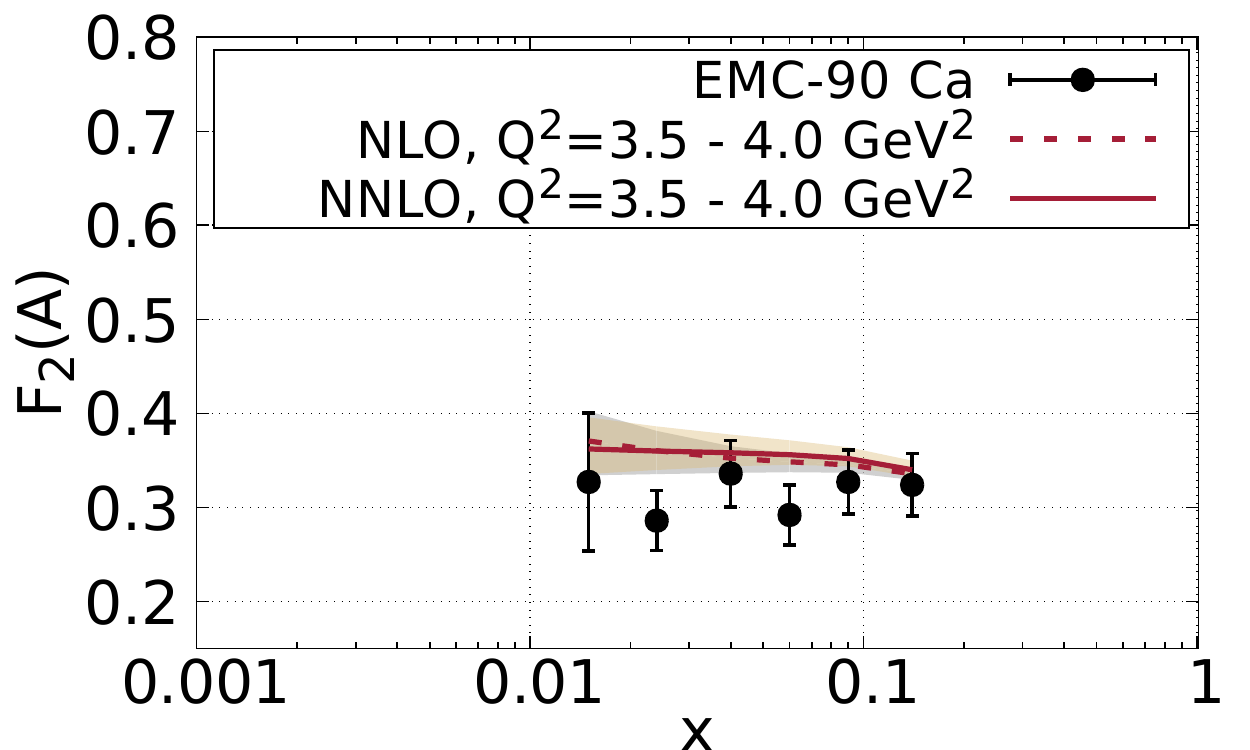}} 
         \subfigure{        
              \includegraphics[width=0.237\textwidth]{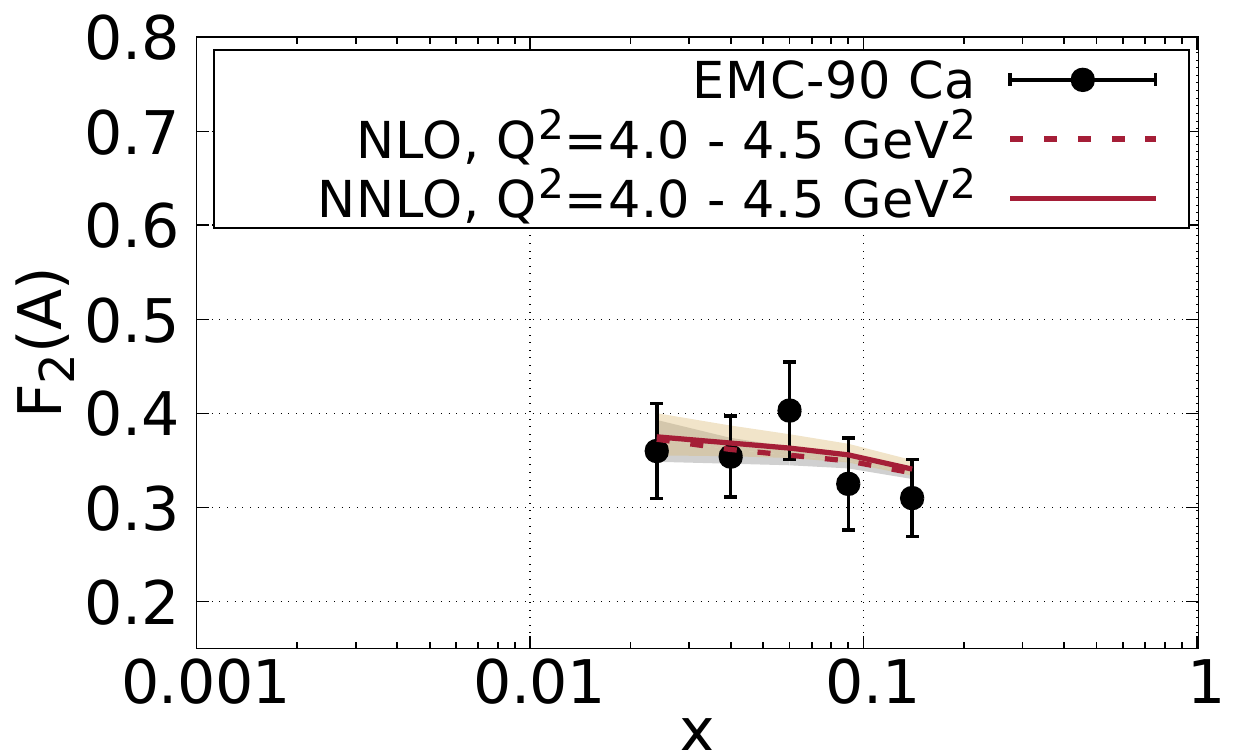}} 
         \subfigure{        
              \includegraphics[width=0.237\textwidth]{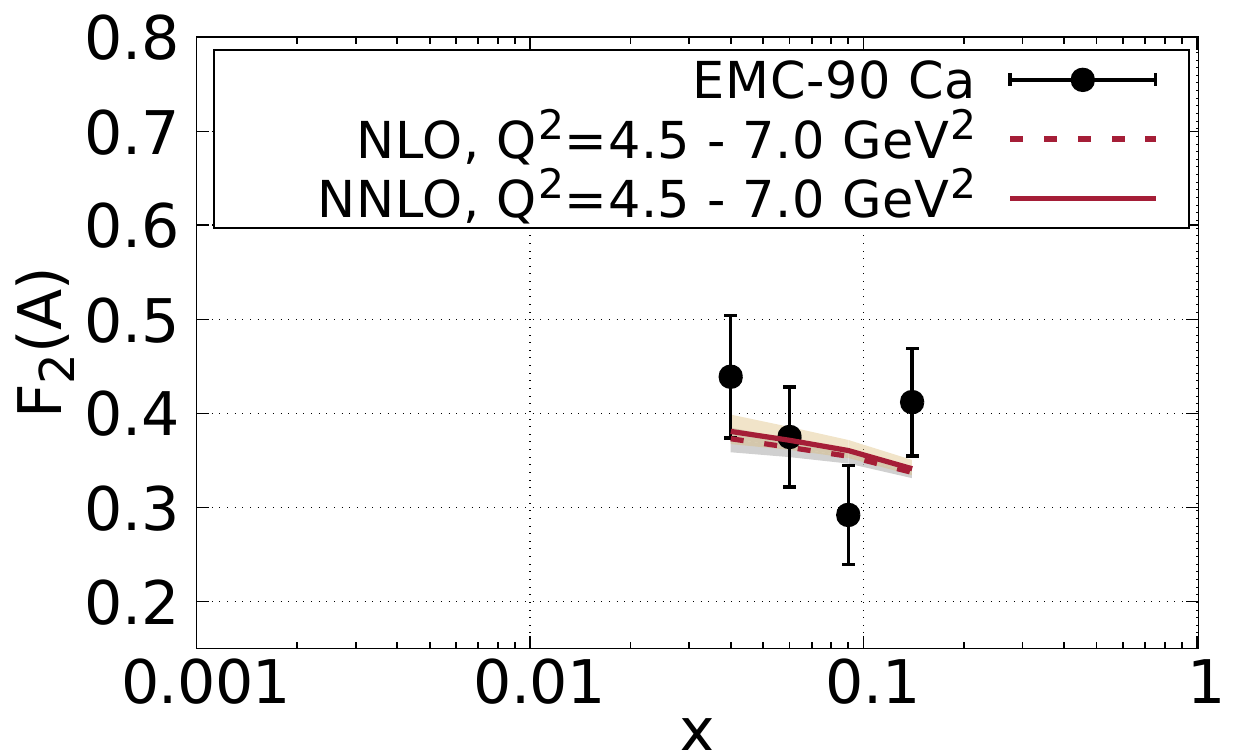}} 
         \subfigure{        
              \includegraphics[width=0.237\textwidth]{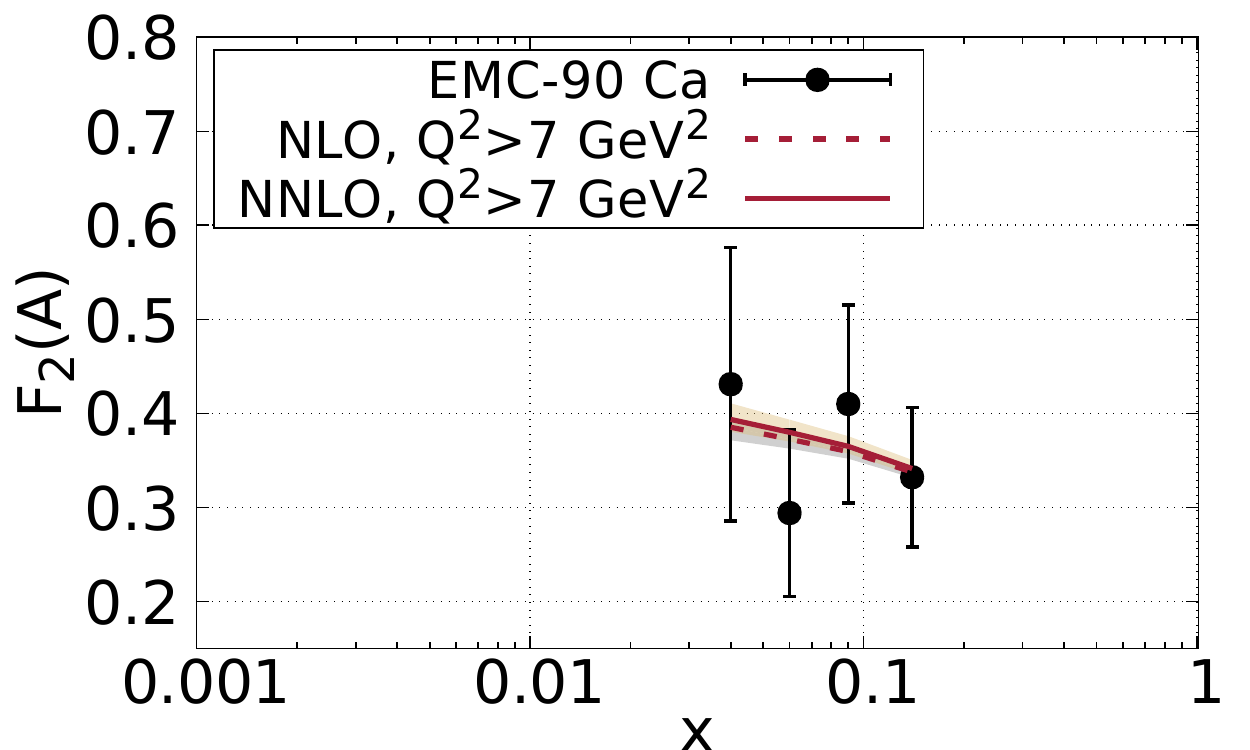}} 
          \subfigure{        
              \includegraphics[width=0.237\textwidth]{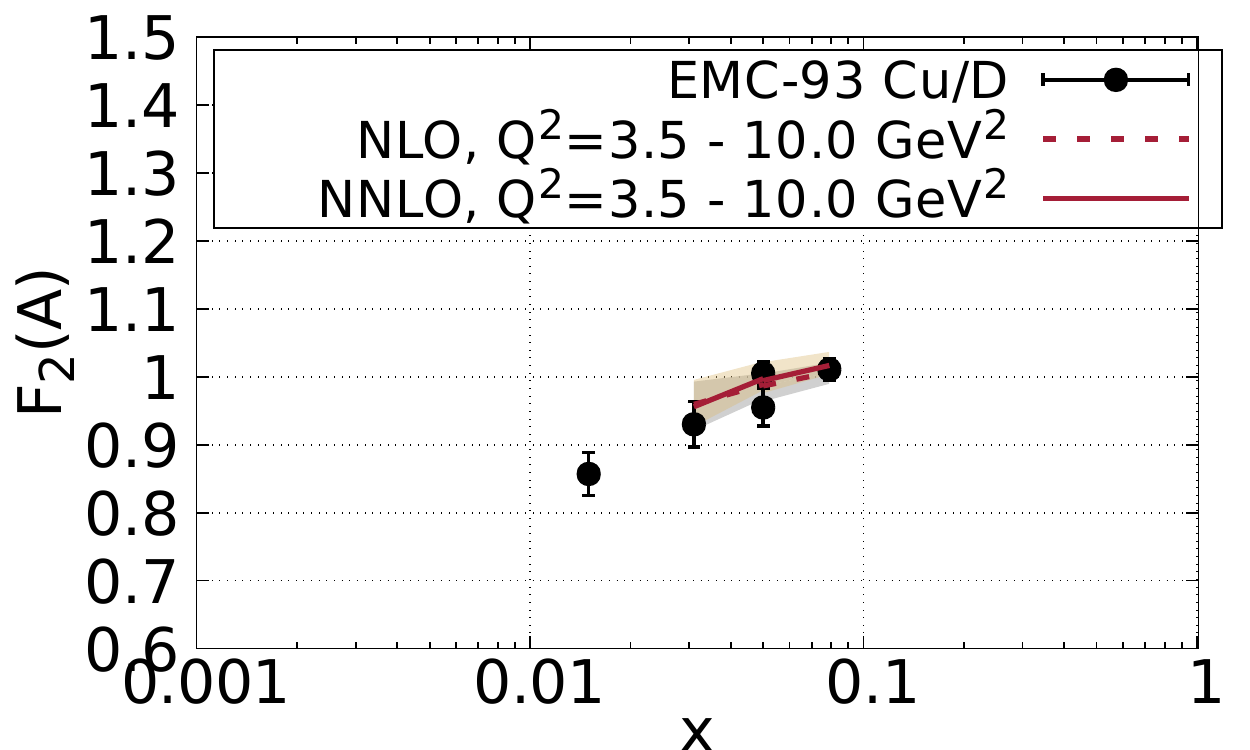}} 
         \subfigure{        
              \includegraphics[width=0.237\textwidth]{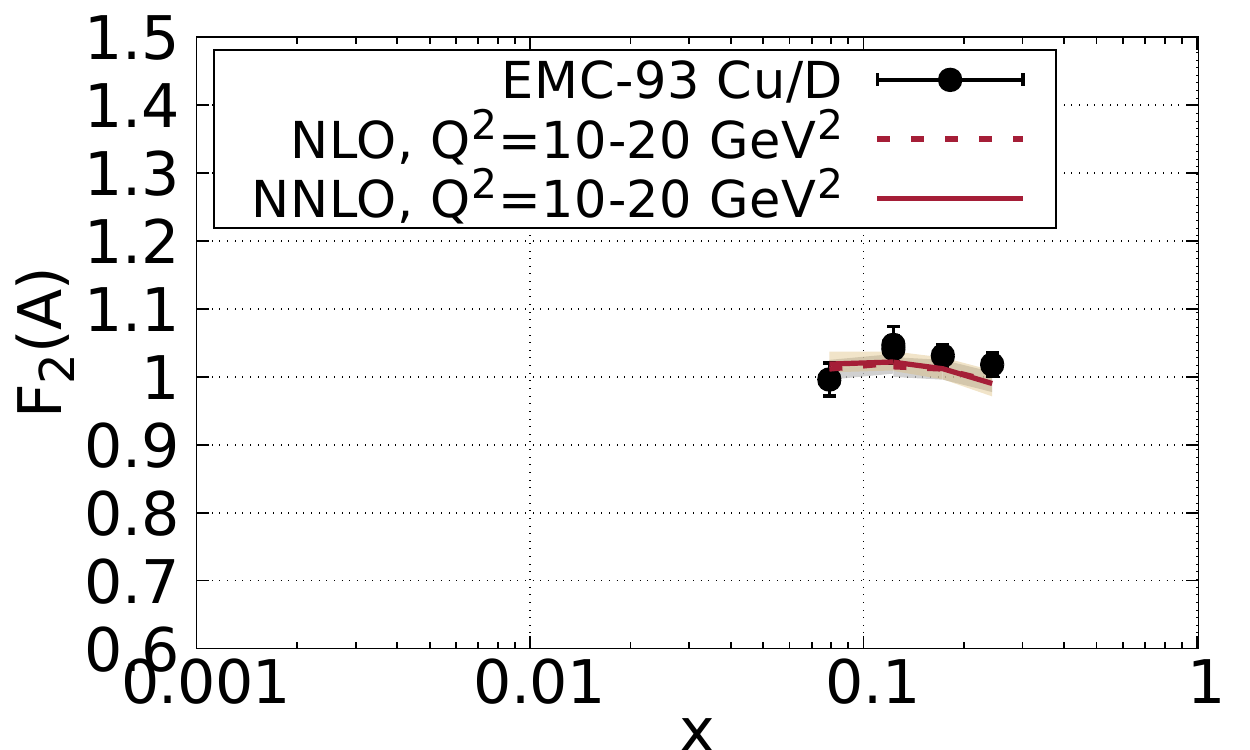}} 
         \subfigure{        
              \includegraphics[width=0.237\textwidth]{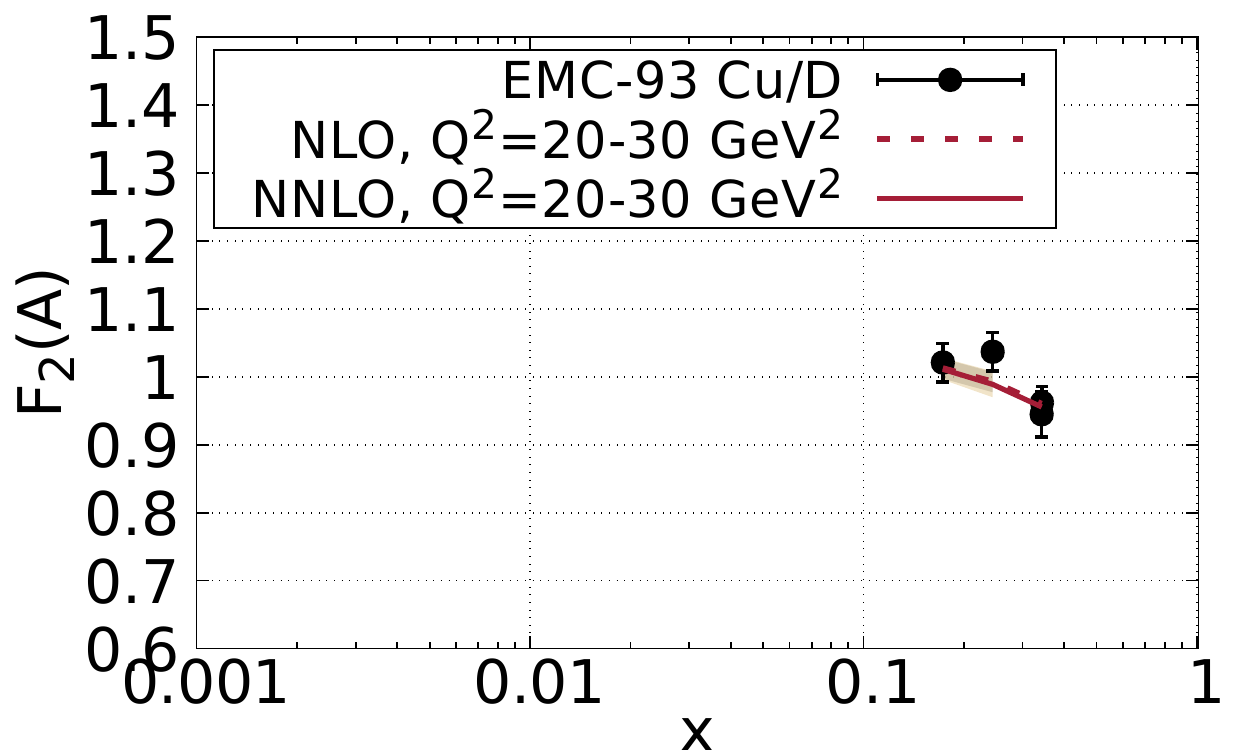}} 
         \subfigure{        
              \includegraphics[width=0.237\textwidth]{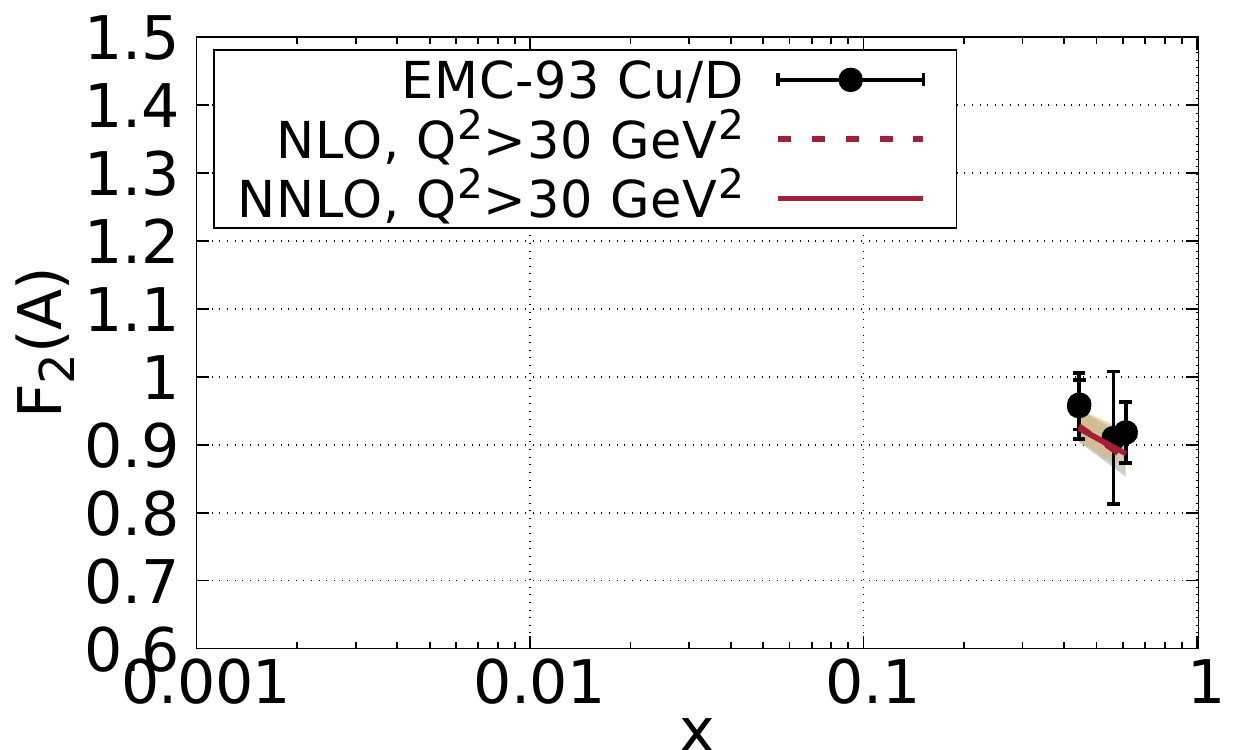}} 
          \subfigure{        
              \includegraphics[width=0.237\textwidth]{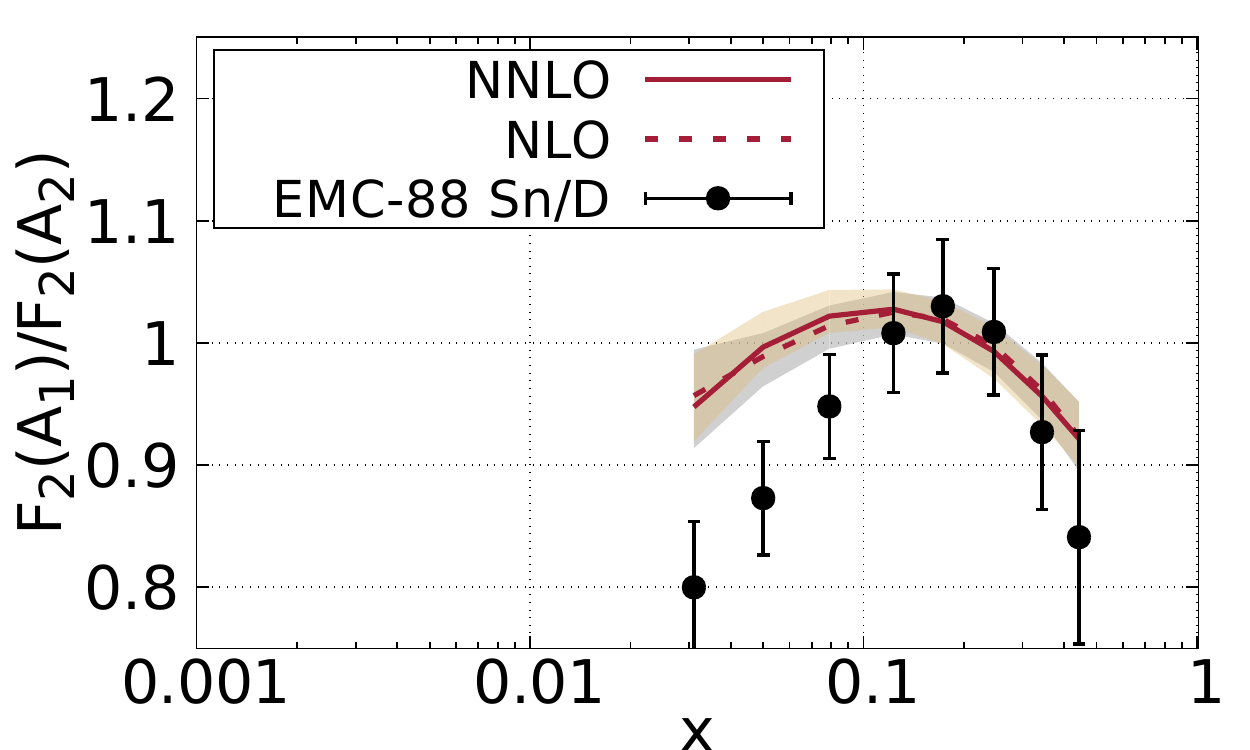}}
          \subfigure{        
              \includegraphics[width=0.237\textwidth]{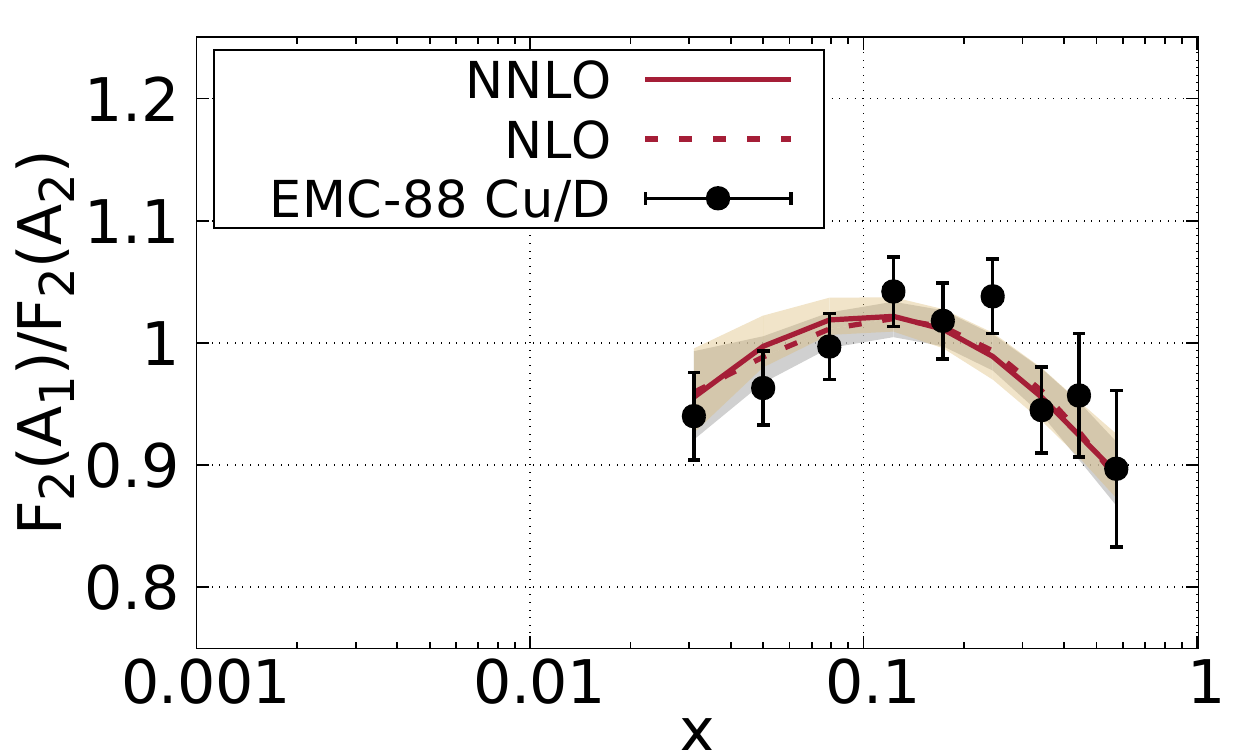}}                                          
          \subfigure{        
              \includegraphics[width=0.237\textwidth]{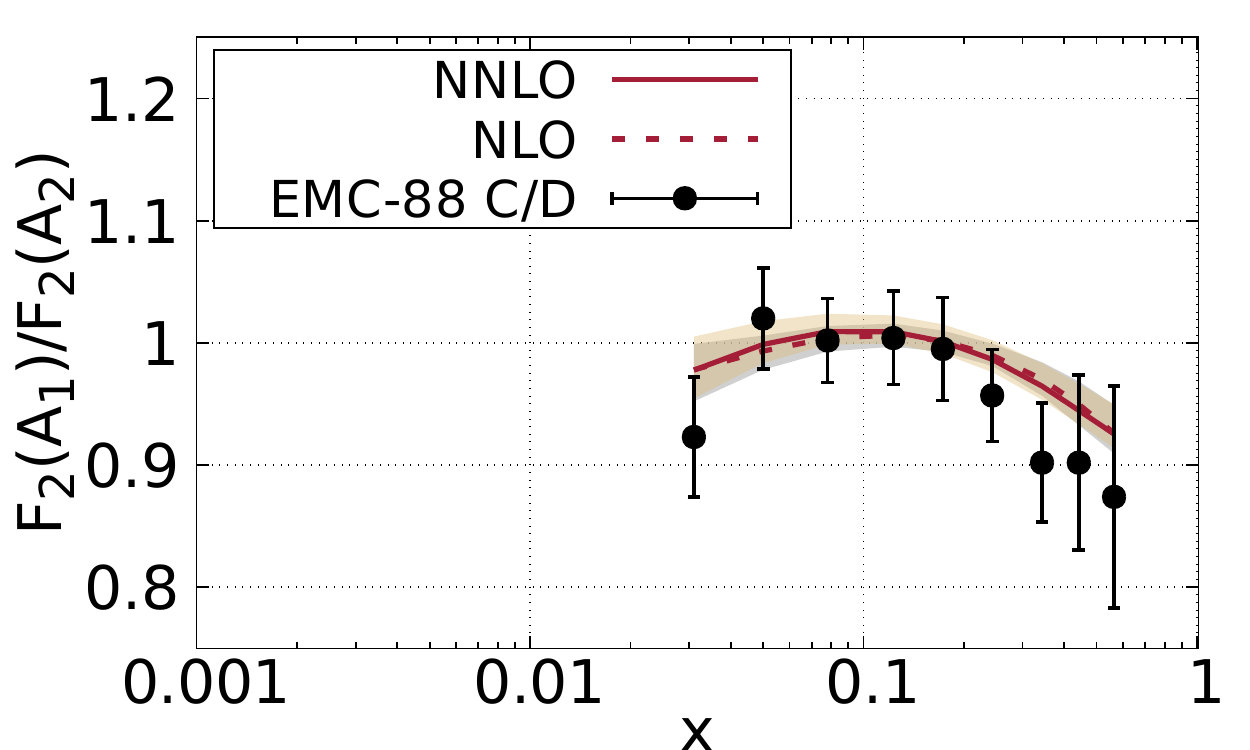}}               
          \end{center} 
    \caption{Comparison to EMC data, first for the structure function $F_2$ at different $Q^2$, and then for different ratios $F_2(A_1)/F_2(A_2)$ measured for nuclei with mass numbers $A_1$ and $A_2$. The calculated quantities are shown at NLO (dashed line, grey error bands) and NNLO (solid line, golden-coloured error bands).}
\label{figEMC}    
    \end{figure*}

\begin{figure*}[htb!]
     \begin{center}
          \subfigure{        
              \includegraphics[width=0.237\textwidth]{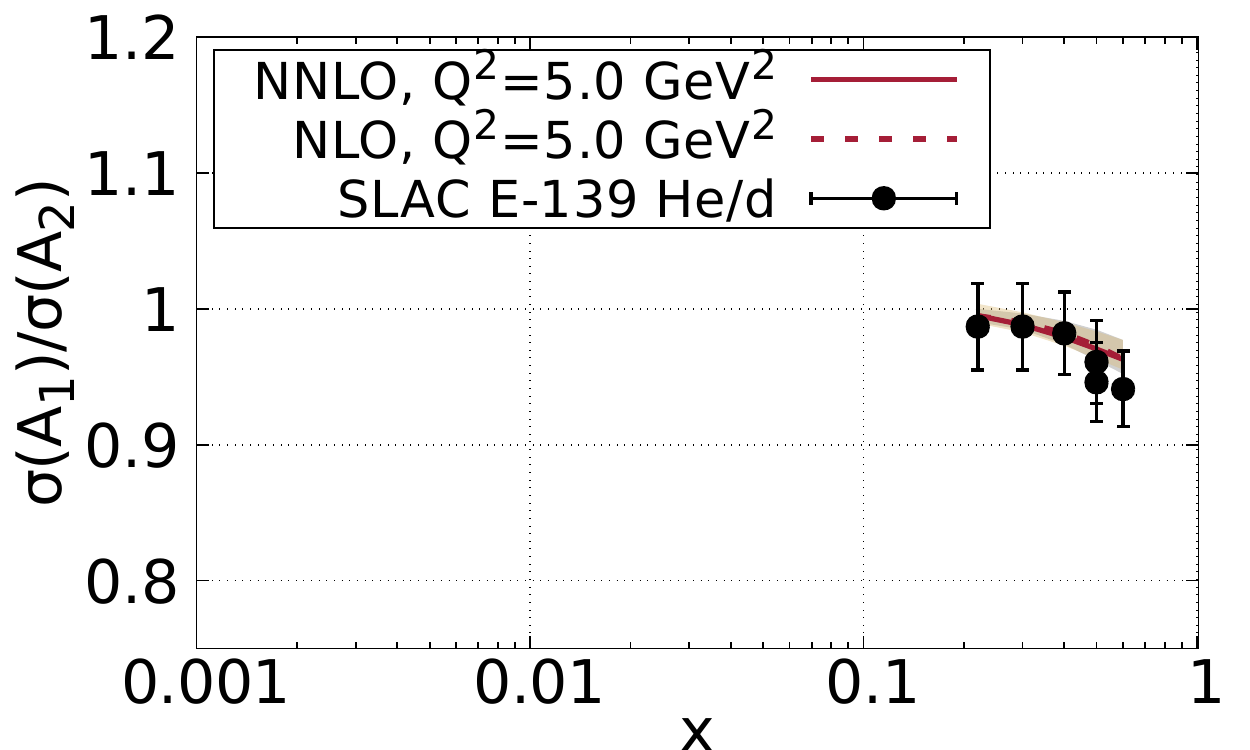}} 
         \subfigure{        
              \includegraphics[width=0.237\textwidth]{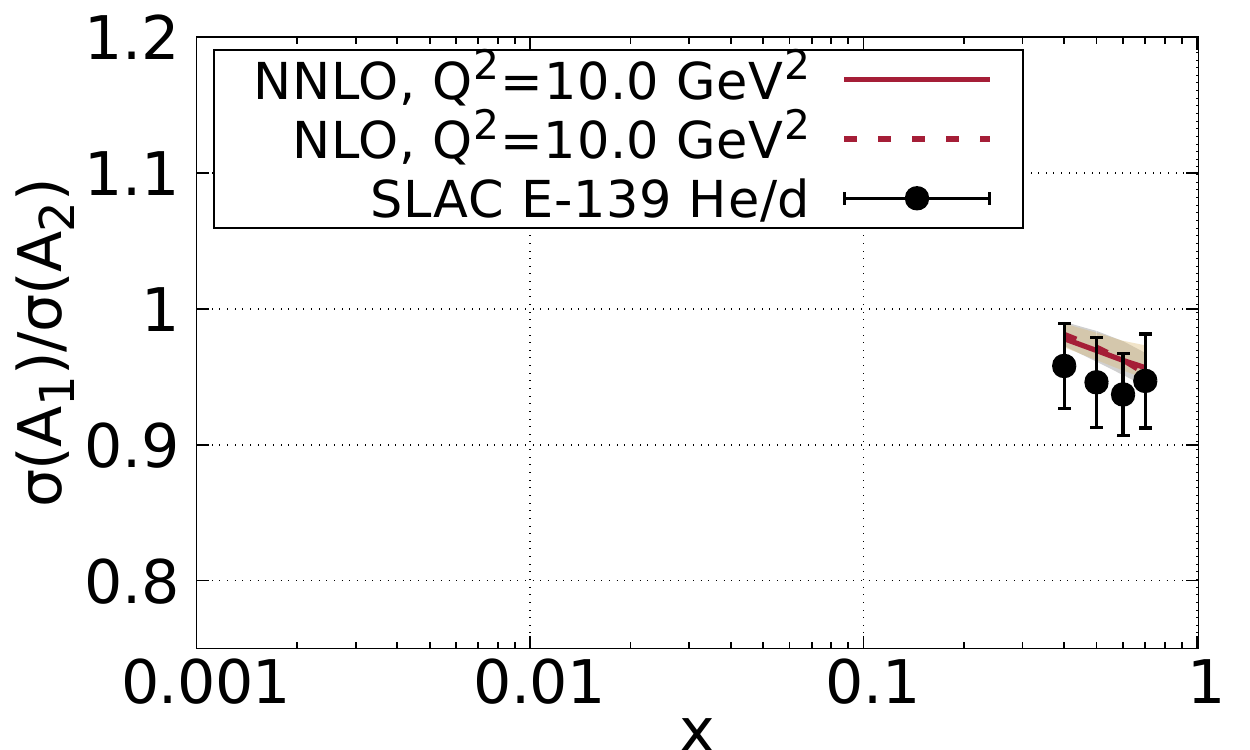}} 
         \subfigure{        
              \includegraphics[width=0.237\textwidth]{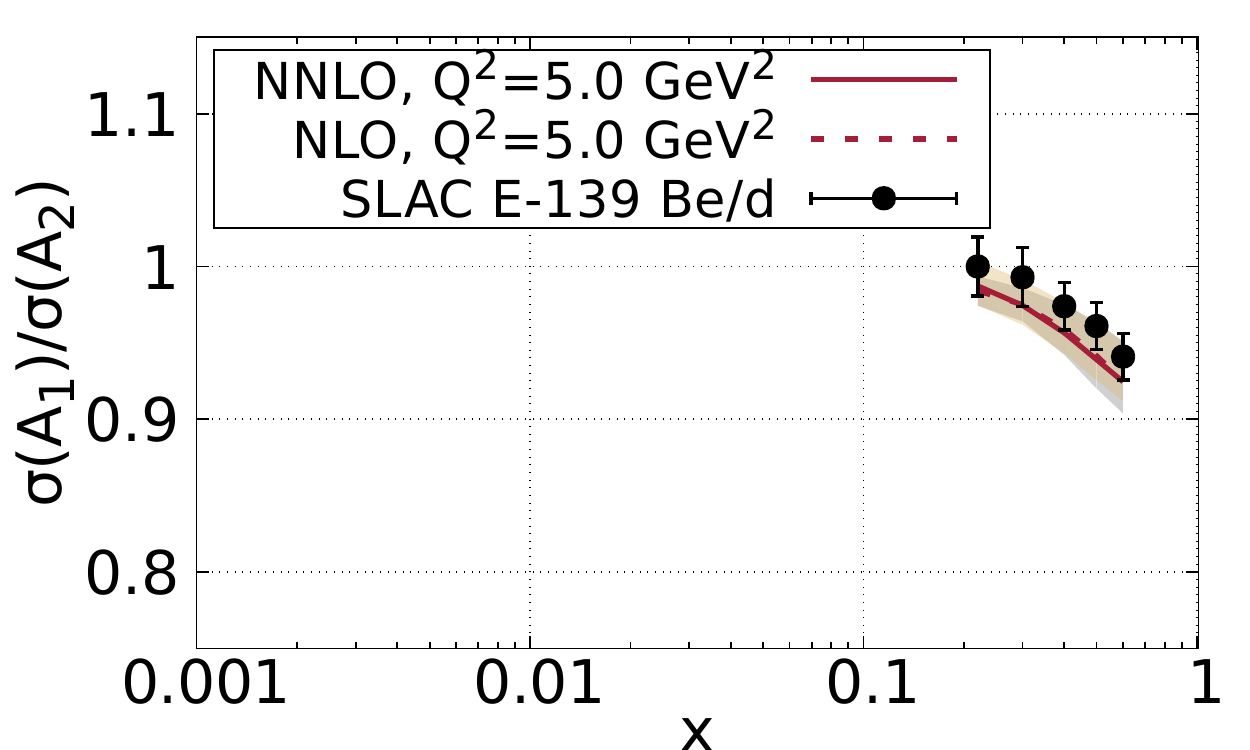}} 
         \subfigure{        
              \includegraphics[width=0.237\textwidth]{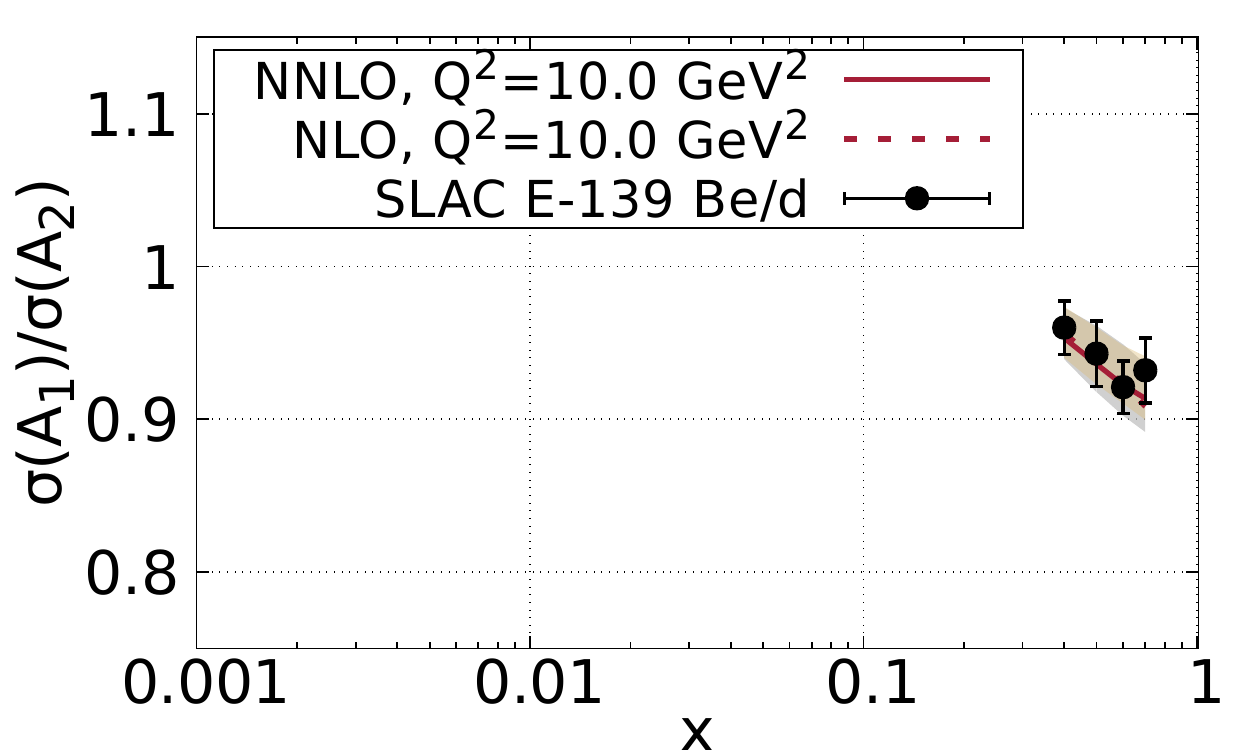}} 
         \subfigure{        
              \includegraphics[width=0.237\textwidth]{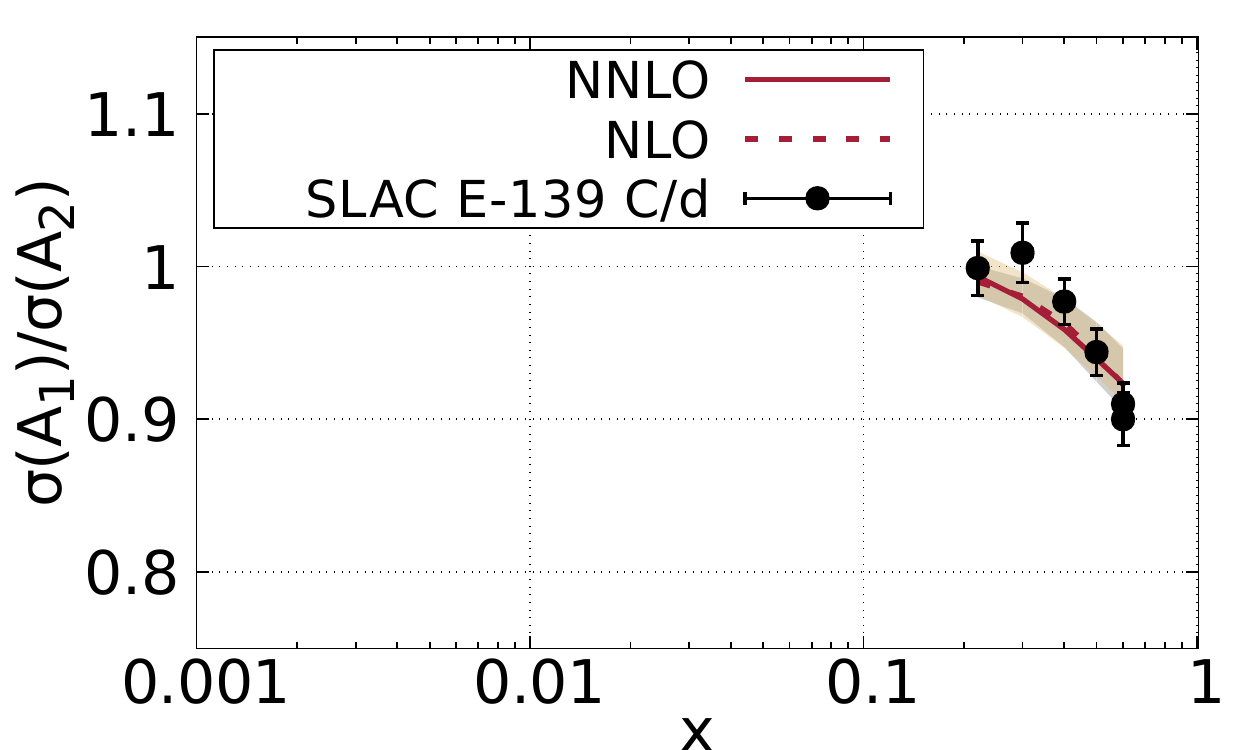}} 
         \subfigure{        
              \includegraphics[width=0.237\textwidth]{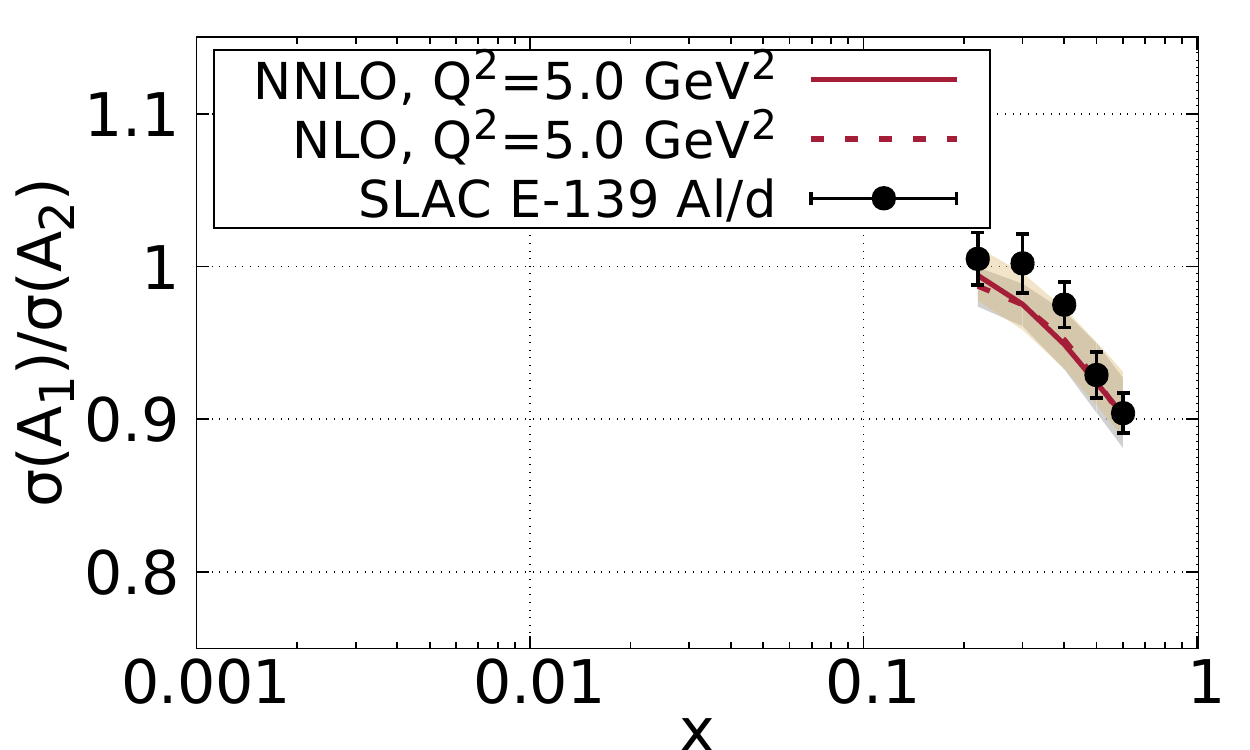}}                             
         \subfigure{        
              \includegraphics[width=0.237\textwidth]{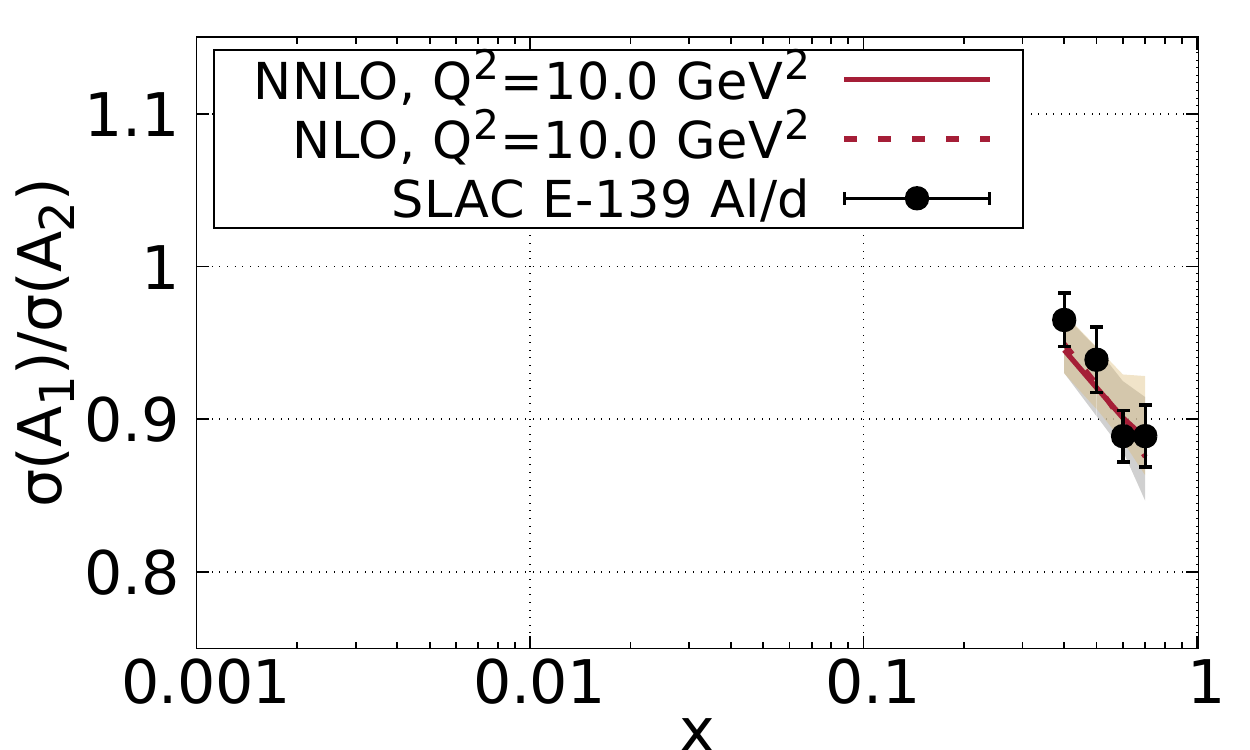}}                             
         \subfigure{        
              \includegraphics[width=0.237\textwidth]{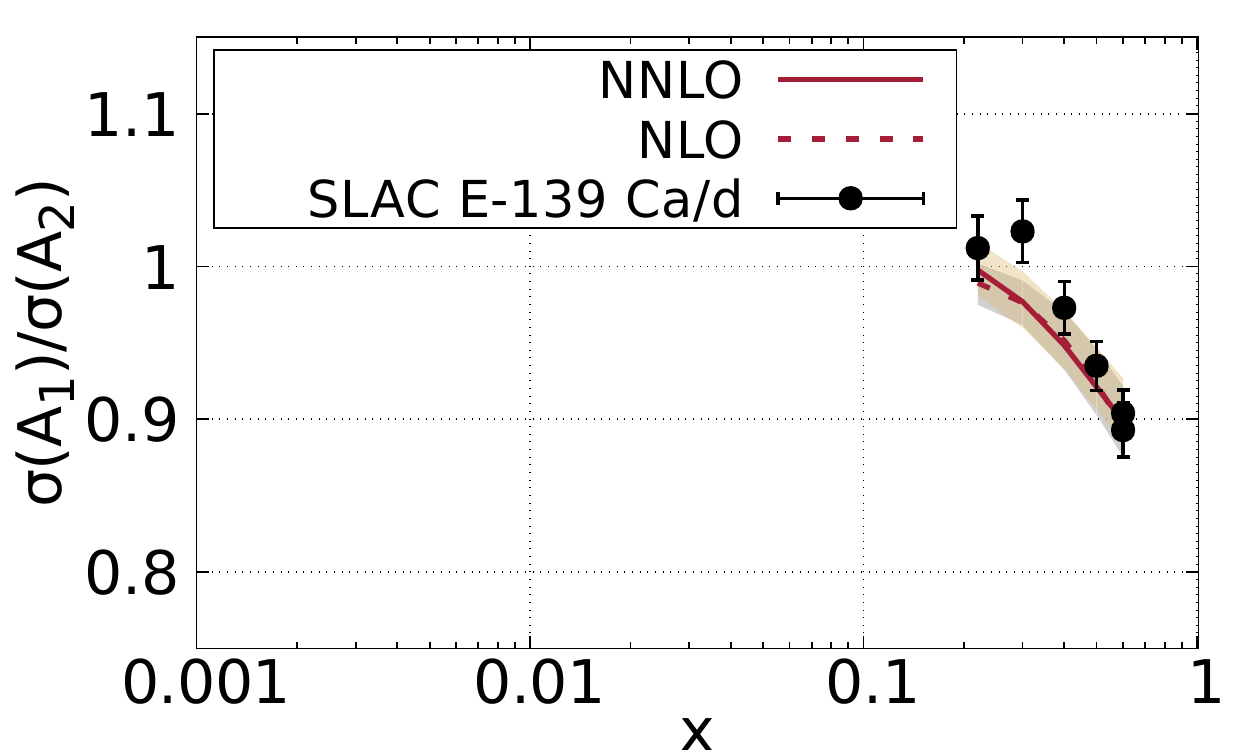}}                             
         \subfigure{        
              \includegraphics[width=0.237\textwidth]{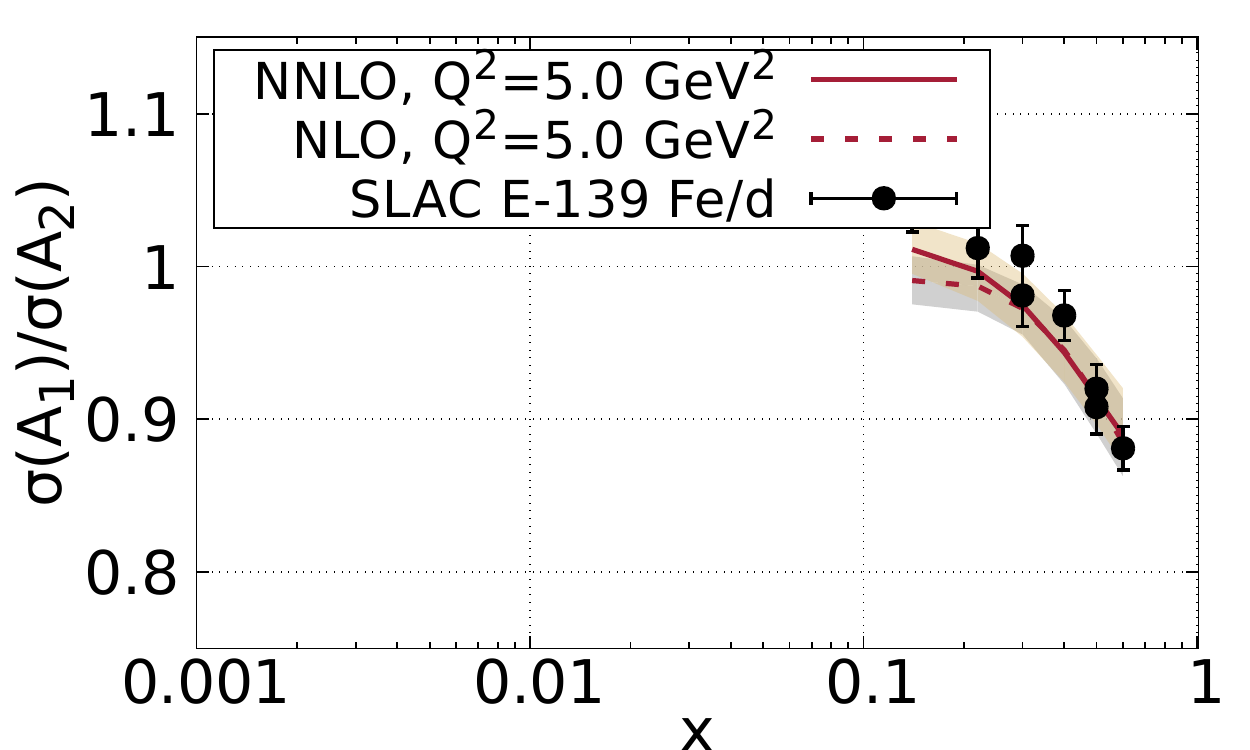}}                             
         \subfigure{        
              \includegraphics[width=0.237\textwidth]{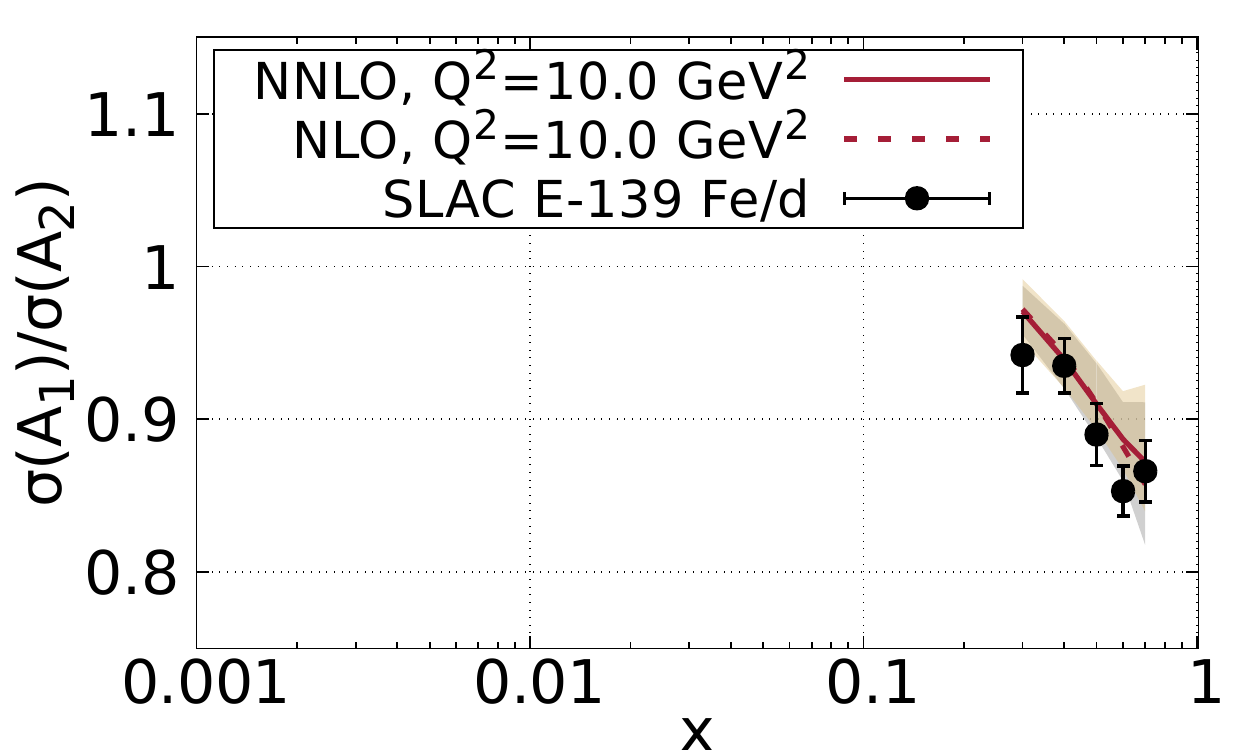}}                             
         \subfigure{        
              \includegraphics[width=0.237\textwidth]{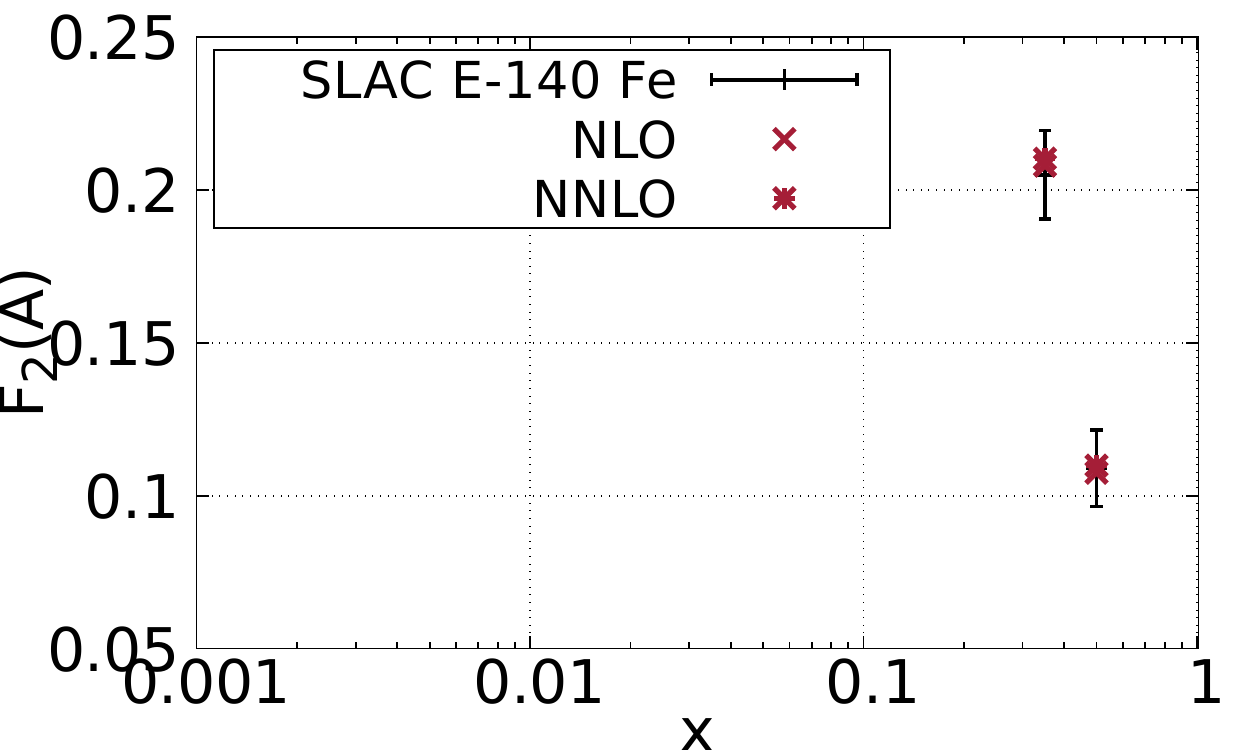}}                             
         \subfigure{        
              \includegraphics[width=0.237\textwidth]{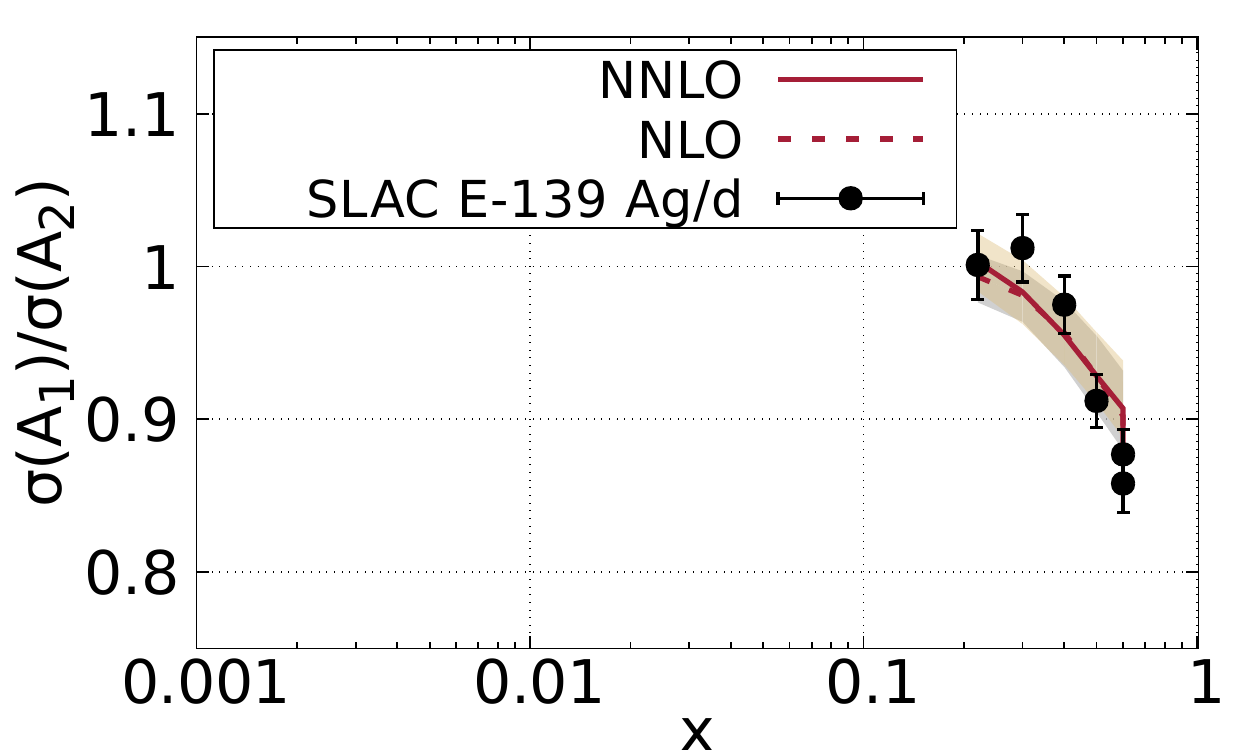}}                             
         \subfigure{        
              \includegraphics[width=0.237\textwidth]{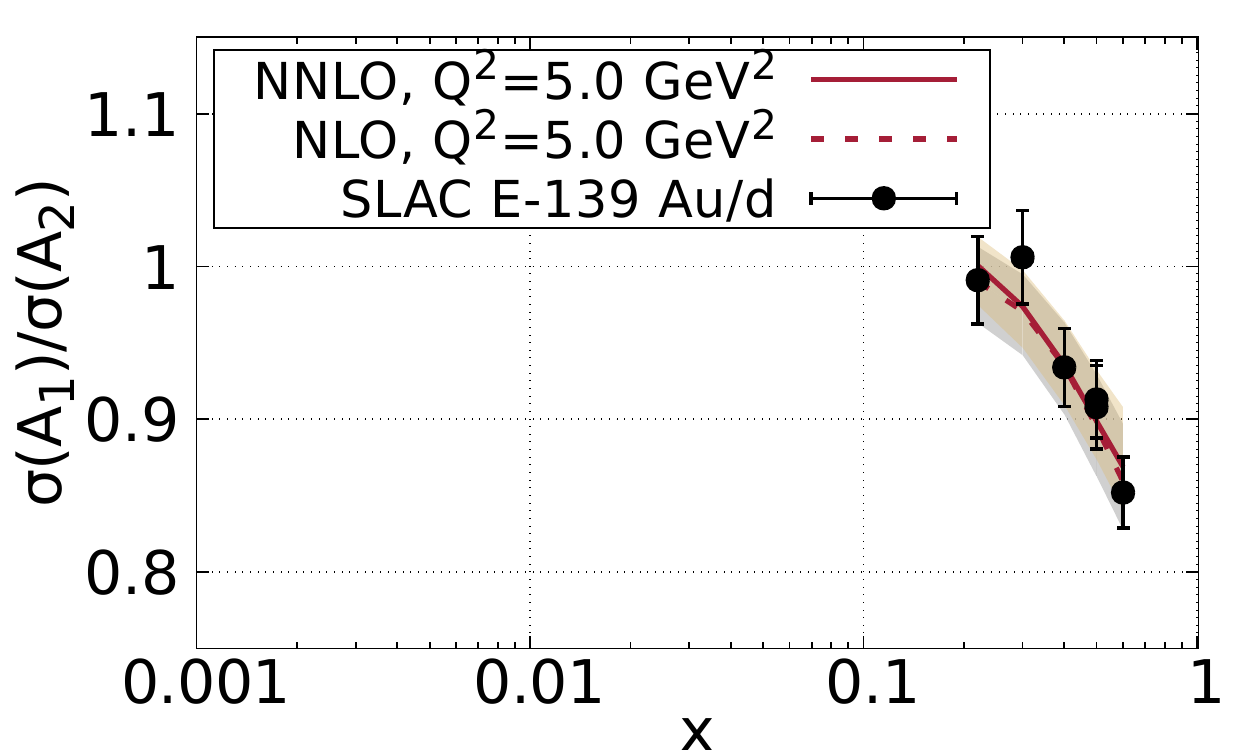}}                             
         \subfigure{        
              \includegraphics[width=0.237\textwidth]{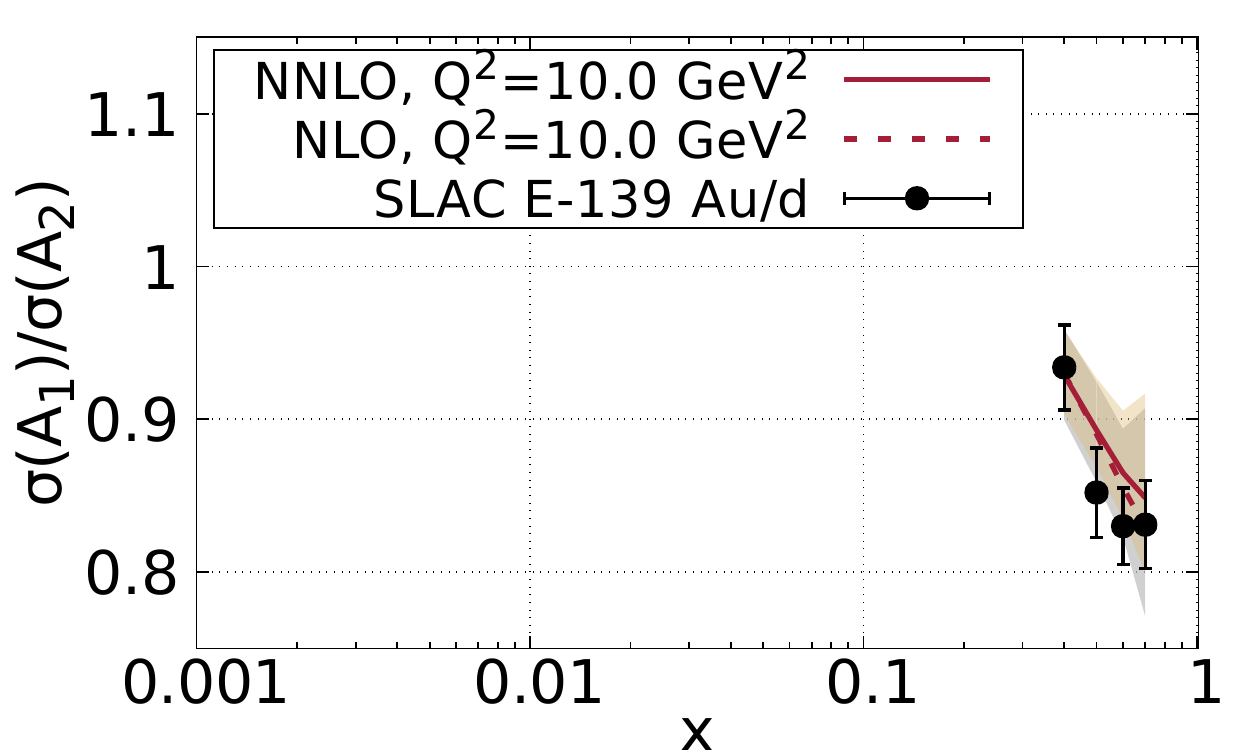}}                             
          \end{center} 
    \caption{Comparison to SLAC data for different ratios of reduced differential cross sections $\sigma(A_1)/\sigma(A_2)$ for nuclei with mass numbers $A_1$ and $A_2$, at different values of $Q^2$ at NLO (dashed line, grey error bands) and NNLO (solid line, golden-coloured error bands).}
\label{figSLAC}    
    \end{figure*}

\begin{figure*}[htb!]
     \begin{center}
          \subfigure{        
              \includegraphics[width=0.237\textwidth]{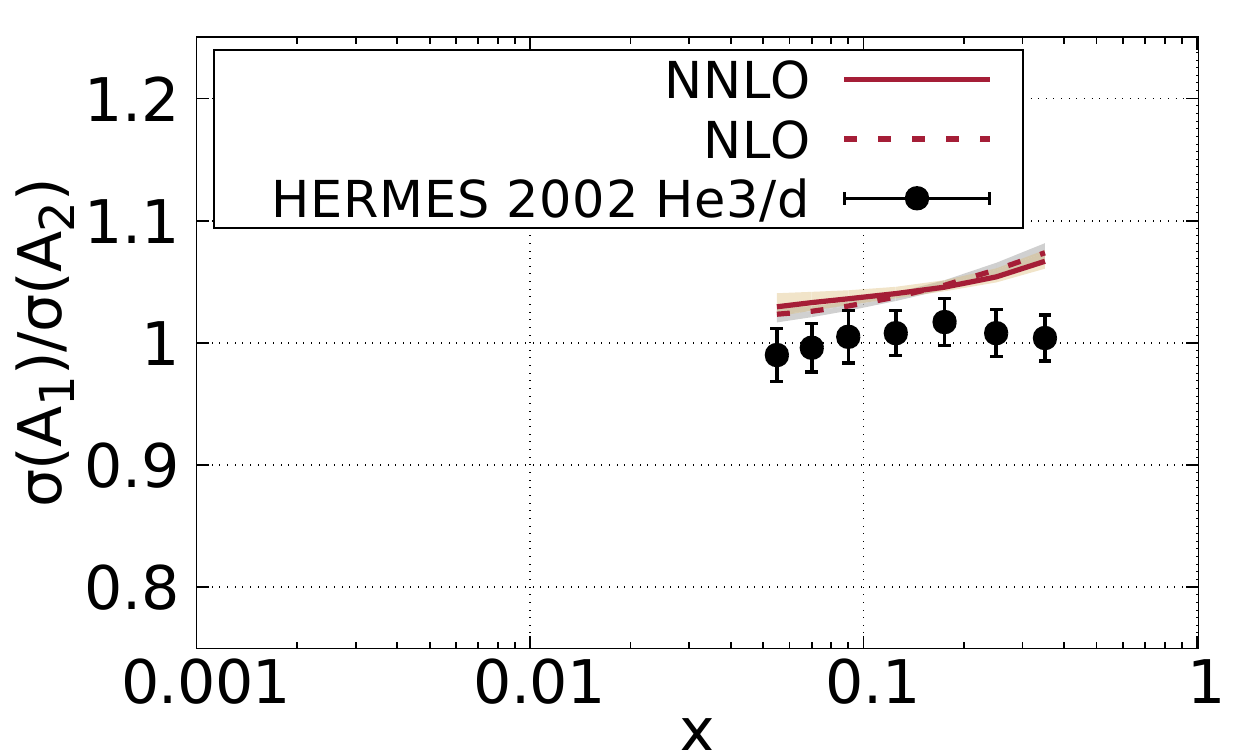}} 
         \subfigure{        
              \includegraphics[width=0.237\textwidth]{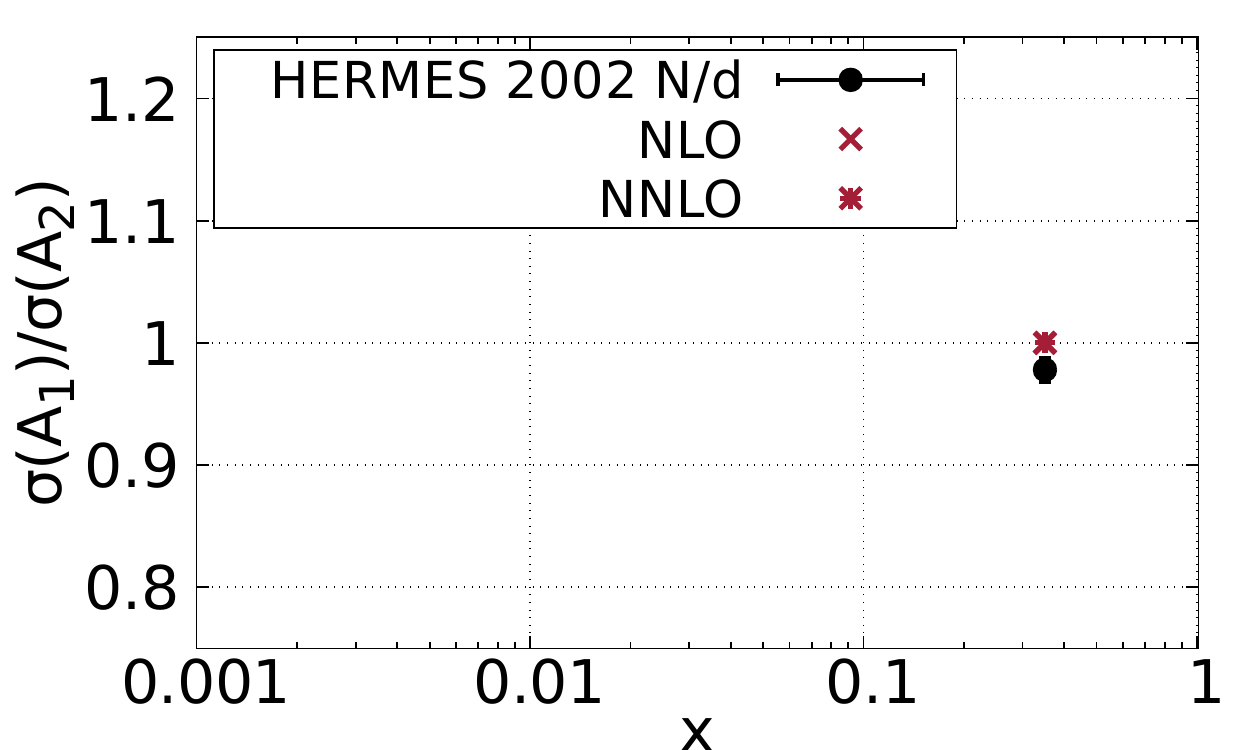}} 
         \subfigure{        
              \includegraphics[width=0.237\textwidth]{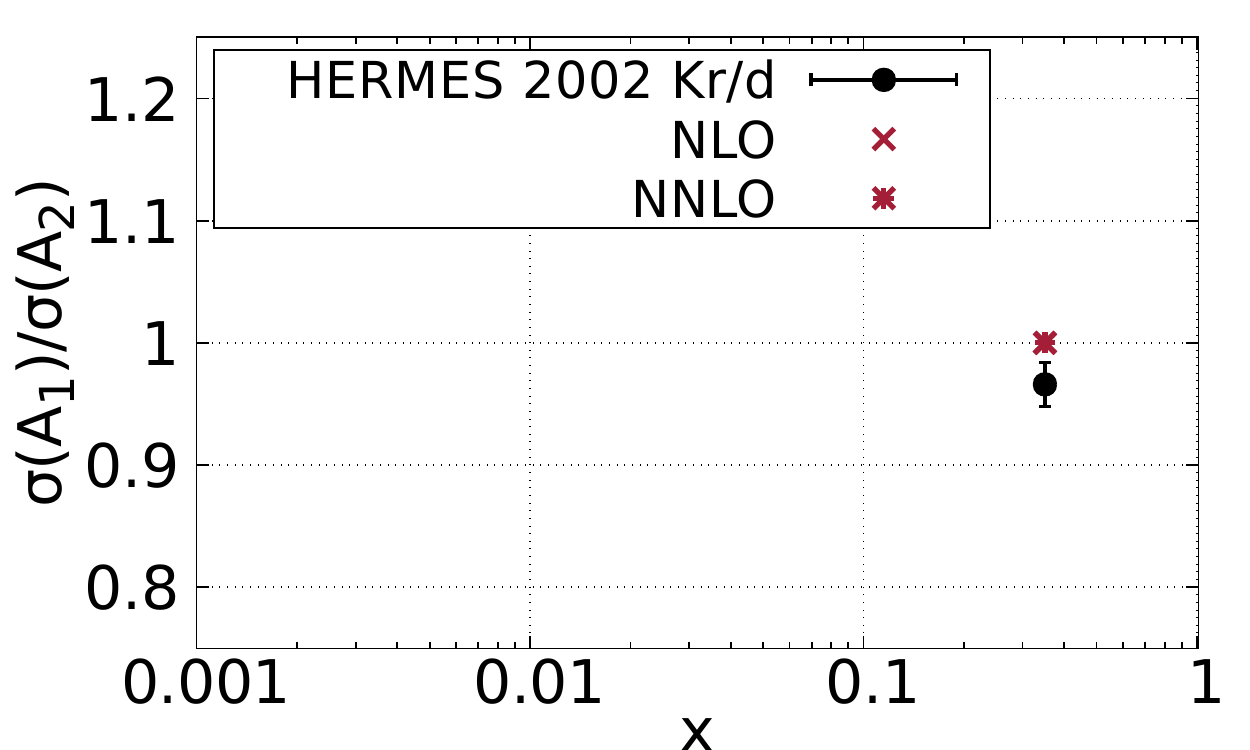}} 
          \end{center} 
    \caption{Comparison to HERMES data for different ratios of reduced differential cross sections $\sigma(A_1)/\sigma(A_2)$ for nuclei with mass numbers $A_1$ and $A_2$, at NLO (dashed line, grey error bands) and NNLO (solid line, golden-coloured error bands).}
\label{figHERMES}    
    \end{figure*}

\begin{figure*}[htb!]
     \begin{center}
          \subfigure{\includegraphics[width=0.325\textwidth]{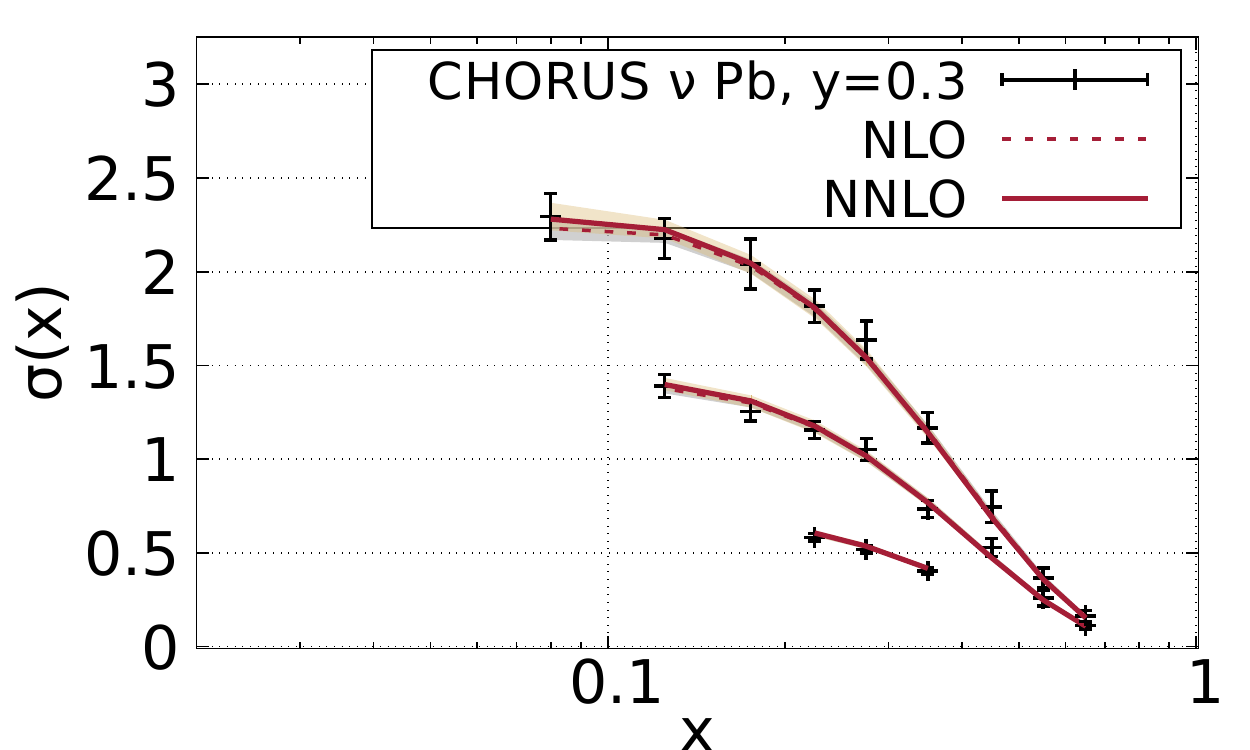}}
          \subfigure{\includegraphics[width=0.325\textwidth]{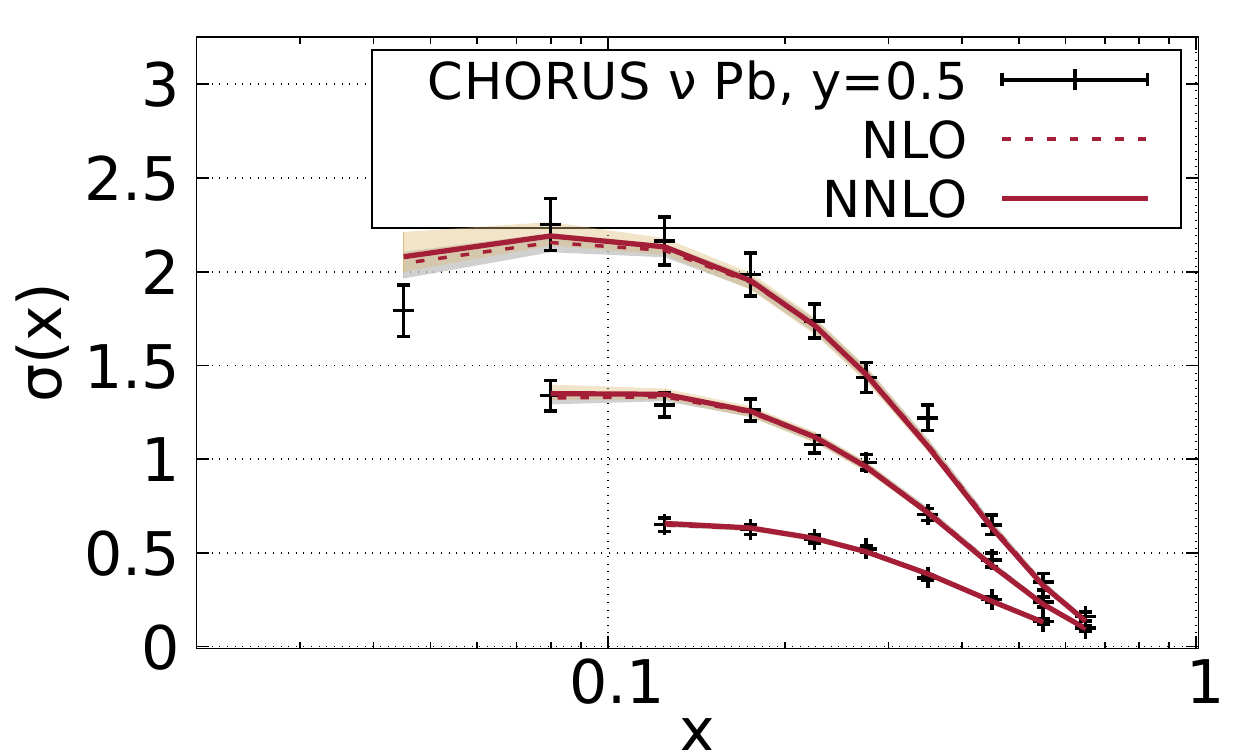}} 
          \subfigure{\includegraphics[width=0.325\textwidth]{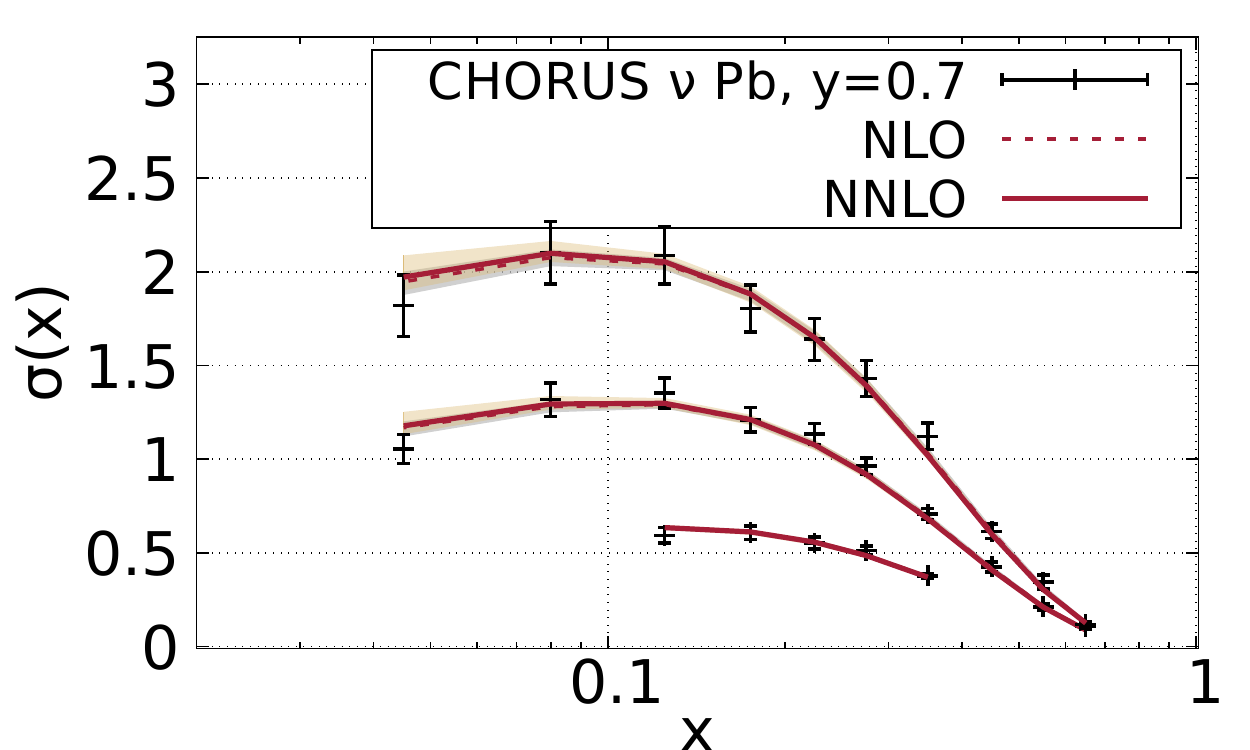}}                      
          \subfigure{\includegraphics[width=0.325\textwidth]{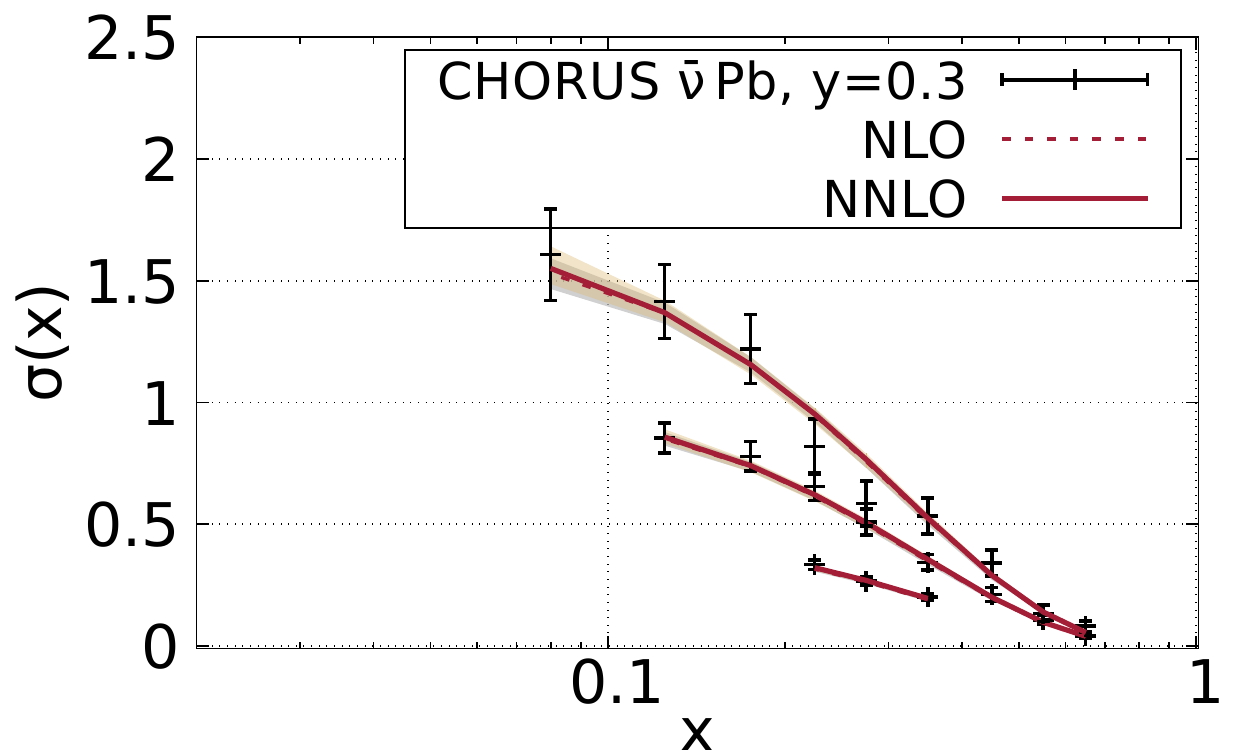}}
          \subfigure{\includegraphics[width=0.325\textwidth]{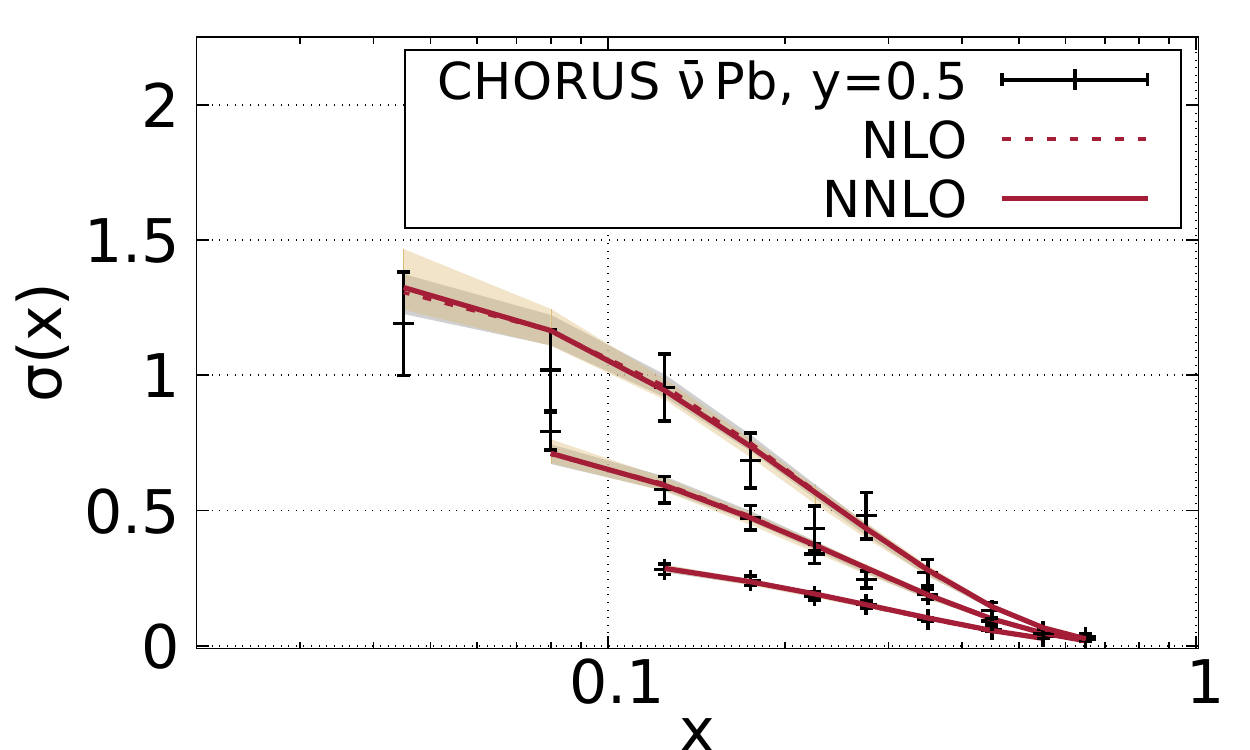}} 
          \subfigure{\includegraphics[width=0.325\textwidth]{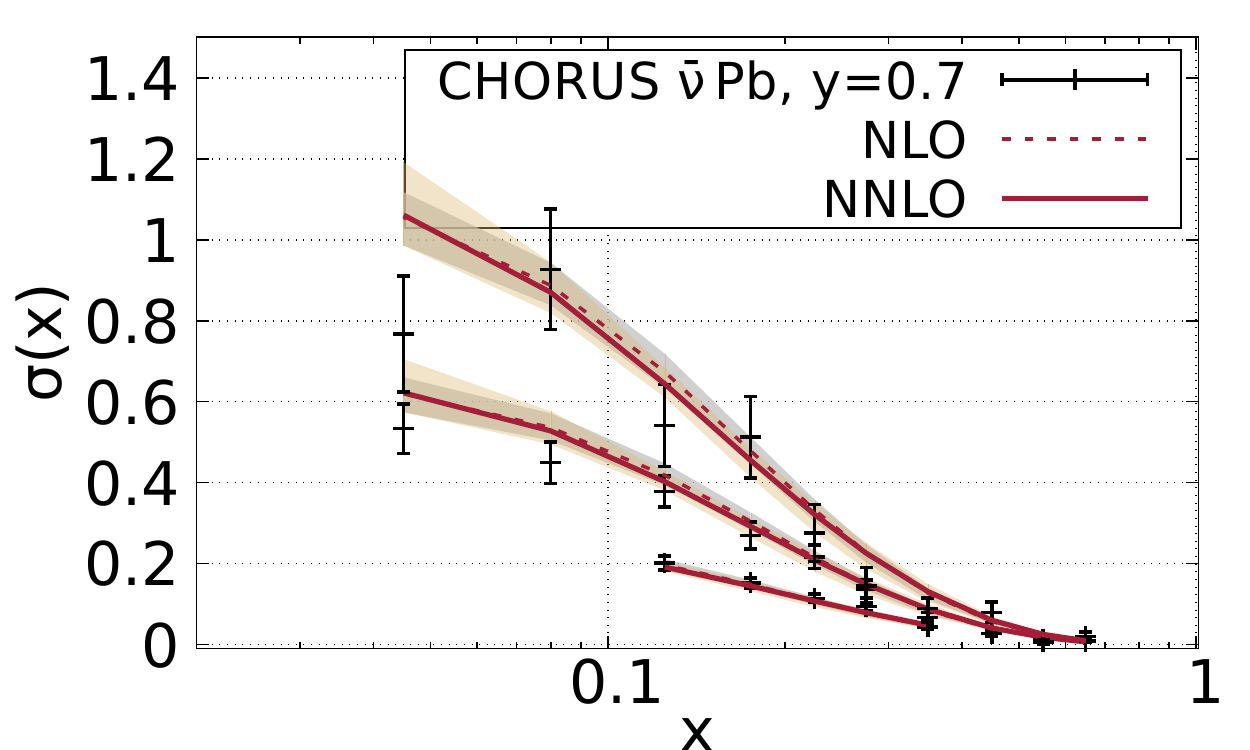}}                      
          \end{center} 
\caption{Sample comparison to selected CHORUS data for CC deeply inelastic scattering on lead (Pb). We show the results for either neutrinos ($\nu$) or antineutrinos ($\bar{\nu}$), for one $y$ value (cf. legend) each at different beam energies ($35$, $70$, $110\,\mathrm{GeV}$). The calculated quantities are shown at NLO (dashed line, grey error bands) and NNLO (solid line, golden-coloured error bands).}
\label{figCHORUS}    
    \end{figure*}

\begin{figure*}[htb!]
     \begin{center}      
          \subfigure{\includegraphics[width=0.325\textwidth]{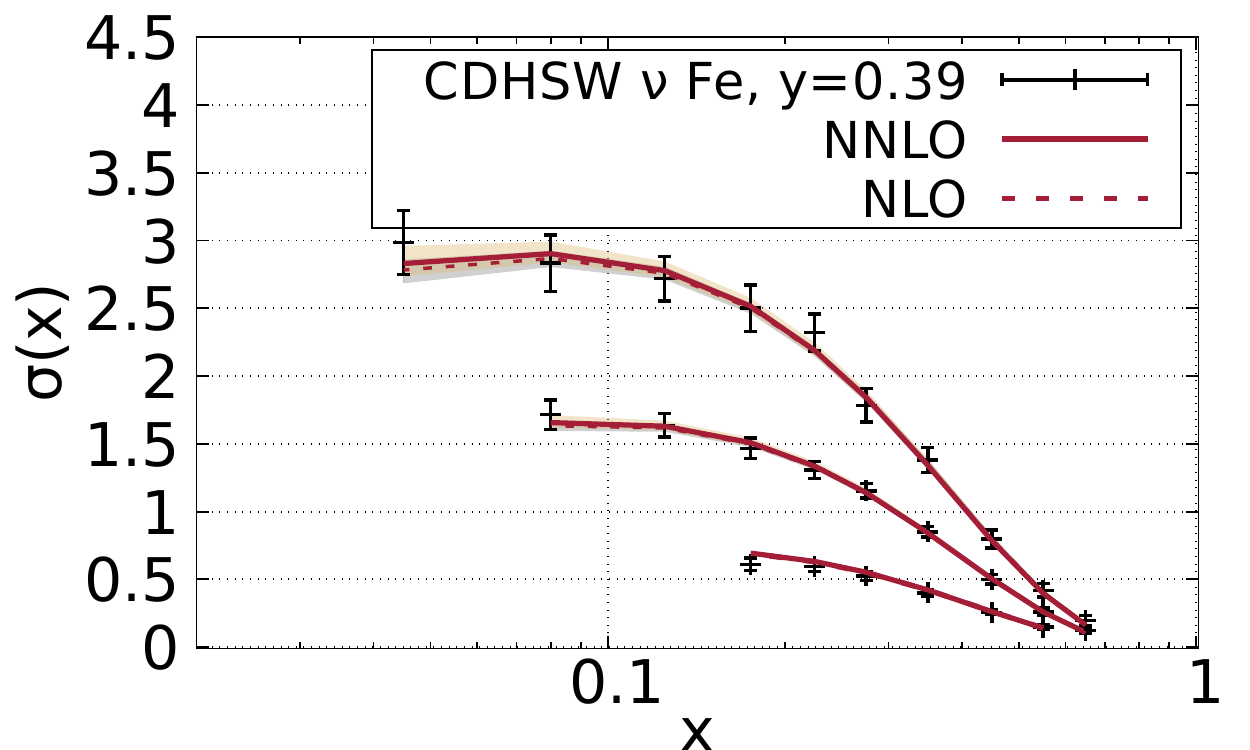}}
          \subfigure{\includegraphics[width=0.325\textwidth]{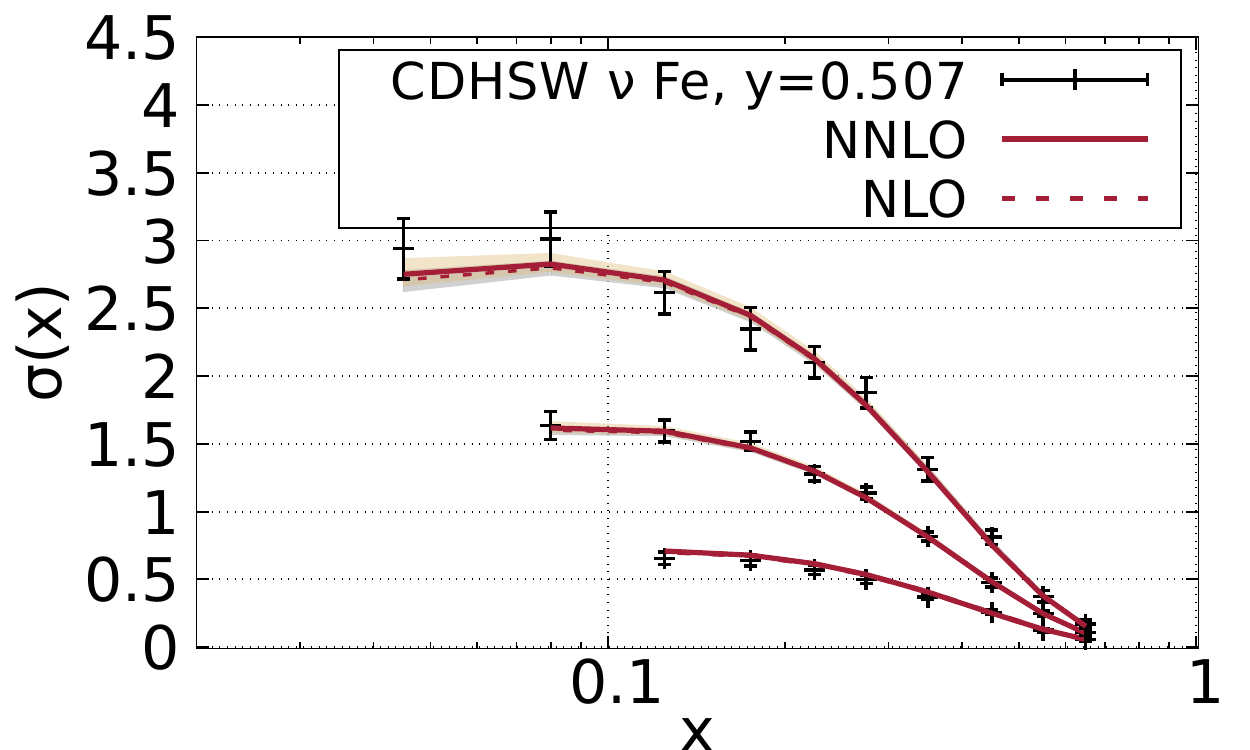}} 
          \subfigure{\includegraphics[width=0.325\textwidth]{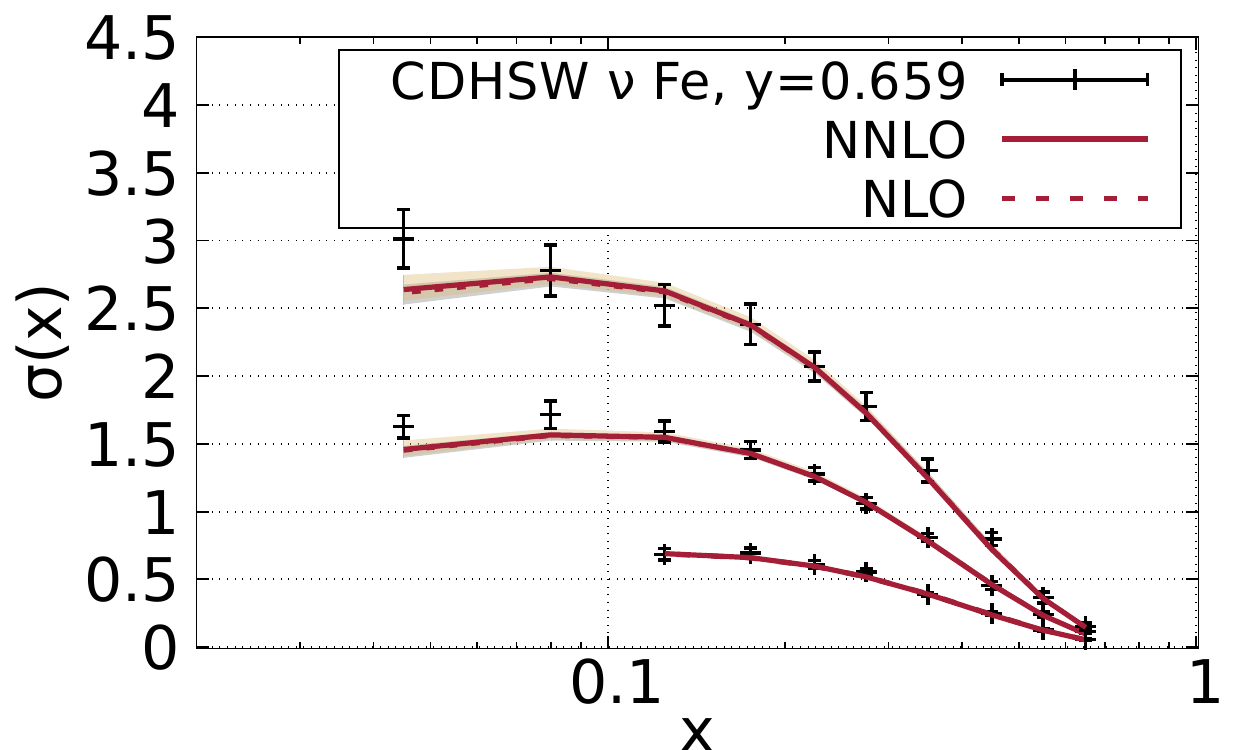}}                      
          \subfigure{\includegraphics[width=0.325\textwidth]{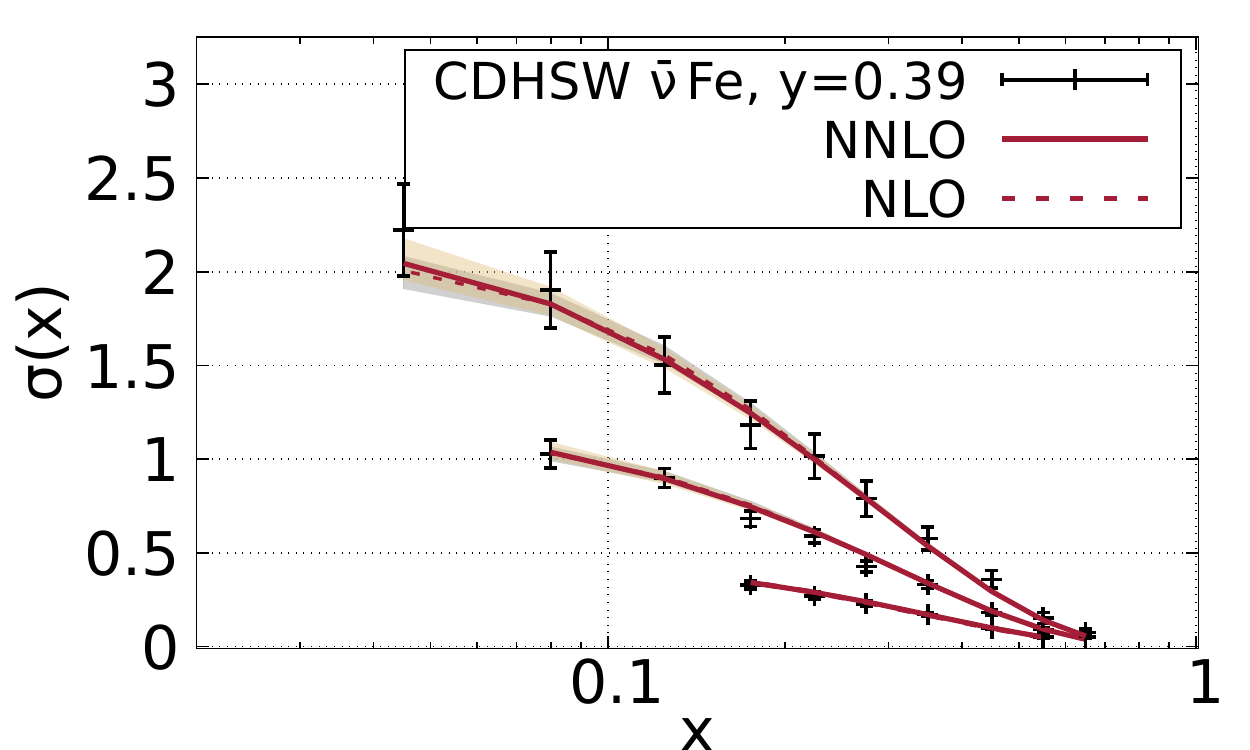}}
          \subfigure{\includegraphics[width=0.325\textwidth]{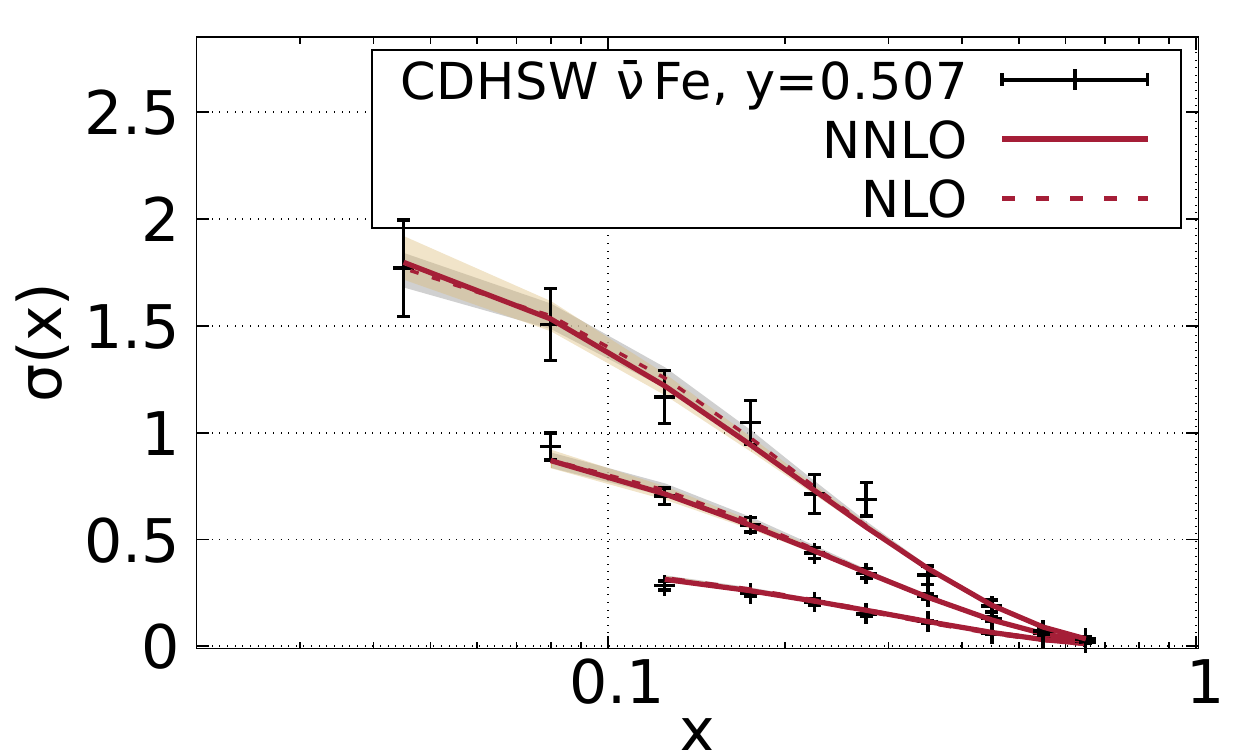}} 
          \subfigure{\includegraphics[width=0.325\textwidth]{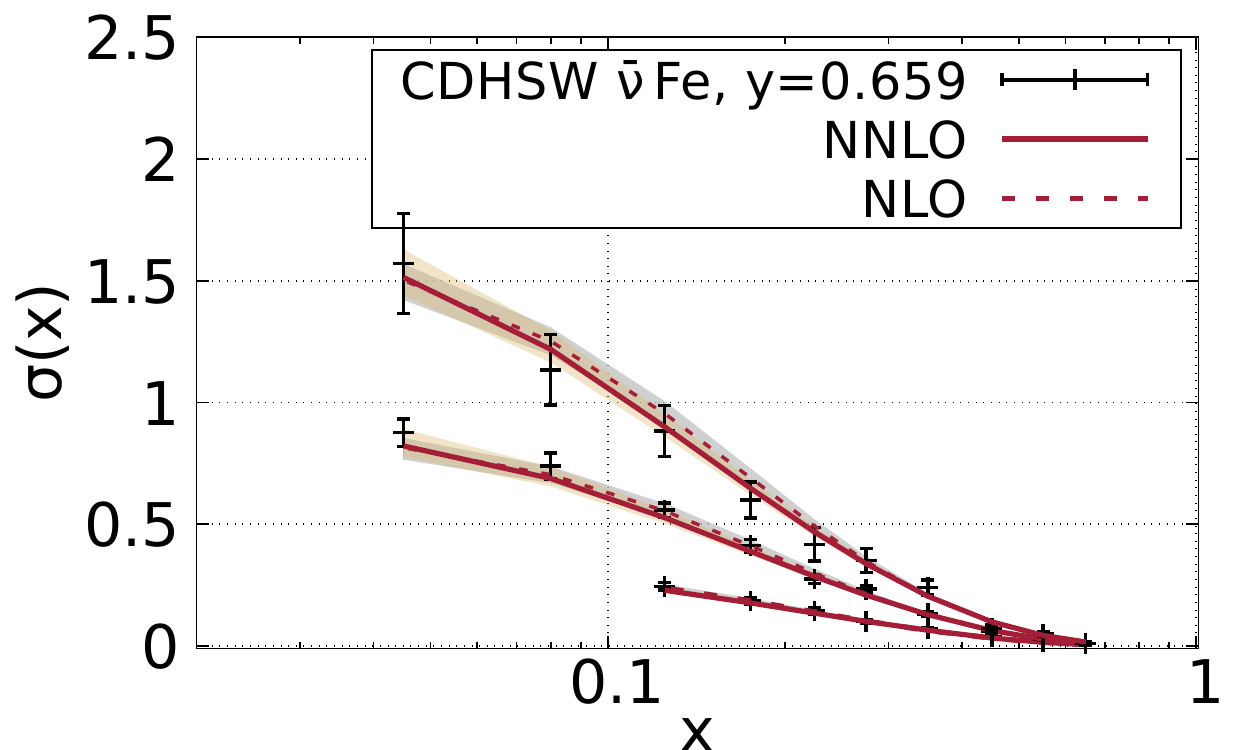}}                      
          \end{center} 
\caption{Sample comparison to selected CDHSW data for CC deeply inelastic scattering on iron (Fe).
We show the results for either neutrinos ($\nu$) or antineutrinos ($\bar{\nu}$), for one $y$ value (cf. legend) each at different beam energies ($38.9$, $85.4$, $144.3\,\mathrm{GeV}$). The calculated quantities are shown at NLO (dashed line, grey error bands) and NNLO (solid line, golden-coloured error bands).}
\label{figCDHSW}    
    \end{figure*}

The optimal set of nPDF parameters is derived by minimizing the $\Chi^2$ as defined in eq.~(\ref{eq-chi2}) by comparing to the measured data presented in table~\ref{tab-expdata}. The resulting cross sections, structure functions and ratios are compared to the data we used in figures \ref{figNMCd}, \ref{figNMC} - \ref{figHERMES} for neutral current DIS processes and in figures \ref{figCHORUS} and \ref{figCDHSW} for charged current DIS processes with neutrinos for a subset of that data.
An overview of the resulting $\Chi^2$ values, divided by the number of data points $N_\mathrm{dp}$, is shown in figure \ref{fig-Chi2} for NLO and NNLO. Values above $\Chi^2/N_{\mathrm{dp}}>3.0$ have been truncated in this graph for better representation, but the actual numbers are shown in table \ref{tab-expdata}. Figure \ref{fig-Chi2} demonstrates that the agreement between the theoretical predictions and the experimental measurements varies between different data sets. For example, the agreement with most of the data published by the NMC collaboration is excellent, whereas the agreement with the HERMES data is clearly not optimal. In this particular example one needs to point out that the number of data points from the HERMES experiment is much smaller than the number of NMC data points, so that the contribution to the total $\Chi^2$ is relatively small for the HERMES data. Apart from the few outlying data sets, the overall agreement is found to be very good, and the total $\Chi^2/N_{\mathrm{dp}}$ is $0.887$ at NLO and $0.862$ at NNLO. Even though some of the data sets are better described at NLO and some at NNLO, the total $\Chi^2$ values are very close at the different orders. The good agreement is also apparent in figures~\ref{figNMCd}--\ref{figCDHSW}. Interestingly, a very good agreement is also achieved for the neutrino data, even though some earlier studies observed difficulties when incorporating these data in a global nPDF analysis \cite{Kovarik:2010uv}. However, as concluded in Ref.~\cite{Paukkunen:2013grz}, this likely because of the tension caused by the NuTeV data which we have not included. 

%

%
     
%
   
%
 
\clearpage

\subsection{Comparisons to other nPDF sets}
\begin{figure}[tb!]
     \begin{center}
          \subfigure{\includegraphics[width=0.235\textwidth]{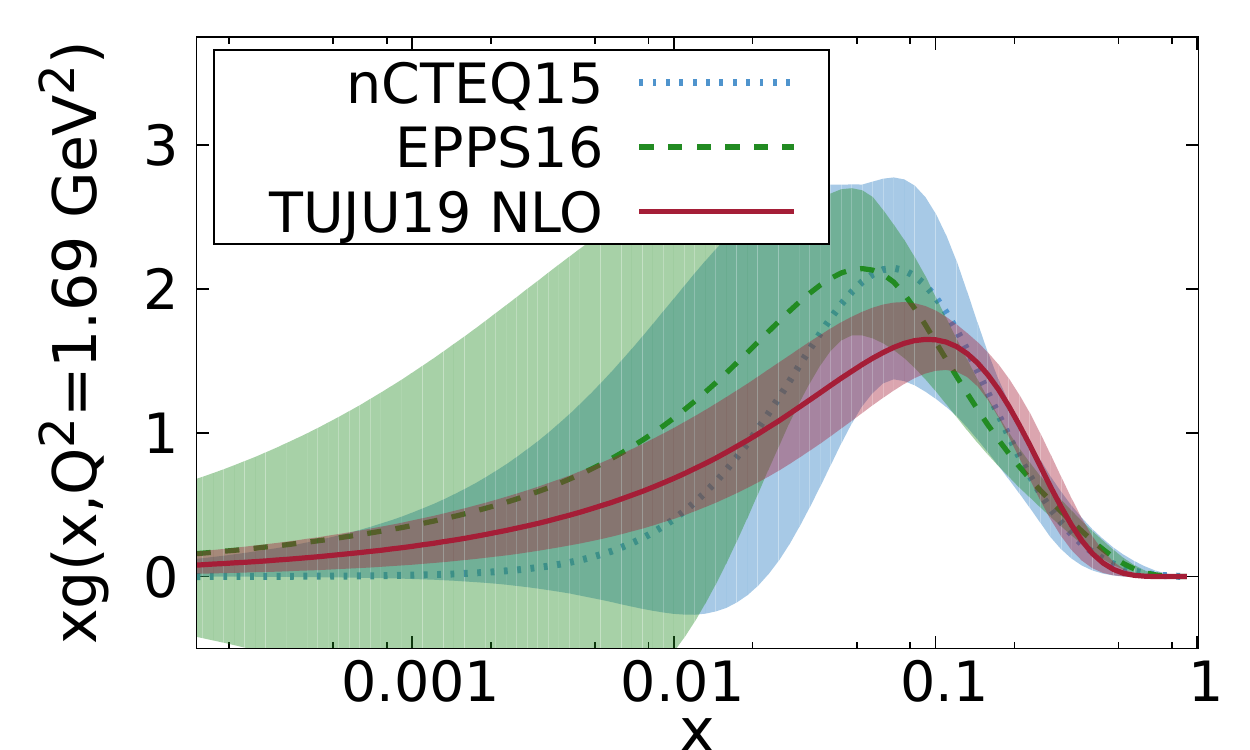}} 
          \subfigure{\includegraphics[width=0.235\textwidth]{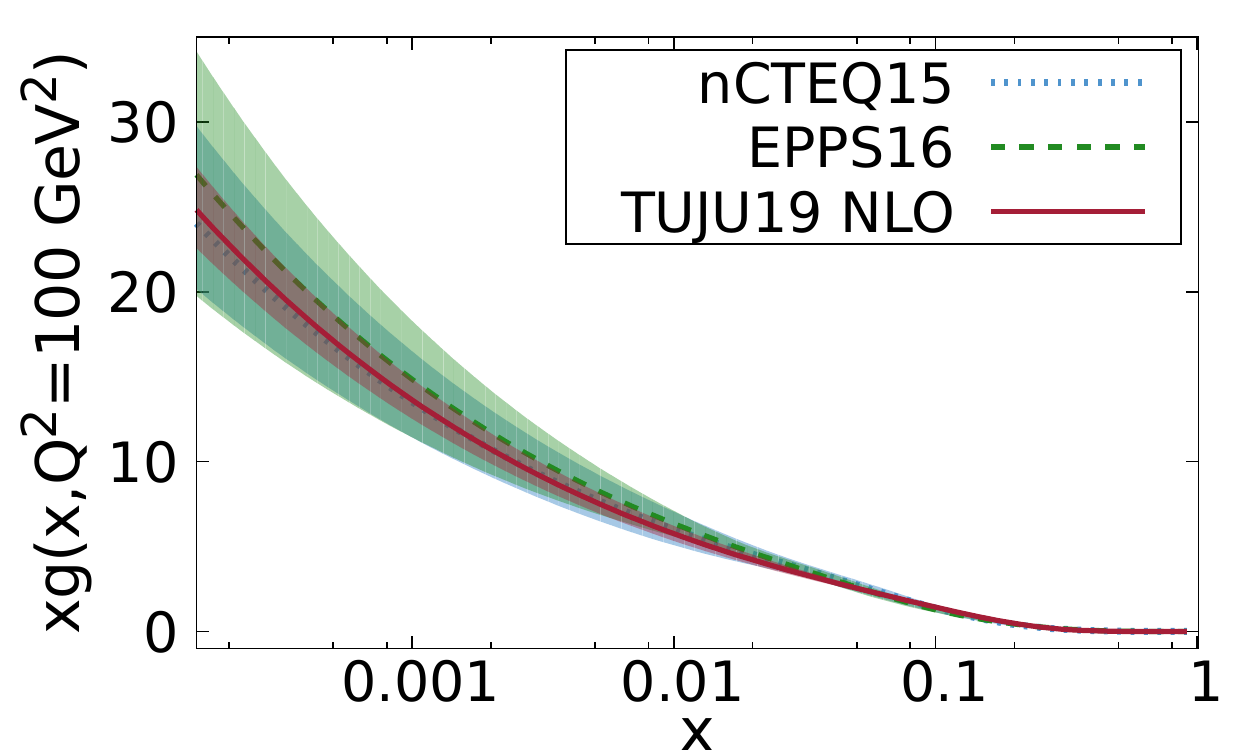}} 
          \subfigure{\includegraphics[width=0.235\textwidth]{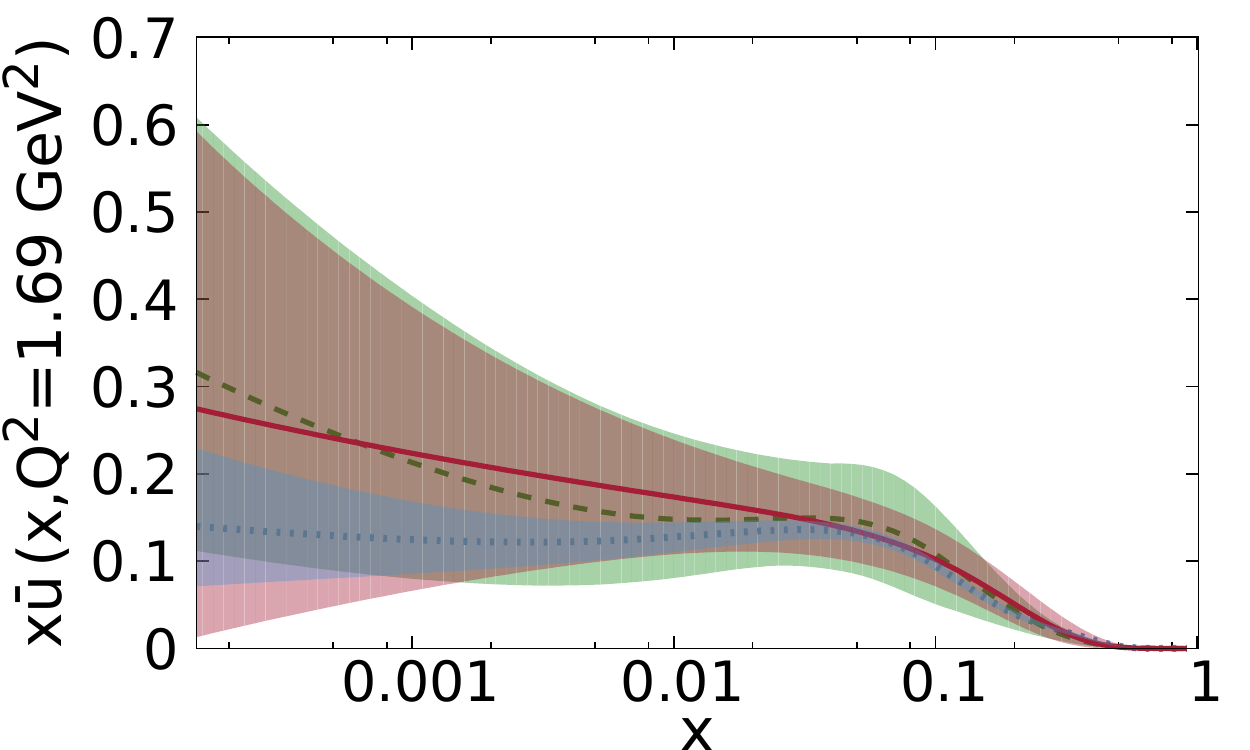}}
          \subfigure{\includegraphics[width=0.235\textwidth]{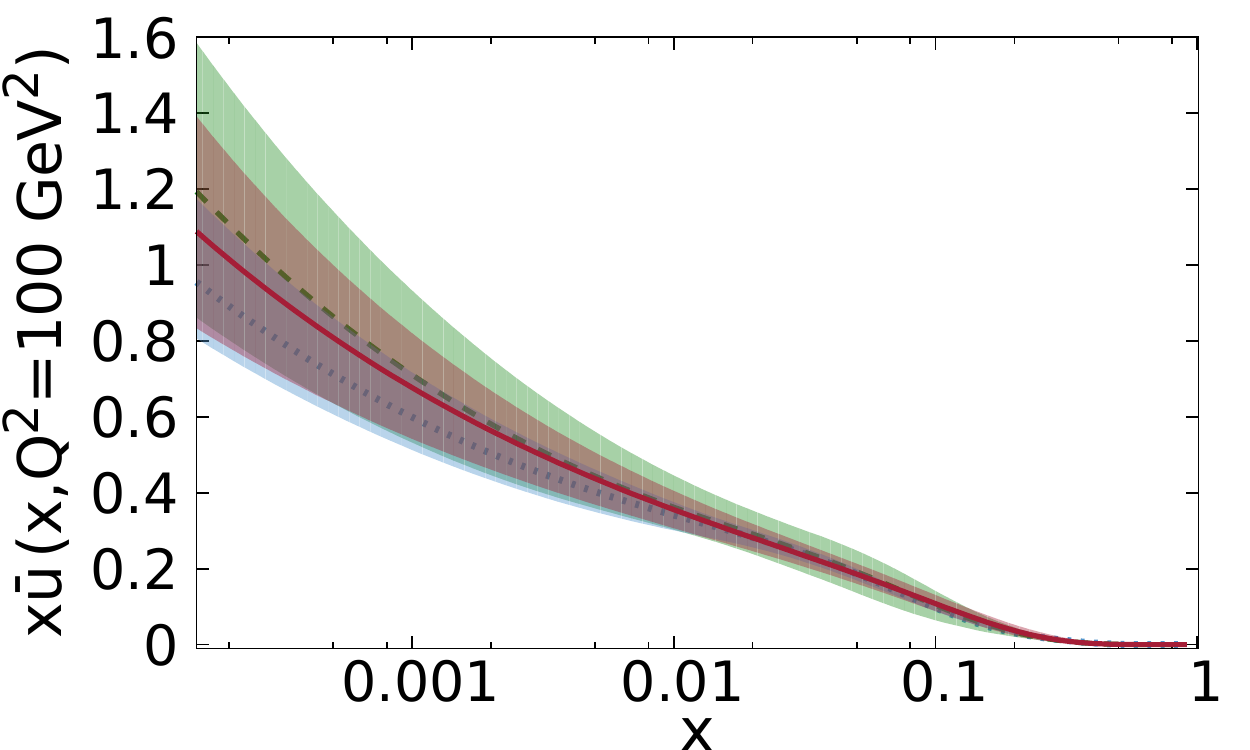}}
          \subfigure{\includegraphics[width=0.235\textwidth]{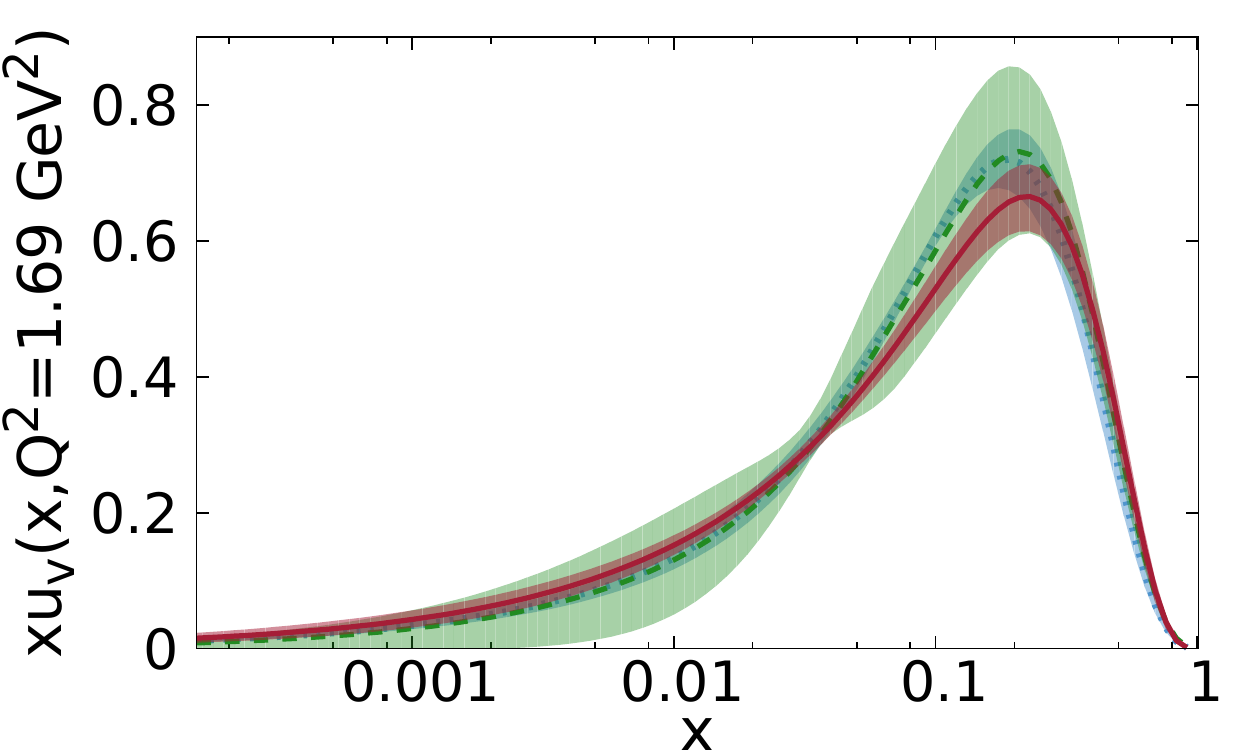}}
          \subfigure{\includegraphics[width=0.235\textwidth]{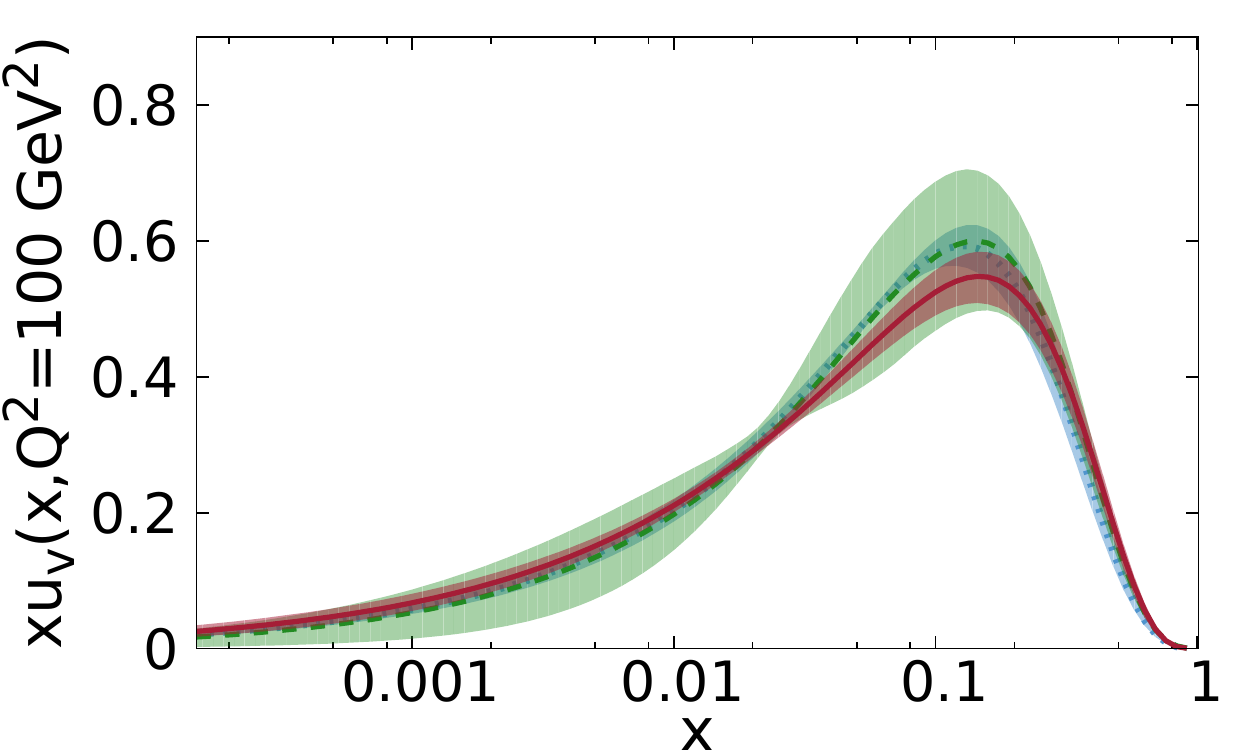}}
          \subfigure{\includegraphics[width=0.235\textwidth]{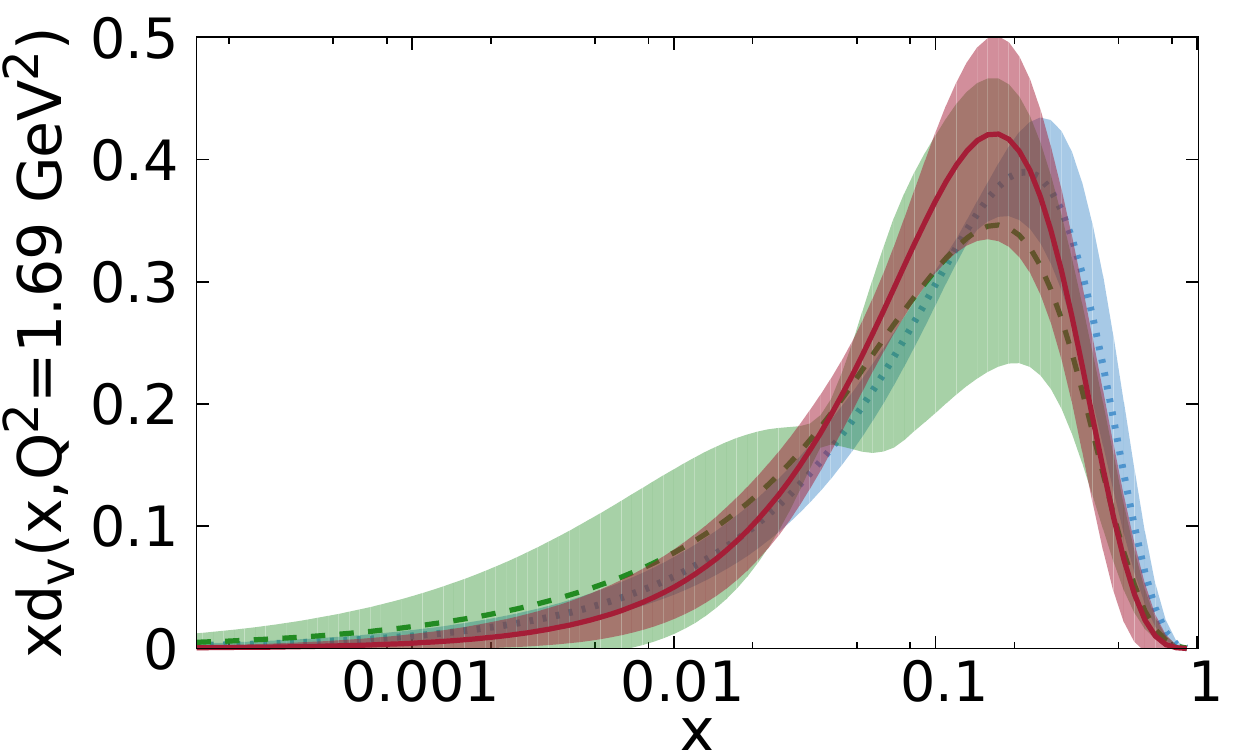}}                         
          \subfigure{\includegraphics[width=0.235\textwidth]{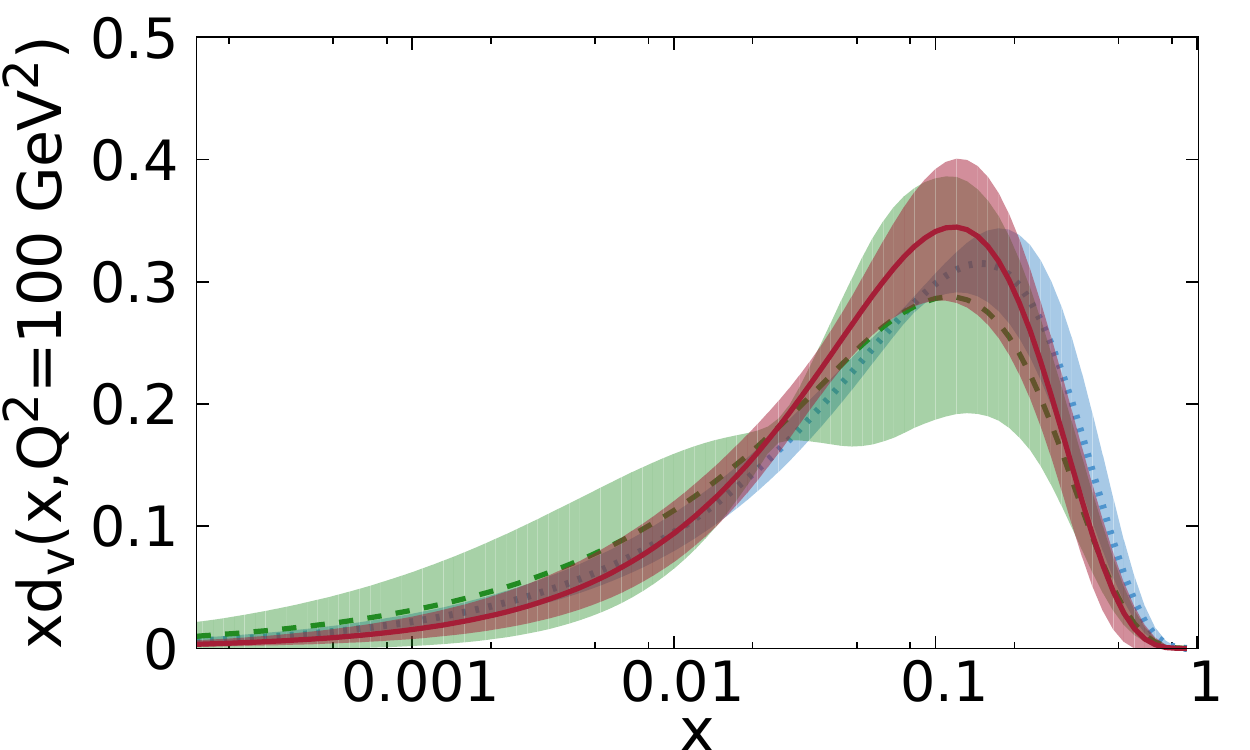}}                         
          \subfigure{\includegraphics[width=0.235\textwidth]{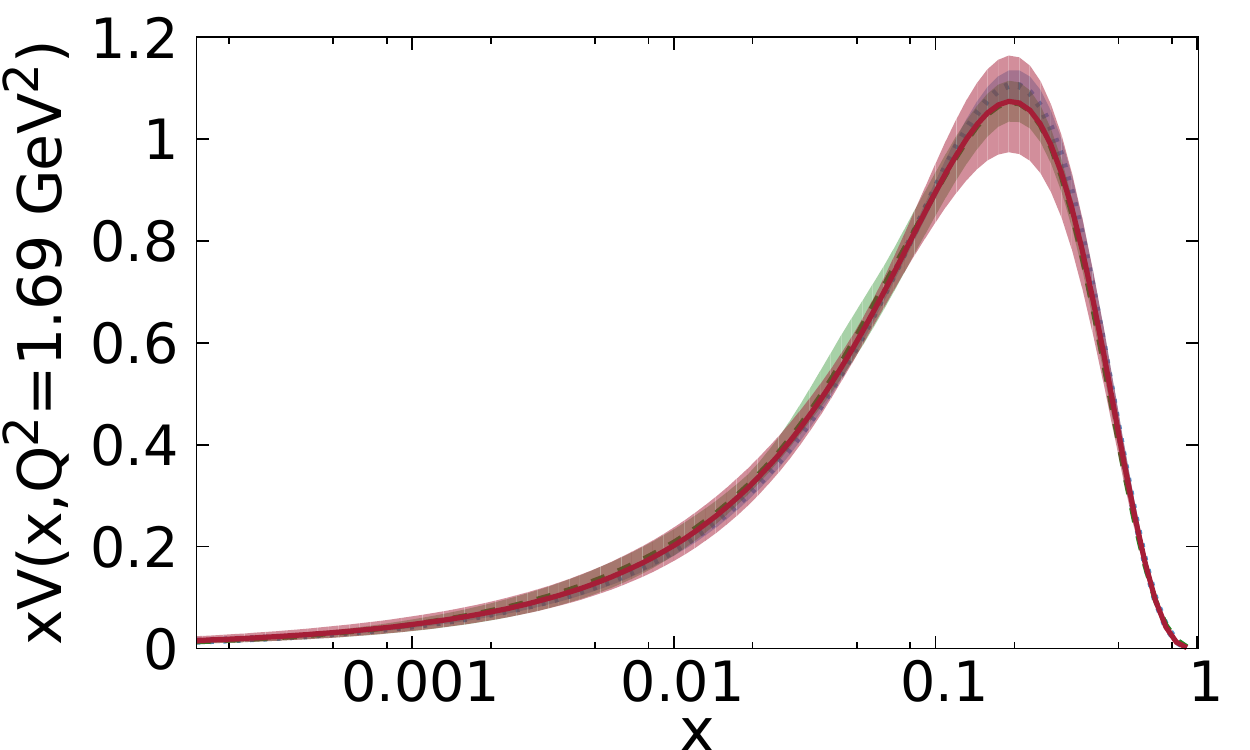}}                         
          \subfigure{\includegraphics[width=0.235\textwidth]{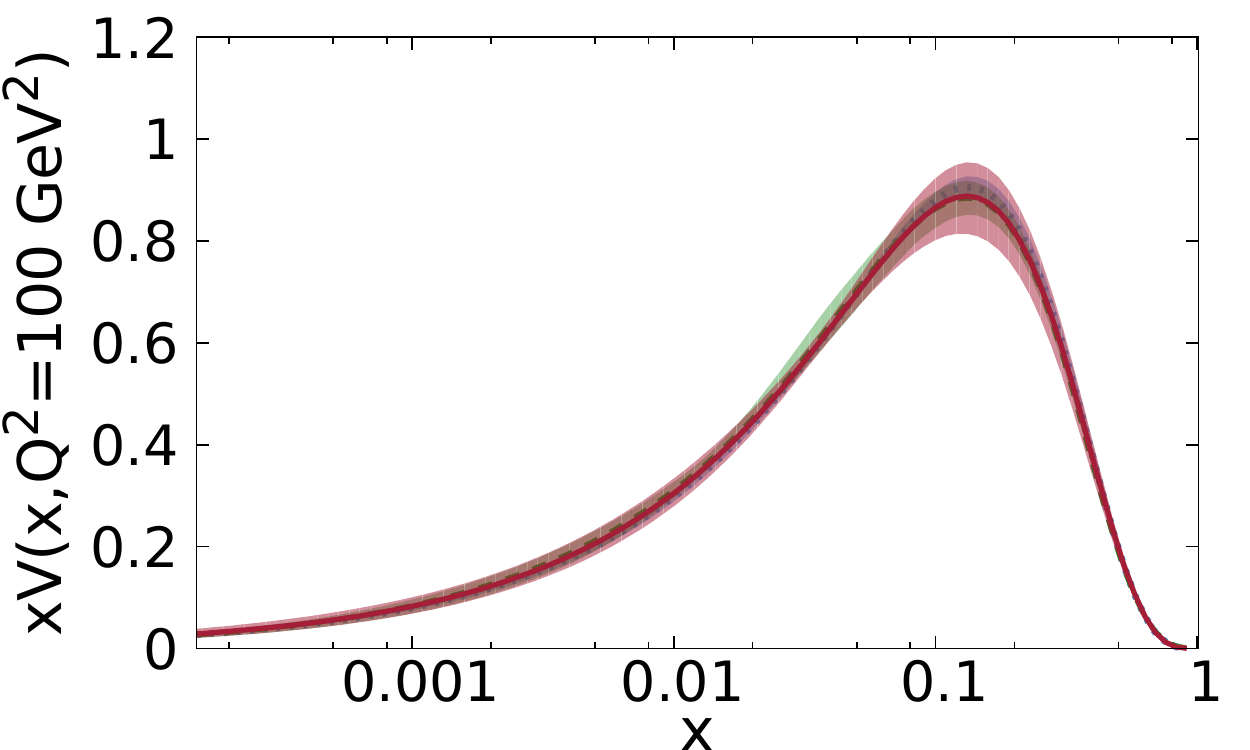}}                         
          \end{center} 
    \caption{Nuclear parton distribution functions TUJU19 in lead at NLO compared to nCTEQ15 \cite{Kovarik:2015cma} and EPPS16 \cite{Eskola:2016oht}, shown at the initial scale $Q_0^2=1.69\,\mathrm{GeV}^2$ and at a higher scale $Q^2=100\,\mathrm{GeV}^2$. The comparison is presented for the distribution functions $xf_i(x,Q^2)$ per parton flavour $i=g,\,\bar{u},\,u_v,\,d_v, \,V$, where $V$ is the sum of valence quarks, in a bound proton in lead.}
\label{fig-nPDF-comparisonQ0-NLO}    
    \end{figure}
Each nPDF analysis is based on a set of assumptions, e.g. the form of the nonperturbative input at the initial scale, the choice of the proton baseline and the kinematical cuts. Therefore, even when based on the same set of data it is not guaranteed that the results will be equivalent. However, some level of agreement -- within the estimated uncertainties -- is expected.
\subsubsection{  Comparison at NLO}

In figure \ref{fig-nPDF-comparisonQ0-NLO} we compare our obtained nPDFs to those of other recent NLO nPDF analyses, nCTEQ15 and EPPS16, at our initial scale $Q^2=1.69~\text{GeV}^2$ and at $Q^2=100~\text{GeV}^2$. The comparisons are shown for $g$, $\bar{u}$, $u_v$, $d_v$ and $V=u_v+d_v$ in a proton bound in lead. For gluons at the initial scale the agreement is not very good, though still well within the uncertainties. Towards higher scales, however, a much better mutual agreement is observed. For sea quarks (here represented by $\bar{u}$) the agreement with the previous analyses is better already at the initial scale, and at $Q^2=100~\text{GeV}^2$ our result is between EPPS16 and nCTEQ15. When comparing the sea-quark distributions one should keep in mind that each analysis has different assumptions for the sea-quark flavour dependence, i.e. in TUJU19 we have assumed $s=\bar{s}=\bar{u}=\bar{d}$, whereas for nCTEQ15 $s=\bar{s}$ and $\bar{u}=\bar{d}$ are connected by an additional factor, and only $s=\bar{s}$ applies for EPPS16. For valence quarks we find that $u_v$ tends to stay below (above) the EPPS16 and nCTEQ15 results at $x\gtrsim 0.03$ ($x\lesssim 0.03$) whereas the opposite behaviour is found for $d_v$. This can be explained by the fact that in the case of nuclear data only a combination of $u_v$ and $d_v$ is probed, and even with the included neutrino data the flavour dependence of valence quarks is not well constrained. Indeed, we find a very good agreement between the three analyses for the sum of valence quarks~$V$.
\begin{figure*}[bt!]
     \begin{center}                           
          \subfigure{\includegraphics[width=0.4\textwidth]{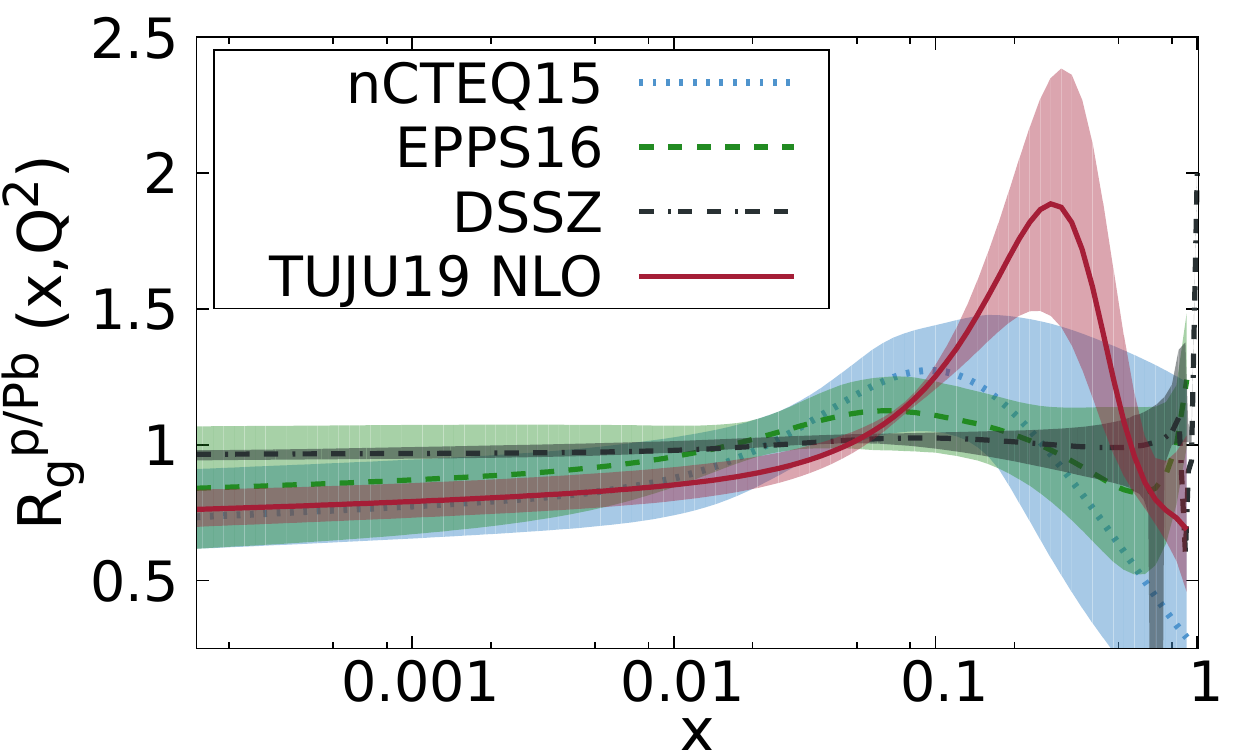}} 
          \subfigure{\includegraphics[width=0.4\textwidth]{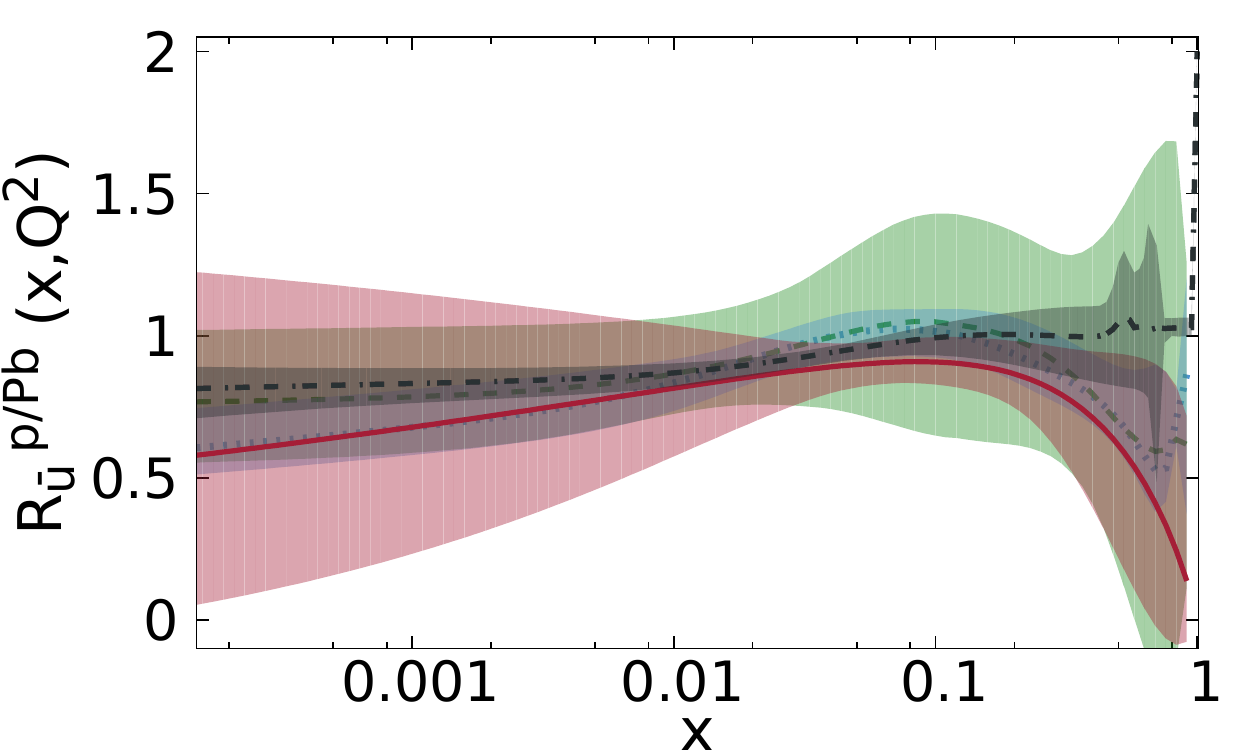}}
          \subfigure{\includegraphics[width=0.4\textwidth]{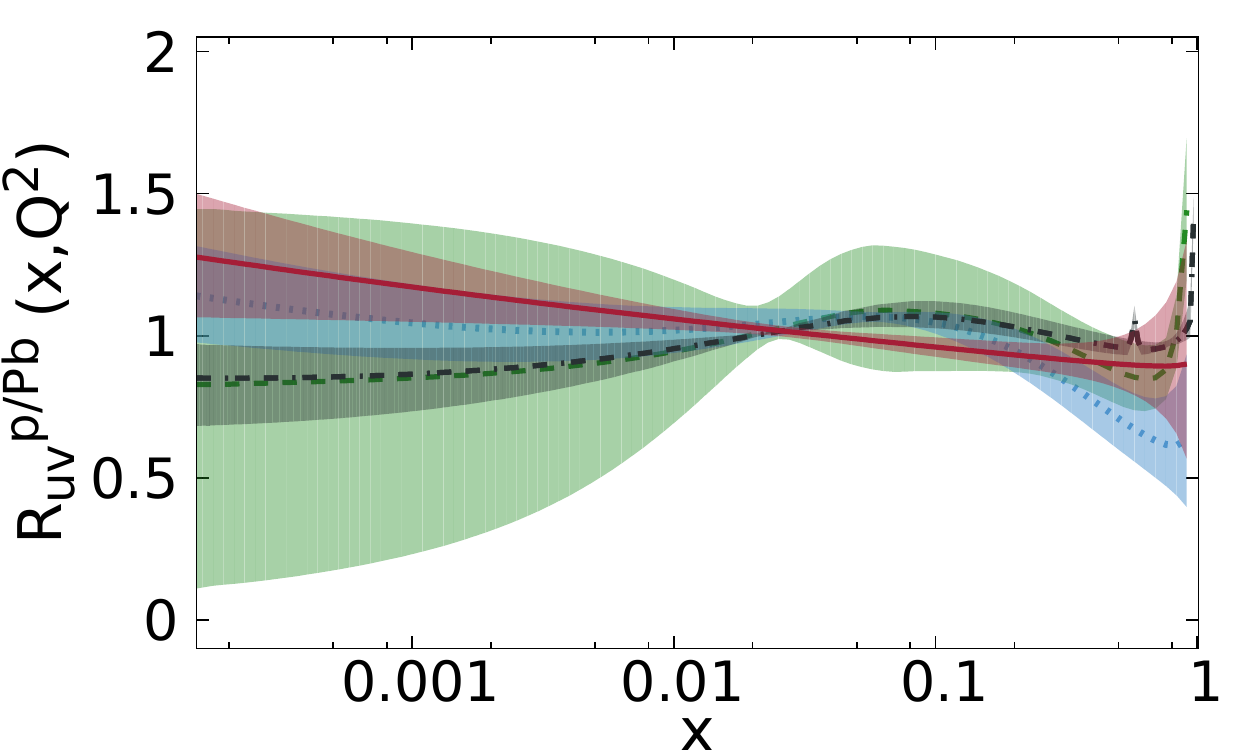}}
          \subfigure{\includegraphics[width=0.4\textwidth]{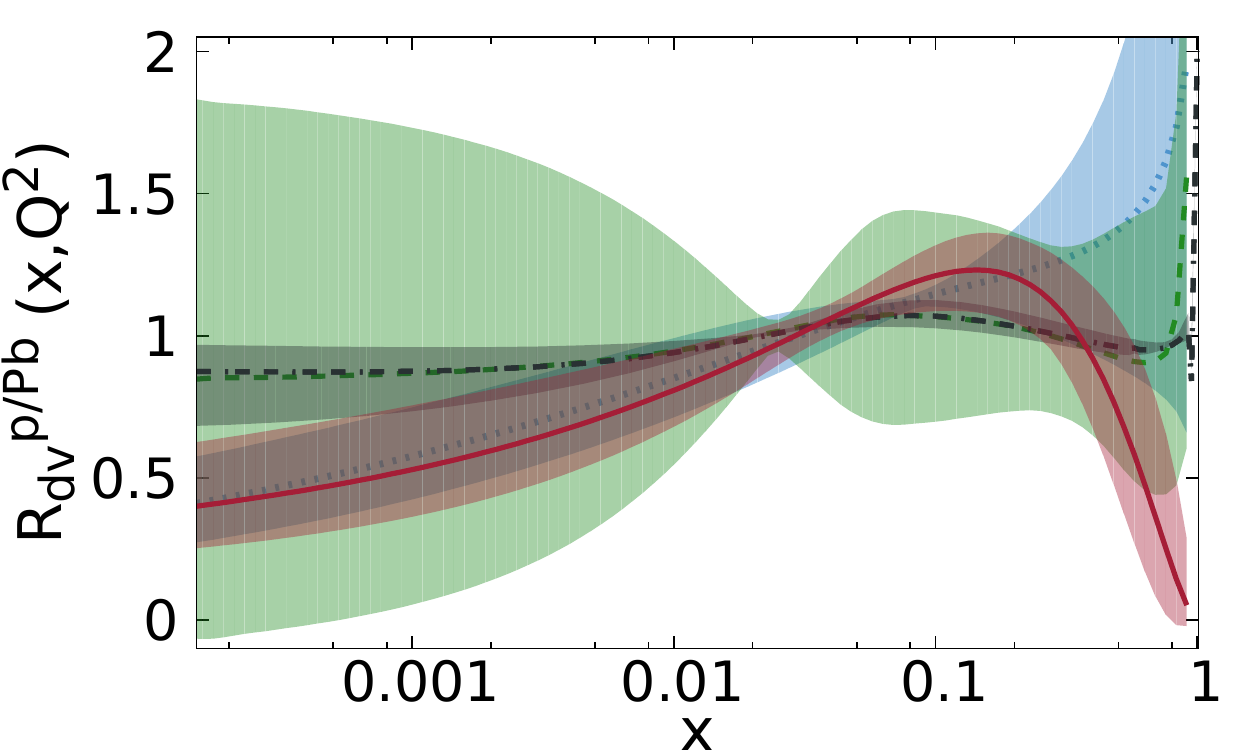}}                                                
          \end{center} 
    \caption{Nuclear parton distribution functions TUJU19 in lead at NLO compared to the nPDF sets nCTEQ15 \cite{Kovarik:2015cma}, EPPS16 \cite{Eskola:2016oht}, and DSSZ \cite{deFlorian:2011fp} shown at the higher scale $Q^2=100\,\mathrm{GeV}^2$. The comparison is presented per parton flavour $i$ for the ratios $R_i^{\mathrm{p/Pb}}$ of PDFs in a proton bound in lead compared to the PDFs in a free proton.}
\label{fig-nPDF-comparisonQ2-NLO}    
    \end{figure*}

The uncertainty bands in our NLO fit are similar to those obtained in the earlier analyses for sea quarks, but for gluons the resulting uncertainties are somewhat smaller. Since both EPPS16 and nCTEQ15 include additional data with some sensitivity to gluons, we conclude that our reduced uncertainties are likely due to the limited number of free parameters in the gluon nPDFs, and the uncertainty due to the lack of data constraints is underestimated. One should note that nCTEQ15 does not provide error sets for the baseline proton PDFs, which partly explains why their uncertainties for the sea quarks (at the initial scale) and the valence quarks tend to be smaller than in EPPS16 and this analysis. The comparisons were generated by using the \textsc{LHAPDF6} library \cite{Buckley:2014ana} and the published grids.

In figure \ref{fig-nPDF-comparisonQ2-NLO} we compare the nuclear modification of the PDFs as defined in eq.~(\ref{eq:Ri}) at $Q^2=100~\text{GeV}^2$. Also comparisons to the DSSZ analysis are included, for which only ratios $R_i^{\mathrm{p/Pb}}$ (eq. \ref{eq:Ri}) were available with error bands\footnote{No LHAPDF6 grids are available for DSSZ.} . In most cases the results are compatible within the estimated uncertainties, though some features stand out. A rather prominent feature of our NLO gluons is the large antishadowing around $x\sim 0.3$. Such a large enhancement is not supported by other analyses which include data sensitive to gluon antishadowing and underlines the need for further data sensitive to such effects. However, the obtained gluon shadowing is in good agreement with EPPS16 and nCTEQ15 results, though with somewhat reduced uncertainty estimates. Only the DSSZ $R_g^{\mathrm{p/Pb}}$ with very mild shadowing is outside the uncertainty bands in this region. For the flavour dependence of the valence quarks we find a similar behaviour as nCTEQ15 where some small-$x$ enhancement and large-$x$ suppression were observed for $u_v$, along with the opposite behaviour for $d_v$. However, when calculating the total valence distribution for a complete nucleon, as shown in figure \ref{fig-nPDF-comparisonQ0-NLO}, we find a good agreement with the other analyses.
\subsubsection{  Comparison at NNLO}
\begin{figure*}[tb!]
     \begin{center}
          \subfigure{\includegraphics[width=0.325\textwidth]{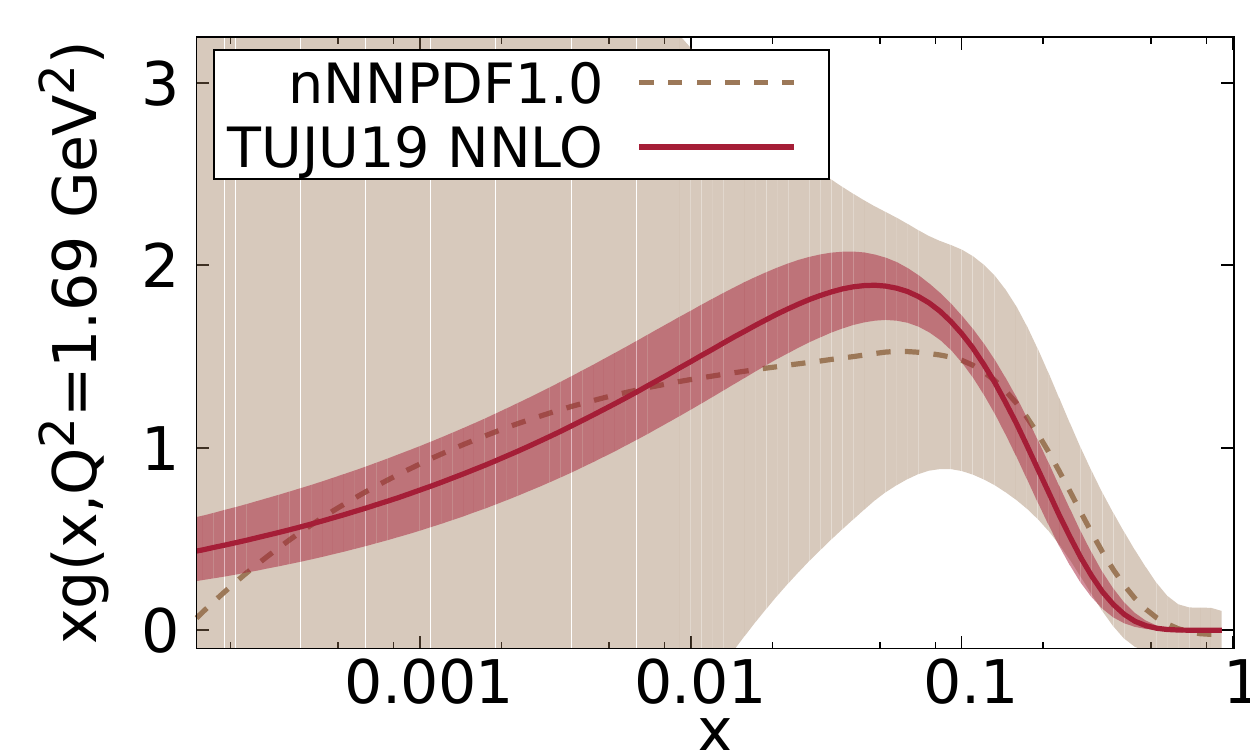}} 
           \subfigure{\includegraphics[width=0.325\textwidth]{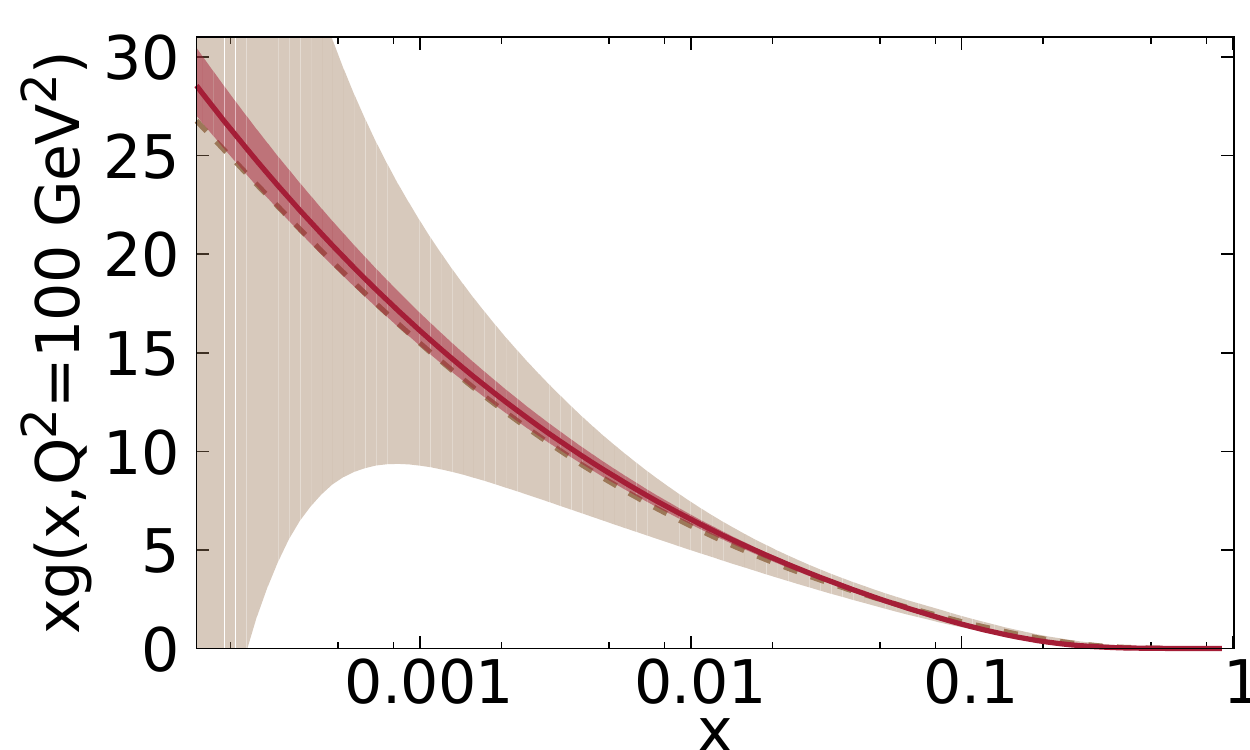}}
           \subfigure{\includegraphics[width=0.325\textwidth]{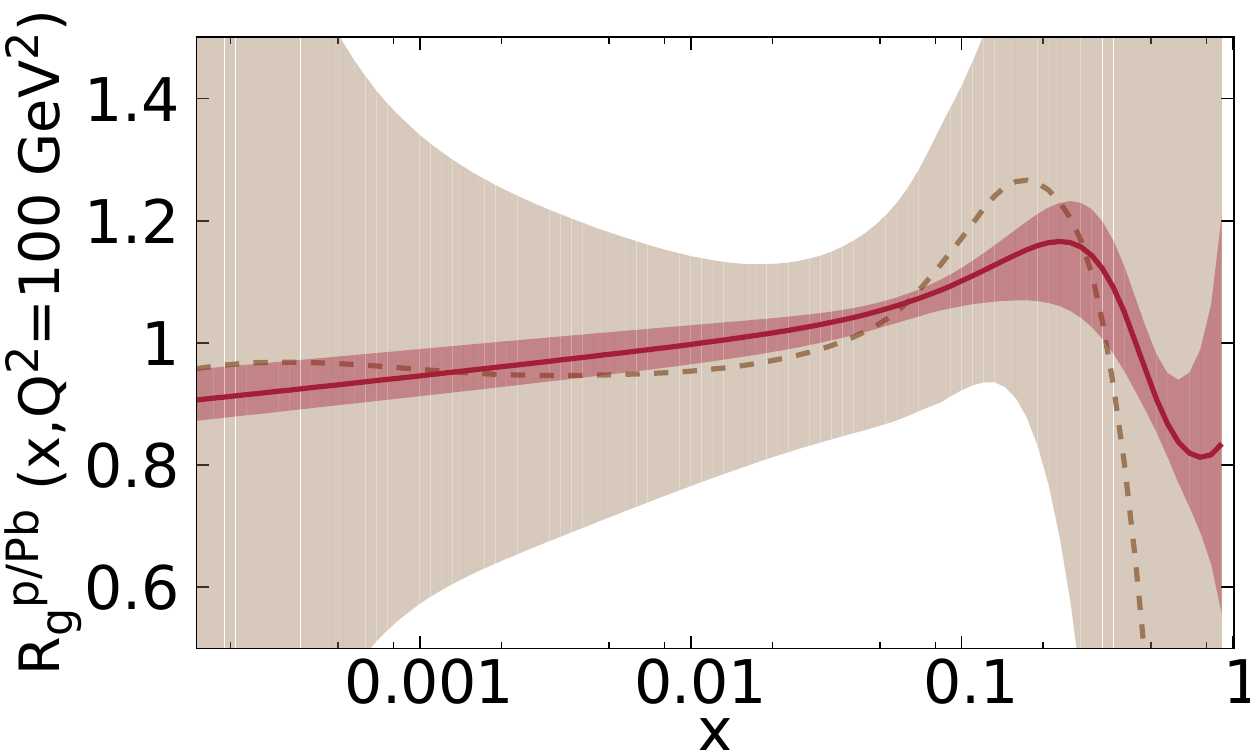}}          
          \subfigure{\includegraphics[width=0.325\textwidth]{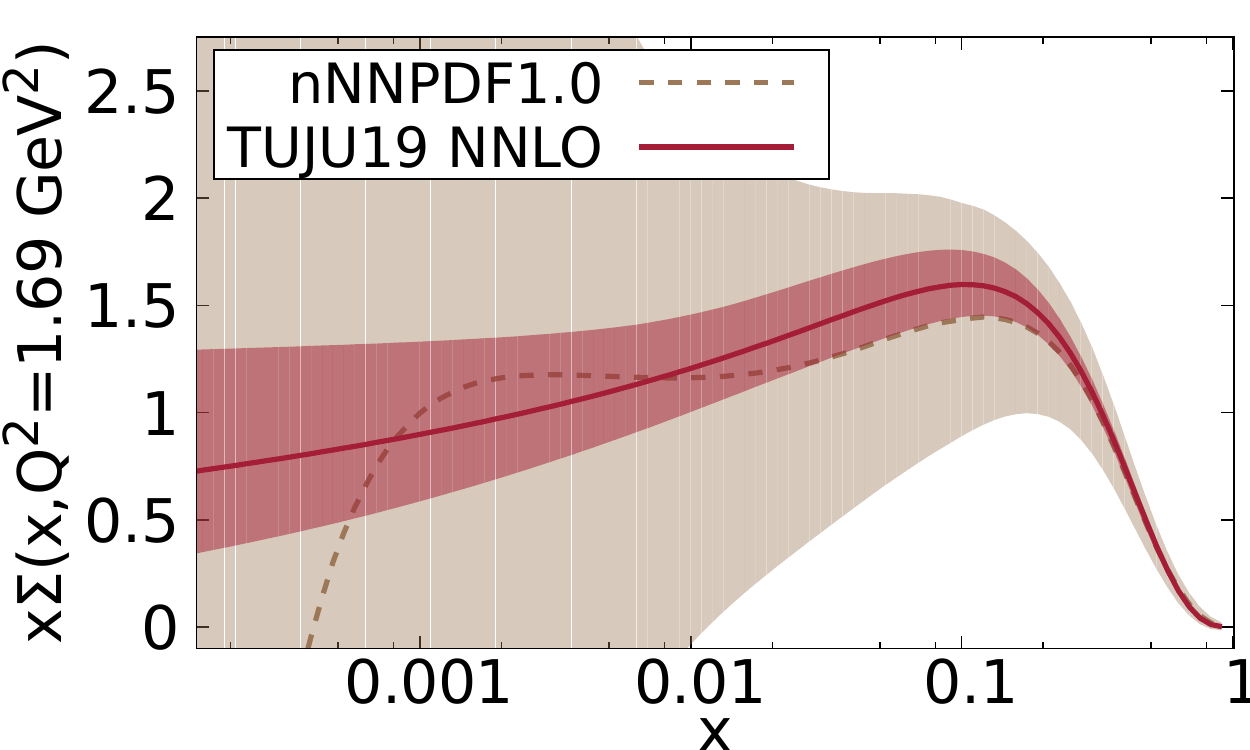}} 
           \subfigure{\includegraphics[width=0.325\textwidth]{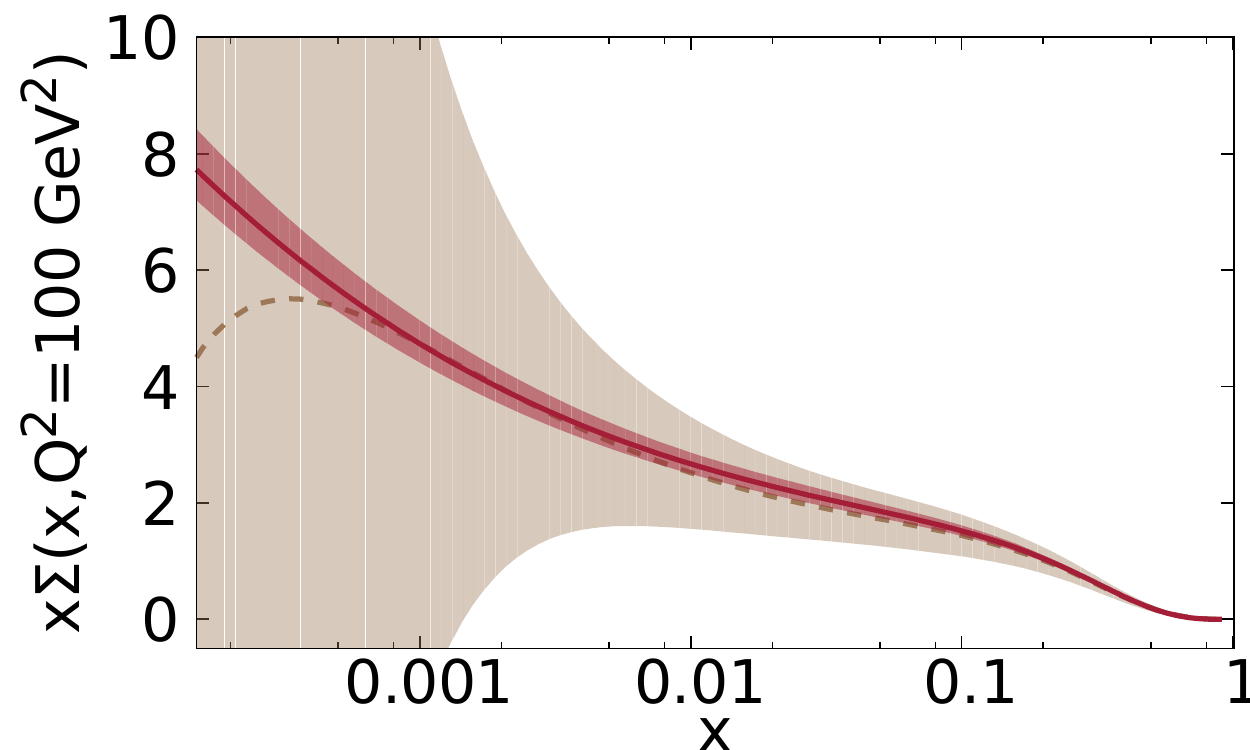}}
           \subfigure{\includegraphics[width=0.325\textwidth]{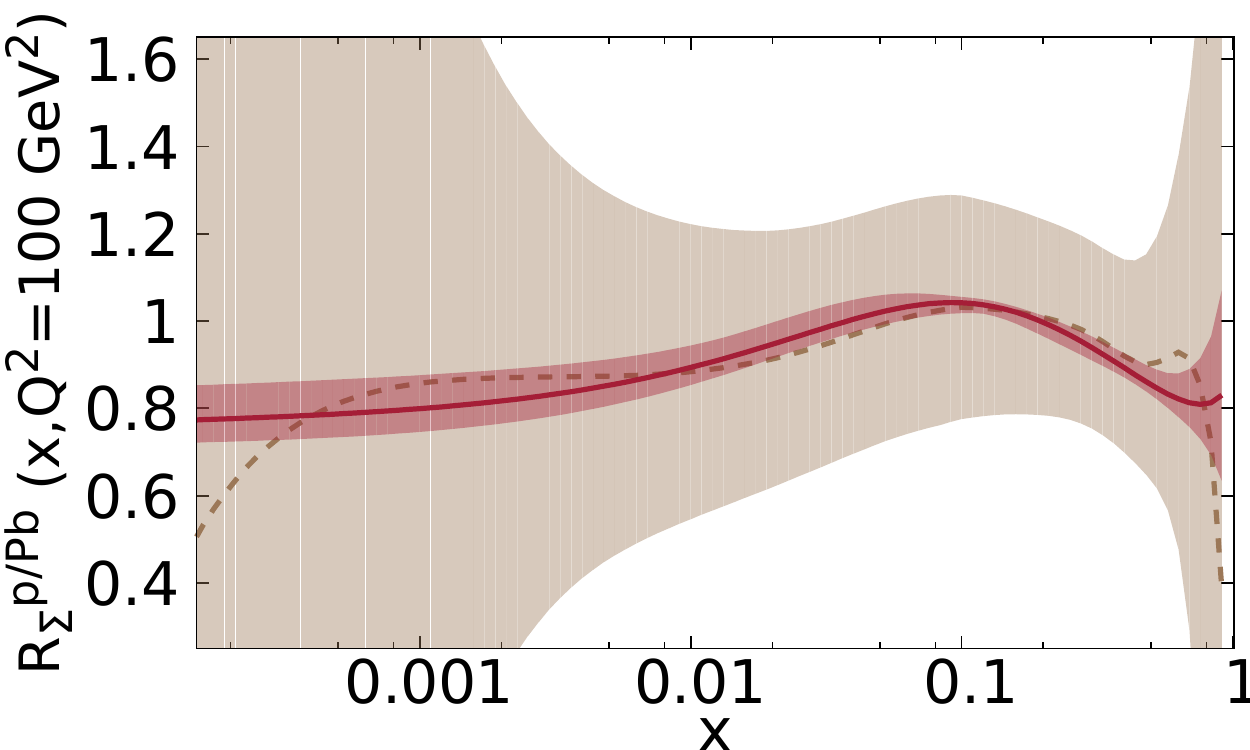}}  
          \end{center} 
    \caption{Nuclear parton distribution functions TUJU19 in lead at NNLO compared to the LHAPDF set nNNPDF1.0 \cite{AbdulKhalek:2019mzd}, shown at our initial scale $Q_0^2=1.69~\text{GeV}^2$ and at a higher scale $Q^2=100\,\mathrm{GeV}^2$ for distribution functions $xf_i$, and at the higher scale $Q^2=100\,\mathrm{GeV}^2$ for the ratios $R_i^{\mathrm{p/Pb}}$ of PDFs in a proton bound in lead compared to PDFs in a free proton. The comparison is presented for the gluon $g$ and for the quark singlet $\Sigma=u+\bar{u}+d+\bar{d}+s+\bar{s}$ in a bound proton in lead.}
\label{fig-nPDF-comparisonQ0-NNLO}    
    \end{figure*}
\begin{figure*}[tb!]
     \begin{center}
          \subfigure{\includegraphics[width=0.325\textwidth]{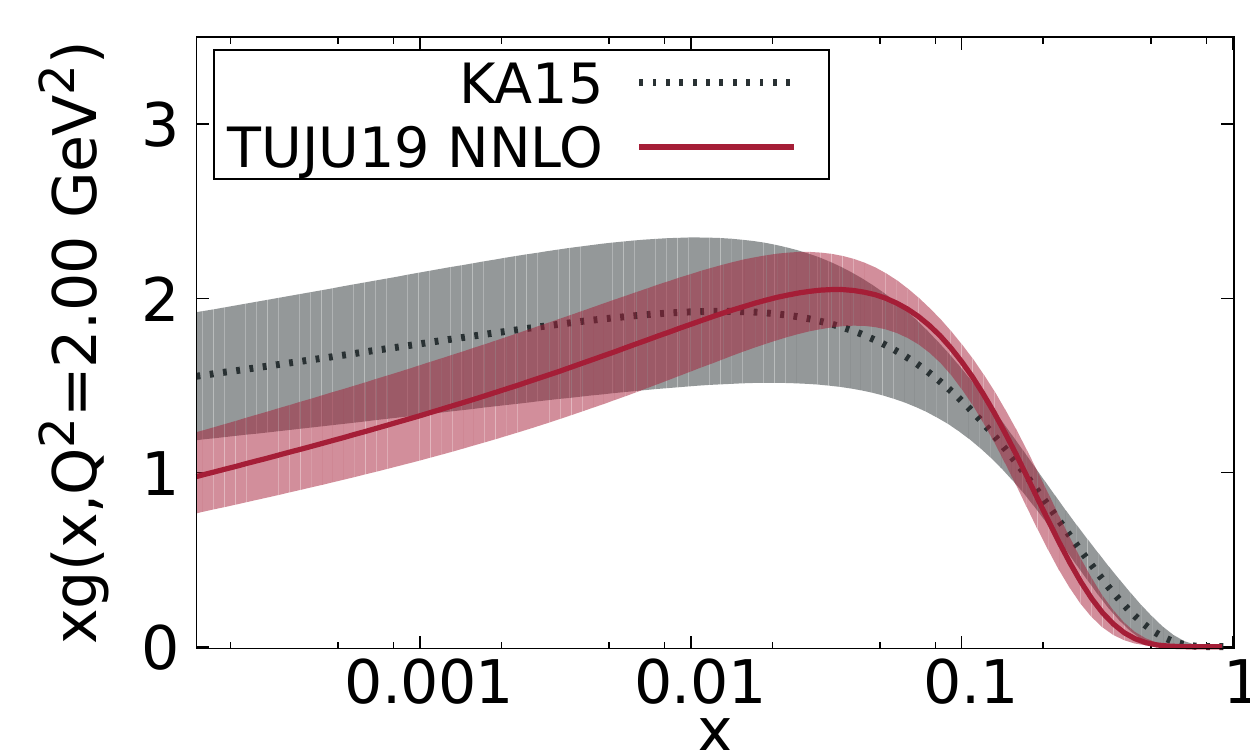}} 
          \subfigure{\includegraphics[width=0.325\textwidth]{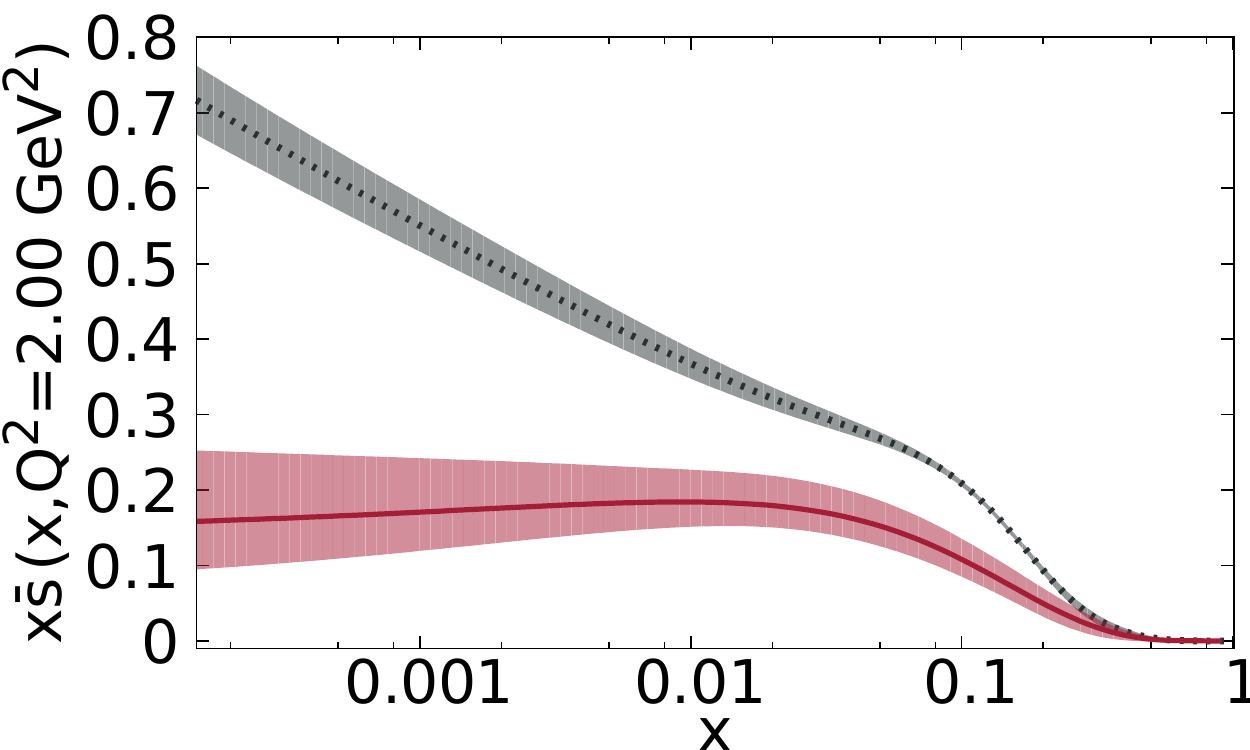}}
          \subfigure{\includegraphics[width=0.325\textwidth]{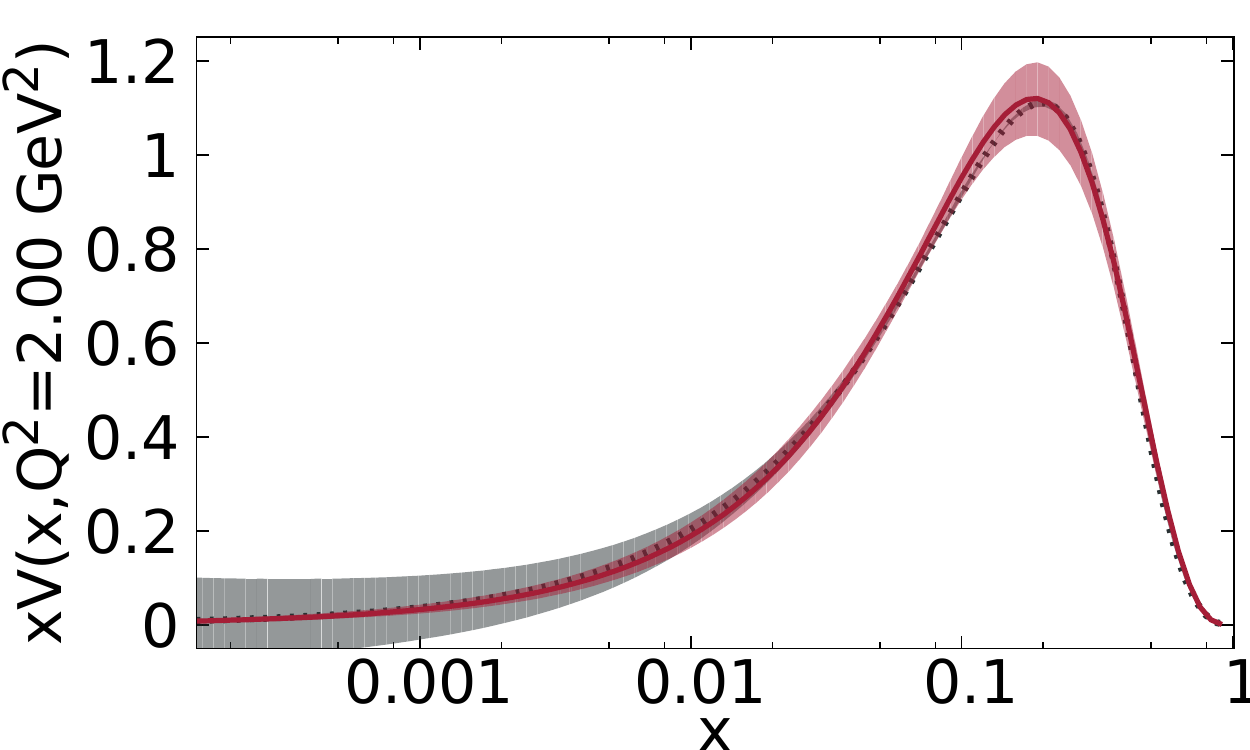}}
          \subfigure{\includegraphics[width=0.325\textwidth]{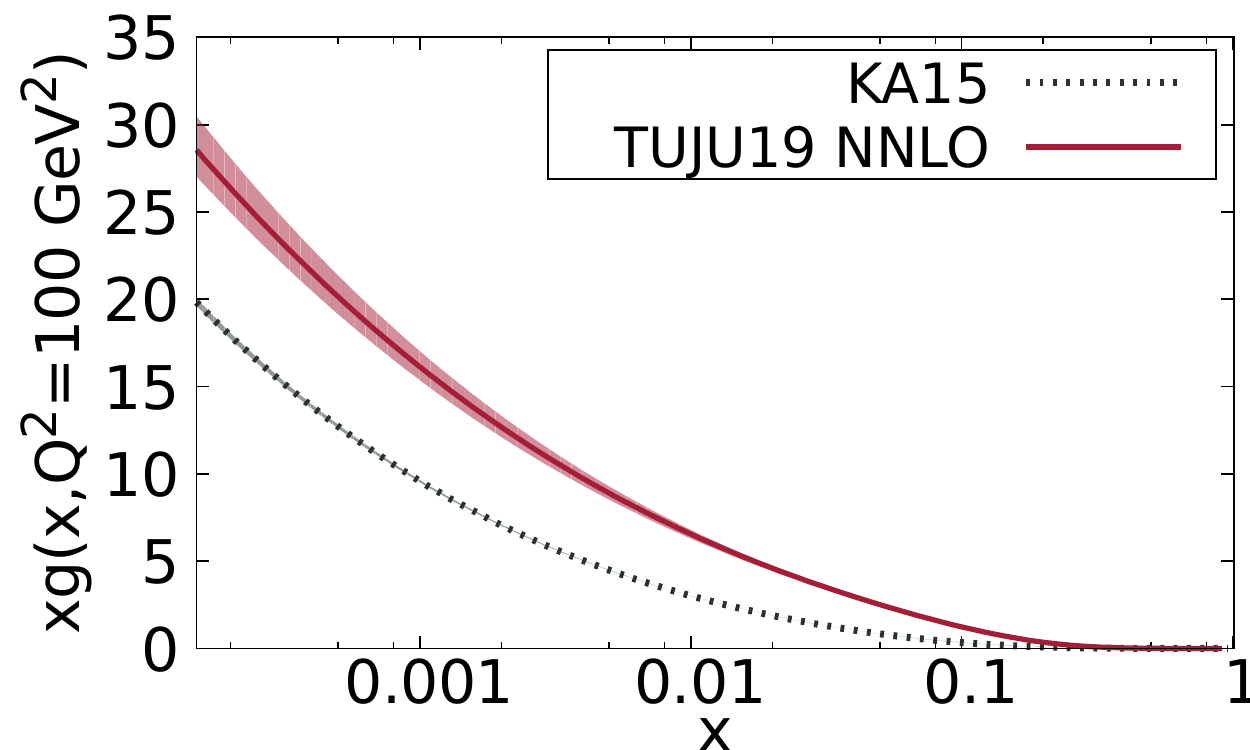}} 
          \subfigure{\includegraphics[width=0.325\textwidth]{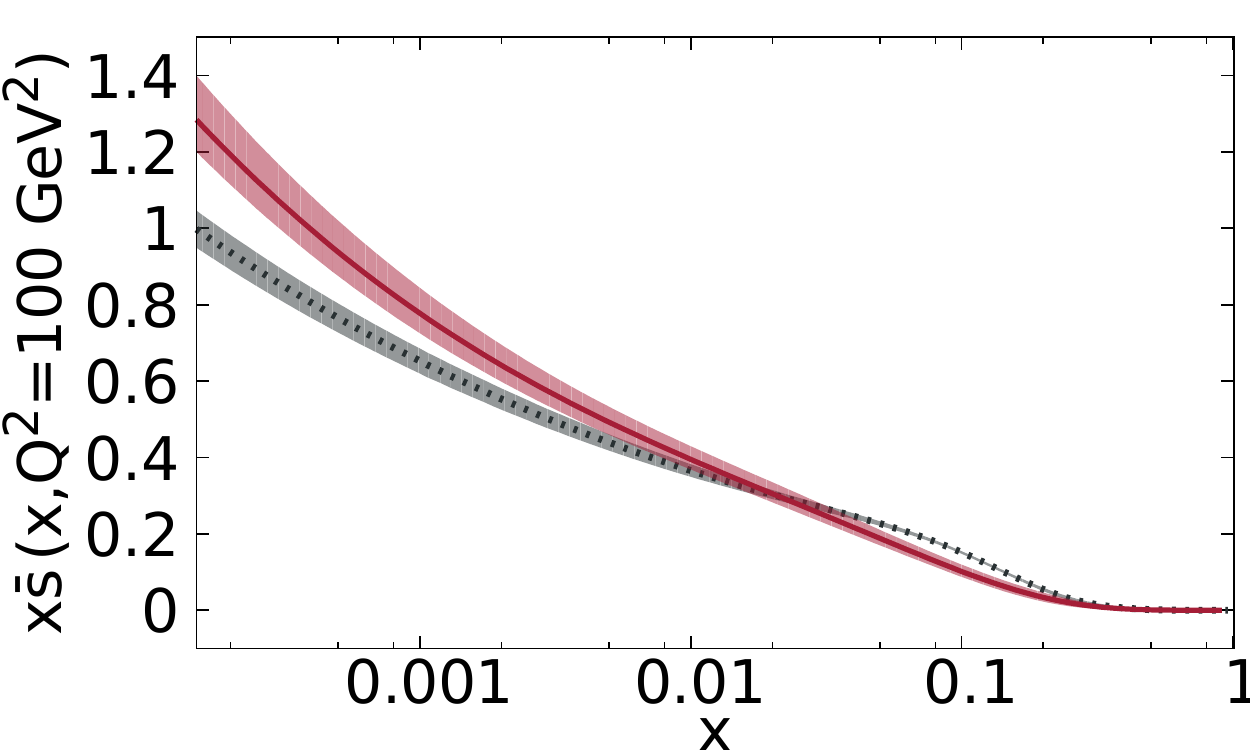}}
          \subfigure{\includegraphics[width=0.325\textwidth]{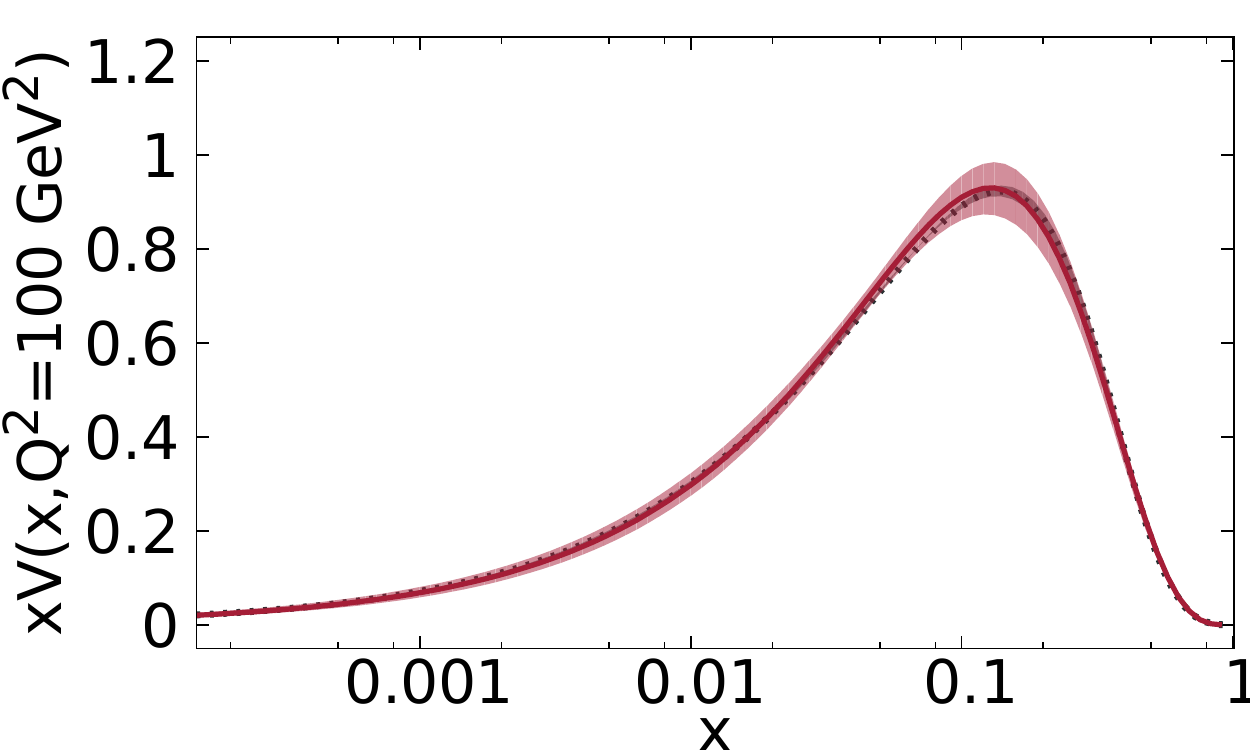}}      
          \end{center} 
    \caption{Nuclear parton distribution functions TUJU19 in lead at NNLO compared to the results by the KA15 group \cite{Khanpour:2016pph}, shown at the initial scale of KA15, $Q_0^2=2.0~\text{GeV}^2$, and at a higher scale $Q^2=100\,\mathrm{GeV}^2$. The comparison is presented for the distribution functions $xf_i(x,Q^2)$ with $i=g,\,\bar{s},\,V$, where $V$ is the sum of valence quarks, in a bound proton in lead.}
\label{fig-nPDF-comparisonQ2-NNLO}    
    \end{figure*}
The comparison of the TUJU19 NNLO nPDF fit to other NNLO nPDF analyses is shown in figure~\ref{fig-nPDF-comparisonQ0-NNLO} for nNNPDF1.0 \cite{AbdulKhalek:2019mzd} and in figure~\ref{fig-nPDF-comparisonQ2-NNLO} for KA15 \cite{Khanpour:2016pph}. The comparison at NNLO is separated into two figures since different information is available for the nNNPDF1.0\footnote{The nNNPDF1.0 LHAPDF6 set is provided with the assumption that $u = d$ and that $\bar{u} = \bar{d} = \bar{s} = s$ to comply with the LHAPDF format, i.e. to provide individual quark flavours. For our comparison, we have used the provided PDFs and LHAPDF uncertainties at the $90\%$ confidence level, keeping in mind that only the quark multiplets $\Sigma$, $T_3$, $T_8$ were determined in the nNNPDF1.0 analysis.} and  KA15\footnote{For the comparison at NNLO the KA15 nPDFs, provided on request by the authors, were also included. The ratios $R^{\mathrm{p/Pb}}$ were not available for this analysis.} analyses. In the case of nNNPDF1.0 we consider lead nPDFs for gluon $g$ and the quark singlet $\Sigma=u+\bar{u}+d+\bar{d}+s+\bar{s}$ (as per \cite{AbdulKhalek:2019mzd}). The comparisons are shown at the two scales $Q_0^2=1.69~\text{GeV}^2$ and $Q^2=100~\text{GeV}^2$ for the distribution functions, and at the higher scale $Q^2=100\,\mathrm{GeV}^2$ for the ratios $R_i^{\mathrm{p/Pb}}$ of PDFs in a proton bound in lead compared to PDFs in a free proton. Even though none of the analyses include data directly sensitive to the gluon distribution, a reasonable behaviour is found for the gluon at the initial scale of this analysis, keeping in mind that our uncertainty bands for the gluon are potentially underestimated, as discussed above. The congruence is improved when going towards higher scales (e.g.~$Q^2=100~\text{GeV}^2$). Furthermore, a very good agreement is observed for the quark singlet $\Sigma$ shown in the lower panels of figure \ref{fig-nPDF-comparisonQ0-NNLO}. The values of $x\Sigma$ are within the error bands at the initial scale, and become even more consistent at the higher scale. The observed deviation in the low-$x$ region ($x < 0.0005$) reflects the lack of low-$x$ constraints by the available nuclear DIS data. We also show the ratios $R^{\mathrm{p/Pb}}$ for TUJU19 compared to nNNPDF1.0 at NNLO in figure \ref{fig-nPDF-comparisonQ0-NNLO}. Again a reasonable shape is found for the gluon nuclear modification, and a very good agreement is visible for the quark singlet, as it is well constrained by the incorporated experimental data. The uncertainties of the nNNPDF1.0 distributions are considerably larger compared to the ones found in our analysis. Indeed, as shown in Ref.~\cite{AbdulKhalek:2019mzd}, the uncertainty bands for a combination of singlet and octet contributions become comparable to those in the earlier works where direct data constraints exist. Still, the nNNPDF1.0 uncertainties grow rapidly towards small-$x$ which can be accounted for by the fact that the applied neural network framework is not as prone to parameterization bias as the traditional Hessian error analysis applied here. 

The nuclear parton distribution functions TUJU19 at NNLO compared to the results of the KA15 group are presented in figure \ref{fig-nPDF-comparisonQ2-NNLO}. Here we consider lead nPDFs for gluons, sea quarks (here $\bar{s}$) and the sum of valence quarks, $V=u_v+d_v$. Again the comparisons are shown at two scales, at the initial scale of KA15, $Q_0^2=2.0\,\mathrm{GeV}^2$, and at $Q^2=100~\text{GeV}^2$. Considering the fact that neither of the analyses includes data directly sensitive to the gluon distribution, a fair agreement is found for the gluon at the initial scales. However, while the agreement between TUJU19 and nNNPDF1.0 remains at higher scales (fig. \ref{fig-nPDF-comparisonQ0-NNLO}), the gluon distribution from KA15 falls below the other two (fig. \ref{fig-nPDF-comparisonQ2-NNLO}) at $Q^2=100~\text{GeV}^2$. The $\bar{s}$ distributions in turn are in a reasonable agreement at higher scales, although at the initial scale the KA15 result is considerably above TUJU19. The total valence distributions from TUJU19 are found to be in very good agreement with those of the KA15 analysis. Apart from the gluon nPDFs at the initial scale, the KA15 uncertainties tend to be very small. This may partly follow from the rather rigid parameterization applied, but it may also be due to the chosen low error tolerance $\Delta \Chi^2=1$.

\section{Summary and outlook}
\label{sec:summary}

We have presented new sets of nPDFs, which we refer to as TUJU19, at NLO and NNLO. Contrary to most previous analyses, our nPDF sets are based on a proton baseline fitted within the same framework, which guarantees consistency throughout the analysis concerning the series of choices on parameter values, assumptions, constraints and kinematical cuts that need to be made when performing a global analysis. The numerical implementation was embedded in the open-source tool \textsc{xFitter}. The source code with all applied modifications required for the treatment of nuclear PDFs will be published, providing a first open-source tool for a global nuclear PDF analysis. The common framework will also enable a simultaneous proton and nuclear PDF analysis in the future.

For the proton PDF baseline analysis DIS data from the HERA, BCDMS and NMC experiments were included. Similarly, the nPDF analysis was based on measurements from fixed-target neutral-current DIS and data from neutrino-nucleus charged-current DIS experiments. The neutrino DIS data were implemented for the first time in a global pQCD analysis at NNLO. The deuteron was treated as a nucleus without neglecting nuclear effects, instead of constructing it as a pure composition of free proton and neutron PDFs as been assumed in several earlier analyses. The resulting cross sections show very good agreement with the included experimental data, both for neutral-current and charged-current DIS processes, as confirmed by the resulting $\Chi^2/N_{\mathrm{dp}}<1.0$ for the nuclear part of the analysis. The comparisons to the existing nPDF sets demonstrate a reasonable agreement within the error bands. The obtained results are consistent with the expectation that due to the consideration of DIS data only the nPDFs for valence quarks are well constrained by the experimental data, whereas gluon and sea quarks are constrained only indirectly by the included data and mostly by DGLAP evolution. The resulting nPDFs will be published in the LHAPDF6 format including uncertainties for both, the proton baseline and the nuclear PDF analysis, derived with a Hessian uncertainty analysis.

As this is the first NNLO nPDF fit within the developed framework, only fixed-target DIS data with lepton and neutrino beams were included. In the future we plan to also add data for other observables for which theoretical calculations at NNLO exist. The fixed-target proton-nucleus DY dilepton data would provide further constraints for the sea quark distributions. Furthermore, the $W^{pm}$ and $Z$ boson production data from p+Pb collisions at the LHC are sensitive to the flavour decomposition and could therefore help to disentangle observed differences in valence quark nuclear modifications. Even after these, direct gluon constraints will remain sparse. Recently it has been shown that such constraints could be obtained from the existing data for dijet and charmed-meson production in p+Pb collisions at the LHC. Further in the future, an electron-ion collider would provide precision data for nPDF analyses. In order to get the best information from these data the highest possible perturbative precision will be required, and we think our NNLO analysis is an important step in this context.

\section*{Acknowledgments}

The authors thank the \textsc{xFitter} development team for providing operative support. In particular Dr.~Alexander Glazov, Dr.~Oleksandr Zenaiev are acknowledged for sharing their expertise on various aspects of the tool and on the details of uncertainty analysis, kindly supported by Prof. Fred Olness. Furthermore, many thanks to Dr.~Valerio Bertone for clarifying details on the implementation of the FONLL scheme and to Dr.~Francesco Giuli for sharing the reference input files. The authors acknowledge discussions and scientific exchange with the other nPDF groups. We thank Dr.~Aleksander Kusina and Prof.~Fred Olness for discussing details in regards to the nCTEQ15 publication, and Dr.~Karol Kovarik for sharing his expertise on the nCTEQ framework. Furthermore, the authors acknowledge discussions, technical input, and comments on the manuscript by Dr.~Hannu Paukkunen. The authors are grateful to Dr.~Hamzeh Khanpour for sharing the KA15 PDFs. We thank Dr.~Pia Zurita for sending the DSSZ PDFs. The authors acknowledge Dr.~Marco Stratmann for valuable discussions and information. The authors acknowledge support by the state of Baden-W\"urttemberg through bwHPC providing the possibility to run the computational calculations on the high-performance cluster. This work was supported in part by the Bundesministerium f\"ur Bildung und Forschung (BMBF) grant 05P18VTCA1, Carl Zeiss Foundation and by the Academy of Finland, project n.o.~308301.

\appendix

\section{PDF parameters}
\label{app-pdf-params}

Here we show the parameters obtained for the proton and nuclear parton distribution functions presented in section \ref{sec-results}. The naming convention corresponds to the PDF parameterization given in equations (\ref{pdf-parameterization}) and (\ref{coeff-A}). Table \ref{tab-results-NLO} provides the NLO parameters, while table \ref{tab-results-NNLO} presents the NNLO ones. 

Some of the parameters were deliberately excluded from the fit. In most cases this means that in the initial version of the analysis procedure those parameters were used, but the obtained parameter value turned out to be very close to zero with very large uncertainty. Thus, that parameter was considered as not required. Alternatively, some of the nuclear parameters were never included as free parameters since the best fit criterion for nuclear PDFs, $\Chi^2\leq 1.0$, could be satisfied by the selected subset of the free parameters.
\begin{center}
\renewcommand{\arraystretch}{1.25}
\begin{table}[h]
\caption{{Values of the NLO fit parameters at the initial scale $Q_0^2=1.69\,\mathrm{GeV^2}$. (SR) means that the normalization for that particular parton is fixed by the sum rules. A dash symbolizes that this parameter was excluded from the fit. Parameter values for the sea quarks, apart from $\bar{u}$, were derived from the applied constraints $\bar{s}=s=\bar{d}=\bar{u}$.}}
\label{tab-results-NLO} 
\scriptsize
\begin{tabular}{lc|lc|lc|lc}
\hline
$g$ & value & $u_v$ & value & $d_v$ & value & $\bar{u}$ & value \\ \hline
\hline
$c^g_{0,0}$& 7.0352 & $c^{u_v}_{0,0}$ & (SR) & $c^{dv}_{0,0}$ & (SR) & $c^{\bar{u}}_{0,0}$& (SR)  \\ 
$c^g_{1,0}$& 0.2871 & $c^{u_v}_{1,0}$ & 0.6046 & $c^{dv}_{1,0}$ & 0.7376 & $c^{\bar{u}}_{1,0}$& -0.1915 \\ 
$c^g_{2,0}$& 14.243 & $c^{u_v}_{2,0}$ & 3.7064 & $c^{dv}_{2,0}$ & 2.9225 & $c^{\bar{u}}_{2,0}$& 7.5403 \\ 
$c^g_{3,0}$& 11.459 & $c^{u_v}_{3,0}$ & 4.6595 & $c^{dv}_{3,0}$ & -0.8736 & $c^{\bar{u}}_{3,0}$& 8.2448 \\ 
$c^g_{4,0}$& - & $c^{u_v}_{4,0}$ & - & $c^{dv}_{4,0}$ & - & $c^{\bar{u}}_{4,0}$& - \\  \hline
$c^g_{1,1}$& -50.064 & $c^{u_v}_{1,1}$ & -0.0616 & $c^{dv}_{1,1}$ & -52.218 & $c^{\bar{u}}_{1,1}$& -7.4250 \\ 
$c^g_{1,2}$& -0.0008 & $c^{u_v}_{1,2}$ & 0.4455 & $c^{dv}_{1,2}$ & -0.1002 & $c^{\bar{u}}_{1,2}$& -0.0021 \\ 
$c^g_{2,1}$& -6.5209 & $c^{u_v}_{2,1}$ & -11.643 & $c^{dv}_{2,1}$ & 3.1722 & $c^{\bar{u}}_{2,1}$& -0.2658 \\ 
$c^g_{2,2}$& 0.2039 & $c^{u_v}_{2,2}$ & 0.0002 & $c^{dv}_{2,2}$ & 0.1336 & $c^{\bar{u}}_{2,2}$& -0.2754 \\ \hline \hline
\end{tabular} 
\end{table}
\end{center}
\begin{center}
\renewcommand{\arraystretch}{1.25}
\begin{table}[h]
\caption{{Same as table \ref{tab-results-NLO}, but at NNLO.}}
\label{tab-results-NNLO} 
\scriptsize
\begin{tabular}{lc|lc|lc|lc}
\hline
$g$ & value & $u_v$ & value & $d_v$ & value & $\bar{u}$ & value \\ \hline
\hline
$c^g_{0,0}$& 6.2654 & $c^{u_v}_{0,0}$ & (SR) & $c^{dv}_{0,0}$ & (SR) & $c^{\bar{u}}_{0,0}$& (SR) \\ 
$c^g_{1,0}$& 0.2712 & $c^{u_v}_{1,0}$ & 0.8060 & $c^{dv}_{1,0}$ & 1.0227 & $c^{\bar{u}}_{1,0}$& -0.1162
\\ 
$c^g_{2,0}$& 11.334 & $c^{u_v}_{2,0}$ & 3.6897 & $c^{dv}_{2,0}$ & 4.2717 & $c^{\bar{u}}_{2,0}$& 7.1632 \\ 
$c^g_{3,0}$& 5.0606 & $c^{u_v}_{3,0}$ & 1.6388 & $c^{dv}_{3,0}$ & -0.6035 & $c^{\bar{u}}_{3,0}$& - \\ 
$c^g_{4,0}$& - & $c^{u_v}_{4,0}$ & - & $c^{dv}_{4,0}$ & - & $c^{\bar{u}}_{4,0}$&  2.4190 \\  \hline
$c^g_{1,1}$& -2.4627 & $c^{u_v}_{1,1}$ & -0.1080 & $c^{dv}_{1,1}$ & -2.8603 & $c^{\bar{u}}_{1,1}$& -3.2213 \\ 
$c^g_{1,2}$& -0.0024 & $c^{u_v}_{1,2}$ & 0.3766 & $c^{dv}_{1,2}$ & -0.0099 & $c^{\bar{u}}_{1,2}$& -0.0123 \\ 
$c^g_{2,1}$& -1.4764 & $c^{u_v}_{2,1}$ & 74.620 & $c^{dv}_{2,1}$ &  1.1235 & $c^{\bar{u}}_{2,1}$& -0.0028 \\ 
$c^g_{2,2}$& 0.3704 & $c^{u_v}_{2,2}$ & -0.0001 & $c^{dv}_{2,2}$ & 0.2357 & $c^{\bar{u}}_{2,2}$& -0.9263 \\ 
\hline \hline
\end{tabular}
\end{table}
\end{center}

\bibliography{Paper-nPDFs}

\begin{thebibliography}{122}%
\makeatletter
\providecommand \@ifxundefined [1]{%
 \@ifx{#1\undefined}
}%
\providecommand \@ifnum [1]{%
 \ifnum #1\expandafter \@firstoftwo
 \else \expandafter \@secondoftwo
 \fi
}%
\providecommand \@ifx [1]{%
 \ifx #1\expandafter \@firstoftwo
 \else \expandafter \@secondoftwo
 \fi
}%
\providecommand \natexlab [1]{#1}%
\providecommand \enquote  [1]{``#1''}%
\providecommand \bibnamefont  [1]{#1}%
\providecommand \bibfnamefont [1]{#1}%
\providecommand \citenamefont [1]{#1}%
\providecommand \href@noop [0]{\@secondoftwo}%
\providecommand \href [0]{\begingroup \@sanitize@url \@href}%
\providecommand \@href[1]{\@@startlink{#1}\@@href}%
\providecommand \@@href[1]{\endgroup#1\@@endlink}%
\providecommand \@sanitize@url [0]{\catcode `\\12\catcode `\$12\catcode
  `\&12\catcode `\#12\catcode `\^12\catcode `\_12\catcode `\%12\relax}%
\providecommand \@@startlink[1]{}%
\providecommand \@@endlink[0]{}%
\providecommand \url  [0]{\begingroup\@sanitize@url \@url }%
\providecommand \@url [1]{\endgroup\@href {#1}{\urlprefix }}%
\providecommand \urlprefix  [0]{URL }%
\providecommand \Eprint [0]{\href }%
\providecommand \doibase [0]{http://dx.doi.org/}%
\providecommand \selectlanguage [0]{\@gobble}%
\providecommand \bibinfo  [0]{\@secondoftwo}%
\providecommand \bibfield  [0]{\@secondoftwo}%
\providecommand \translation [1]{[#1]}%
\providecommand \BibitemOpen [0]{}%
\providecommand \bibitemStop [0]{}%
\providecommand \bibitemNoStop [0]{.\EOS\space}%
\providecommand \EOS [0]{\spacefactor3000\relax}%
\providecommand \BibitemShut  [1]{\csname bibitem#1\endcsname}%
\let\auto@bib@innerbib\@empty
\bibitem [{\citenamefont {Collins}\ \emph {et~al.}(1989)\citenamefont
  {Collins}, \citenamefont {Soper},\ and\ \citenamefont
  {Sterman}}]{Collins:1989gx}%
  \BibitemOpen
  \bibfield  {author} {\bibinfo {author} {\bibfnamefont {J.~C.}\ \bibnamefont
  {Collins}}, \bibinfo {author} {\bibfnamefont {D.~E.}\ \bibnamefont {Soper}},
  \ and\ \bibinfo {author} {\bibfnamefont {G.~F.}\ \bibnamefont {Sterman}},\
  }\href {\doibase 10.1142/9789814503266_0001} {\bibfield  {journal} {\bibinfo
  {journal} {Adv. Ser. Direct. High Energy Phys.}\ }\textbf {\bibinfo {volume}
  {5}},\ \bibinfo {pages} {1} (\bibinfo {year} {1989})},\ \Eprint
  {http://arxiv.org/abs/hep-ph/0409313} {arXiv:hep-ph/0409313 [hep-ph]}
  \BibitemShut {NoStop}%
\bibitem [{\citenamefont {Kovařík}\ \emph {et~al.}(2019)\citenamefont
  {Kovařík}, \citenamefont {Nadolsky},\ and\ \citenamefont
  {Soper}}]{Kovarik:2019xvh}%
  \BibitemOpen
  \bibfield  {author} {\bibinfo {author} {\bibfnamefont {K.}~\bibnamefont
  {Kovařík}}, \bibinfo {author} {\bibfnamefont {P.~M.}\ \bibnamefont
  {Nadolsky}}, \ and\ \bibinfo {author} {\bibfnamefont {D.~E.}\ \bibnamefont
  {Soper}},\ }\href@noop {} {\  (\bibinfo {year} {2019})},\ \Eprint
  {http://arxiv.org/abs/1905.06957} {arXiv:1905.06957 [hep-ph]} \BibitemShut
  {NoStop}%
\bibitem [{\citenamefont {Dokshitzer}\ \emph {et~al.}(1991)\citenamefont
  {Dokshitzer}, \citenamefont {Khoze}, \citenamefont {Mueller},\ and\
  \citenamefont {Troian}}]{Dokshitzer:1991wu}%
  \BibitemOpen
  \bibfield  {author} {\bibinfo {author} {\bibfnamefont {Y.~L.}\ \bibnamefont
  {Dokshitzer}}, \bibinfo {author} {\bibfnamefont {V.~A.}\ \bibnamefont
  {Khoze}}, \bibinfo {author} {\bibfnamefont {A.~H.}\ \bibnamefont {Mueller}},
  \ and\ \bibinfo {author} {\bibfnamefont {S.~I.}\ \bibnamefont {Troian}},\
  }\href@noop {} {\emph {\bibinfo {title} {{Basics of perturbative QCD}}}}\
  (\bibinfo  {publisher} {Ed. Frontieres},\ \bibinfo {year} {1991})\BibitemShut
  {NoStop}%
\bibitem [{\citenamefont {Brock}\ \emph {et~al.}(1995)\citenamefont {Brock}
  \emph {et~al.}}]{Brock:1993sz}%
  \BibitemOpen
  \bibfield  {author} {\bibinfo {author} {\bibfnamefont {R.}~\bibnamefont
  {Brock}} \emph {et~al.} (\bibinfo {collaboration} {CTEQ}),\ }\href {\doibase
  10.1103/RevModPhys.67.157} {\bibfield  {journal} {\bibinfo  {journal} {Rev.
  Mod. Phys.}\ }\textbf {\bibinfo {volume} {67}},\ \bibinfo {pages} {157}
  (\bibinfo {year} {1995})}\BibitemShut {NoStop}%
\bibitem [{\citenamefont {Lipatov}(1975)}]{Lipatov:1974qm}%
  \BibitemOpen
  \bibfield  {author} {\bibinfo {author} {\bibfnamefont {L.~N.}\ \bibnamefont
  {Lipatov}},\ }\href@noop {} {\bibfield  {journal} {\bibinfo  {journal} {Sov.
  J. Nucl. Phys.}\ }\textbf {\bibinfo {volume} {20}},\ \bibinfo {pages} {94}
  (\bibinfo {year} {1975})},\ \bibinfo {note} {[Yad.
  Fiz.20,181(1974)]}\BibitemShut {NoStop}%
\bibitem [{\citenamefont {Gribov}\ and\ \citenamefont
  {Lipatov}(1972)}]{Gribov:1972ri}%
  \BibitemOpen
  \bibfield  {author} {\bibinfo {author} {\bibfnamefont {V.~N.}\ \bibnamefont
  {Gribov}}\ and\ \bibinfo {author} {\bibfnamefont {L.~N.}\ \bibnamefont
  {Lipatov}},\ }\href@noop {} {\bibfield  {journal} {\bibinfo  {journal} {Sov.
  J. Nucl. Phys.}\ }\textbf {\bibinfo {volume} {15}},\ \bibinfo {pages} {438}
  (\bibinfo {year} {1972})},\ \bibinfo {note} {[Yad.
  Fiz.15,781(1972)]}\BibitemShut {NoStop}%
\bibitem [{\citenamefont {Altarelli}\ and\ \citenamefont
  {Parisi}(1977)}]{Altarelli:1977zs}%
  \BibitemOpen
  \bibfield  {author} {\bibinfo {author} {\bibfnamefont {G.}~\bibnamefont
  {Altarelli}}\ and\ \bibinfo {author} {\bibfnamefont {G.}~\bibnamefont
  {Parisi}},\ }\href {\doibase 10.1016/0550-3213(77)90384-4} {\bibfield
  {journal} {\bibinfo  {journal} {Nucl. Phys.}\ }\textbf {\bibinfo {volume}
  {B126}},\ \bibinfo {pages} {298} (\bibinfo {year} {1977})}\BibitemShut
  {NoStop}%
\bibitem [{\citenamefont {Dokshitzer}(1977)}]{Dokshitzer:1977sg}%
  \BibitemOpen
  \bibfield  {author} {\bibinfo {author} {\bibfnamefont {Y.~L.}\ \bibnamefont
  {Dokshitzer}},\ }\href@noop {} {\bibfield  {journal} {\bibinfo  {journal}
  {Sov. Phys. JETP}\ }\textbf {\bibinfo {volume} {46}},\ \bibinfo {pages} {641}
  (\bibinfo {year} {1977})},\ \bibinfo {note} {[Zh. Eksp. Teor.
  Fiz.73,1216(1977)]}\BibitemShut {NoStop}%
\bibitem [{\citenamefont {Abramowicz}\ \emph {et~al.}(2015)\citenamefont
  {Abramowicz} \emph {et~al.}}]{Abramowicz:2015mha}%
  \BibitemOpen
  \bibfield  {author} {\bibinfo {author} {\bibfnamefont {H.}~\bibnamefont
  {Abramowicz}} \emph {et~al.} (\bibinfo {collaboration} {H1, ZEUS}),\ }\href
  {\doibase 10.1140/epjc/s10052-015-3710-4} {\bibfield  {journal} {\bibinfo
  {journal} {Eur. Phys. J.}\ }\textbf {\bibinfo {volume} {C75}},\ \bibinfo
  {pages} {580} (\bibinfo {year} {2015})},\ \Eprint
  {http://arxiv.org/abs/1506.06042} {arXiv:1506.06042 [hep-ex]} \BibitemShut
  {NoStop}%
\bibitem [{\citenamefont {Gao}\ \emph {et~al.}(2018)\citenamefont {Gao},
  \citenamefont {Harland-Lang},\ and\ \citenamefont {Rojo}}]{Gao:2017yyd}%
  \BibitemOpen
  \bibfield  {author} {\bibinfo {author} {\bibfnamefont {J.}~\bibnamefont
  {Gao}}, \bibinfo {author} {\bibfnamefont {L.}~\bibnamefont {Harland-Lang}}, \
  and\ \bibinfo {author} {\bibfnamefont {J.}~\bibnamefont {Rojo}},\ }\href
  {\doibase 10.1016/j.physrep.2018.03.002} {\bibfield  {journal} {\bibinfo
  {journal} {Phys. Rept.}\ }\textbf {\bibinfo {volume} {742}},\ \bibinfo
  {pages} {1} (\bibinfo {year} {2018})},\ \Eprint
  {http://arxiv.org/abs/1709.04922} {arXiv:1709.04922 [hep-ph]} \BibitemShut
  {NoStop}%
\bibitem [{\citenamefont {Armesto}(2006)}]{Armesto:2006ph}%
  \BibitemOpen
  \bibfield  {author} {\bibinfo {author} {\bibfnamefont {N.}~\bibnamefont
  {Armesto}},\ }\href {\doibase 10.1088/0954-3899/32/11/R01} {\bibfield
  {journal} {\bibinfo  {journal} {J. Phys.}\ }\textbf {\bibinfo {volume}
  {G32}},\ \bibinfo {pages} {R367} (\bibinfo {year} {2006})},\ \Eprint
  {http://arxiv.org/abs/hep-ph/0604108} {arXiv:hep-ph/0604108 [hep-ph]}
  \BibitemShut {NoStop}%
\bibitem [{\citenamefont {de~Florian}\ \emph {et~al.}(2012)\citenamefont
  {de~Florian}, \citenamefont {Sassot}, \citenamefont {Zurita},\ and\
  \citenamefont {Stratmann}}]{deFlorian:2011fp}%
  \BibitemOpen
  \bibfield  {author} {\bibinfo {author} {\bibfnamefont {D.}~\bibnamefont
  {de~Florian}}, \bibinfo {author} {\bibfnamefont {R.}~\bibnamefont {Sassot}},
  \bibinfo {author} {\bibfnamefont {P.}~\bibnamefont {Zurita}}, \ and\ \bibinfo
  {author} {\bibfnamefont {M.}~\bibnamefont {Stratmann}},\ }\href {\doibase
  10.1103/PhysRevD.85.074028} {\bibfield  {journal} {\bibinfo  {journal} {Phys.
  Rev.}\ }\textbf {\bibinfo {volume} {D85}},\ \bibinfo {pages} {074028}
  (\bibinfo {year} {2012})},\ \Eprint {http://arxiv.org/abs/1112.6324}
  {arXiv:1112.6324 [hep-ph]} \BibitemShut {NoStop}%
\bibitem [{\citenamefont {Adler}\ \emph {et~al.}(2007)\citenamefont {Adler}
  \emph {et~al.}}]{Adler:2006wg}%
  \BibitemOpen
  \bibfield  {author} {\bibinfo {author} {\bibfnamefont {S.~S.}\ \bibnamefont
  {Adler}} \emph {et~al.} (\bibinfo {collaboration} {PHENIX}),\ }\href
  {\doibase 10.1103/PhysRevLett.98.172302} {\bibfield  {journal} {\bibinfo
  {journal} {Phys. Rev. Lett.}\ }\textbf {\bibinfo {volume} {98}},\ \bibinfo
  {pages} {172302} (\bibinfo {year} {2007})},\ \Eprint
  {http://arxiv.org/abs/nucl-ex/0610036} {arXiv:nucl-ex/0610036 [nucl-ex]}
  \BibitemShut {NoStop}%
\bibitem [{\citenamefont {Eskola}\ \emph {et~al.}(2008)\citenamefont {Eskola},
  \citenamefont {Paukkunen},\ and\ \citenamefont {Salgado}}]{Eskola:2008ca}%
  \BibitemOpen
  \bibfield  {author} {\bibinfo {author} {\bibfnamefont {K.~J.}\ \bibnamefont
  {Eskola}}, \bibinfo {author} {\bibfnamefont {H.}~\bibnamefont {Paukkunen}}, \
  and\ \bibinfo {author} {\bibfnamefont {C.~A.}\ \bibnamefont {Salgado}},\
  }\href {\doibase 10.1088/1126-6708/2008/07/102} {\bibfield  {journal}
  {\bibinfo  {journal} {JHEP}\ }\textbf {\bibinfo {volume} {07}},\ \bibinfo
  {pages} {102} (\bibinfo {year} {2008})},\ \Eprint
  {http://arxiv.org/abs/0802.0139} {arXiv:0802.0139 [hep-ph]} \BibitemShut
  {NoStop}%
\bibitem [{\citenamefont {Eskola}\ \emph {et~al.}(2009)\citenamefont {Eskola},
  \citenamefont {Paukkunen},\ and\ \citenamefont {Salgado}}]{Eskola:2009uj}%
  \BibitemOpen
  \bibfield  {author} {\bibinfo {author} {\bibfnamefont {K.~J.}\ \bibnamefont
  {Eskola}}, \bibinfo {author} {\bibfnamefont {H.}~\bibnamefont {Paukkunen}}, \
  and\ \bibinfo {author} {\bibfnamefont {C.~A.}\ \bibnamefont {Salgado}},\
  }\href {\doibase 10.1088/1126-6708/2009/04/065} {\bibfield  {journal}
  {\bibinfo  {journal} {JHEP}\ }\textbf {\bibinfo {volume} {04}},\ \bibinfo
  {pages} {065} (\bibinfo {year} {2009})},\ \Eprint
  {http://arxiv.org/abs/0902.4154} {arXiv:0902.4154 [hep-ph]} \BibitemShut
  {NoStop}%
\bibitem [{\citenamefont {Kovarik}\ \emph {et~al.}(2016)\citenamefont {Kovarik}
  \emph {et~al.}}]{Kovarik:2015cma}%
  \BibitemOpen
  \bibfield  {author} {\bibinfo {author} {\bibfnamefont {K.}~\bibnamefont
  {Kovarik}} \emph {et~al.},\ }\href {\doibase 10.1103/PhysRevD.93.085037}
  {\bibfield  {journal} {\bibinfo  {journal} {Phys. Rev.}\ }\textbf {\bibinfo
  {volume} {D93}},\ \bibinfo {pages} {085037} (\bibinfo {year} {2016})},\
  \Eprint {http://arxiv.org/abs/1509.00792} {arXiv:1509.00792 [hep-ph]}
  \BibitemShut {NoStop}%
\bibitem [{\citenamefont {Eskola}\ \emph {et~al.}(2017)\citenamefont {Eskola},
  \citenamefont {Paakkinen}, \citenamefont {Paukkunen},\ and\ \citenamefont
  {Salgado}}]{Eskola:2016oht}%
  \BibitemOpen
  \bibfield  {author} {\bibinfo {author} {\bibfnamefont {K.~J.}\ \bibnamefont
  {Eskola}}, \bibinfo {author} {\bibfnamefont {P.}~\bibnamefont {Paakkinen}},
  \bibinfo {author} {\bibfnamefont {H.}~\bibnamefont {Paukkunen}}, \ and\
  \bibinfo {author} {\bibfnamefont {C.~A.}\ \bibnamefont {Salgado}},\ }\href
  {\doibase 10.1140/epjc/s10052-017-4725-9} {\bibfield  {journal} {\bibinfo
  {journal} {Eur. Phys. J.}\ }\textbf {\bibinfo {volume} {C77}},\ \bibinfo
  {pages} {163} (\bibinfo {year} {2017})},\ \Eprint
  {http://arxiv.org/abs/1612.05741} {arXiv:1612.05741 [hep-ph]} \BibitemShut
  {NoStop}%
\bibitem [{\citenamefont {Khachatryan}\ \emph {et~al.}(2015)\citenamefont
  {Khachatryan} \emph {et~al.}}]{Khachatryan:2015hha}%
  \BibitemOpen
  \bibfield  {author} {\bibinfo {author} {\bibfnamefont {V.}~\bibnamefont
  {Khachatryan}} \emph {et~al.} (\bibinfo {collaboration} {CMS}),\ }\href
  {\doibase 10.1016/j.physletb.2015.09.057} {\bibfield  {journal} {\bibinfo
  {journal} {Phys. Lett.}\ }\textbf {\bibinfo {volume} {B750}},\ \bibinfo
  {pages} {565} (\bibinfo {year} {2015})},\ \Eprint
  {http://arxiv.org/abs/1503.05825} {arXiv:1503.05825 [nucl-ex]} \BibitemShut
  {NoStop}%
\bibitem [{\citenamefont {Khachatryan}\ \emph {et~al.}(2016)\citenamefont
  {Khachatryan} \emph {et~al.}}]{Khachatryan:2015pzs}%
  \BibitemOpen
  \bibfield  {author} {\bibinfo {author} {\bibfnamefont {V.}~\bibnamefont
  {Khachatryan}} \emph {et~al.} (\bibinfo {collaboration} {CMS}),\ }\href
  {\doibase 10.1016/j.physletb.2016.05.044} {\bibfield  {journal} {\bibinfo
  {journal} {Phys. Lett.}\ }\textbf {\bibinfo {volume} {B759}},\ \bibinfo
  {pages} {36} (\bibinfo {year} {2016})},\ \Eprint
  {http://arxiv.org/abs/1512.06461} {arXiv:1512.06461 [hep-ex]} \BibitemShut
  {NoStop}%
\bibitem [{\citenamefont {Aad}\ \emph {et~al.}(2015)\citenamefont {Aad} \emph
  {et~al.}}]{Aad:2015gta}%
  \BibitemOpen
  \bibfield  {author} {\bibinfo {author} {\bibfnamefont {G.}~\bibnamefont
  {Aad}} \emph {et~al.} (\bibinfo {collaboration} {ATLAS}),\ }\href {\doibase
  10.1103/PhysRevC.92.044915} {\bibfield  {journal} {\bibinfo  {journal} {Phys.
  Rev.}\ }\textbf {\bibinfo {volume} {C92}},\ \bibinfo {pages} {044915}
  (\bibinfo {year} {2015})},\ \Eprint {http://arxiv.org/abs/1507.06232}
  {arXiv:1507.06232 [hep-ex]} \BibitemShut {NoStop}%
\bibitem [{\citenamefont {Chatrchyan}\ \emph {et~al.}(2014)\citenamefont
  {Chatrchyan} \emph {et~al.}}]{Chatrchyan:2014hqa}%
  \BibitemOpen
  \bibfield  {author} {\bibinfo {author} {\bibfnamefont {S.}~\bibnamefont
  {Chatrchyan}} \emph {et~al.} (\bibinfo {collaboration} {CMS}),\ }\href
  {\doibase 10.1140/epjc/s10052-014-2951-y} {\bibfield  {journal} {\bibinfo
  {journal} {Eur. Phys. J.}\ }\textbf {\bibinfo {volume} {C74}},\ \bibinfo
  {pages} {2951} (\bibinfo {year} {2014})},\ \Eprint
  {http://arxiv.org/abs/1401.4433} {arXiv:1401.4433 [nucl-ex]} \BibitemShut
  {NoStop}%
\bibitem [{\citenamefont {Eskola}\ \emph
  {et~al.}(2019{\natexlab{a}})\citenamefont {Eskola}, \citenamefont
  {Paakkinen},\ and\ \citenamefont {Paukkunen}}]{Eskola:2019dui}%
  \BibitemOpen
  \bibfield  {author} {\bibinfo {author} {\bibfnamefont {K.~J.}\ \bibnamefont
  {Eskola}}, \bibinfo {author} {\bibfnamefont {P.}~\bibnamefont {Paakkinen}}, \
  and\ \bibinfo {author} {\bibfnamefont {H.}~\bibnamefont {Paukkunen}},\
  }\href@noop {} {\bibfield  {journal} {\bibinfo  {journal} {Eur. Phys. J.}\
  }\textbf {\bibinfo {volume} {C79}},\ \bibinfo {pages} {511} (\bibinfo {year}
  {2019}{\natexlab{a}})},\ \Eprint {http://arxiv.org/abs/1903.09832}
  {arXiv:1903.09832 [hep-ph]} \BibitemShut {NoStop}%
\bibitem [{\citenamefont {Zenaiev}\ \emph {et~al.}(2015)\citenamefont {Zenaiev}
  \emph {et~al.}}]{Zenaiev:2015rfa}%
  \BibitemOpen
  \bibfield  {author} {\bibinfo {author} {\bibfnamefont {O.}~\bibnamefont
  {Zenaiev}} \emph {et~al.} (\bibinfo {collaboration} {PROSA}),\ }\href
  {\doibase 10.1140/epjc/s10052-015-3618-z} {\bibfield  {journal} {\bibinfo
  {journal} {Eur. Phys. J.}\ }\textbf {\bibinfo {volume} {C75}},\ \bibinfo
  {pages} {396} (\bibinfo {year} {2015})},\ \Eprint
  {http://arxiv.org/abs/1503.04581} {arXiv:1503.04581 [hep-ph]} \BibitemShut
  {NoStop}%
\bibitem [{\citenamefont {Cacciari}\ \emph {et~al.}(2015)\citenamefont
  {Cacciari}, \citenamefont {Mangano},\ and\ \citenamefont
  {Nason}}]{Cacciari:2015fta}%
  \BibitemOpen
  \bibfield  {author} {\bibinfo {author} {\bibfnamefont {M.}~\bibnamefont
  {Cacciari}}, \bibinfo {author} {\bibfnamefont {M.~L.}\ \bibnamefont
  {Mangano}}, \ and\ \bibinfo {author} {\bibfnamefont {P.}~\bibnamefont
  {Nason}},\ }\href {\doibase 10.1140/epjc/s10052-015-3814-x} {\bibfield
  {journal} {\bibinfo  {journal} {Eur. Phys. J.}\ }\textbf {\bibinfo {volume}
  {C75}},\ \bibinfo {pages} {610} (\bibinfo {year} {2015})},\ \Eprint
  {http://arxiv.org/abs/1507.06197} {arXiv:1507.06197 [hep-ph]} \BibitemShut
  {NoStop}%
\bibitem [{\citenamefont {Gauld}\ and\ \citenamefont
  {Rojo}(2017)}]{Gauld:2016kpd}%
  \BibitemOpen
  \bibfield  {author} {\bibinfo {author} {\bibfnamefont {R.}~\bibnamefont
  {Gauld}}\ and\ \bibinfo {author} {\bibfnamefont {J.}~\bibnamefont {Rojo}},\
  }\href {\doibase 10.1103/PhysRevLett.118.072001} {\bibfield  {journal}
  {\bibinfo  {journal} {Phys. Rev. Lett.}\ }\textbf {\bibinfo {volume} {118}},\
  \bibinfo {pages} {072001} (\bibinfo {year} {2017})},\ \Eprint
  {http://arxiv.org/abs/1610.09373} {arXiv:1610.09373 [hep-ph]} \BibitemShut
  {NoStop}%
\bibitem [{\citenamefont {De~Oliveira}\ \emph {et~al.}(2018)\citenamefont
  {De~Oliveira}, \citenamefont {Martin},\ and\ \citenamefont
  {Ryskin}}]{deOliveira:2017lop}%
  \BibitemOpen
  \bibfield  {author} {\bibinfo {author} {\bibfnamefont {E.~G.}\ \bibnamefont
  {De~Oliveira}}, \bibinfo {author} {\bibfnamefont {A.~D.}\ \bibnamefont
  {Martin}}, \ and\ \bibinfo {author} {\bibfnamefont {M.~G.}\ \bibnamefont
  {Ryskin}},\ }\href {\doibase 10.1103/PhysRevD.97.074021} {\bibfield
  {journal} {\bibinfo  {journal} {Phys. Rev.}\ }\textbf {\bibinfo {volume}
  {D97}},\ \bibinfo {pages} {074021} (\bibinfo {year} {2018})},\ \Eprint
  {http://arxiv.org/abs/1712.06834} {arXiv:1712.06834 [hep-ph]} \BibitemShut
  {NoStop}%
\bibitem [{\citenamefont {Eskola}\ \emph
  {et~al.}(2019{\natexlab{b}})\citenamefont {Eskola}, \citenamefont {Helenius},
  \citenamefont {Paakkinen},\ and\ \citenamefont {Paukkunen}}]{Eskola:2019bgf}%
  \BibitemOpen
  \bibfield  {author} {\bibinfo {author} {\bibfnamefont {K.~J.}\ \bibnamefont
  {Eskola}}, \bibinfo {author} {\bibfnamefont {I.}~\bibnamefont {Helenius}},
  \bibinfo {author} {\bibfnamefont {P.}~\bibnamefont {Paakkinen}}, \ and\
  \bibinfo {author} {\bibfnamefont {H.}~\bibnamefont {Paukkunen}},\ }\href@noop
  {} {\  (\bibinfo {year} {2019}{\natexlab{b}})},\ \Eprint
  {http://arxiv.org/abs/1906.02512} {arXiv:1906.02512 [hep-ph]} \BibitemShut
  {NoStop}%
\bibitem [{\citenamefont {Helenius}\ \emph {et~al.}(2014)\citenamefont
  {Helenius}, \citenamefont {Eskola},\ and\ \citenamefont
  {Paukkunen}}]{Helenius:2014qla}%
  \BibitemOpen
  \bibfield  {author} {\bibinfo {author} {\bibfnamefont {I.}~\bibnamefont
  {Helenius}}, \bibinfo {author} {\bibfnamefont {K.~J.}\ \bibnamefont
  {Eskola}}, \ and\ \bibinfo {author} {\bibfnamefont {H.}~\bibnamefont
  {Paukkunen}},\ }\href {\doibase 10.1007/JHEP09(2014)138} {\bibfield
  {journal} {\bibinfo  {journal} {JHEP}\ }\textbf {\bibinfo {volume} {09}},\
  \bibinfo {pages} {138} (\bibinfo {year} {2014})},\ \Eprint
  {http://arxiv.org/abs/1406.1689} {arXiv:1406.1689 [hep-ph]} \BibitemShut
  {NoStop}%
\bibitem [{\citenamefont {Peitzmann}(2018)}]{Peitzmann:2018kie}%
  \BibitemOpen
  \bibfield  {author} {\bibinfo {author} {\bibfnamefont {T.}~\bibnamefont
  {Peitzmann}} (\bibinfo {collaboration} {ALICE}),\ }\bibfield  {booktitle}
  {\emph {\bibinfo {booktitle} {{Proceedings, 9th International Conference on
  Hard and Electromagnetic Probes of High-Energy Nuclear Collisions: Hard
  Probes 2018 (HP2018): Aix-Les-Bains, France, October 1-5, 2018}}},\ }\href
  {\doibase 10.22323/1.345.0062} {\bibfield  {journal} {\bibinfo  {journal}
  {PoS}\ }\textbf {\bibinfo {volume} {HardProbes2018}},\ \bibinfo {pages} {062}
  (\bibinfo {year} {2018})},\ \Eprint {http://arxiv.org/abs/1812.07246}
  {arXiv:1812.07246 [hep-ex]} \BibitemShut {NoStop}%
\bibitem [{\citenamefont {Harland-Lang}\ \emph {et~al.}(2015)\citenamefont
  {Harland-Lang}, \citenamefont {Martin}, \citenamefont {Motylinski},\ and\
  \citenamefont {Thorne}}]{Harland-Lang:2014zoa}%
  \BibitemOpen
  \bibfield  {author} {\bibinfo {author} {\bibfnamefont {L.~A.}\ \bibnamefont
  {Harland-Lang}}, \bibinfo {author} {\bibfnamefont {A.~D.}\ \bibnamefont
  {Martin}}, \bibinfo {author} {\bibfnamefont {P.}~\bibnamefont {Motylinski}},
  \ and\ \bibinfo {author} {\bibfnamefont {R.~S.}\ \bibnamefont {Thorne}},\
  }\href {\doibase 10.1140/epjc/s10052-015-3397-6} {\bibfield  {journal}
  {\bibinfo  {journal} {Eur. Phys. J.}\ }\textbf {\bibinfo {volume} {C75}},\
  \bibinfo {pages} {204} (\bibinfo {year} {2015})},\ \Eprint
  {http://arxiv.org/abs/1412.3989} {arXiv:1412.3989 [hep-ph]} \BibitemShut
  {NoStop}%
\bibitem [{\citenamefont {Dulat}\ \emph {et~al.}(2016)\citenamefont {Dulat},
  \citenamefont {Hou}, \citenamefont {Gao}, \citenamefont {Guzzi},
  \citenamefont {Huston}, \citenamefont {Nadolsky}, \citenamefont {Pumplin},
  \citenamefont {Schmidt}, \citenamefont {Stump},\ and\ \citenamefont
  {Yuan}}]{Dulat:2015mca}%
  \BibitemOpen
  \bibfield  {author} {\bibinfo {author} {\bibfnamefont {S.}~\bibnamefont
  {Dulat}}, \bibinfo {author} {\bibfnamefont {T.-J.}\ \bibnamefont {Hou}},
  \bibinfo {author} {\bibfnamefont {J.}~\bibnamefont {Gao}}, \bibinfo {author}
  {\bibfnamefont {M.}~\bibnamefont {Guzzi}}, \bibinfo {author} {\bibfnamefont
  {J.}~\bibnamefont {Huston}}, \bibinfo {author} {\bibfnamefont
  {P.}~\bibnamefont {Nadolsky}}, \bibinfo {author} {\bibfnamefont
  {J.}~\bibnamefont {Pumplin}}, \bibinfo {author} {\bibfnamefont
  {C.}~\bibnamefont {Schmidt}}, \bibinfo {author} {\bibfnamefont
  {D.}~\bibnamefont {Stump}}, \ and\ \bibinfo {author} {\bibfnamefont {C.~P.}\
  \bibnamefont {Yuan}},\ }\href {\doibase 10.1103/PhysRevD.93.033006}
  {\bibfield  {journal} {\bibinfo  {journal} {Phys. Rev.}\ }\textbf {\bibinfo
  {volume} {D93}},\ \bibinfo {pages} {033006} (\bibinfo {year} {2016})},\
  \Eprint {http://arxiv.org/abs/1506.07443} {arXiv:1506.07443 [hep-ph]}
  \BibitemShut {NoStop}%
\bibitem [{\citenamefont {Ball}\ \emph {et~al.}(2017)\citenamefont {Ball} \emph
  {et~al.}}]{Ball:2017nwa}%
  \BibitemOpen
  \bibfield  {author} {\bibinfo {author} {\bibfnamefont {R.~D.}\ \bibnamefont
  {Ball}} \emph {et~al.} (\bibinfo {collaboration} {NNPDF}),\ }\href {\doibase
  10.1140/epjc/s10052-017-5199-5} {\bibfield  {journal} {\bibinfo  {journal}
  {Eur. Phys. J.}\ }\textbf {\bibinfo {volume} {C77}},\ \bibinfo {pages} {663}
  (\bibinfo {year} {2017})},\ \Eprint {http://arxiv.org/abs/1706.00428}
  {arXiv:1706.00428 [hep-ph]} \BibitemShut {NoStop}%
\bibitem [{\citenamefont {Alekhin}\ \emph {et~al.}(2017)\citenamefont
  {Alekhin}, \citenamefont {Bl\"{u}mlein}, \citenamefont {Moch},\ and\
  \citenamefont {Placakyte}}]{Alekhin:2017kpj}%
  \BibitemOpen
  \bibfield  {author} {\bibinfo {author} {\bibfnamefont {S.}~\bibnamefont
  {Alekhin}}, \bibinfo {author} {\bibfnamefont {J.}~\bibnamefont
  {Bl\"{u}mlein}}, \bibinfo {author} {\bibfnamefont {S.}~\bibnamefont {Moch}},
  \ and\ \bibinfo {author} {\bibfnamefont {R.}~\bibnamefont {Placakyte}},\
  }\href {\doibase 10.1103/PhysRevD.96.014011} {\bibfield  {journal} {\bibinfo
  {journal} {Phys. Rev.}\ }\textbf {\bibinfo {volume} {D96}},\ \bibinfo {pages}
  {014011} (\bibinfo {year} {2017})},\ \Eprint
  {http://arxiv.org/abs/1701.05838} {arXiv:1701.05838 [hep-ph]} \BibitemShut
  {NoStop}%
\bibitem [{\citenamefont {Vogt}\ \emph {et~al.}(2004)\citenamefont {Vogt},
  \citenamefont {Moch},\ and\ \citenamefont {Vermaseren}}]{Vogt:2004mw}%
  \BibitemOpen
  \bibfield  {author} {\bibinfo {author} {\bibfnamefont {A.}~\bibnamefont
  {Vogt}}, \bibinfo {author} {\bibfnamefont {S.}~\bibnamefont {Moch}}, \ and\
  \bibinfo {author} {\bibfnamefont {J.~A.~M.}\ \bibnamefont {Vermaseren}},\
  }\href {\doibase 10.1016/j.nuclphysb.2004.04.024} {\bibfield  {journal}
  {\bibinfo  {journal} {Nucl. Phys.}\ }\textbf {\bibinfo {volume} {B691}},\
  \bibinfo {pages} {129} (\bibinfo {year} {2004})},\ \Eprint
  {http://arxiv.org/abs/hep-ph/0404111} {arXiv:hep-ph/0404111 [hep-ph]}
  \BibitemShut {NoStop}%
\bibitem [{\citenamefont {Moch}\ \emph {et~al.}(2004)\citenamefont {Moch},
  \citenamefont {Vermaseren},\ and\ \citenamefont {Vogt}}]{Moch:2004pa}%
  \BibitemOpen
  \bibfield  {author} {\bibinfo {author} {\bibfnamefont {S.}~\bibnamefont
  {Moch}}, \bibinfo {author} {\bibfnamefont {J.~A.~M.}\ \bibnamefont
  {Vermaseren}}, \ and\ \bibinfo {author} {\bibfnamefont {A.}~\bibnamefont
  {Vogt}},\ }\href {\doibase 10.1016/j.nuclphysb.2004.03.030} {\bibfield
  {journal} {\bibinfo  {journal} {Nucl. Phys.}\ }\textbf {\bibinfo {volume}
  {B688}},\ \bibinfo {pages} {101} (\bibinfo {year} {2004})},\ \Eprint
  {http://arxiv.org/abs/hep-ph/0403192} {arXiv:hep-ph/0403192 [hep-ph]}
  \BibitemShut {NoStop}%
\bibitem [{\citenamefont {Boughezal}\ \emph {et~al.}(2017)\citenamefont
  {Boughezal}, \citenamefont {Campbell}, \citenamefont {Ellis}, \citenamefont
  {Focke}, \citenamefont {Giele}, \citenamefont {Liu}, \citenamefont
  {Petriello},\ and\ \citenamefont {Williams}}]{Boughezal:2016wmq}%
  \BibitemOpen
  \bibfield  {author} {\bibinfo {author} {\bibfnamefont {R.}~\bibnamefont
  {Boughezal}}, \bibinfo {author} {\bibfnamefont {J.~M.}\ \bibnamefont
  {Campbell}}, \bibinfo {author} {\bibfnamefont {R.~K.}\ \bibnamefont {Ellis}},
  \bibinfo {author} {\bibfnamefont {C.}~\bibnamefont {Focke}}, \bibinfo
  {author} {\bibfnamefont {W.}~\bibnamefont {Giele}}, \bibinfo {author}
  {\bibfnamefont {X.}~\bibnamefont {Liu}}, \bibinfo {author} {\bibfnamefont
  {F.}~\bibnamefont {Petriello}}, \ and\ \bibinfo {author} {\bibfnamefont
  {C.}~\bibnamefont {Williams}},\ }\href {\doibase
  10.1140/epjc/s10052-016-4558-y} {\bibfield  {journal} {\bibinfo  {journal}
  {Eur. Phys. J.}\ }\textbf {\bibinfo {volume} {C77}},\ \bibinfo {pages} {7}
  (\bibinfo {year} {2017})},\ \Eprint {http://arxiv.org/abs/1605.08011}
  {arXiv:1605.08011 [hep-ph]} \BibitemShut {NoStop}%
\bibitem [{\citenamefont {Currie}\ \emph {et~al.}(2017)\citenamefont {Currie},
  \citenamefont {Glover},\ and\ \citenamefont {Pires}}]{Currie:2016bfm}%
  \BibitemOpen
  \bibfield  {author} {\bibinfo {author} {\bibfnamefont {J.}~\bibnamefont
  {Currie}}, \bibinfo {author} {\bibfnamefont {E.~W.~N.}\ \bibnamefont
  {Glover}}, \ and\ \bibinfo {author} {\bibfnamefont {J.}~\bibnamefont
  {Pires}},\ }\href {\doibase 10.1103/PhysRevLett.118.072002} {\bibfield
  {journal} {\bibinfo  {journal} {Phys. Rev. Lett.}\ }\textbf {\bibinfo
  {volume} {118}},\ \bibinfo {pages} {072002} (\bibinfo {year} {2017})},\
  \Eprint {http://arxiv.org/abs/1611.01460} {arXiv:1611.01460 [hep-ph]}
  \BibitemShut {NoStop}%
\bibitem [{\citenamefont {Campbell}\ \emph {et~al.}(2017)\citenamefont
  {Campbell}, \citenamefont {Ellis},\ and\ \citenamefont
  {Williams}}]{Campbell:2016lzl}%
  \BibitemOpen
  \bibfield  {author} {\bibinfo {author} {\bibfnamefont {J.~M.}\ \bibnamefont
  {Campbell}}, \bibinfo {author} {\bibfnamefont {R.~K.}\ \bibnamefont {Ellis}},
  \ and\ \bibinfo {author} {\bibfnamefont {C.}~\bibnamefont {Williams}},\
  }\href {\doibase 10.1103/PhysRevLett.118.222001} {\bibfield  {journal}
  {\bibinfo  {journal} {Phys. Rev. Lett.}\ }\textbf {\bibinfo {volume} {118}},\
  \bibinfo {pages} {222001} (\bibinfo {year} {2017})},\ \Eprint
  {http://arxiv.org/abs/1612.04333} {arXiv:1612.04333 [hep-ph]} \BibitemShut
  {NoStop}%
\bibitem [{\citenamefont {Gaunt}\ \emph {et~al.}(2015)\citenamefont {Gaunt},
  \citenamefont {Stahlhofen}, \citenamefont {Tackmann},\ and\ \citenamefont
  {Walsh}}]{Gaunt:2015pea}%
  \BibitemOpen
  \bibfield  {author} {\bibinfo {author} {\bibfnamefont {J.}~\bibnamefont
  {Gaunt}}, \bibinfo {author} {\bibfnamefont {M.}~\bibnamefont {Stahlhofen}},
  \bibinfo {author} {\bibfnamefont {F.~J.}\ \bibnamefont {Tackmann}}, \ and\
  \bibinfo {author} {\bibfnamefont {J.~R.}\ \bibnamefont {Walsh}},\ }\href
  {\doibase 10.1007/JHEP09(2015)058} {\bibfield  {journal} {\bibinfo  {journal}
  {JHEP}\ }\textbf {\bibinfo {volume} {09}},\ \bibinfo {pages} {058} (\bibinfo
  {year} {2015})},\ \Eprint {http://arxiv.org/abs/1505.04794} {arXiv:1505.04794
  [hep-ph]} \BibitemShut {NoStop}%
\bibitem [{\citenamefont {Czakon}\ \emph {et~al.}(2016)\citenamefont {Czakon},
  \citenamefont {Fiedler}, \citenamefont {Heymes},\ and\ \citenamefont
  {Mitov}}]{Czakon:2016ckf}%
  \BibitemOpen
  \bibfield  {author} {\bibinfo {author} {\bibfnamefont {M.}~\bibnamefont
  {Czakon}}, \bibinfo {author} {\bibfnamefont {P.}~\bibnamefont {Fiedler}},
  \bibinfo {author} {\bibfnamefont {D.}~\bibnamefont {Heymes}}, \ and\ \bibinfo
  {author} {\bibfnamefont {A.}~\bibnamefont {Mitov}},\ }\href {\doibase
  10.1007/JHEP05(2016)034} {\bibfield  {journal} {\bibinfo  {journal} {JHEP}\
  }\textbf {\bibinfo {volume} {05}},\ \bibinfo {pages} {034} (\bibinfo {year}
  {2016})},\ \Eprint {http://arxiv.org/abs/1601.05375} {arXiv:1601.05375
  [hep-ph]} \BibitemShut {NoStop}%
\bibitem [{\citenamefont {Grazzini}\ \emph {et~al.}(2018)\citenamefont
  {Grazzini}, \citenamefont {Kallweit},\ and\ \citenamefont
  {Wiesemann}}]{Grazzini:2017mhc}%
  \BibitemOpen
  \bibfield  {author} {\bibinfo {author} {\bibfnamefont {M.}~\bibnamefont
  {Grazzini}}, \bibinfo {author} {\bibfnamefont {S.}~\bibnamefont {Kallweit}},
  \ and\ \bibinfo {author} {\bibfnamefont {M.}~\bibnamefont {Wiesemann}},\
  }\href {\doibase 10.1140/epjc/s10052-018-5771-7} {\bibfield  {journal}
  {\bibinfo  {journal} {Eur. Phys. J.}\ }\textbf {\bibinfo {volume} {C78}},\
  \bibinfo {pages} {537} (\bibinfo {year} {2018})},\ \Eprint
  {http://arxiv.org/abs/1711.06631} {arXiv:1711.06631 [hep-ph]} \BibitemShut
  {NoStop}%
\bibitem [{\citenamefont {Chen}\ \emph {et~al.}(2019)\citenamefont {Chen},
  \citenamefont {Gehrmann}, \citenamefont {Glover}, \citenamefont {Hoefer},\
  and\ \citenamefont {Huss}}]{Chen:2019zmr}%
  \BibitemOpen
  \bibfield  {author} {\bibinfo {author} {\bibfnamefont {X.}~\bibnamefont
  {Chen}}, \bibinfo {author} {\bibfnamefont {T.}~\bibnamefont {Gehrmann}},
  \bibinfo {author} {\bibfnamefont {N.}~\bibnamefont {Glover}}, \bibinfo
  {author} {\bibfnamefont {M.}~\bibnamefont {Hoefer}}, \ and\ \bibinfo {author}
  {\bibfnamefont {A.}~\bibnamefont {Huss}},\ }\href@noop {} {\bibfield
  {journal} {\bibinfo  {journal} {Submitted to: J. High Energy Phys.}\ }
  (\bibinfo {year} {2019})},\ \Eprint {http://arxiv.org/abs/1904.01044}
  {arXiv:1904.01044 [hep-ph]} \BibitemShut {NoStop}%
\bibitem [{\citenamefont {Caola}\ \emph {et~al.}(2019)\citenamefont {Caola},
  \citenamefont {Melnikov},\ and\ \citenamefont {Röntsch}}]{Caola:2019nzf}%
  \BibitemOpen
  \bibfield  {author} {\bibinfo {author} {\bibfnamefont {F.}~\bibnamefont
  {Caola}}, \bibinfo {author} {\bibfnamefont {K.}~\bibnamefont {Melnikov}}, \
  and\ \bibinfo {author} {\bibfnamefont {R.}~\bibnamefont {Röntsch}},\ }\href
  {\doibase 10.1140/epjc/s10052-019-6880-7} {\  (\bibinfo {year} {2019}),\
  10.1140/epjc/s10052-019-6880-7},\ \Eprint {http://arxiv.org/abs/1902.02081}
  {arXiv:1902.02081 [hep-ph]} \BibitemShut {NoStop}%
\bibitem [{\citenamefont {Khanpour}\ and\ \citenamefont
  {Atashbar~Tehrani}(2016)}]{Khanpour:2016pph}%
  \BibitemOpen
  \bibfield  {author} {\bibinfo {author} {\bibfnamefont {H.}~\bibnamefont
  {Khanpour}}\ and\ \bibinfo {author} {\bibfnamefont {S.}~\bibnamefont
  {Atashbar~Tehrani}},\ }\href {\doibase 10.1103/PhysRevD.93.014026} {\bibfield
   {journal} {\bibinfo  {journal} {Phys. Rev.}\ }\textbf {\bibinfo {volume}
  {D93}},\ \bibinfo {pages} {014026} (\bibinfo {year} {2016})},\ \Eprint
  {http://arxiv.org/abs/1601.00939} {arXiv:1601.00939 [hep-ph]} \BibitemShut
  {NoStop}%
\bibitem [{\citenamefont {Abdul~Khalek}\ \emph {et~al.}(2019)\citenamefont
  {Abdul~Khalek}, \citenamefont {Ethier},\ and\ \citenamefont
  {Rojo}}]{AbdulKhalek:2019mzd}%
  \BibitemOpen
  \bibfield  {author} {\bibinfo {author} {\bibfnamefont {R.}~\bibnamefont
  {Abdul~Khalek}}, \bibinfo {author} {\bibfnamefont {J.~J.}\ \bibnamefont
  {Ethier}}, \ and\ \bibinfo {author} {\bibfnamefont {J.}~\bibnamefont {Rojo}}
  (\bibinfo {collaboration} {NNPDF}),\ }\href {\doibase
  10.1140/epjc/s10052-019-6983-1} {\bibfield  {journal} {\bibinfo  {journal}
  {Eur. Phys. J.}\ }\textbf {\bibinfo {volume} {C79}},\ \bibinfo {pages} {471}
  (\bibinfo {year} {2019})},\ \Eprint {http://arxiv.org/abs/1904.00018}
  {arXiv:1904.00018 [hep-ph]} \BibitemShut {NoStop}%
\bibitem [{\citenamefont {Forte}\ \emph {et~al.}(2002)\citenamefont {Forte},
  \citenamefont {Garrido}, \citenamefont {Latorre},\ and\ \citenamefont
  {Piccione}}]{Forte:2002fg}%
  \BibitemOpen
  \bibfield  {author} {\bibinfo {author} {\bibfnamefont {S.}~\bibnamefont
  {Forte}}, \bibinfo {author} {\bibfnamefont {L.}~\bibnamefont {Garrido}},
  \bibinfo {author} {\bibfnamefont {J.~I.}\ \bibnamefont {Latorre}}, \ and\
  \bibinfo {author} {\bibfnamefont {A.}~\bibnamefont {Piccione}},\ }\href
  {\doibase 10.1088/1126-6708/2002/05/062} {\bibfield  {journal} {\bibinfo
  {journal} {JHEP}\ }\textbf {\bibinfo {volume} {05}},\ \bibinfo {pages} {062}
  (\bibinfo {year} {2002})},\ \Eprint {http://arxiv.org/abs/hep-ph/0204232}
  {arXiv:hep-ph/0204232 [hep-ph]} \BibitemShut {NoStop}%
\bibitem [{\citenamefont {Ball}\ \emph {et~al.}(2009)\citenamefont {Ball},
  \citenamefont {Del~Debbio}, \citenamefont {Forte}, \citenamefont {Guffanti},
  \citenamefont {Latorre}, \citenamefont {Piccione}, \citenamefont {Rojo},\
  and\ \citenamefont {Ubiali}}]{Ball:2008by}%
  \BibitemOpen
  \bibfield  {author} {\bibinfo {author} {\bibfnamefont {R.~D.}\ \bibnamefont
  {Ball}}, \bibinfo {author} {\bibfnamefont {L.}~\bibnamefont {Del~Debbio}},
  \bibinfo {author} {\bibfnamefont {S.}~\bibnamefont {Forte}}, \bibinfo
  {author} {\bibfnamefont {A.}~\bibnamefont {Guffanti}}, \bibinfo {author}
  {\bibfnamefont {J.~I.}\ \bibnamefont {Latorre}}, \bibinfo {author}
  {\bibfnamefont {A.}~\bibnamefont {Piccione}}, \bibinfo {author}
  {\bibfnamefont {J.}~\bibnamefont {Rojo}}, \ and\ \bibinfo {author}
  {\bibfnamefont {M.}~\bibnamefont {Ubiali}} (\bibinfo {collaboration}
  {NNPDF}),\ }\href {\doibase 10.1016/j.nuclphysb.2008.09.037,
  10.1016/j.nuclphysb.2009.02.027} {\bibfield  {journal} {\bibinfo  {journal}
  {Nucl. Phys.}\ }\textbf {\bibinfo {volume} {B809}},\ \bibinfo {pages} {1}
  (\bibinfo {year} {2009})},\ \bibinfo {note} {[Erratum: Nucl.
  Phys.B816,293(2009)]},\ \Eprint {http://arxiv.org/abs/0808.1231}
  {arXiv:0808.1231 [hep-ph]} \BibitemShut {NoStop}%
\bibitem [{\citenamefont {Zenaiev}(2016)}]{Zenaiev:2016jnq}%
  \BibitemOpen
  \bibfield  {author} {\bibinfo {author} {\bibfnamefont {O.}~\bibnamefont
  {Zenaiev}} (\bibinfo {collaboration} {xFitter team}),\ }\href {\doibase
  10.22323/1.265.0033} {\bibfield  {journal} {\bibinfo  {journal} {PoS}\
  }\textbf {\bibinfo {volume} {DIS2016}},\ \bibinfo {pages} {033} (\bibinfo
  {year} {2016})}\BibitemShut {NoStop}%
\bibitem [{\citenamefont {Bertone}\ \emph {et~al.}(2018)\citenamefont {Bertone}
  \emph {et~al.}}]{Bertone:2017tig}%
  \BibitemOpen
  \bibfield  {author} {\bibinfo {author} {\bibfnamefont {V.}~\bibnamefont
  {Bertone}} \emph {et~al.} (\bibinfo {collaboration} {xFitter Developers'
  Team}),\ }\href {\doibase 10.22323/1.297.0203} {\bibfield  {journal}
  {\bibinfo  {journal} {PoS}\ }\textbf {\bibinfo {volume} {DIS2017}},\ \bibinfo
  {pages} {203} (\bibinfo {year} {2018})},\ \Eprint
  {http://arxiv.org/abs/1709.01151} {arXiv:1709.01151 [hep-ph]} \BibitemShut
  {NoStop}%
\bibitem [{\citenamefont {Alekhin}\ \emph {et~al.}(2015)\citenamefont {Alekhin}
  \emph {et~al.}}]{Alekhin:2014irh}%
  \BibitemOpen
  \bibfield  {author} {\bibinfo {author} {\bibfnamefont {S.}~\bibnamefont
  {Alekhin}} \emph {et~al.},\ }\href {\doibase 10.1140/epjc/s10052-015-3480-z}
  {\bibfield  {journal} {\bibinfo  {journal} {Eur. Phys. J.}\ }\textbf
  {\bibinfo {volume} {C75}},\ \bibinfo {pages} {304} (\bibinfo {year}
  {2015})},\ \Eprint {http://arxiv.org/abs/1410.4412} {arXiv:1410.4412
  [hep-ph]} \BibitemShut {NoStop}%
\bibitem [{\citenamefont {Roberts}(1994)}]{Roberts:1990ww}%
  \BibitemOpen
  \bibfield  {author} {\bibinfo {author} {\bibfnamefont {R.~G.}\ \bibnamefont
  {Roberts}},\ }\href {\doibase 10.1017/CBO9780511564062} {\emph {\bibinfo
  {title} {{The Structure of the proton: Deep inelastic scattering}}}},\
  Cambridge Monographs on Mathematical Physics\ (\bibinfo  {publisher}
  {Cambridge University Press},\ \bibinfo {year} {1994})\BibitemShut {NoStop}%
\bibitem [{\citenamefont {Sterman~at al.}(2001)}]{sterman_handbook_2001}%
  \BibitemOpen
  \bibfield  {author} {\bibinfo {author} {\bibfnamefont {G.}~\bibnamefont
  {Sterman~at al.}},\ }\href
  {http://www.e-booksdirectory.com/details.php?ebook=5644} {\emph {\bibinfo
  {title} {Handbook of {Perturbative} {QCD}}}}\ (\bibinfo  {publisher} {CTEQ},\
  \bibinfo {year} {2001})\BibitemShut {NoStop}%
\bibitem [{\citenamefont {Patrignani}\ \emph {et~al.}(2016)\citenamefont
  {Patrignani} \emph {et~al.}}]{Patrignani:2016xqp}%
  \BibitemOpen
  \bibfield  {author} {\bibinfo {author} {\bibfnamefont {C.}~\bibnamefont
  {Patrignani}} \emph {et~al.} (\bibinfo {collaboration} {Particle Data
  Group}),\ }\href {\doibase 10.1088/1674-1137/40/10/100001} {\bibfield
  {journal} {\bibinfo  {journal} {Chin. Phys.}\ }\textbf {\bibinfo {volume}
  {C40}},\ \bibinfo {pages} {100001} (\bibinfo {year} {2016})}\BibitemShut
  {NoStop}%
\bibitem [{\citenamefont {van Neerven}\ and\ \citenamefont
  {Vogt}(2000{\natexlab{a}})}]{vanNeerven:1999ca}%
  \BibitemOpen
  \bibfield  {author} {\bibinfo {author} {\bibfnamefont {W.~L.}\ \bibnamefont
  {van Neerven}}\ and\ \bibinfo {author} {\bibfnamefont {A.}~\bibnamefont
  {Vogt}},\ }\href {\doibase 10.1016/S0550-3213(99)00668-9} {\bibfield
  {journal} {\bibinfo  {journal} {Nucl. Phys.}\ }\textbf {\bibinfo {volume}
  {B568}},\ \bibinfo {pages} {263} (\bibinfo {year} {2000}{\natexlab{a}})},\
  \Eprint {http://arxiv.org/abs/hep-ph/9907472} {arXiv:hep-ph/9907472 [hep-ph]}
  \BibitemShut {NoStop}%
\bibitem [{\citenamefont {van Neerven}\ and\ \citenamefont
  {Vogt}(2000{\natexlab{b}})}]{vanNeerven:2000uj}%
  \BibitemOpen
  \bibfield  {author} {\bibinfo {author} {\bibfnamefont {W.~L.}\ \bibnamefont
  {van Neerven}}\ and\ \bibinfo {author} {\bibfnamefont {A.}~\bibnamefont
  {Vogt}},\ }\href {\doibase 10.1016/S0550-3213(00)00480-6} {\bibfield
  {journal} {\bibinfo  {journal} {Nucl. Phys.}\ }\textbf {\bibinfo {volume}
  {B588}},\ \bibinfo {pages} {345} (\bibinfo {year} {2000}{\natexlab{b}})},\
  \Eprint {http://arxiv.org/abs/hep-ph/0006154} {arXiv:hep-ph/0006154 [hep-ph]}
  \BibitemShut {NoStop}%
\bibitem [{\citenamefont {Vermaseren}\ \emph {et~al.}(2005)\citenamefont
  {Vermaseren}, \citenamefont {Vogt},\ and\ \citenamefont
  {Moch}}]{Vermaseren:2005qc}%
  \BibitemOpen
  \bibfield  {author} {\bibinfo {author} {\bibfnamefont {J.~A.~M.}\
  \bibnamefont {Vermaseren}}, \bibinfo {author} {\bibfnamefont
  {A.}~\bibnamefont {Vogt}}, \ and\ \bibinfo {author} {\bibfnamefont
  {S.}~\bibnamefont {Moch}},\ }\href {\doibase 10.1016/j.nuclphysb.2005.06.020}
  {\bibfield  {journal} {\bibinfo  {journal} {Nucl. Phys.}\ }\textbf {\bibinfo
  {volume} {B724}},\ \bibinfo {pages} {3} (\bibinfo {year} {2005})},\ \Eprint
  {http://arxiv.org/abs/hep-ph/0504242} {arXiv:hep-ph/0504242 [hep-ph]}
  \BibitemShut {NoStop}%
\bibitem [{\citenamefont {Tanabashi}\ \emph {et~al.}(2018)\citenamefont
  {Tanabashi} \emph {et~al.}}]{Tanabashi:2018oca}%
  \BibitemOpen
  \bibfield  {author} {\bibinfo {author} {\bibfnamefont {M.}~\bibnamefont
  {Tanabashi}} \emph {et~al.} (\bibinfo {collaboration} {Particle Data
  Group}),\ }\href {\doibase 10.1103/PhysRevD.98.030001} {\bibfield  {journal}
  {\bibinfo  {journal} {Phys. Rev.}\ }\textbf {\bibinfo {volume} {D98}},\
  \bibinfo {pages} {030001} (\bibinfo {year} {2018})}\BibitemShut {NoStop}%
\bibitem [{\citenamefont {Accardi}\ \emph {et~al.}(2016)\citenamefont {Accardi}
  \emph {et~al.}}]{Accardi:2016ndt}%
  \BibitemOpen
  \bibfield  {author} {\bibinfo {author} {\bibfnamefont {A.}~\bibnamefont
  {Accardi}} \emph {et~al.},\ }\href {\doibase 10.1140/epjc/s10052-016-4285-4}
  {\bibfield  {journal} {\bibinfo  {journal} {Eur. Phys. J.}\ }\textbf
  {\bibinfo {volume} {C76}},\ \bibinfo {pages} {471} (\bibinfo {year}
  {2016})},\ \Eprint {http://arxiv.org/abs/1603.08906} {arXiv:1603.08906
  [hep-ph]} \BibitemShut {NoStop}%
\bibitem [{\citenamefont {Alekhin}()}]{ABM-scheme}%
  \BibitemOpen
  \bibfield  {author} {\bibinfo {author} {\bibfnamefont {S.}~\bibnamefont
  {Alekhin}},\ }\href {http://www.zeuthen.desy.de/$\sim$alekhin/OPENQCDRAD}
  {\enquote {\bibinfo {title} {Openqcdrad, a program description and the code
  are available via: http://www.zeuthen.desy.de/$\sim$alekhin/openqcdrad},}\
  }\BibitemShut {NoStop}%
\bibitem [{\citenamefont {Alekhin}\ and\ \citenamefont
  {Moch}(2011)}]{Alekhin:2010sv}%
  \BibitemOpen
  \bibfield  {author} {\bibinfo {author} {\bibfnamefont {S.}~\bibnamefont
  {Alekhin}}\ and\ \bibinfo {author} {\bibfnamefont {S.}~\bibnamefont {Moch}},\
  }\href {\doibase 10.1016/j.physletb.2011.04.026} {\bibfield  {journal}
  {\bibinfo  {journal} {Phys. Lett.}\ }\textbf {\bibinfo {volume} {B699}},\
  \bibinfo {pages} {345} (\bibinfo {year} {2011})},\ \Eprint
  {http://arxiv.org/abs/1011.5790} {arXiv:1011.5790 [hep-ph]} \BibitemShut
  {NoStop}%
\bibitem [{\citenamefont {Aivazis}\ \emph {et~al.}(1994)\citenamefont
  {Aivazis}, \citenamefont {Collins}, \citenamefont {Olness},\ and\
  \citenamefont {Tung}}]{Aivazis:1993pi}%
  \BibitemOpen
  \bibfield  {author} {\bibinfo {author} {\bibfnamefont {M.~A.~G.}\
  \bibnamefont {Aivazis}}, \bibinfo {author} {\bibfnamefont {J.~C.}\
  \bibnamefont {Collins}}, \bibinfo {author} {\bibfnamefont {F.~I.}\
  \bibnamefont {Olness}}, \ and\ \bibinfo {author} {\bibfnamefont {W.-K.}\
  \bibnamefont {Tung}},\ }\href {\doibase 10.1103/PhysRevD.50.3102} {\bibfield
  {journal} {\bibinfo  {journal} {Phys. Rev.}\ }\textbf {\bibinfo {volume}
  {D50}},\ \bibinfo {pages} {3102} (\bibinfo {year} {1994})},\ \Eprint
  {http://arxiv.org/abs/hep-ph/9312319} {arXiv:hep-ph/9312319 [hep-ph]}
  \BibitemShut {NoStop}%
\bibitem [{\citenamefont {Kr\"{a}mer}\ \emph {et~al.}(2000)\citenamefont
  {Kr\"{a}mer}, \citenamefont {Olness},\ and\ \citenamefont
  {Soper}}]{Kramer:2000hn}%
  \BibitemOpen
  \bibfield  {author} {\bibinfo {author} {\bibfnamefont {M.}~\bibnamefont
  {Kr\"{a}mer}}, \bibinfo {author} {\bibfnamefont {F.~I.}\ \bibnamefont
  {Olness}}, \ and\ \bibinfo {author} {\bibfnamefont {D.~E.}\ \bibnamefont
  {Soper}},\ }\href {\doibase 10.1103/PhysRevD.62.096007} {\bibfield  {journal}
  {\bibinfo  {journal} {Phys. Rev.}\ }\textbf {\bibinfo {volume} {D62}},\
  \bibinfo {pages} {096007} (\bibinfo {year} {2000})},\ \Eprint
  {http://arxiv.org/abs/hep-ph/0003035} {arXiv:hep-ph/0003035 [hep-ph]}
  \BibitemShut {NoStop}%
\bibitem [{\citenamefont {Tung}\ \emph {et~al.}(2002)\citenamefont {Tung},
  \citenamefont {Kretzer},\ and\ \citenamefont {Schmidt}}]{Tung:2001mv}%
  \BibitemOpen
  \bibfield  {author} {\bibinfo {author} {\bibfnamefont {W.-K.}\ \bibnamefont
  {Tung}}, \bibinfo {author} {\bibfnamefont {S.}~\bibnamefont {Kretzer}}, \
  and\ \bibinfo {author} {\bibfnamefont {C.}~\bibnamefont {Schmidt}},\ }\href
  {\doibase 10.1088/0954-3899/28/5/321} {\bibfield  {journal} {\bibinfo
  {journal} {J. Phys.}\ }\textbf {\bibinfo {volume} {G28}},\ \bibinfo {pages}
  {983} (\bibinfo {year} {2002})},\ \Eprint
  {http://arxiv.org/abs/hep-ph/0110247} {arXiv:hep-ph/0110247 [hep-ph]}
  \BibitemShut {NoStop}%
\bibitem [{\citenamefont {Kretzer}\ \emph {et~al.}(2004)\citenamefont
  {Kretzer}, \citenamefont {Lai}, \citenamefont {Olness},\ and\ \citenamefont
  {Tung}}]{Kretzer:2003it}%
  \BibitemOpen
  \bibfield  {author} {\bibinfo {author} {\bibfnamefont {S.}~\bibnamefont
  {Kretzer}}, \bibinfo {author} {\bibfnamefont {H.~L.}\ \bibnamefont {Lai}},
  \bibinfo {author} {\bibfnamefont {F.~I.}\ \bibnamefont {Olness}}, \ and\
  \bibinfo {author} {\bibfnamefont {W.~K.}\ \bibnamefont {Tung}},\ }\href
  {\doibase 10.1103/PhysRevD.69.114005} {\bibfield  {journal} {\bibinfo
  {journal} {Phys. Rev.}\ }\textbf {\bibinfo {volume} {D69}},\ \bibinfo {pages}
  {114005} (\bibinfo {year} {2004})},\ \Eprint
  {http://arxiv.org/abs/hep-ph/0307022} {arXiv:hep-ph/0307022 [hep-ph]}
  \BibitemShut {NoStop}%
\bibitem [{\citenamefont {Thorne}\ and\ \citenamefont
  {Roberts}(1998)}]{Thorne:1997ga}%
  \BibitemOpen
  \bibfield  {author} {\bibinfo {author} {\bibfnamefont {R.~S.}\ \bibnamefont
  {Thorne}}\ and\ \bibinfo {author} {\bibfnamefont {R.~G.}\ \bibnamefont
  {Roberts}},\ }\href {\doibase 10.1103/PhysRevD.57.6871} {\bibfield  {journal}
  {\bibinfo  {journal} {Phys. Rev.}\ }\textbf {\bibinfo {volume} {D57}},\
  \bibinfo {pages} {6871} (\bibinfo {year} {1998})},\ \Eprint
  {http://arxiv.org/abs/hep-ph/9709442} {arXiv:hep-ph/9709442 [hep-ph]}
  \BibitemShut {NoStop}%
\bibitem [{\citenamefont {Thorne}(2006)}]{Thorne:2006qt}%
  \BibitemOpen
  \bibfield  {author} {\bibinfo {author} {\bibfnamefont {R.~S.}\ \bibnamefont
  {Thorne}},\ }\href {\doibase 10.1103/PhysRevD.73.054019} {\bibfield
  {journal} {\bibinfo  {journal} {Phys. Rev.}\ }\textbf {\bibinfo {volume}
  {D73}},\ \bibinfo {pages} {054019} (\bibinfo {year} {2006})},\ \Eprint
  {http://arxiv.org/abs/hep-ph/0601245} {arXiv:hep-ph/0601245 [hep-ph]}
  \BibitemShut {NoStop}%
\bibitem [{\citenamefont {Thorne}(2012)}]{Thorne:2012az}%
  \BibitemOpen
  \bibfield  {author} {\bibinfo {author} {\bibfnamefont {R.~S.}\ \bibnamefont
  {Thorne}},\ }\href {\doibase 10.1103/PhysRevD.86.074017} {\bibfield
  {journal} {\bibinfo  {journal} {Phys. Rev.}\ }\textbf {\bibinfo {volume}
  {D86}},\ \bibinfo {pages} {074017} (\bibinfo {year} {2012})},\ \Eprint
  {http://arxiv.org/abs/1201.6180} {arXiv:1201.6180 [hep-ph]} \BibitemShut
  {NoStop}%
\bibitem [{\citenamefont {Cacciari}\ \emph {et~al.}(1998)\citenamefont
  {Cacciari}, \citenamefont {Greco},\ and\ \citenamefont
  {Nason}}]{Cacciari:1998it}%
  \BibitemOpen
  \bibfield  {author} {\bibinfo {author} {\bibfnamefont {M.}~\bibnamefont
  {Cacciari}}, \bibinfo {author} {\bibfnamefont {M.}~\bibnamefont {Greco}}, \
  and\ \bibinfo {author} {\bibfnamefont {P.}~\bibnamefont {Nason}},\ }\href
  {\doibase 10.1088/1126-6708/1998/05/007} {\bibfield  {journal} {\bibinfo
  {journal} {JHEP}\ }\textbf {\bibinfo {volume} {05}},\ \bibinfo {pages} {007}
  (\bibinfo {year} {1998})},\ \Eprint {http://arxiv.org/abs/hep-ph/9803400}
  {arXiv:hep-ph/9803400 [hep-ph]} \BibitemShut {NoStop}%
\bibitem [{\citenamefont {Forte}\ \emph {et~al.}(2010)\citenamefont {Forte},
  \citenamefont {Laenen}, \citenamefont {Nason},\ and\ \citenamefont
  {Rojo}}]{Forte:2010ta}%
  \BibitemOpen
  \bibfield  {author} {\bibinfo {author} {\bibfnamefont {S.}~\bibnamefont
  {Forte}}, \bibinfo {author} {\bibfnamefont {E.}~\bibnamefont {Laenen}},
  \bibinfo {author} {\bibfnamefont {P.}~\bibnamefont {Nason}}, \ and\ \bibinfo
  {author} {\bibfnamefont {J.}~\bibnamefont {Rojo}},\ }\href {\doibase
  10.1016/j.nuclphysb.2010.03.014} {\bibfield  {journal} {\bibinfo  {journal}
  {Nucl. Phys.}\ }\textbf {\bibinfo {volume} {B834}},\ \bibinfo {pages} {116}
  (\bibinfo {year} {2010})},\ \Eprint {http://arxiv.org/abs/1001.2312}
  {arXiv:1001.2312 [hep-ph]} \BibitemShut {NoStop}%
\bibitem [{\citenamefont {Bertone}\ \emph {et~al.}(2014)\citenamefont
  {Bertone}, \citenamefont {Carrazza},\ and\ \citenamefont
  {Rojo}}]{Bertone:2013vaa}%
  \BibitemOpen
  \bibfield  {author} {\bibinfo {author} {\bibfnamefont {V.}~\bibnamefont
  {Bertone}}, \bibinfo {author} {\bibfnamefont {S.}~\bibnamefont {Carrazza}}, \
  and\ \bibinfo {author} {\bibfnamefont {J.}~\bibnamefont {Rojo}},\ }\href
  {\doibase 10.1016/j.cpc.2014.03.007} {\bibfield  {journal} {\bibinfo
  {journal} {Comput. Phys. Commun.}\ }\textbf {\bibinfo {volume} {185}},\
  \bibinfo {pages} {1647} (\bibinfo {year} {2014})},\ \Eprint
  {http://arxiv.org/abs/1310.1394} {arXiv:1310.1394 [hep-ph]} \BibitemShut
  {NoStop}%
\bibitem [{\citenamefont {Boros}\ \emph {et~al.}(1999)\citenamefont {Boros},
  \citenamefont {Londergan},\ and\ \citenamefont {Thomas}}]{Boros:1998es}%
  \BibitemOpen
  \bibfield  {author} {\bibinfo {author} {\bibfnamefont {C.}~\bibnamefont
  {Boros}}, \bibinfo {author} {\bibfnamefont {J.~T.}\ \bibnamefont
  {Londergan}}, \ and\ \bibinfo {author} {\bibfnamefont {A.~W.}\ \bibnamefont
  {Thomas}},\ }\href {\doibase 10.1103/PhysRevD.59.074021} {\bibfield
  {journal} {\bibinfo  {journal} {Phys. Rev.}\ }\textbf {\bibinfo {volume}
  {D59}},\ \bibinfo {pages} {074021} (\bibinfo {year} {1999})},\ \Eprint
  {http://arxiv.org/abs/hep-ph/9810220} {arXiv:hep-ph/9810220 [hep-ph]}
  \BibitemShut {NoStop}%
\bibitem [{\citenamefont {Saito}\ \emph {et~al.}(2000)\citenamefont {Saito},
  \citenamefont {Boros}, \citenamefont {Tsushima}, \citenamefont {Bissey},
  \citenamefont {Afnan},\ and\ \citenamefont {Thomas}}]{Saito:2000fx}%
  \BibitemOpen
  \bibfield  {author} {\bibinfo {author} {\bibfnamefont {K.}~\bibnamefont
  {Saito}}, \bibinfo {author} {\bibfnamefont {C.}~\bibnamefont {Boros}},
  \bibinfo {author} {\bibfnamefont {K.}~\bibnamefont {Tsushima}}, \bibinfo
  {author} {\bibfnamefont {F.~R.~P.}\ \bibnamefont {Bissey}}, \bibinfo {author}
  {\bibfnamefont {I.~R.}\ \bibnamefont {Afnan}}, \ and\ \bibinfo {author}
  {\bibfnamefont {A.~W.}\ \bibnamefont {Thomas}},\ }\href {\doibase
  10.1016/S0370-2693(00)01166-7} {\bibfield  {journal} {\bibinfo  {journal}
  {Phys. Lett.}\ }\textbf {\bibinfo {volume} {B493}},\ \bibinfo {pages} {288}
  (\bibinfo {year} {2000})},\ \Eprint {http://arxiv.org/abs/nucl-th/0008017}
  {arXiv:nucl-th/0008017 [nucl-th]} \BibitemShut {NoStop}%
\bibitem [{\citenamefont {Londergan}\ \emph {et~al.}(2005)\citenamefont
  {Londergan}, \citenamefont {Murdock},\ and\ \citenamefont
  {Thomas}}]{Londergan:2005ht}%
  \BibitemOpen
  \bibfield  {author} {\bibinfo {author} {\bibfnamefont {J.~T.}\ \bibnamefont
  {Londergan}}, \bibinfo {author} {\bibfnamefont {D.~P.}\ \bibnamefont
  {Murdock}}, \ and\ \bibinfo {author} {\bibfnamefont {A.~W.}\ \bibnamefont
  {Thomas}},\ }\href {\doibase 10.1103/PhysRevD.72.036010} {\bibfield
  {journal} {\bibinfo  {journal} {Phys. Rev.}\ }\textbf {\bibinfo {volume}
  {D72}},\ \bibinfo {pages} {036010} (\bibinfo {year} {2005})},\ \Eprint
  {http://arxiv.org/abs/hep-ph/0507029} {arXiv:hep-ph/0507029 [hep-ph]}
  \BibitemShut {NoStop}%
\bibitem [{\citenamefont {Martin}\ \emph {et~al.}(2004)\citenamefont {Martin},
  \citenamefont {Roberts}, \citenamefont {Stirling},\ and\ \citenamefont
  {Thorne}}]{Martin:2003sk}%
  \BibitemOpen
  \bibfield  {author} {\bibinfo {author} {\bibfnamefont {A.~D.}\ \bibnamefont
  {Martin}}, \bibinfo {author} {\bibfnamefont {R.~G.}\ \bibnamefont {Roberts}},
  \bibinfo {author} {\bibfnamefont {W.~J.}\ \bibnamefont {Stirling}}, \ and\
  \bibinfo {author} {\bibfnamefont {R.~S.}\ \bibnamefont {Thorne}},\ }\href
  {\doibase 10.1140/epjc/s2004-01825-2} {\bibfield  {journal} {\bibinfo
  {journal} {Eur. Phys. J.}\ }\textbf {\bibinfo {volume} {C35}},\ \bibinfo
  {pages} {325} (\bibinfo {year} {2004})},\ \Eprint
  {http://arxiv.org/abs/hep-ph/0308087} {arXiv:hep-ph/0308087 [hep-ph]}
  \BibitemShut {NoStop}%
\bibitem [{\citenamefont {Wang}\ \emph {et~al.}(2016)\citenamefont {Wang},
  \citenamefont {Thomas},\ and\ \citenamefont {Young}}]{Wang:2015msk}%
  \BibitemOpen
  \bibfield  {author} {\bibinfo {author} {\bibfnamefont {X.~G.}\ \bibnamefont
  {Wang}}, \bibinfo {author} {\bibfnamefont {A.~W.}\ \bibnamefont {Thomas}}, \
  and\ \bibinfo {author} {\bibfnamefont {R.~D.}\ \bibnamefont {Young}},\ }\href
  {\doibase 10.1016/j.physletb.2015.12.062} {\bibfield  {journal} {\bibinfo
  {journal} {Phys. Lett.}\ }\textbf {\bibinfo {volume} {B753}},\ \bibinfo
  {pages} {595} (\bibinfo {year} {2016})},\ \Eprint
  {http://arxiv.org/abs/1512.04139} {arXiv:1512.04139 [nucl-th]} \BibitemShut
  {NoStop}%
\bibitem [{\citenamefont {Pumplin}\ \emph
  {et~al.}(2001{\natexlab{a}})\citenamefont {Pumplin}, \citenamefont {Stump},
  \citenamefont {Brock}, \citenamefont {Casey}, \citenamefont {Huston},
  \citenamefont {Kalk}, \citenamefont {Lai},\ and\ \citenamefont
  {Tung}}]{Pumplin:2001ct}%
  \BibitemOpen
  \bibfield  {author} {\bibinfo {author} {\bibfnamefont {J.}~\bibnamefont
  {Pumplin}}, \bibinfo {author} {\bibfnamefont {D.}~\bibnamefont {Stump}},
  \bibinfo {author} {\bibfnamefont {R.}~\bibnamefont {Brock}}, \bibinfo
  {author} {\bibfnamefont {D.}~\bibnamefont {Casey}}, \bibinfo {author}
  {\bibfnamefont {J.}~\bibnamefont {Huston}}, \bibinfo {author} {\bibfnamefont
  {J.}~\bibnamefont {Kalk}}, \bibinfo {author} {\bibfnamefont {H.~L.}\
  \bibnamefont {Lai}}, \ and\ \bibinfo {author} {\bibfnamefont {W.~K.}\
  \bibnamefont {Tung}},\ }\href {\doibase 10.1103/PhysRevD.65.014013}
  {\bibfield  {journal} {\bibinfo  {journal} {Phys. Rev.}\ }\textbf {\bibinfo
  {volume} {D65}},\ \bibinfo {pages} {014013} (\bibinfo {year}
  {2001}{\natexlab{a}})},\ \Eprint {http://arxiv.org/abs/hep-ph/0101032}
  {arXiv:hep-ph/0101032 [hep-ph]} \BibitemShut {NoStop}%
\bibitem [{\citenamefont {Pumplin}\ \emph
  {et~al.}(2001{\natexlab{b}})\citenamefont {Pumplin}, \citenamefont {Stump},\
  and\ \citenamefont {Tung}}]{Pumplin:2000vx}%
  \BibitemOpen
  \bibfield  {author} {\bibinfo {author} {\bibfnamefont {J.}~\bibnamefont
  {Pumplin}}, \bibinfo {author} {\bibfnamefont {D.~R.}\ \bibnamefont {Stump}},
  \ and\ \bibinfo {author} {\bibfnamefont {W.~K.}\ \bibnamefont {Tung}},\
  }\href {\doibase 10.1103/PhysRevD.65.014011} {\bibfield  {journal} {\bibinfo
  {journal} {Phys. Rev.}\ }\textbf {\bibinfo {volume} {D65}},\ \bibinfo {pages}
  {014011} (\bibinfo {year} {2001}{\natexlab{b}})},\ \Eprint
  {http://arxiv.org/abs/hep-ph/0008191} {arXiv:hep-ph/0008191 [hep-ph]}
  \BibitemShut {NoStop}%
\bibitem [{\citenamefont {Giele}\ and\ \citenamefont
  {Keller}(1998)}]{Giele:1998gw}%
  \BibitemOpen
  \bibfield  {author} {\bibinfo {author} {\bibfnamefont {W.~T.}\ \bibnamefont
  {Giele}}\ and\ \bibinfo {author} {\bibfnamefont {S.}~\bibnamefont {Keller}},\
  }\href {\doibase 10.1103/PhysRevD.58.094023} {\bibfield  {journal} {\bibinfo
  {journal} {Phys. Rev.}\ }\textbf {\bibinfo {volume} {D58}},\ \bibinfo {pages}
  {094023} (\bibinfo {year} {1998})},\ \Eprint
  {http://arxiv.org/abs/hep-ph/9803393} {arXiv:hep-ph/9803393 [hep-ph]}
  \BibitemShut {NoStop}%
\bibitem [{\citenamefont {Giele}\ \emph {et~al.}(2001)\citenamefont {Giele},
  \citenamefont {Keller},\ and\ \citenamefont {Kosower}}]{Giele:2001mr}%
  \BibitemOpen
  \bibfield  {author} {\bibinfo {author} {\bibfnamefont {W.~T.}\ \bibnamefont
  {Giele}}, \bibinfo {author} {\bibfnamefont {S.~A.}\ \bibnamefont {Keller}}, \
  and\ \bibinfo {author} {\bibfnamefont {D.~A.}\ \bibnamefont {Kosower}},\
  }\href@noop {} {\  (\bibinfo {year} {2001})},\ \Eprint
  {http://arxiv.org/abs/hep-ph/0104052} {arXiv:hep-ph/0104052 [hep-ph]}
  \BibitemShut {NoStop}%
\bibitem [{\citenamefont {Hartland}\ \emph {et~al.}(2019)\citenamefont
  {Hartland}, \citenamefont {Maltoni}, \citenamefont {Nocera}, \citenamefont
  {Rojo}, \citenamefont {Slade}, \citenamefont {Vryonidou},\ and\ \citenamefont
  {Zhang}}]{Hartland:2019bjb}%
  \BibitemOpen
  \bibfield  {author} {\bibinfo {author} {\bibfnamefont {N.~P.}\ \bibnamefont
  {Hartland}}, \bibinfo {author} {\bibfnamefont {F.}~\bibnamefont {Maltoni}},
  \bibinfo {author} {\bibfnamefont {E.~R.}\ \bibnamefont {Nocera}}, \bibinfo
  {author} {\bibfnamefont {J.}~\bibnamefont {Rojo}}, \bibinfo {author}
  {\bibfnamefont {E.}~\bibnamefont {Slade}}, \bibinfo {author} {\bibfnamefont
  {E.}~\bibnamefont {Vryonidou}}, \ and\ \bibinfo {author} {\bibfnamefont
  {C.}~\bibnamefont {Zhang}},\ }\href {\doibase 10.1007/JHEP04(2019)100}
  {\bibfield  {journal} {\bibinfo  {journal} {JHEP}\ }\textbf {\bibinfo
  {volume} {04}},\ \bibinfo {pages} {100} (\bibinfo {year} {2019})},\ \Eprint
  {http://arxiv.org/abs/1901.05965} {arXiv:1901.05965 [hep-ph]} \BibitemShut
  {NoStop}%
\bibitem [{\citenamefont {Stump}\ \emph {et~al.}(2001)\citenamefont {Stump},
  \citenamefont {Pumplin}, \citenamefont {Brock}, \citenamefont {Casey},
  \citenamefont {Huston}, \citenamefont {Kalk}, \citenamefont {Lai},\ and\
  \citenamefont {Tung}}]{Stump:2001gu}%
  \BibitemOpen
  \bibfield  {author} {\bibinfo {author} {\bibfnamefont {D.}~\bibnamefont
  {Stump}}, \bibinfo {author} {\bibfnamefont {J.}~\bibnamefont {Pumplin}},
  \bibinfo {author} {\bibfnamefont {R.}~\bibnamefont {Brock}}, \bibinfo
  {author} {\bibfnamefont {D.}~\bibnamefont {Casey}}, \bibinfo {author}
  {\bibfnamefont {J.}~\bibnamefont {Huston}}, \bibinfo {author} {\bibfnamefont
  {J.}~\bibnamefont {Kalk}}, \bibinfo {author} {\bibfnamefont {H.~L.}\
  \bibnamefont {Lai}}, \ and\ \bibinfo {author} {\bibfnamefont {W.~K.}\
  \bibnamefont {Tung}},\ }\href {\doibase 10.1103/PhysRevD.65.014012}
  {\bibfield  {journal} {\bibinfo  {journal} {Phys. Rev.}\ }\textbf {\bibinfo
  {volume} {D65}},\ \bibinfo {pages} {014012} (\bibinfo {year} {2001})},\
  \Eprint {http://arxiv.org/abs/hep-ph/0101051} {arXiv:hep-ph/0101051 [hep-ph]}
  \BibitemShut {NoStop}%
\bibitem [{\citenamefont {Martin}\ \emph {et~al.}(2003)\citenamefont {Martin},
  \citenamefont {Roberts}, \citenamefont {Stirling},\ and\ \citenamefont
  {Thorne}}]{Martin:2002aw}%
  \BibitemOpen
  \bibfield  {author} {\bibinfo {author} {\bibfnamefont {A.~D.}\ \bibnamefont
  {Martin}}, \bibinfo {author} {\bibfnamefont {R.~G.}\ \bibnamefont {Roberts}},
  \bibinfo {author} {\bibfnamefont {W.~J.}\ \bibnamefont {Stirling}}, \ and\
  \bibinfo {author} {\bibfnamefont {R.~S.}\ \bibnamefont {Thorne}},\ }\href
  {\doibase 10.1140/epjc/s2003-01196-2} {\bibfield  {journal} {\bibinfo
  {journal} {Eur. Phys. J.}\ }\textbf {\bibinfo {volume} {C28}},\ \bibinfo
  {pages} {455} (\bibinfo {year} {2003})},\ \Eprint
  {http://arxiv.org/abs/hep-ph/0211080} {arXiv:hep-ph/0211080 [hep-ph]}
  \BibitemShut {NoStop}%
\bibitem [{\citenamefont {Hirai}\ \emph {et~al.}(2007)\citenamefont {Hirai},
  \citenamefont {Kumano},\ and\ \citenamefont {Nagai}}]{Hirai:2007sx}%
  \BibitemOpen
  \bibfield  {author} {\bibinfo {author} {\bibfnamefont {M.}~\bibnamefont
  {Hirai}}, \bibinfo {author} {\bibfnamefont {S.}~\bibnamefont {Kumano}}, \
  and\ \bibinfo {author} {\bibfnamefont {T.~H.}\ \bibnamefont {Nagai}},\ }\href
  {\doibase 10.1103/PhysRevC.76.065207} {\bibfield  {journal} {\bibinfo
  {journal} {Phys. Rev.}\ }\textbf {\bibinfo {volume} {C76}},\ \bibinfo {pages}
  {065207} (\bibinfo {year} {2007})},\ \Eprint {http://arxiv.org/abs/0709.3038}
  {arXiv:0709.3038 [hep-ph]} \BibitemShut {NoStop}%
\bibitem [{xfi()}]{xfitter_link}%
  \BibitemOpen
  \href {https://www.xfitter.org/xFitter/} {\enquote {\bibinfo {title}
  {xfitter: https://www.xfitter.org/xfitter/},}\ }\BibitemShut {NoStop}%
\bibitem [{\citenamefont {James}\ and\ \citenamefont
  {Roos}(1975)}]{James:1975dr}%
  \BibitemOpen
  \bibfield  {author} {\bibinfo {author} {\bibfnamefont {F.}~\bibnamefont
  {James}}\ and\ \bibinfo {author} {\bibfnamefont {M.}~\bibnamefont {Roos}},\
  }\href {\doibase 10.1016/0010-4655(75)90039-9} {\bibfield  {journal}
  {\bibinfo  {journal} {Comput. Phys. Commun.}\ }\textbf {\bibinfo {volume}
  {10}},\ \bibinfo {pages} {343} (\bibinfo {year} {1975})}\BibitemShut
  {NoStop}%
\bibitem [{\citenamefont {Lazzaro}\ and\ \citenamefont
  {Moneta}(2010)}]{Lazzaro:2010zza}%
  \BibitemOpen
  \bibfield  {author} {\bibinfo {author} {\bibfnamefont {A.}~\bibnamefont
  {Lazzaro}}\ and\ \bibinfo {author} {\bibfnamefont {L.}~\bibnamefont
  {Moneta}},\ }\href {\doibase 10.1088/1742-6596/219/4/042044} {\bibfield
  {journal} {\bibinfo  {journal} {J. Phys. Conf. Ser.}\ }\textbf {\bibinfo
  {volume} {219}},\ \bibinfo {pages} {042044} (\bibinfo {year}
  {2010})}\BibitemShut {NoStop}%
\bibitem [{\citenamefont {Botje}(2011)}]{Botje:2010ay}%
  \BibitemOpen
  \bibfield  {author} {\bibinfo {author} {\bibfnamefont {M.}~\bibnamefont
  {Botje}},\ }\href {\doibase 10.1016/j.cpc.2010.10.020} {\bibfield  {journal}
  {\bibinfo  {journal} {Comput. Phys. Commun.}\ }\textbf {\bibinfo {volume}
  {182}},\ \bibinfo {pages} {490} (\bibinfo {year} {2011})},\ \Eprint
  {http://arxiv.org/abs/1005.1481} {arXiv:1005.1481 [hep-ph]} \BibitemShut
  {NoStop}%
\bibitem [{\citenamefont {Carli}\ \emph {et~al.}(2010)\citenamefont {Carli},
  \citenamefont {Clements}, \citenamefont {Cooper-Sarkar}, \citenamefont
  {Gwenlan}, \citenamefont {Salam}, \citenamefont {Siegert}, \citenamefont
  {Starovoitov},\ and\ \citenamefont {Sutton}}]{Carli:2010rw}%
  \BibitemOpen
  \bibfield  {author} {\bibinfo {author} {\bibfnamefont {T.}~\bibnamefont
  {Carli}}, \bibinfo {author} {\bibfnamefont {D.}~\bibnamefont {Clements}},
  \bibinfo {author} {\bibfnamefont {A.}~\bibnamefont {Cooper-Sarkar}}, \bibinfo
  {author} {\bibfnamefont {C.}~\bibnamefont {Gwenlan}}, \bibinfo {author}
  {\bibfnamefont {G.~P.}\ \bibnamefont {Salam}}, \bibinfo {author}
  {\bibfnamefont {F.}~\bibnamefont {Siegert}}, \bibinfo {author} {\bibfnamefont
  {P.}~\bibnamefont {Starovoitov}}, \ and\ \bibinfo {author} {\bibfnamefont
  {M.}~\bibnamefont {Sutton}},\ }\href {\doibase
  10.1140/epjc/s10052-010-1255-0} {\bibfield  {journal} {\bibinfo  {journal}
  {Eur. Phys. J.}\ }\textbf {\bibinfo {volume} {C66}},\ \bibinfo {pages} {503}
  (\bibinfo {year} {2010})},\ \Eprint {http://arxiv.org/abs/0911.2985}
  {arXiv:0911.2985 [hep-ph]} \BibitemShut {NoStop}%
\bibitem [{\citenamefont {Luszczak}\ and\ \citenamefont
  {Kowalski}(2014)}]{Luszczak:2013rxa}%
  \BibitemOpen
  \bibfield  {author} {\bibinfo {author} {\bibfnamefont {A.}~\bibnamefont
  {Luszczak}}\ and\ \bibinfo {author} {\bibfnamefont {H.}~\bibnamefont
  {Kowalski}},\ }\href {\doibase 10.1103/PhysRevD.89.074051} {\bibfield
  {journal} {\bibinfo  {journal} {Phys. Rev.}\ }\textbf {\bibinfo {volume}
  {D89}},\ \bibinfo {pages} {074051} (\bibinfo {year} {2014})},\ \Eprint
  {http://arxiv.org/abs/1312.4060} {arXiv:1312.4060 [hep-ph]} \BibitemShut
  {NoStop}%
\bibitem [{\citenamefont {Luszczak}\ and\ \citenamefont
  {Kowalski}(2017)}]{Luszczak:2016bxd}%
  \BibitemOpen
  \bibfield  {author} {\bibinfo {author} {\bibfnamefont {A.}~\bibnamefont
  {Luszczak}}\ and\ \bibinfo {author} {\bibfnamefont {H.}~\bibnamefont
  {Kowalski}},\ }\href {\doibase 10.1103/PhysRevD.95.014030} {\bibfield
  {journal} {\bibinfo  {journal} {Phys. Rev.}\ }\textbf {\bibinfo {volume}
  {D95}},\ \bibinfo {pages} {014030} (\bibinfo {year} {2017})},\ \Eprint
  {http://arxiv.org/abs/1611.10100} {arXiv:1611.10100 [hep-ph]} \BibitemShut
  {NoStop}%
\bibitem [{\citenamefont {Ball}\ \emph {et~al.}(2018)\citenamefont {Ball},
  \citenamefont {Bertone}, \citenamefont {Bonvini}, \citenamefont {Marzani},
  \citenamefont {Rojo},\ and\ \citenamefont {Rottoli}}]{Ball:2017otu}%
  \BibitemOpen
  \bibfield  {author} {\bibinfo {author} {\bibfnamefont {R.~D.}\ \bibnamefont
  {Ball}}, \bibinfo {author} {\bibfnamefont {V.}~\bibnamefont {Bertone}},
  \bibinfo {author} {\bibfnamefont {M.}~\bibnamefont {Bonvini}}, \bibinfo
  {author} {\bibfnamefont {S.}~\bibnamefont {Marzani}}, \bibinfo {author}
  {\bibfnamefont {J.}~\bibnamefont {Rojo}}, \ and\ \bibinfo {author}
  {\bibfnamefont {L.}~\bibnamefont {Rottoli}},\ }\href {\doibase
  10.1140/epjc/s10052-018-5774-4} {\bibfield  {journal} {\bibinfo  {journal}
  {Eur. Phys. J.}\ }\textbf {\bibinfo {volume} {C78}},\ \bibinfo {pages} {321}
  (\bibinfo {year} {2018})},\ \Eprint {http://arxiv.org/abs/1710.05935}
  {arXiv:1710.05935 [hep-ph]} \BibitemShut {NoStop}%
\bibitem [{\citenamefont {Abdolmaleki}\ \emph {et~al.}(2018)\citenamefont
  {Abdolmaleki} \emph {et~al.}}]{Abdolmaleki:2018jln}%
  \BibitemOpen
  \bibfield  {author} {\bibinfo {author} {\bibfnamefont {H.}~\bibnamefont
  {Abdolmaleki}} \emph {et~al.} (\bibinfo {collaboration} {xFitter Developers'
  Team}),\ }\href {\doibase 10.1140/epjc/s10052-018-6090-8} {\bibfield
  {journal} {\bibinfo  {journal} {Eur. Phys. J.}\ }\textbf {\bibinfo {volume}
  {C78}},\ \bibinfo {pages} {621} (\bibinfo {year} {2018})},\ \Eprint
  {http://arxiv.org/abs/1802.00064} {arXiv:1802.00064 [hep-ph]} \BibitemShut
  {NoStop}%
\bibitem [{\citenamefont {Adloff}\ \emph {et~al.}(2003)\citenamefont {Adloff}
  \emph {et~al.}}]{Adloff:2003uh}%
  \BibitemOpen
  \bibfield  {author} {\bibinfo {author} {\bibfnamefont {C.}~\bibnamefont
  {Adloff}} \emph {et~al.} (\bibinfo {collaboration} {H1}),\ }\href {\doibase
  10.1140/epjc/s2003-01257-6} {\bibfield  {journal} {\bibinfo  {journal} {Eur.
  Phys. J.}\ }\textbf {\bibinfo {volume} {C30}},\ \bibinfo {pages} {1}
  (\bibinfo {year} {2003})},\ \Eprint {http://arxiv.org/abs/hep-ex/0304003}
  {arXiv:hep-ex/0304003 [hep-ex]} \BibitemShut {NoStop}%
\bibitem [{\citenamefont {Benvenuti}\ \emph {et~al.}(1989)\citenamefont
  {Benvenuti} \emph {et~al.}}]{Benvenuti:1989rh}%
  \BibitemOpen
  \bibfield  {author} {\bibinfo {author} {\bibfnamefont {A.~C.}\ \bibnamefont
  {Benvenuti}} \emph {et~al.} (\bibinfo {collaboration} {BCDMS}),\ }\href
  {\doibase 10.1016/0370-2693(89)91637-7} {\bibfield  {journal} {\bibinfo
  {journal} {Phys. Lett.}\ }\textbf {\bibinfo {volume} {B223}},\ \bibinfo
  {pages} {485} (\bibinfo {year} {1989})}\BibitemShut {NoStop}%
\bibitem [{\citenamefont {Arneodo}\ \emph {et~al.}(1997)\citenamefont {Arneodo}
  \emph {et~al.}}]{Arneodo:1996qe}%
  \BibitemOpen
  \bibfield  {author} {\bibinfo {author} {\bibfnamefont {M.}~\bibnamefont
  {Arneodo}} \emph {et~al.} (\bibinfo {collaboration} {New Muon}),\ }\href
  {\doibase 10.1016/S0550-3213(96)00538-X} {\bibfield  {journal} {\bibinfo
  {journal} {Nucl. Phys.}\ }\textbf {\bibinfo {volume} {B483}},\ \bibinfo
  {pages} {3} (\bibinfo {year} {1997})},\ \Eprint
  {http://arxiv.org/abs/hep-ph/9610231} {arXiv:hep-ph/9610231 [hep-ph]}
  \BibitemShut {NoStop}%
\bibitem [{\citenamefont {Arneodo}\ \emph {et~al.}(1990)\citenamefont {Arneodo}
  \emph {et~al.}}]{Arneodo:1989sy}%
  \BibitemOpen
  \bibfield  {author} {\bibinfo {author} {\bibfnamefont {M.}~\bibnamefont
  {Arneodo}} \emph {et~al.} (\bibinfo {collaboration} {European Muon}),\ }\href
  {\doibase 10.1016/0550-3213(90)90221-X} {\bibfield  {journal} {\bibinfo
  {journal} {Nucl. Phys.}\ }\textbf {\bibinfo {volume} {B333}},\ \bibinfo
  {pages} {1} (\bibinfo {year} {1990})}\BibitemShut {NoStop}%
\bibitem [{\citenamefont {Airapetian}\ \emph {et~al.}(2002)\citenamefont
  {Airapetian} \emph {et~al.}}]{Airapetian:2002fx}%
  \BibitemOpen
  \bibfield  {author} {\bibinfo {author} {\bibfnamefont {A.}~\bibnamefont
  {Airapetian}} \emph {et~al.} (\bibinfo {collaboration} {HERMES}),\
  }\href@noop {} {\  (\bibinfo {year} {2002})},\ \Eprint
  {http://arxiv.org/abs/hep-ex/0210068} {arXiv:hep-ex/0210068 [hep-ex]}
  \BibitemShut {NoStop}%
\bibitem [{\citenamefont {Amaudruz}\ \emph {et~al.}(1995)\citenamefont
  {Amaudruz} \emph {et~al.}}]{Amaudruz:1995tq}%
  \BibitemOpen
  \bibfield  {author} {\bibinfo {author} {\bibfnamefont {P.}~\bibnamefont
  {Amaudruz}} \emph {et~al.} (\bibinfo {collaboration} {New Muon}),\ }\href
  {\doibase 10.1016/0550-3213(94)00023-9} {\bibfield  {journal} {\bibinfo
  {journal} {Nucl. Phys.}\ }\textbf {\bibinfo {volume} {B441}},\ \bibinfo
  {pages} {3} (\bibinfo {year} {1995})},\ \Eprint
  {http://arxiv.org/abs/hep-ph/9503291} {arXiv:hep-ph/9503291 [hep-ph]}
  \BibitemShut {NoStop}%
\bibitem [{\citenamefont {Gomez}\ \emph {et~al.}(1994)\citenamefont {Gomez}
  \emph {et~al.}}]{Gomez:1993ri}%
  \BibitemOpen
  \bibfield  {author} {\bibinfo {author} {\bibfnamefont {J.}~\bibnamefont
  {Gomez}} \emph {et~al.},\ }\href {\doibase 10.1103/PhysRevD.49.4348}
  {\bibfield  {journal} {\bibinfo  {journal} {Phys. Rev.}\ }\textbf {\bibinfo
  {volume} {D49}},\ \bibinfo {pages} {4348} (\bibinfo {year}
  {1994})}\BibitemShut {NoStop}%
\bibitem [{\citenamefont {Arneodo}\ \emph {et~al.}(1995)\citenamefont {Arneodo}
  \emph {et~al.}}]{Arneodo:1995cs}%
  \BibitemOpen
  \bibfield  {author} {\bibinfo {author} {\bibfnamefont {M.}~\bibnamefont
  {Arneodo}} \emph {et~al.} (\bibinfo {collaboration} {New Muon}),\ }\href
  {\doibase 10.1016/0550-3213(95)00023-2} {\bibfield  {journal} {\bibinfo
  {journal} {Nucl. Phys.}\ }\textbf {\bibinfo {volume} {B441}},\ \bibinfo
  {pages} {12} (\bibinfo {year} {1995})},\ \Eprint
  {http://arxiv.org/abs/hep-ex/9504002} {arXiv:hep-ex/9504002 [hep-ex]}
  \BibitemShut {NoStop}%
\bibitem [{\citenamefont {Arneodo}\ \emph
  {et~al.}(1996{\natexlab{a}})\citenamefont {Arneodo} \emph
  {et~al.}}]{Arneodo:1996rv}%
  \BibitemOpen
  \bibfield  {author} {\bibinfo {author} {\bibfnamefont {M.}~\bibnamefont
  {Arneodo}} \emph {et~al.} (\bibinfo {collaboration} {New Muon}),\ }\href
  {\doibase 10.1016/S0550-3213(96)90117-0} {\bibfield  {journal} {\bibinfo
  {journal} {Nucl. Phys.}\ }\textbf {\bibinfo {volume} {B481}},\ \bibinfo
  {pages} {3} (\bibinfo {year} {1996}{\natexlab{a}})}\BibitemShut {NoStop}%
\bibitem [{\citenamefont {Adams}\ \emph {et~al.}(1995)\citenamefont {Adams}
  \emph {et~al.}}]{Adams:1995is}%
  \BibitemOpen
  \bibfield  {author} {\bibinfo {author} {\bibfnamefont {M.~R.}\ \bibnamefont
  {Adams}} \emph {et~al.} (\bibinfo {collaboration} {E665}),\ }\href {\doibase
  10.1007/BF01624583} {\bibfield  {journal} {\bibinfo  {journal} {Z. Phys.}\
  }\textbf {\bibinfo {volume} {C67}},\ \bibinfo {pages} {403} (\bibinfo {year}
  {1995})},\ \Eprint {http://arxiv.org/abs/hep-ex/9505006}
  {arXiv:hep-ex/9505006 [hep-ex]} \BibitemShut {NoStop}%
\bibitem [{\citenamefont {Ashman}\ \emph {et~al.}(1988)\citenamefont {Ashman}
  \emph {et~al.}}]{Ashman:1988bf}%
  \BibitemOpen
  \bibfield  {author} {\bibinfo {author} {\bibfnamefont {J.}~\bibnamefont
  {Ashman}} \emph {et~al.} (\bibinfo {collaboration} {European Muon}),\ }\href
  {\doibase 10.1016/0370-2693(88)91872-2} {\bibfield  {journal} {\bibinfo
  {journal} {Phys. Lett.}\ }\textbf {\bibinfo {volume} {B202}},\ \bibinfo
  {pages} {603} (\bibinfo {year} {1988})}\BibitemShut {NoStop}%
\bibitem [{\citenamefont {Dasu}\ \emph {et~al.}(1994)\citenamefont {Dasu} \emph
  {et~al.}}]{Dasu:1993vk}%
  \BibitemOpen
  \bibfield  {author} {\bibinfo {author} {\bibfnamefont {S.}~\bibnamefont
  {Dasu}} \emph {et~al.},\ }\href {\doibase 10.1103/PhysRevD.49.5641}
  {\bibfield  {journal} {\bibinfo  {journal} {Phys. Rev.}\ }\textbf {\bibinfo
  {volume} {D49}},\ \bibinfo {pages} {5641} (\bibinfo {year}
  {1994})}\BibitemShut {NoStop}%
\bibitem [{\citenamefont {Berge}\ \emph {et~al.}(1991)\citenamefont {Berge}
  \emph {et~al.}}]{Berge:1989hr}%
  \BibitemOpen
  \bibfield  {author} {\bibinfo {author} {\bibfnamefont {J.~P.}\ \bibnamefont
  {Berge}} \emph {et~al.},\ }\href {\doibase 10.1007/BF01555493} {\bibfield
  {journal} {\bibinfo  {journal} {Z. Phys.}\ }\textbf {\bibinfo {volume}
  {C49}},\ \bibinfo {pages} {187} (\bibinfo {year} {1991})}\BibitemShut
  {NoStop}%
\bibitem [{\citenamefont {Ashman}\ \emph {et~al.}(1993)\citenamefont {Ashman}
  \emph {et~al.}}]{Ashman:1992kv}%
  \BibitemOpen
  \bibfield  {author} {\bibinfo {author} {\bibfnamefont {J.}~\bibnamefont
  {Ashman}} \emph {et~al.} (\bibinfo {collaboration} {European Muon}),\ }\href
  {\doibase 10.1007/BF01565050} {\bibfield  {journal} {\bibinfo  {journal} {Z.
  Phys.}\ }\textbf {\bibinfo {volume} {C57}},\ \bibinfo {pages} {211} (\bibinfo
  {year} {1993})}\BibitemShut {NoStop}%
\bibitem [{\citenamefont {Arneodo}\ \emph
  {et~al.}(1996{\natexlab{b}})\citenamefont {Arneodo} \emph
  {et~al.}}]{Arneodo:1996ru}%
  \BibitemOpen
  \bibfield  {author} {\bibinfo {author} {\bibfnamefont {M.}~\bibnamefont
  {Arneodo}} \emph {et~al.} (\bibinfo {collaboration} {New Muon}),\ }\href
  {\doibase 10.1016/S0550-3213(96)90119-4} {\bibfield  {journal} {\bibinfo
  {journal} {Nucl. Phys.}\ }\textbf {\bibinfo {volume} {B481}},\ \bibinfo
  {pages} {23} (\bibinfo {year} {1996}{\natexlab{b}})}\BibitemShut {NoStop}%
\bibitem [{\citenamefont {Adams}\ \emph {et~al.}(1992)\citenamefont {Adams}
  \emph {et~al.}}]{Adams:1992nf}%
  \BibitemOpen
  \bibfield  {author} {\bibinfo {author} {\bibfnamefont {M.~R.}\ \bibnamefont
  {Adams}} \emph {et~al.} (\bibinfo {collaboration} {E665}),\ }\href {\doibase
  10.1103/PhysRevLett.68.3266} {\bibfield  {journal} {\bibinfo  {journal}
  {Phys. Rev. Lett.}\ }\textbf {\bibinfo {volume} {68}},\ \bibinfo {pages}
  {3266} (\bibinfo {year} {1992})}\BibitemShut {NoStop}%
\bibitem [{\citenamefont {Onengut}\ \emph {et~al.}(2006)\citenamefont {Onengut}
  \emph {et~al.}}]{Onengut:2005kv}%
  \BibitemOpen
  \bibfield  {author} {\bibinfo {author} {\bibfnamefont {G.}~\bibnamefont
  {Onengut}} \emph {et~al.} (\bibinfo {collaboration} {CHORUS}),\ }\href
  {\doibase 10.1016/j.physletb.2005.10.062} {\bibfield  {journal} {\bibinfo
  {journal} {Phys. Lett.}\ }\textbf {\bibinfo {volume} {B632}},\ \bibinfo
  {pages} {65} (\bibinfo {year} {2006})}\BibitemShut {NoStop}%
\bibitem [{\citenamefont {Tzanov}\ \emph {et~al.}(2006)\citenamefont {Tzanov}
  \emph {et~al.}}]{Tzanov:2005kr}%
  \BibitemOpen
  \bibfield  {author} {\bibinfo {author} {\bibfnamefont {M.}~\bibnamefont
  {Tzanov}} \emph {et~al.} (\bibinfo {collaboration} {NuTeV}),\ }\href
  {\doibase 10.1103/PhysRevD.74.012008} {\bibfield  {journal} {\bibinfo
  {journal} {Phys. Rev.}\ }\textbf {\bibinfo {volume} {D74}},\ \bibinfo {pages}
  {012008} (\bibinfo {year} {2006})},\ \Eprint
  {http://arxiv.org/abs/hep-ex/0509010} {arXiv:hep-ex/0509010 [hep-ex]}
  \BibitemShut {NoStop}%
\bibitem [{\citenamefont {MacFarlane}\ \emph {et~al.}(1984)\citenamefont
  {MacFarlane} \emph {et~al.}}]{MacFarlane:1983ax}%
  \BibitemOpen
  \bibfield  {author} {\bibinfo {author} {\bibfnamefont {D.}~\bibnamefont
  {MacFarlane}} \emph {et~al.},\ }\href {\doibase 10.1007/BF01572534}
  {\bibfield  {journal} {\bibinfo  {journal} {Z. Phys.}\ }\textbf {\bibinfo
  {volume} {C26}},\ \bibinfo {pages} {1} (\bibinfo {year} {1984})}\BibitemShut
  {NoStop}%
\bibitem [{\citenamefont {Owens}\ \emph {et~al.}(2007)\citenamefont {Owens},
  \citenamefont {Huston}, \citenamefont {Keppel}, \citenamefont {Kuhlmann},
  \citenamefont {Morfin}, \citenamefont {Olness}, \citenamefont {Pumplin},\
  and\ \citenamefont {Stump}}]{Owens:2007kp}%
  \BibitemOpen
  \bibfield  {author} {\bibinfo {author} {\bibfnamefont {J.~F.}\ \bibnamefont
  {Owens}}, \bibinfo {author} {\bibfnamefont {J.}~\bibnamefont {Huston}},
  \bibinfo {author} {\bibfnamefont {C.~E.}\ \bibnamefont {Keppel}}, \bibinfo
  {author} {\bibfnamefont {S.}~\bibnamefont {Kuhlmann}}, \bibinfo {author}
  {\bibfnamefont {J.~G.}\ \bibnamefont {Morfin}}, \bibinfo {author}
  {\bibfnamefont {F.}~\bibnamefont {Olness}}, \bibinfo {author} {\bibfnamefont
  {J.}~\bibnamefont {Pumplin}}, \ and\ \bibinfo {author} {\bibfnamefont
  {D.}~\bibnamefont {Stump}},\ }\href {\doibase 10.1103/PhysRevD.75.054030}
  {\bibfield  {journal} {\bibinfo  {journal} {Phys. Rev.}\ }\textbf {\bibinfo
  {volume} {D75}},\ \bibinfo {pages} {054030} (\bibinfo {year} {2007})},\
  \Eprint {http://arxiv.org/abs/hep-ph/0702159} {arXiv:hep-ph/0702159 [HEP-PH]}
  \BibitemShut {NoStop}%
\bibitem [{\citenamefont {Schienbein}\ \emph {et~al.}(2008)\citenamefont
  {Schienbein}, \citenamefont {Yu}, \citenamefont {Keppel}, \citenamefont
  {Morfin}, \citenamefont {Olness},\ and\ \citenamefont
  {Owens}}]{Schienbein:2007fs}%
  \BibitemOpen
  \bibfield  {author} {\bibinfo {author} {\bibfnamefont {I.}~\bibnamefont
  {Schienbein}}, \bibinfo {author} {\bibfnamefont {J.~Y.}\ \bibnamefont {Yu}},
  \bibinfo {author} {\bibfnamefont {C.}~\bibnamefont {Keppel}}, \bibinfo
  {author} {\bibfnamefont {J.~G.}\ \bibnamefont {Morfin}}, \bibinfo {author}
  {\bibfnamefont {F.}~\bibnamefont {Olness}}, \ and\ \bibinfo {author}
  {\bibfnamefont {J.~F.}\ \bibnamefont {Owens}},\ }\href {\doibase
  10.1103/PhysRevD.77.054013} {\bibfield  {journal} {\bibinfo  {journal} {Phys.
  Rev.}\ }\textbf {\bibinfo {volume} {D77}},\ \bibinfo {pages} {054013}
  (\bibinfo {year} {2008})},\ \Eprint {http://arxiv.org/abs/0710.4897}
  {arXiv:0710.4897 [hep-ph]} \BibitemShut {NoStop}%
\bibitem [{\citenamefont {Schienbein}\ \emph {et~al.}(2009)\citenamefont
  {Schienbein}, \citenamefont {Yu}, \citenamefont {Kovarik}, \citenamefont
  {Keppel}, \citenamefont {Morfin}, \citenamefont {Olness},\ and\ \citenamefont
  {Owens}}]{Schienbein:2009kk}%
  \BibitemOpen
  \bibfield  {author} {\bibinfo {author} {\bibfnamefont {I.}~\bibnamefont
  {Schienbein}}, \bibinfo {author} {\bibfnamefont {J.~Y.}\ \bibnamefont {Yu}},
  \bibinfo {author} {\bibfnamefont {K.}~\bibnamefont {Kovarik}}, \bibinfo
  {author} {\bibfnamefont {C.}~\bibnamefont {Keppel}}, \bibinfo {author}
  {\bibfnamefont {J.~G.}\ \bibnamefont {Morfin}}, \bibinfo {author}
  {\bibfnamefont {F.}~\bibnamefont {Olness}}, \ and\ \bibinfo {author}
  {\bibfnamefont {J.~F.}\ \bibnamefont {Owens}},\ }\href {\doibase
  10.1103/PhysRevD.80.094004} {\bibfield  {journal} {\bibinfo  {journal} {Phys.
  Rev.}\ }\textbf {\bibinfo {volume} {D80}},\ \bibinfo {pages} {094004}
  (\bibinfo {year} {2009})},\ \Eprint {http://arxiv.org/abs/0907.2357}
  {arXiv:0907.2357 [hep-ph]} \BibitemShut {NoStop}%
\bibitem [{\citenamefont {Kovarik}\ \emph {et~al.}(2011)\citenamefont
  {Kovarik}, \citenamefont {Schienbein}, \citenamefont {Olness}, \citenamefont
  {Yu}, \citenamefont {Keppel}, \citenamefont {Morfin}, \citenamefont {Owens},\
  and\ \citenamefont {Stavreva}}]{Kovarik:2010uv}%
  \BibitemOpen
  \bibfield  {author} {\bibinfo {author} {\bibfnamefont {K.}~\bibnamefont
  {Kovarik}}, \bibinfo {author} {\bibfnamefont {I.}~\bibnamefont {Schienbein}},
  \bibinfo {author} {\bibfnamefont {F.~I.}\ \bibnamefont {Olness}}, \bibinfo
  {author} {\bibfnamefont {J.~Y.}\ \bibnamefont {Yu}}, \bibinfo {author}
  {\bibfnamefont {C.}~\bibnamefont {Keppel}}, \bibinfo {author} {\bibfnamefont
  {J.~G.}\ \bibnamefont {Morfin}}, \bibinfo {author} {\bibfnamefont {J.~F.}\
  \bibnamefont {Owens}}, \ and\ \bibinfo {author} {\bibfnamefont
  {T.}~\bibnamefont {Stavreva}},\ }\href {\doibase
  10.1103/PhysRevLett.106.122301} {\bibfield  {journal} {\bibinfo  {journal}
  {Phys. Rev. Lett.}\ }\textbf {\bibinfo {volume} {106}},\ \bibinfo {pages}
  {122301} (\bibinfo {year} {2011})},\ \Eprint {http://arxiv.org/abs/1012.0286}
  {arXiv:1012.0286 [hep-ph]} \BibitemShut {NoStop}%
\bibitem [{\citenamefont {Paukkunen}\ and\ \citenamefont
  {Salgado}(2010)}]{Paukkunen:2010hb}%
  \BibitemOpen
  \bibfield  {author} {\bibinfo {author} {\bibfnamefont {H.}~\bibnamefont
  {Paukkunen}}\ and\ \bibinfo {author} {\bibfnamefont {C.~A.}\ \bibnamefont
  {Salgado}},\ }\href {\doibase 10.1007/JHEP07(2010)032} {\bibfield  {journal}
  {\bibinfo  {journal} {JHEP}\ }\textbf {\bibinfo {volume} {07}},\ \bibinfo
  {pages} {032} (\bibinfo {year} {2010})},\ \Eprint
  {http://arxiv.org/abs/1004.3140} {arXiv:1004.3140 [hep-ph]} \BibitemShut
  {NoStop}%
\bibitem [{\citenamefont {Paukkunen}\ and\ \citenamefont
  {Salgado}(2013)}]{Paukkunen:2013grz}%
  \BibitemOpen
  \bibfield  {author} {\bibinfo {author} {\bibfnamefont {H.}~\bibnamefont
  {Paukkunen}}\ and\ \bibinfo {author} {\bibfnamefont {C.~A.}\ \bibnamefont
  {Salgado}},\ }\href {\doibase 10.1103/PhysRevLett.110.212301} {\bibfield
  {journal} {\bibinfo  {journal} {Phys. Rev. Lett.}\ }\textbf {\bibinfo
  {volume} {110}},\ \bibinfo {pages} {212301} (\bibinfo {year} {2013})},\
  \Eprint {http://arxiv.org/abs/1302.2001} {arXiv:1302.2001 [hep-ph]}
  \BibitemShut {NoStop}%
\bibitem [{\citenamefont {Amaudruz}\ \emph {et~al.}(1992)\citenamefont
  {Amaudruz} \emph {et~al.}}]{Amaudruz:1991nw}%
  \BibitemOpen
  \bibfield  {author} {\bibinfo {author} {\bibfnamefont {P.}~\bibnamefont
  {Amaudruz}} \emph {et~al.} (\bibinfo {collaboration} {New Muon}),\ }\href
  {\doibase 10.1016/0550-3213(92)90227-3} {\bibfield  {journal} {\bibinfo
  {journal} {Nucl. Phys.}\ }\textbf {\bibinfo {volume} {B371}},\ \bibinfo
  {pages} {3} (\bibinfo {year} {1992})}\BibitemShut {NoStop}%
\bibitem [{\citenamefont {Botje}(2002)}]{Botje:2001fx}%
  \BibitemOpen
  \bibfield  {author} {\bibinfo {author} {\bibfnamefont {M.}~\bibnamefont
  {Botje}},\ }\href {\doibase 10.1088/0954-3899/28/5/305} {\bibfield  {journal}
  {\bibinfo  {journal} {J. Phys.}\ }\textbf {\bibinfo {volume} {G28}},\
  \bibinfo {pages} {779} (\bibinfo {year} {2002})},\ \Eprint
  {http://arxiv.org/abs/hep-ph/0110123} {arXiv:hep-ph/0110123 [hep-ph]}
  \BibitemShut {NoStop}%
\bibitem [{\citenamefont {Aaron}\ \emph {et~al.}(2009)\citenamefont {Aaron}
  \emph {et~al.}}]{Aaron:2009bp}%
  \BibitemOpen
  \bibfield  {author} {\bibinfo {author} {\bibfnamefont {F.~D.}\ \bibnamefont
  {Aaron}} \emph {et~al.} (\bibinfo {collaboration} {H1}),\ }\href {\doibase
  10.1140/epjc/s10052-009-1128-6} {\bibfield  {journal} {\bibinfo  {journal}
  {Eur. Phys. J.}\ }\textbf {\bibinfo {volume} {C63}},\ \bibinfo {pages} {625}
  (\bibinfo {year} {2009})},\ \Eprint {http://arxiv.org/abs/0904.0929}
  {arXiv:0904.0929 [hep-ex]} \BibitemShut {NoStop}%
\bibitem [{\citenamefont {Bodek}\ \emph {et~al.}(1983)\citenamefont {Bodek}
  \emph {et~al.}}]{Bodek:1983ec}%
  \BibitemOpen
  \bibfield  {author} {\bibinfo {author} {\bibfnamefont {A.}~\bibnamefont
  {Bodek}} \emph {et~al.},\ }\href {\doibase 10.1103/PhysRevLett.51.534}
  {\bibfield  {journal} {\bibinfo  {journal} {Phys. Rev. Lett.}\ }\textbf
  {\bibinfo {volume} {51}},\ \bibinfo {pages} {534} (\bibinfo {year}
  {1983})}\BibitemShut {NoStop}%
\bibitem [{\citenamefont {Buckley}\ \emph {et~al.}(2015)\citenamefont
  {Buckley}, \citenamefont {Ferrando}, \citenamefont {Lloyd}, \citenamefont
  {Nordstr\"{o}m}, \citenamefont {Page}, \citenamefont {R\"{u}fenacht},
  \citenamefont {Sch\"{o}nherr},\ and\ \citenamefont {Watt}}]{Buckley:2014ana}%
  \BibitemOpen
  \bibfield  {author} {\bibinfo {author} {\bibfnamefont {A.}~\bibnamefont
  {Buckley}}, \bibinfo {author} {\bibfnamefont {J.}~\bibnamefont {Ferrando}},
  \bibinfo {author} {\bibfnamefont {S.}~\bibnamefont {Lloyd}}, \bibinfo
  {author} {\bibfnamefont {K.}~\bibnamefont {Nordstr\"{o}m}}, \bibinfo {author}
  {\bibfnamefont {B.}~\bibnamefont {Page}}, \bibinfo {author} {\bibfnamefont
  {M.}~\bibnamefont {R\"{u}fenacht}}, \bibinfo {author} {\bibfnamefont
  {M.}~\bibnamefont {Sch\"{o}nherr}}, \ and\ \bibinfo {author} {\bibfnamefont
  {G.}~\bibnamefont {Watt}},\ }\href {\doibase 10.1140/epjc/s10052-015-3318-8}
  {\bibfield  {journal} {\bibinfo  {journal} {Eur. Phys. J.}\ }\textbf
  {\bibinfo {volume} {C75}},\ \bibinfo {pages} {132} (\bibinfo {year}
  {2015})},\ \Eprint {http://arxiv.org/abs/1412.7420} {arXiv:1412.7420
  [hep-ph]} \BibitemShut {NoStop}%
\end{thebibliography}%

\end{document}